\documentclass[12pt]{article}

\usepackage[a4paper,margin=1in]{geometry}
\usepackage{setspace}
\usepackage{amsmath,amssymb,amsfonts,amsthm,bm}
\usepackage{mathtools}
\usepackage{bbm}
\usepackage{enumitem}
\usepackage{natbib}
\usepackage{comment}
\usepackage{graphicx}
\usepackage{booktabs}
\usepackage{longtable}
\usepackage{array}
\usepackage{algorithm}
\usepackage{algpseudocode}
\usepackage[labelfont=bf,labelsep=period,font=small]{caption}
\usepackage{xcolor}
\usepackage[normalem]{ulem}

\graphicspath{{figures/}}

\setlist[enumerate,1]{label=(\roman*)}

\theoremstyle{plain}
\newtheorem{theorem}{Theorem}
\newtheorem{proposition}{Proposition}
\newtheorem{lemma}{Lemma}
\newtheorem{corollary}{Corollary}
\theoremstyle{definition}
\newtheorem{assumption}{Assumption}
\newtheorem{definition}{Definition}
\theoremstyle{remark}
\newtheorem{remark}{Remark}

\DeclareMathOperator{\argmin}{argmin}

\newcommand{\R}{\mathbb{R}}
\newcommand{\E}{\mathbb{E}}
\newcommand{\Var}{\mathbb{V}\mathrm{ar}}
\newcommand{\Cov}{\mathbb{C}\mathrm{ov}}
\newcommand{\1}{\mathbbm{1}}
\newcommand{\norm}[1]{\left\lVert #1 \right\rVert}
\newcommand{\maxnorm}[1]{\left\lVert #1 \right\rVert_{\max}}
\newcommand{\mnorm}[1]{\left\lVert #1 \right\rVert_{L_\infty}}
\newcommand{\op}{o_{\mathbb{P}}}
\newcommand{\Op}{O_{\mathbb{P}}}
\newcommand{\inner}[2]{\left\langle #1,#2 \right\rangle}

\begin{document}

\title{Bias-Aware External-Model-Assisted Inference in High-Dimensional Regression}
\author{Hongzhe Zhang, Hanxuan Ye, Hongzhe Li\\
University of Pennsylvania}
\date{\today}
\maketitle

\begin{abstract}
{
We study high-dimensional semi-supervised linear regression, in which a small labelled sample is paired with unlabelled covariates and an external model of unknown quality.  Such a model carries information about the target coefficients along two complementary channels: a linear signal in the span of the covariates, and a residual orthogonal to that span.  The prevailing framework, prediction-powered inference, reads only the residual.  When the predictor is linear that channel is empty, so the correction is self-cancelling: prediction-powered inference and its variants reduce to ordinary least squares on the labelled data, and no first-order efficiency is gained.  We propose the Debiased External-model-Assisted Lasso (DEAL), which reads the other channel, routing the external model and the unlabelled covariates into the variance of a debiased estimator.  We establish a coordinate-wise central limit theorem whose variance adapts to the model's accuracy.  A bias-aware, data-driven shrinkage discards the model when it is uninformative: its accuracy affects only the width of the interval, never the coverage.  The guarantee holds at the best linear approximation of the target when the truth is not linear, and equally when the predictor is non-linear.  It survives covariate shift in the unlabelled design; where the unlabelled pool dominates, a shift-aware adjustment also recovers the efficiency the shift would otherwise cost.  A hybrid that reads both channels dominates prediction-powered inference at a common budget.
}
\end{abstract}

\textit{Keywords:} covariate shift, debiased Lasso, high-dimensional inference, model misspecification, prediction-powered inference, semi-supervised learning, transfer learning.

\section{Introduction}\label{sec:intro}

A recurring feature of modern statistical practice is an asymmetry of information.  The covariates of interest are recorded in abundance, while the gold-standard response is scarce, because it requires expert annotation, a wet-lab assay, a clinical follow-up, or a costly experiment.  At the same time a machine-learned predictor can impute the missing response cheaply but imperfectly.  Such a predictor is often trained on an external population, and is increasingly a foundation model queried off the shelf, one whose calibration the analyst can neither control nor audit.  The predictor need not even share the analyst's covariates; its own feature set might be text, images, or a separate assay, and it is seen only through the values it returns on the target units.  The inferential problem is to exploit such a predictor to sharpen conclusions about a target parameter without inheriting its bias.

{Prediction-powered inference (PPI) \citep{angelopoulos2023prediction} is the leading framework for this task, building on earlier post-prediction-inference proposals \citep{wang2020methods, motwani2023revisiting}.  It forms an estimator from the model's predictions on the unlabelled covariates, then corrects it with a \emph{rectifier}, a labelled-sample estimate of the error of that prediction-only fit, so that validity holds whatever the model's quality.  Later work adds a power-tuned rectifier weight \citep[PPI++,][]{angelopoulos2023ppipp}, cross-fitting \citep{zrnic2024crossppi}, and active label acquisition \citep{zrnic2024active}.}

{Beneath these variants lies a choice of which information to use.  A model's predictions carry information about the target coefficients along two channels.  One is a linear signal in the span of the covariates: projected onto the design, the predictions read the coefficients themselves, up to noise.  The other is the residual left orthogonal to that span.  The rectifier reads only the residual channel, recovering what the labelled design cannot already reproduce.  For linear regression this is a structural limit.  When the predictor is linear the residual channel is empty, and the correction cancels: PPI reduces to target-only ordinary least squares whatever the predictor's coefficients (Proposition~\ref{prop:ppi-cancellation}), and optimally-tuned PPI++ coincides with it.  A near-oracle but non-linear predictor can even inflate the per-coordinate variance, as the rectifier injects noise.  Any genuine efficiency must therefore come from the other channel: the design-spanned signal the rectifier is built to subtract.}

{We develop such a procedure.  The Debiased External-model-Assisted Lasso (DEAL) reads the channel that the rectifier discards; we call it the variance channel, after the use the procedure makes of it.  It observes a small labelled target sample $(X_0,y_0)$, unlabelled target covariates $\tilde X$, and an external model of unknown quality.  Its construction has two elements.  First, it routes the model and the covariates into the \emph{variance} of a debiased estimator, not the mean of a rectified one: the model imputes pseudo-labels on $\tilde X$, a Lasso is fitted on the stacked labelled and pseudo-labelled data, and a debiasing step follows.    Second, a \emph{bias-aware} shrinkage keeps this safe when the model is poor.  It treats the correction as a single noisy reading of an unknown bias, and shrinks it by a data-driven amount: reverting to target-only debiased Lasso when the model is uninformative, and standing aside when it is near-perfect.  At a common unlabelled budget the intervals are never longer than those of target-only debiased Lasso, and strictly shorter when the external model is informative.  Under a linear predictor the prediction-powered family collapses to that same benchmark; the intervals therefore beat the prediction-powered ones as well.}
{The proposal draws on, and departs from, three lines of work.  The first is semi-supervised inference, in which a large unlabelled covariate sample sharpens estimation of a low-dimensional functional---the mean \citep{zhang2019semisupervised}, the linear-regression coefficient \citep{chakrabortty2018efficient, azriel2022semisupervised}---carried to high-dimensional mean estimation by \citet{zhang2022semisupervised} and to optimal and safe high-dimensional estimation by \citet{deng2024optimal}.  A second branch of that literature imputes the unobserved response itself from auxiliary proxies: the surrogate index of \citet{athey2019surrogate} imputes long-term outcomes from short-term surrogates, and \citet{kallus2025surrogate} characterise the efficiency such imputation can buy.  Those methods draw on the unlabelled covariates alone, or on surrogates whose validity rests on surrogacy or missingness assumptions; we admit an external model of arbitrary quality and replace such assumptions with a bias-aware safeguard.  The fixed-$p$ efficiency ceiling of \citet{azriel2022semisupervised} does not bind us: our gain draws on the external model, information that the unlabelled covariates alone do not carry.
The second is the debiased-Lasso machinery of \citet{vandegeer2014asymptotic}, \citet{javanmard2014confidence}, and \citet{zhang2014confidence}, on which our debiasing step is built.  Within that family, the correlation-adjusted debiased Lasso of \citet{celentano2024correlation} is the closest semi-supervised member: its unlabelled samples enter solely through the covariate model of the debiasing step, in a proportional-asymptotics regime; we work in the sparse regime and route an external model, not the covariates alone, into the enlarged design.  The third is inference with model-generated (synthetic) data, the setting closest to ours.  \citet{keret2025glm} show that in low-dimensional generalised linear models a misspecified linear regression on AI-generated data, combined with summary statistics from the original sample, restores root-$n$ inference; we share the misspecified-projection device but work in high dimensions and target coordinate-wise intervals.  \citet{rezaei2025highdim} study how the covariance shift of synthetic training data governs generalisation and propose covariance matching; their object is predictive risk, whereas ours is valid inference under covariate shift.}

{Recent refinements of prediction-powered inference, such as POP-Inf \citep{miao2025postprediction} and stratified prediction-powered inference \citep{fisch2024stratified}, remain rectifier-based and inherit the linear-predictor cancellation of Proposition~\ref{prop:ppi-cancellation}.  The efficiency here is drawn instead from imputation: the external model imputes pseudo-labels on the unlabelled covariates, enlarging the design the debiased estimator inverts, and DEAL carries that mechanism to coordinate-wise inference in high dimensions under the bias-aware safeguard.}
{
Building on the two-channel view, the paper's contributions are fourfold.  First, DEAL and its central limit theory: under sub-Gaussian design we establish coordinate-wise asymptotic normality, with an explicit variance that adapts to the model's quality and yields valid confidence intervals (Section~\ref{sec:stacking}).  Second, validity beyond a correctly specified model: the intervals remain valid for the best linear approximation when the target is misspecified or the predictor non-linear (Section~\ref{sec:robust}).  Third, robustness without trust in the model: a bias-aware shrinkage reverts to target-only debiased Lasso when the model is uninformative, and coverage is preserved under covariate shift in the unlabelled design; where the unlabelled pool dominates, a shift-aware adjustment recovers the efficiency the shift would otherwise cost (Sections~\ref{sec:method} and~\ref{sec:robust}).  Fourth, the two channels combine: a hybrid reading both is never worse than optimally-tuned PPI++ at a common budget, and strictly better wherever a non-linear predictor leaves design-orthogonal signal (Section~\ref{sec:hybrid}).

Two devices make the procedure deployable: a data-driven choice of the unlabelled sample size (Section~\ref{sec:choose-N}), which balances the external model's bias against the sampling noise the unlabelled rows remove, and the shift-aware adjustment of the debiasing step.  The theory rests on non-asymptotic $\ell_1$ and prediction-error bounds for the stacked Lasso.  We corroborate it with a Monte Carlo study and a five-application real-data portfolio (astronomy, materials chemistry, and oncology, one of the three oncology applications imputed by a large language model), across which the intervals tighten, while a no-information check confirms that a predictor uncorrelated with the response yields no spurious gain (Section~\ref{sec:experiments}).
}

{The remainder of the paper is organised as follows.  Section~\ref{sec:setup} sets up the problem, Section~\ref{sec:method} the method, Section~\ref{sec:stacking} the theory, Section~\ref{sec:robust} the comparison with prediction-powered inference and the robustness analysis, Section~\ref{sec:experiments} the experiments, and Section~\ref{sec:discussion} concludes.  Notation is indexed in Section~\ref{app:notation} of the supplementary material.}

\section{Problem setup and assumptions}\label{sec:setup}

\subsection{Motivation}\label{sec:motivations}
{The Introduction described the rectifier's structural limit for a linear predictor; we now state it exactly.}  For ordinary least-squares (OLS) regression coefficients the PPI estimator
takes the form
\begin{equation}
    \hat\beta^{\mathrm{PP}}
    = \big(\tilde X^{\top}\tilde X\big)^{-1}\tilde X^{\top} f(\tilde X)
      - \big(X_0^{\top}X_0\big)^{-1} X_0^{\top}\big(f(X_0) - y_0\big),
    \label{eq:ppi-ols}
\end{equation}
in which $f$ is the external model, the first term is the prediction-only regression fit on the $N$ rows of unlabelled
covariates $\tilde X$ and the second is the rectifier estimated on the $n_0$ labelled observations $(X_0,y_0)$.
{The rectifier secures validity: on the labelled sample it removes exactly the error the prediction-only fit would otherwise incur.  The following proposition shows that when both the regression function and the predictor are linear, this same rectifier removes the contribution of $f$ entirely, collapsing the estimator onto ordinary least squares.}

\begin{proposition}[Cancellation of the PPI rectifier under a linear model and predictor]\label{prop:ppi-cancellation}
Suppose the target response follows the linear model \eqref{eq:target-model},
$y_0 = X_0\beta^\star + \varepsilon_0$ with $\E[\varepsilon_0\mid X_0]=0$ and
$\Var(\varepsilon_0\mid X_0)=\sigma^2 I_{n_0}$, and suppose the predictor is linear,
$f(X)=X\beta_f$ for some fixed $\beta_f\in\R^p$. Assume the Gram matrices
$X_0^{\top}X_0$ and $\tilde X^{\top}\tilde X$ are invertible. Then the PPI estimator
\eqref{eq:ppi-ols} satisfies
\[
    \hat\beta^{\mathrm{PP}}
    = \beta^\star + \big(X_0^{\top}X_0\big)^{-1} X_0^{\top}\varepsilon_0
    = \hat\beta_{\mathrm{OLS}}
\]
identically, for every value of $\beta_f$; in particular $\Var(\hat\beta^{\mathrm{PP}}\mid X_0)=\sigma^2\big(X_0^{\top}X_0\big)^{-1}$, the ordinary least-squares variance.
\end{proposition}

{The cancellation is not an artefact of the unit rectifier weight.  PPI++ \citep{angelopoulos2023ppipp} replaces that unit weight by a data-driven $\omega\in\R$ applied to the whole prediction-powered increment, $\hat\beta_{\mathrm{OLS}}+\omega(\hat\beta^{\mathrm{PP}}-\hat\beta_{\mathrm{OLS}})$, so that $\omega=0$ returns the target-only ordinary least-squares fit and the procedure never underperforms it.  Under a linear predictor, however, Proposition~\ref{prop:ppi-cancellation} gives $\hat\beta^{\mathrm{PP}}=\hat\beta_{\mathrm{OLS}}$ identically: the increment is exactly zero, and PPI++ collapses onto ordinary least squares for every $\omega$, whatever the tuning.  In its rectified-loss form, the estimating equation of Section~\ref{app:ppi-comparison} of the supplementary material, the increment is instead the two-sample covariance fluctuation $\omega\,[\omega\hat\Sigma_{\mathrm u}+(1-\omega)\hat\Sigma_0]^{-1}(\hat\Sigma_{\mathrm u}-\hat\Sigma_0)(\beta_f-\hat\beta_{\mathrm{OLS}})$, which Proposition~\ref{prop:ppipp-equivalence} shows to be asymptotically negligible for a consistent external estimator.  When $f$ is non-linear, however close to the regression function, the unit-weight rectifier can even inject noise, inflating the per-coordinate variance above the ordinary least-squares floor $\sigma^2(\Sigma^{-1})_{jj}/n_0$, with $\Sigma$ the population covariance of the covariates.  Proposition~\ref{prop:ppi-cancellation} is the low-dimensional shadow of the high-dimensional equivalence of Proposition~\ref{prop:ppipp-equivalence}: at a common unlabelled budget, neither PPI nor optimally-tuned PPI++ improves on target-only debiased Lasso under a linear predictor.}

{The construction developed in Section~\ref{sec:method} instead uses the unlabelled covariates and the external estimator to lower this per-coordinate variance.}

\subsection{Data structure and parameter of interest}

The construction operates on two target data blocks and one external input; we fix these, and the target parameter, here.

We observe:
\begin{itemize}
    \item Target-domain labelled data, comprising an inference sample $(X_0,y_0)$ with $X_0\in\R^{n_0\times p}$ and $y_0\in\R^{n_0}$, on which every display below is written, and a tuning subsample $(X_{\mathrm{tun}},y_{\mathrm{tun}})$ of size $n_{\mathrm{tun}}\asymp n_0$, used to select the shrinkage level of Section~\ref{sec:proxy-shrinkage},
    \item Target-domain unlabelled covariates $\tilde X\in\R^{N\times p}$, the $N$ rows used being drawn from an available pool of $N_{\mathrm{avail}}\ge N$ rows (the choice of $N$ is made in Section~\ref{sec:choose-N}),
    \item {An external estimator $\hat\beta_{\mathrm{ext}}\in\R^p$, possibly built from labelled data in related domains.}
\end{itemize}

The target-domain response obeys the linear model
\begin{equation}
    y_0 = X_0 \beta^\star + \varepsilon_0,\qquad \E[\varepsilon_0\mid X_0]=0,\quad
    \Var(\varepsilon_0\mid X_0)=\sigma^2 I_{n_0}
    \label{eq:target-model}
\end{equation}
for an unknown sparse regression vector $\beta^\star\in\R^p$.  The primary target is low-dimensional inference on individual coordinates $\beta^\star_j$, $1\le j\le p$, in a high-dimensional regime with $\log p/n_0\to0$ and $p$ possibly comparable to or exceeding $n_0$, under sparsity.

We assume the labelled and unlabelled blocks share the population covariance
\[
    \Sigma := \E\left[\frac{1}{n_0}X_0^\top X_0\right]
    = \E\left[\frac{1}{N}\tilde X^\top \tilde X\right],
\]
and denote the sample covariances
\[
    \hat\Sigma_0 := \frac{1}{n_0} X_0^\top X_0,\qquad
    \hat\Sigma_{\mathrm u} := \frac{1}{N} \tilde X^\top \tilde X,\qquad
    \hat\Sigma_{\mathrm{stk}} := \frac{1}{n_0+N}\,(X_0;\tilde X)^\top (X_0;\tilde X).
\]
For a matrix $A=(a_{jk})$, we write
\[
    \maxnorm{A} := \max_{1\le j,k\le p}|a_{jk}|,
    \qquad
    \mnorm{A} := \max_{1\le j\le p} \sum_{k=1}^p |a_{jk}|
\]
for its max-entry norm and row-sum norm, respectively.

{Write $\mu(X):=\E[Y\mid X]$ for the conditional mean under the target distribution $P_0$, and $\Sigma_0$ for the population covariance of the labelled block (equal to the common $\Sigma$ absent covariate shift).  The unlabelled-to-labelled ratio is $\kappa:=N/n_0$.}
\begin{definition}[Estimands]\label{def:estimands}
Under the linear model \eqref{eq:target-model} the inferential target is the coefficient vector $\beta^\star$.  Without the linearity assumption it is the population least-squares projection
\begin{equation}\label{eq:beta-proj}
    \beta^\star_{\mathrm{proj}} \;:=\; \argmin_{\beta\in\R^p}\,\E_{P_0}\bigl[(Y-X^\top\beta)^2\bigr] \;=\; \Sigma_0^{-1}\,\E_{P_0}[X\,\mu(X)],
\end{equation}
which coincides with $\beta^\star$ under linear truth.
\end{definition}

\subsection{High-dimensional regime and sparsity}

We work in a high-dimensional regime in which the dimension and all sample sizes grow jointly.  Throughout the theoretical analysis, an asymptotic statement refers to the sequence limit
\begin{equation}\label{eq:regime}
\begin{gathered}
    n_0\to\infty,\qquad
    p=p_{n_0}\to\infty \ \text{ with } \ \frac{\log p}{n_0}\to 0,\\
    N=N_{n_0}\to\infty \ \text{ with } \ \frac{\log p}{N}\to 0
    \ \text{ and } \ \kappa_n:=\frac{N}{n_0}\to\kappa_\infty\in[0,\infty),
\end{gathered}
\end{equation}
with the sparsities and the external sample size growing only as permitted by the rate conditions imposed below.  The phrase ``with probability tending to one'' and the symbols $\op(\cdot)$, $\Op(\cdot)$ always refer to this joint limit, and every constant is uniform along the sequence unless its dependence is stated.  In particular the unlabelled budget is at most proportional to the labelled size ($\kappa_n$ bounded), the regime in which the core theory operates; the covariate-shift result of Corollary~\ref{cor:shift-restore}, stated at $\kappa_n\to\infty$, is the one exception.  Let
\[
    s := \norm{\beta^\star}_0
\]
be the sparsity of the target parameter.  We impose the following standard assumptions.

\begin{assumption}[Design and noise]\label{ass:design}
\emph{(a) Design.}  The rows of $(X_0;\tilde X)$ are i.i.d.\ sub-Gaussian with mean zero and covariance $\Sigma$ satisfying
\[
    0 < c_\Sigma \le \lambda_{\min}(\Sigma) \le \lambda_{\max}(\Sigma) \le C_\Sigma < \infty,
    \qquad
    \mnorm{\Sigma} \le C_{\Sigma,1} < \infty,
\]
for constants $c_\Sigma, C_\Sigma, C_{\Sigma,1}$ not depending on $(n_0,N,p)$, and with sub-Gaussian norms bounded uniformly along the sequence \eqref{eq:regime}.  The row-norm bound holds, for example, under banded, autoregressive, or otherwise row-summable covariance structures.

\emph{(b) Noise.}  The noise variables $\varepsilon_{0,i}$ are i.i.d., mean-zero, sub-Gaussian with variance $\sigma^2$ and sub-Gaussian norm bounded by a constant, and independent of the covariates $(X_0;\tilde X)$.
\end{assumption}

\begin{assumption}[Restricted eigenvalue]\label{ass:rsc}
There exists $\phi_0>0$ such that, with probability tending to one, the sample covariance $\hat\Sigma_0$ satisfies the restricted eigenvalue condition: for all vectors $\delta\in\R^p$ obeying
\[
    \norm{\delta_{S^c}}_1 \le 3 \norm{\delta_S}_1,\qquad S:=\mathrm{supp}(\beta^\star),
\]
one has
\[
    \delta^\top \hat\Sigma_0\, \delta \ge \phi_0 \norm{\delta}_2^2.
\]
The same condition, on the same cone, holds for $\hat\Sigma_{\mathrm{stk}}$ with a constant $\phi_{\mathrm{stk}}>0$.  We write $\phi:=\min(\phi_0,\phi_{\mathrm{stk}})$ where no distinction is needed, so every bound displayed in terms of $\phi$ is valid for both blocks.  The general-predictor reduction of Section~\ref{sec:general-predictors} imposes the same condition on one further block, with a constant $\phi_{\mathrm{lin}}>0$, defined at its point of use there.
\end{assumption}

\begin{assumption}[Sparsity scaling]\label{ass:sparsity}
The sparsity and dimension satisfy
\[
   s \frac{\log p}{n_0} \to 0.
\]
\end{assumption}

Assumptions~\ref{ass:design}--\ref{ass:sparsity} are standard in the analysis of high-dimensional Lasso and debiased Lasso estimators.

\subsection{External estimator}

{The external estimator $\hat\beta_{\mathrm{ext}}$ is a black-box input: we place no assumption on its quality and impose only a rate bound on its error.}

\begin{assumption}[External-estimator rate]\label{ass:external}


The error of the external estimator,
$\Delta := \hat\beta_{\mathrm{ext}}-\beta^\star$, obeys, with probability tending to one, the rate bound
\[
    \norm{\Delta}_1 \le a_1,\qquad
    \norm{\Sigma^{1/2}\Delta}_2 \le a_2,
\]
for deterministic sequences $(a_1,a_2)$; the second bound is in the population prediction norm, the form the analysis consumes; the plug-in $\hat a_2$ of Section~\ref{sec:choose-N} bounds it in practice.
\end{assumption}

{The bias-aware shrinkage of Section~\ref{sec:method} adapts to the size of $(a_1,a_2)$ automatically, so one analysis covers the relevant regimes by specialising the rate.  Three recur: an external estimator at the target-only Lasso rate $a_1\asymp s\sigma\sqrt{(\log p)/n_0}/\phi$; a near-perfect estimator with $a_1,a_2=\op(n_0^{-1/2})$; and an estimator built from auxiliary labelled data of effective size $n_A$ (in the numerical study of Section~\ref{sec:experiments}, the size of the external sample) with $a_1,a_2\asymp s_A\sqrt{(\log p)/n_A}$, where $s_A$ is the sparsity of the coefficient fitted on the auxiliary data.  Each is introduced at its point of use.}

\begin{assumption}[External estimator independent of the inference noise]\label{ass:external-indep}
For the formal analysis, $\hat\beta_{\mathrm{ext}}$ is independent of the inference-stage data $(X_0,\varepsilon_0,\tilde X)$: it may use auxiliary data and held-out target covariates, but none of the target labelled responses, the labelled covariates $X_0$, or the unlabelled rows $\tilde X$ that enter the correction, stacked refit, and debiasing.  The same exclusion covers the tuning responses that select the shrinkage level (Section~\ref{sec:proxy-shrinkage}) and, where the control-variate hybrid of Section~\ref{sec:hybrid} is in use, its coordinate weight $\hat\omega_j$, introduced there: conditional on the tuning covariates, $\hat\beta_{\mathrm{ext}}$ is independent of the tuning-sample noise.  In the general-predictor reduction of Section~\ref{sec:general-predictors} this holds by construction, the linearisation being fitted on a separate held-out block.
\end{assumption}

{We develop the construction for an external estimator in coefficient form, $\hat\beta_{\mathrm{ext}}\in\R^p$, purely for concreteness; any measurable predictor is admitted through the linearisation of Section~\ref{sec:general-predictors}, and a predictor $\hat\mu(X,W)$ that also reads \emph{auxiliary} covariates $W$ lying outside the design through Section~\ref{sec:aux-cov} of the supplementary material.}

\subsection{Debiasing matrices of Javanmard--Montanari type}

Following the debiased Lasso literature, we construct the debiasing matrices, approximate inverses of the sample covariances, via convex feasibility problems of Javanmard--Montanari (JM) type. For each coordinate $j\in\{1,\dots,p\}$, define $m_{1j}\in\R^p$ as a solution to
\begin{equation}
    m_{1j} \in \argmin_{m\in\R^p}
    m^\top \hat\Sigma_0 m
    \quad\text{subject to}\quad
    \norm{\hat\Sigma_0 m - e_j}_\infty \le \mu_1,
    \label{eq:JM-M1}
\end{equation}
where $e_j$ is the $j$th standard basis vector and $\mu_1$ is a tolerance of order
\[
    \mu_1 \asymp \sqrt{\frac{\log p}{n_0}}.
\]
{Let $M_1:=[m_{11},\dots,m_{1p}]^\top\in\R^{p\times p}$ stack these vectors as its rows, so that $e_j^\top M_1=m_{1j}^\top$.}

Analogously, we define $M_2:=[m_{21},\dots,m_{2p}]^\top$, again stacked by rows, by solving a JM problem with the stacked covariance $\hat\Sigma_{\mathrm{stk}}$, with a tolerance $\mu_2\asymp\sqrt{(\log p)/(n_0+N)}$:
\begin{equation}
    m_{2j} \in \argmin_{m\in\R^p}
    m^\top \hat\Sigma_{\mathrm{stk}} m
    \quad\text{subject to}\quad
    \norm{\hat\Sigma_{\mathrm{stk}} m - e_j}_\infty \le \mu_2.
    \label{eq:JM-M2}
\end{equation}

The next assumption summarises the stability properties of $M_1$ and $M_2$ that we use.

\begin{assumption}[JM feasibility and matrix-norm control]\label{ass:JM}
With probability tending to one, the JM matrices satisfy
\[
    \maxnorm{M_1\hat\Sigma_0 - I_p} \le \mu_1,\qquad
    \maxnorm{M_2\hat\Sigma_{\mathrm{stk}} - I_p} \le \mu_2,
\]
\[
    \mnorm{M_1} + \mnorm{M_2} = \Op(1).
\]
\end{assumption}

The max-entry control is what enters the bias-envelope and debiasing-remainder bounds, whereas the row-sum norms of $M_1$ and $M_2$ enter the score and pseudo-label calculations, paired with the covariance row norm of Assumption~\ref{ass:design} (Lemma~\ref{lem:bias-transfer} of the supplementary material).  Under row sparsity of the precision matrix these conditions hold for the standard constructions \citep{javanmard2014confidence}.

\begin{remark}[Precision constructions and the variance-matching condition]\label{rem:jm-scope}
{Assumption~\ref{ass:JM} is imposed on any pair $(M_1,M_2)$ with the stated properties.  The convex programmes \eqref{eq:JM-M1}--\eqref{eq:JM-M2} are one construction meeting it; the nodewise-Lasso construction of Section~\ref{sec:refine} of the supplementary material (used in the numerical study and every application) is another.  The two displays already deliver, for each fixed coordinate $j$, bounded coordinate-wise variance proxies, $e_j^\top M_2\Sigma M_2^\top e_j=\Op(1)$ and $e_j^\top M_1\hat\Sigma_0M_1^\top e_j=\Op(1)$.  The stronger \emph{variance-matching} limit $e_j^\top M_2\Sigma M_2^\top e_j\to(\Sigma^{-1})_{jj}$, under which the variance induced by the estimated precision matches that of the true precision and the asymptotic variance in Theorem~\ref{thm:CLT} takes its exact quadratic form, is not imposed here; it is discharged with the asymptotic theory, for both constructions, in Lemma~\ref{lem:m2-efficiency} and Remark~\ref{rem:sparsityfree-discharge} of the supplementary material, at the sparsity and design conditions recorded there.}
\end{remark}

Assumptions~\ref{ass:design}--\ref{ass:JM} are maintained throughout.  Design- and model-specific conditions are introduced at their point of use: the tuning-regularity condition{ collected in Section~\ref{app:shrinkage}}, the Gaussian-design conditions of Section~\ref{sec:refine}, and the misspecification and predictor-projection conditions of Section~\ref{sec:robust}.
{A consolidated notation index is provided in Section~\ref{app:notation}.}

\section[The Debiased External-model-Assisted Lasso (DEAL)]{The Debiased External-model-Assisted\\ Lasso (DEAL)}\label{sec:method}

This section presents the complete four-stage construction and develops the bias-aware initialisation step; the pseudo-label stacking and final debiasing steps are analysed in Section~\ref{sec:stacking}.

\subsection{Algorithmic pipeline}

The four stages are a bias-aware initialiser, pseudo-label imputation, a stacked Lasso, and a final debiasing.  {Recall from Section~\ref{sec:setup} that the labelled data are split into a tuning subsample $(X_{\mathrm{tun}},y_{\mathrm{tun}})$ and an inference sample $(X_0,y_0)$ of size $n_0$.  The inference sample carries the one-step correction, the stacked-Lasso refit, and the final debiasing; the tuning subsample selects the scalar shrinkage level $\hat t$.}


\emph{Stage~1 (bias-aware initialiser).}  {We form a one-step correction $C$ to the external estimator and add back a data-driven fraction $\hat t\in[0,1]$ of it.  The fraction trades the bias that $C$ removes against the noise it adds; it is constructed on the tuning subsample (Section~\ref{sec:proxy-shrinkage}) and shown consistent, in the supplementary material, for the tuning-sample counterpart of the ideal level $t^\star$ of that section.}
With $M_1$ the Javanmard--Montanari matrix \eqref{eq:JM-M1} and $\hat\Sigma_0$ the labelled sample covariance of Section~\ref{sec:setup}, form the one-step correction $C:=n_0^{-1}M_1 X_0^\top(y_0-X_0\hat\beta_{\mathrm{ext}})$. The bias-aware initialiser is
\begin{equation}
    \tilde\beta^{\mathrm{init}} := \hat\beta_{\mathrm{ext}} + \hat t\, C.
    \label{eq:init-bias-aware}
\end{equation}
The level $\hat t=1$ applies the full one-step correction and returns the target-only debiased fit, whereas $\hat t=0$ trusts the external estimator outright; throughout, \emph{shrinkage} refers to shrinking the correction $C$, so a smaller $\hat t$ is a stronger shrinkage, and $\hat t\to0$ is the regime an accurate external estimator earns.

\emph{Stage~2 (pseudo-label imputation).}  Using the unlabelled covariates $\tilde X$ and the bias-aware initialiser
$\tilde\beta^{\mathrm{init}}$, we construct pseudo-labels
\[
    \tilde f := \tilde X\, \tilde\beta^{\mathrm{init}}\in\R^N.
\]
These play the role of imputed responses for the unlabelled covariates.

\emph{Stage~3 (stacked Lasso).}  We form the stacked design and response
\[
    X_{\mathrm{stk}} := 
    \begin{pmatrix}
        X_0\\
        \tilde X
    \end{pmatrix}\in\R^{(n_0+N)\times p},\qquad
    y_{\mathrm{stk}} := 
    \begin{pmatrix}
        y_0\\
        \tilde f
    \end{pmatrix}\in\R^{n_0+N},
\]
and compute a Lasso estimator
\begin{equation}
    \hat\beta := \argmin_{\beta\in\R^p}
    \left\{
        \frac{1}{2(n_0+N)}\norm{y_{\mathrm{stk}} - X_{\mathrm{stk}}\beta}_2^2
        + \lambda \norm{\beta}_1
    \right\},
    \label{eq:stacked-lasso}
\end{equation}
for a tuning parameter $\lambda$ to be specified below.

\emph{Stage~4 (final debiasing).}  We construct the debiased estimator using $M_2$:
\begin{equation}
    \tilde\beta := \hat\beta + \frac{1}{n_0+N}M_2 X_{\mathrm{stk}}^\top (y_{\mathrm{stk}} - X_{\mathrm{stk}}\hat\beta).
    \label{eq:final-debias}
\end{equation}
The final debiasing in \eqref{eq:final-debias} inverts a design of $n_0+N$ rows rather than the $n_0$ rows a target-only debiased Lasso sees: the leading noise term is normalised by $n_0+N$, and the debiasing matrix $M_2$, read off the stacked sample covariance $\hat\Sigma_{\mathrm{stk}}$, is sharpened by the larger row count.  How much of this pseudo-labelled enlargement enters the variance is governed by the Stage-1 shrinkage $\hat t$, because the pseudo-labels carry a $\hat t$-fraction of the labelled correction's noise back into the stacked fit.  The resulting variance is given explicitly in \eqref{eq:idealized-var} and Corollary~\ref{cor:idealized-efficiency}.
Our main theorems show that, under appropriate choices of $(\mu_2,\lambda)$,
assumptions on $\hat\beta_{\mathrm{ext}}$, and an explicit bias condition for the
bias-aware initialiser, each coordinate $\tilde\beta_j$ is asymptotically normal with explicit variance.  The data-driven choice of the unlabelled sample size is developed in Section~\ref{sec:choose-N}.  The complete practical procedure, including the plug-in variance, is summarised in Algorithm~\ref{alg:deal} (Section~\ref{sec:implementation} of the supplementary material).

\subsection{Why shrink: the one-step-correction decomposition}

This subsection motivates the Stage~1 shrinkage $\hat t$; Section~\ref{sec:proxy-shrinkage} then constructs the data-driven selector.

{Recall the one-step correction $C$ of Stage~1,}
\begin{equation}
    C := \frac{1}{n_0} M_1 X_0^\top (y_0 - X_0\hat\beta_{\mathrm{ext}})\in\R^p.
    \label{eq:C-def}
\end{equation}

{With $\Delta=\hat\beta_{\mathrm{ext}}-\beta^\star$ the external error of Assumption~\ref{ass:external}, define}
\[
    \bm{b} := -M_1\hat\Sigma_0\,\Delta\in\R^p,\qquad
    Z_1 := \frac{1}{n_0}M_1 X_0^\top \varepsilon_0\in\R^p.
\]
{Then}
\[
    C = \bm{b} + Z_1,
\]
{and conditionally on $X_0$, $Z_1$ has mean zero and covariance}
\[
    V := \Var(Z_1\mid X_0) = \frac{\sigma^2}{n_0} M_1 \hat\Sigma_0 M_1^\top.
\]
{(This is Lemma~\ref{lem:C-decomposition} of the supplementary material.)}

{Thus} $C$ decomposes into signal and noise, $C=\bm{b}+Z_1$, with $\bm{b}$ carrying the external bias $\Delta$.  If $\hat\beta_{\mathrm{ext}}$ is already close to $\beta^\star$, then $\bm{b}$ is small and
$C$ is dominated by the noise $Z_1$; adding $C$ then primarily injects variance.
Conversely, when $\hat\beta_{\mathrm{ext}}$ is biased, the bias term may be large and $C$
can usefully correct towards $\beta^\star$.  Our aim is to adapt between these regimes while
keeping the selected shrinkage level independent of the inference-sample noise.

\subsection{From proxy risk to the data-driven shrinkage selector}\label{sec:proxy-shrinkage}

We now make the shrinkage concrete: first the ideal level that minimises the bias--variance trade-off, then a selector that estimates it from the tuning sample alone.

If one applies a generic shrinkage level $t\in[0,1]$ to a correction of the form
$C=\bm{b}+Z_1$, then
\[
    \hat\beta_{\mathrm{ext}} + tC - \beta^\star
    = (1-t)\Delta + t(I_p-M_1\hat\Sigma_0)\Delta + tZ_1.
\]
Under the proxy approximation $M_1\hat\Sigma_0\approx I_p$, which removes the middle term, the corresponding conditional
quadratic risk is the \emph{proxy risk}
\[
    R_{\mathrm{proxy}}(t)
    := (t-1)^2 \norm{\bm{b}}_2^2 + t^2 \mathrm{tr}(V).
\]

\begin{proposition}[Proxy oracle shrinkage]\label{prop:proxy-risk}
The proxy risk $R_{\mathrm{proxy}}(t)$ is minimised over $t\in[0,1]$ at
\[
    t^\star = \frac{\norm{\bm{b}}_2^2}{\norm{\bm{b}}_2^2 + \mathrm{tr}(V)}\in[0,1].
\]
In particular, if $\bm{b}=0$, then $t^\star=0$; if $\norm{\bm{b}}_2^2\gg \mathrm{tr}(V)$, then
$t^\star\to 1$.
\end{proposition}

To obtain a practical and inferentially valid selector, we estimate the shrinkage level on
the separate tuning subsample only.
{The two tuning blocks below have fixed proportions.}
In the two-block tuning split, divide $(X_{\mathrm{tun}},y_{\mathrm{tun}})$ into two
disjoint blocks $(X_{\mathrm{tun},a},y_{\mathrm{tun},a})$ and
$(X_{\mathrm{tun},b},y_{\mathrm{tun},b})$ of sizes
$n_{\mathrm{tun},a}+n_{\mathrm{tun},b}=n_{\mathrm{tun}}$, construct the corresponding JM
matrices $M_{1,\mathrm{tun}}^{(a)}$ and $M_{1,\mathrm{tun}}^{(b)}$ and block sample covariances $\hat\Sigma_{\mathrm{tun},c}:=n_{\mathrm{tun},c}^{-1}X_{\mathrm{tun},c}^\top X_{\mathrm{tun},c}$ for $c\in\{a,b\}$, and form
\begin{align*}
    C_{\mathrm{tun}}^{(a)} &:= \frac{1}{n_{\mathrm{tun},a}}
    M_{1,\mathrm{tun}}^{(a)}X_{\mathrm{tun},a}^\top
    (y_{\mathrm{tun},a}-X_{\mathrm{tun},a}\hat\beta_{\mathrm{ext}}),\\
    C_{\mathrm{tun}}^{(b)} &:= \frac{1}{n_{\mathrm{tun},b}}
    M_{1,\mathrm{tun}}^{(b)}X_{\mathrm{tun},b}^\top
    (y_{\mathrm{tun},b}-X_{\mathrm{tun},b}\hat\beta_{\mathrm{ext}}).
\end{align*}
Set
\[
    \hat B := \inner{C_{\mathrm{tun}}^{(a)}}{C_{\mathrm{tun}}^{(b)}},
    \qquad
    \hat T := \frac{\hat\sigma_{\mathrm{tun}}^2}{n_{\mathrm{tun}}}
    \,\mathrm{tr}\!\bigl(M_{1,\mathrm{tun}}\hat\Sigma_{\mathrm{tun}}M_{1,\mathrm{tun}}^\top\bigr),
\]
{where $\hat\Sigma_{\mathrm{tun}}:=n_{\mathrm{tun}}^{-1}X_{\mathrm{tun}}^\top X_{\mathrm{tun}}$ is the pooled tuning-sample covariance, $M_{1,\mathrm{tun}}$ the Javanmard--Montanari matrix~\eqref{eq:JM-M1} formed from $\hat\Sigma_{\mathrm{tun}}$ with tolerance $\mu_{1,\mathrm{tun}}\asymp\sqrt{(\log p)/n_{\mathrm{tun}}}$, and $\hat\sigma^2_{\mathrm{tun}}$ the scaled-Lasso noise-variance estimate \citep{sun2012scaled} computed on the pooled tuning sample,} and define
\begin{equation}
    \hat t := \Pi_{[0,1]}\left(
        \frac{\hat B}{\hat B+\hat T}
    \right)
    \quad\text{on }\{\hat B+\hat T>0\},
    \qquad
    \hat t := 0
    \quad\text{on }\{\hat B+\hat T\le 0\}.
    \label{eq:t-hat}
\end{equation}
{The second clause makes the selector defined on the whole sample space and equal to the positive-part form $\hat t=\hat B_+/(\hat B_++\hat T)$; the elementary verification is given in Section~\ref{app:shrinkage}.}
{The two-block inner product $\hat B$ targets the reproducible part of the bias energy: the two tuning blocks being independent, their correction noises cancel in expectation, leaving the cross-block product of the two bias components, $B_{\mathrm{tun}}$, the tuning-sample analogue of the bias energy $\norm{\bm{b}}_2^2$ (Section~\ref{app:shrinkage}).  A companion statistic $\hat T$ estimates $\mathrm{tr}(V_{\mathrm{tun}})$, the analogue of $\mathrm{tr}(V)$ for the covariance $V_{\mathrm{tun}}$ of the tuning-sample correction noise; $\hat t$ is then the empirical counterpart of the tuning-scale oracle $t^\star_{\mathrm{tun}}$, the tuning-sample analogue of the proxy oracle $t^\star$ of Proposition~\ref{prop:proxy-risk}.  That $t^\star_{\mathrm{tun}}$ coincides with the inference-scale oracle $t^\star$ is not claimed; the question bears on the efficiency limit alone, not on coverage (Remark~\ref{rem:t-hat-scope} of Section~\ref{app:shrinkage}).  In form a signal-to-signal-plus-noise ratio, $\hat t$ is the James--Stein / empirical-Bayes shrinkage level \citep{james1961estimation,efron1973stein} specialised to the bias-correction channel.}
By construction, $\hat t$ is measurable with respect to the tuning sample, the target-domain
covariates, and $\hat\beta_{\mathrm{ext}}$; since $\hat\beta_{\mathrm{ext}}$ is itself independent of the
inference-stage data, and of the tuning noise conditionally on the tuning covariates
(Assumption~\ref{ass:external-indep}), $\hat t$ is independent of the inference-sample noise
$\varepsilon_0$ conditional on the covariates.
{$\hat t$ is consistent for the tuning-scale oracle under the regularity conditions collected in Section~\ref{app:shrinkage} (Lemma~\ref{lem:t-hat-consistency} there); no rate is claimed at this point.}

\subsection{The bias-aware initialiser and its remainder}\label{sec:init-decomp}

With the shrinkage level fixed, we record the bias-aware initialiser it produces and the remainder the theory must control.  Write $\bar\mu_1:=\mu_1\asymp\sqrt{(\log p)/n_0}$ for the Javanmard--Montanari tolerance of $M_1$; by Assumption~\ref{ass:JM},
\[
    \maxnorm{M_1\hat\Sigma_0 - I_p} \le \bar\mu_1,
    \qquad
    \mnorm{M_1} = \Op(1).
\]

{With $\bm{b}:=-M_1\hat\Sigma_0\Delta$ and $Z_1:=n_0^{-1}M_1X_0^\top\varepsilon_0$ as above, the bias-aware initialiser satisfies the exact decomposition $\tilde\beta^{\mathrm{init}}-\beta^\star=r(\hat t)+\hat t\,Z_1$, with $Z_1$ conditionally mean zero given the covariates (Lemma~\ref{lem:oof-decomposition} of the supplementary material, Section~\ref{app:proofs}), where}
\begin{equation}
    r(t) := \Delta - t\,M_1\hat\Sigma_0\Delta = (I_p - t\,M_1\hat\Sigma_0)\Delta.
    \label{eq:r-oof}
\end{equation}
%

The quantity $\norm{r(t)}_\infty$ will be controlled explicitly in
Section~\ref{sec:theory-bias}.

{
\subsection{General predictors}\label{sec:general-predictors}

So far the external estimator has been a coefficient vector $\hat\beta_{\mathrm{ext}}\in\R^p$.  DEAL is not restricted to this form.  Let $\hat\mu:\R^p\to\R$ be any external predictor of the response, written $f$ in Section~\ref{sec:motivations} and now unrestricted in form.  It may be a gradient-boosted model, a neural network, or a language model queried off the shelf, and a possibly poor estimate of the conditional mean $\mu$.  A single linearisation step reduces it to a coefficient.  Write its population projection onto the design and the orthogonal residual it leaves as
\begin{equation}\label{eq:lab-projection}
\begin{gathered}
    \beta^{\mathrm{proj}}_0[\hat\mu] \;:=\; \Sigma_0^{-1}\,\E_{P_0}[X\,\hat\mu(X)],
    \qquad
    \nu[\hat\mu](X) \;:=\; \hat\mu(X) - X^\top\beta^{\mathrm{proj}}_0[\hat\mu],\\
    \delta_{\mathrm{lin}}[\hat\mu] \;:=\; \beta^{\mathrm{proj}}_0[\hat\mu] - \beta^\star_{\mathrm{proj}},
\end{gathered}
\end{equation}
so that $\hat\mu(X)=X^\top\beta^{\mathrm{proj}}_0[\hat\mu]+\nu[\hat\mu](X)$ with $\E_{P_0}[X\,\nu[\hat\mu](X)]=0$---the two channels of the Introduction, here for a single predictor.  To obtain a working coefficient, split off from the unlabelled pool a block $X_{\mathrm u}^{(\mathrm{lin})}$ of $n_{\mathrm{lin}}$ rows, withheld thereafter from the stacking, and regress the predictor's values on the design:
\begin{equation}\label{eq:lin-lasso}
    \hat\beta_{\mathrm{ext}}^{\mathrm{lin}} \;:=\; \argmin_{\beta\in\R^p}\;\frac{1}{2 n_{\mathrm{lin}}}\,\bigl\|\hat\mu(X_{\mathrm u}^{(\mathrm{lin})}) - X_{\mathrm u}^{(\mathrm{lin})}\beta\bigr\|_2^2 \;+\; \lambda^{\mathrm{lin}}\|\beta\|_1,
\end{equation}
with $\lambda^{\mathrm{lin}}\asymp\sqrt{\log p / n_{\mathrm{lin}}}$, or ordinary least squares ($\lambda^{\mathrm{lin}}=0$) when $n_{\mathrm{lin}}\gg p$.  Then run Stages~1--4 with $\hat\beta_{\mathrm{ext}}\leftarrow\hat\beta_{\mathrm{ext}}^{\mathrm{lin}}$.  The split does two jobs.  It keeps $\hat\beta_{\mathrm{ext}}^{\mathrm{lin}}$ independent of the stacking block, the inference-stage exclusion that Assumption~\ref{ass:external-indep} demands, and it supplies the fresh rows on which the linearisation rate of Lemma~\ref{lem:lin-rate} is established.  For a linear predictor the residual $\nu[\hat\mu]$ vanishes and the step returns $\hat\beta_{\mathrm{ext}}$ unchanged.  This reduction inherits DEAL's central limit theorem.  Its regularity conditions and the comparison with prediction-powered inference are developed in Section~\ref{sec:misspec-nonlinear-labeler}; in that comparison the residual $\nu[\hat\mu]$ is what matters.

The reduction invokes the restricted-eigenvalue condition of Assumption~\ref{ass:rsc} on one further block: the linearisation-block sample covariance $\hat\Sigma_{\mathrm{lin}}:=n_{\mathrm{lin}}^{-1}X_{\mathrm u}^{(\mathrm{lin})\top}X_{\mathrm u}^{(\mathrm{lin})}$ satisfies it on the cone of $S_{\mathrm{lin}}:=\mathrm{supp}(\beta^{\mathrm{proj}}_0[\hat\mu])$, with a constant $\phi_{\mathrm{lin}}>0$; where the reduction is in use, $\phi$ is read as $\min(\phi_0,\phi_{\mathrm{stk}},\phi_{\mathrm{lin}})$.
}

\section{Asymptotic theory for DEAL}\label{sec:stacking}\label{sec:theory-bias}

This section develops the large-sample theory of the four-stage estimator of Section~\ref{sec:method}: the estimation error of the stacked-Lasso refit, the bias control for the bias-aware initialiser, and the coordinate-wise central limit theorem (CLT) with its adaptive variance.

\subsection{The effective score and the stacked Lasso rate}

Define
\[
    \xi := \tilde f - \tilde X\beta^\star = \tilde X\,(\tilde\beta^{\mathrm{init}} - \beta^\star)
    = \tilde X r(\hat t) + \hat t\, \tilde X Z_1,
\]
the effective pseudo-label noise, the last equality being the decomposition of Section~\ref{sec:init-decomp}.  The pseudo-labels constructed in Stage~2 are then
\[
    \tilde f = \tilde X\,\tilde\beta^{\mathrm{init}}
    = \tilde X\beta^\star + \xi.
\]
Combining labelled and pseudo-labelled samples, the score driving the stacked Lasso is
\[
    g := \frac{1}{n_0+N}\,X_{\mathrm{stk}}^\top
    (y_{\mathrm{stk}} - X_{\mathrm{stk}}\beta^\star).
\]
Substituting the pseudo-label noise,
\[
    g
    = \frac{1}{n_0+N}\left(
        X_0^\top\varepsilon_0 + \tilde X^\top \xi
    \right).
\]
We decompose the score into its labelled part and two pseudo-label parts, driven by the correction noise $Z_1$ (subscript G) and by the residual bias $r(\hat t)$ (subscript R):
\[
    g = g_{\mathrm{lab}} + g_{\mathrm{pl,G}} + g_{\mathrm{pl,R}},
\]
where
\begin{align*}
    g_{\mathrm{lab}} &:= \frac{1}{n_0+N} X_0^\top\varepsilon_0,\\
    g_{\mathrm{pl,G}} &:= \frac{\hat t}{n_0+N}\tilde X^\top \tilde X Z_1
    = \frac{\hat t N}{n_0+N} \hat\Sigma_{\mathrm u} Z_1,\\
    g_{\mathrm{pl,R}} &:= \frac{1}{n_0+N}\tilde X^\top \tilde X r(\hat t)
    = \frac{N}{n_0+N} \hat\Sigma_{\mathrm u} r(\hat t).
\end{align*}

The bounds on the effective score $g$ and the resulting stacked-Lasso rate below are direct analogues of the standard Lasso theory, with additional terms reflecting the pseudo-label noise.

{Under Assumptions~\ref{ass:design}, \ref{ass:external-indep}, and \ref{ass:JM}, for $\hat t\in[0,1]$ there exist constants $C_1,C_2,C_3>0$ such that, with probability tending to one (Lemma~\ref{lem:score} of the supplementary material, Section~\ref{app:proofs}),}
\begin{align*}
    \norm{g}_\infty \le{}&
    C_1\sigma\sqrt{\frac{\log p}{n_0+N}}
    + C_2\,\frac{N}{n_0+N}\,\hat t\,\Bigl(1+\sqrt{\tfrac{p}{N}}\Bigr)\sqrt{\frac{\log p}{n_0}}\\
    &+ C_3\,\frac{N}{n_0+N}\,\Bigl(\norm{r(\hat t)}_\infty + \sqrt{\tfrac{\log p}{N}}\,\norm{\Sigma^{1/2}r(\hat t)}_2\Bigr).
\end{align*}
The factor $1+\sqrt{p/N}$ and the second entry of the bias bracket are the price of averaging the pseudo-label channel over the raw unlabelled covariance; both reduce to constants once $N\gtrsim p\log p$, and are carried explicitly otherwise.

Because $\norm{r(\hat t)}_\infty$ depends on the unknown target parameter, we state the
stacked-Lasso rate in terms of a computable surrogate upper bound. Lemma~\ref{lem:bias-envelope} {of the supplementary material (Section~\ref{app:proofs})}, with Assumption~\ref{ass:external}, supplies an explicit bound $b_n$---the generic-design surrogate $b_n^{\mathrm{gen}}$ of \eqref{eq:bn-gen}---for which
\begin{equation}
    \Pr\!\left(\norm{r(\hat t)}_\infty + \sqrt{\tfrac{\log p}{N}}\,\norm{\Sigma^{1/2}r(\hat t)}_2 \le b_n\right)\to 1.
    \label{eq:bn-surrogate}
\end{equation}
The quantity so bounded is the \emph{composite envelope} of the imputation bias, its $\ell_\infty$ and weighted-$\ell_2$ entries combined; the stacked Lasso may then be tuned against $b_n$ rather than against the unknown
envelope itself.

\begin{theorem}[Stacked Lasso rate]\label{thm:stacked-lasso}
Let $\hat\beta$ be the stacked Lasso estimator \eqref{eq:stacked-lasso} with tuning parameter
\[
    \lambda \ge K_0\left\{
        \sigma\sqrt{\frac{\log p}{n_0+N}}
        + \frac{N}{n_0+N}\hat t\Bigl(1+\sqrt{\tfrac{p}{N}}\Bigr)\sqrt{\frac{\log p}{n_0}}
        + \frac{N}{n_0+N}b_n
    \right\}
\]
for a sufficiently large constant $K_0$, where $b_n$ satisfies \eqref{eq:bn-surrogate}.
Suppose Assumptions~\ref{ass:design}, \ref{ass:rsc}, \ref{ass:JM}, and
\ref{ass:external-indep} hold, and that $\Delta$ satisfies $\norm{\Delta}_1\le a_1$.
Then, with probability tending to one,
\begin{align*}
    \norm{\hat\beta - \beta^\star}_1 &\le \frac{4s}{\phi}\lambda,\\
    \frac{1}{n_0+N}\norm{X_{\mathrm{stk}}(\hat\beta - \beta^\star)}_2^2
    &= (\hat\beta - \beta^\star)^\top \hat\Sigma_{\mathrm{stk}}(\hat\beta - \beta^\star)
    \le {\frac{9s}{\phi}\lambda^2}.
\end{align*}
\end{theorem}

The stacked Lasso thus attains the usual $\ell_1$ and prediction rates in $\lambda$ once $\lambda$ dominates the three terms of its tuning bound: the labelled noise, the pseudo-label noise, and the surrogate bias $b_n$.  For later use we record the generic-design surrogate; two sharper surrogates, available under the Gaussian design, are recorded in Section~\ref{sec:refine}.  Under a general design, if $\Pr(\norm{\Delta}_1\le a_1)\to 1$, then Lemma~\ref{lem:bias-envelope} {of the supplementary material} implies that
\begin{equation}
    b_n^{\mathrm{gen}}
    :=
    a_1\left\{\bar\mu_1 + (1-\hat t)(1+\bar\mu_1)\right\}
    + a_1\sqrt{C_{\Sigma,1}\,\tfrac{\log p}{N}\,\bigl\{(1-\hat t)+\hat t\,p\,\bar\mu_1\bigr\}\bigl\{\bar\mu_1+(1-\hat t)(1+\bar\mu_1)\bigr\}}
    \label{eq:bn-gen}
\end{equation}
is a valid surrogate upper bound for the composite envelope of \eqref{eq:bn-surrogate}, the two terms covering its $\ell_\infty$ and weighted-$\ell_2$ entries respectively.

{We now turn to the asymptotic normality of the final debiased estimator $\tilde\beta$ in \eqref{eq:final-debias}, beginning with the bias control for the bias-aware initialiser.}

\subsection{Bias control for the bias-aware initialiser}

We first quantify the residual bias $r(t)$ appearing in the bias-aware initialiser.  Recall
\[
    r(t) = \Delta - t\,M_1\hat\Sigma_0\Delta.
\]


{Lemma~\ref{lem:bias-envelope} of the supplementary material (Section~\ref{app:proofs}) shows that whenever the JM matrix satisfies $\maxnorm{M_1\hat\Sigma_0 - I_p}\le \bar\mu_1$, then for any $t\in[0,1]$,}
\[
    \norm{r(t)}_\infty \le \bigl(\bar\mu_1 + (1-t)(1+\bar\mu_1)\bigr)\,\norm{\Delta}_1.
\]
{Thus as $t\to 1$ the bias term contracts at rate $O(\bar\mu_1\norm{\Delta}_1)$, while as $t\to 0$ it is dominated by $\norm{\Delta}_1$.  At the target-only Lasso rate of Assumption~\ref{ass:external}---written $a_1\asymp s\lambda_0$, where $\lambda_0:=\sigma\sqrt{(\log p)/n_0}/\phi$ is the target-only Lasso tuning rate---}
\[
    \norm{r(1)}_\infty
    =
    \Op\!\left(\bar\mu_1 s\lambda_0\right).
\]
Hence, at full correction $t=1$, a sufficient condition for the bias to be negligible at the final debiasing scale is
\[
    \sqrt{n_0+N}\,\bar\mu_1\norm{\Delta}_1 \to 0.
\]
More generally, the practical shrinkage level $\hat t$ must satisfy
\[
    \sqrt{n_0+N}\,\bigl(\bar\mu_1 + (1-\hat t)(1+\bar\mu_1)\bigr)\norm{\Delta}_1 \to 0
\]
in order for the bias term to be asymptotically negligible; $N$ is chosen from the data (Section~\ref{sec:choose-N}).


\subsection{CLT for the final debiased estimator}

We now state the main asymptotic normality result for the final debiased estimator.
Define the noise-amplification factor
\[
    A_n := I_p + \frac{\hat t\,N}{n_0}\hat\Sigma_{\mathrm u}M_1.
\]
It collects the two routes by which the labelled noise $\varepsilon_0$ reaches the final estimator: directly, through the labelled block, and through the $\hat t$-fraction carried into the pseudo-labels.  Because the shrinkage level $\hat t$ is selected on the separate tuning subsample, $A_n$ is independent of the inference-sample noise conditional on the covariates and the tuning data.

{The proof of Theorem~\ref{thm:CLT} uses the tuning-stage shrinkage only through its independence from the inference-sample noise; its limit $\hat t\to_{\mathbb P}t_0$ enters the identification of the variance in Section~\ref{sec:idealized-efficiency} and, at a rate, the misspecification analysis (proof of Theorem~\ref{thm:CLT-misspec}), and we state it as an explicit condition.  It is discharged in two parts in Section~\ref{app:shrinkage}: the estimation half, $|\hat t-t^\star_{\mathrm{tun}}|=\op\bigl((n_0+N)^{-1/4}\bigr)$, is proved there (Proposition~\ref{prop:that-rate}), leaving the stabilisation of the tuning-stage bias-to-noise ratio, a functional of the design and the external error alone.}

\begin{assumption}[{Shrinkage-rate condition (T)}]\label{ass:condT}
{The tuning-stage shrinkage obeys $|\hat t-t_0|=\op\bigl((n_0+N)^{-1/4}\bigr)$, where $t_0\in[0,1]$ denotes the in-probability limit of $\hat t$.}
\end{assumption}
Table~\ref{tab:condT-regimes} records where condition~(T) is discharged and where it is carried as an assumption, together with the shrinkage limit $t_0$ in each regime.

\begin{table}[!ht]
\centering
\caption{\textit{Status of the shrinkage-rate condition~(T) across the external-estimator regimes.}  Entries marked ``proved'' are discharged in Proposition~\ref{prop:condT-regimes} of the supplementary material; $t_0$ is the in-probability limit of $\hat t$.  $\rho_n:=B_{\mathrm{tun}}/\mathrm{tr}(V_{\mathrm{tun}})$ is the tuning-block bias-to-noise ratio of Proposition~\ref{prop:condT-regimes}.}
\label{tab:condT-regimes}
\small
\begin{tabular}{@{}p{4.8cm}p{1.7cm}p{5.3cm}@{}}
\toprule
Regime & $t_0$ & Condition~(T) \\
\midrule
Near-perfect external estimator & $0$ & proved (Prop.~\ref{prop:condT-regimes}(i)) \\
\addlinespace[2pt]
Target-only Lasso rate; auxiliary-data estimator & $0$ & proved under auxiliary scaling (Prop.~\ref{prop:condT-regimes}(ii),(iii)) \\
\addlinespace[2pt]
Dominant bias & $1$ & proved when $\rho_n\gg(n_0+N)^{1/4}$ (Prop.~\ref{prop:condT-regimes}(iv)) \\
\addlinespace[2pt]
Intermediate & $\in(0,1)$ & assumed (Prop.~\ref{prop:condT-regimes}(v)) \\
\addlinespace[2pt]
Misspecified tuning model & $\in[0,1]$ & assumed; open question (Rem.~\ref{rem:condT-scope}; Thm.~\ref{thm:CLT-misspec}) \\
\bottomrule
\end{tabular}
\end{table}

\begin{theorem}[Asymptotic linearity and coordinate-wise CLT]\label{thm:CLT}
Suppose Assumptions~\ref{ass:design}, \ref{ass:rsc}, \ref{ass:JM},
\ref{ass:external}, and \ref{ass:external-indep} hold.  Let $\hat\beta$ and $\tilde\beta$
be defined by \eqref{eq:stacked-lasso} and \eqref{eq:final-debias}, with the
tuning-sample shrinkage level $\hat t$ from \eqref{eq:t-hat}.  Fix a coordinate $j$, and let
\[
    \tau_{j,n}^2
    :=
    \frac{\sigma^2 n_0}{n_0+N}\,
    e_j^\top M_2 A_n\hat\Sigma_0 A_n^\top M_2^\top e_j.
\]
Assume that
\begin{enumerate}
    \item $\sqrt{n_0+N}\,\Bigl(\norm{r(\hat t)}_\infty + \sqrt{\tfrac{\log p}{N}}\,\norm{\Sigma^{1/2}r(\hat t)}_2\Bigr) \to 0$ in probability,
    \item $\sqrt{n_0+N}\,\mu_2 s\lambda \to 0$,
    \item $\tau_{j,n}^2 \to \tau_j^2\in(0,\infty)$ in probability, and
    \item the Lindeberg condition
    \[
        \frac{\max_{1\le i\le n_0}
            \bigl|x_{0,i}^\top A_n^\top M_2^\top e_j\bigr|}
        {\left\{\sum_{i=1}^{n_0}
            \bigl(x_{0,i}^\top A_n^\top M_2^\top e_j\bigr)^2\right\}^{1/2}}
        \to 0
    \]
    holds in probability.
\end{enumerate}
Then
\[
    \sqrt{n_0+N}\,(\tilde\beta_j - \beta^\star_j)
    =
    \frac{1}{\sqrt{n_0+N}}
    \sum_{i=1}^{n_0}
    \bigl(x_{0,i}^\top A_n^\top M_2^\top e_j\bigr)\,\varepsilon_{0,i}
    + \op(1),
\]
and consequently
\[
    \sqrt{n_0+N}\,(\tilde\beta_j - \beta^\star_j)
    \;\Rightarrow\; \mathcal{N}(0,\tau_j^2).
\]
\end{theorem}

Theorem~\ref{thm:CLT} imposes four conditions beyond the standing assumptions.  Condition~(i) is the bias requirement: the residual bias $r(\hat t)$ of the bias-aware initialiser, in its composite $\ell_\infty$ and weighted-$\ell_2$ form, must be negligible at the $\sqrt{n_0+N}$ scale; (ii) makes the debiasing remainder, the product of the tolerance $\mu_2$ and the stacked-Lasso $\ell_1$ rate, negligible at the same scale; (iii) asks the variance to stabilise at a positive finite limit; (iv) is Lindeberg's condition.  Condition~(iii) is treated in two cases.  For the nodewise precision used in practice it is an assumption: Assumption~\ref{ass:rowsparse} below supplies its variance-matching part, the limit $e_j^\top M_2\Sigma M_2^\top e_j\to(\Sigma^{-1})_{jj}$ (Lemma~\ref{lem:m2-efficiency} of the supplementary material), at the price of row sparsity of the precision matrix, $s_\Omega\log p=o(\sqrt{n_0+N})$ with $s_\Omega$ defined there, and the full stabilisation of $\tau_{j,n}^2$ remains assumed (Remark~\ref{rem:nodewise-iii-hypothesis} of the supplementary material).  For a Gaussian design with known precision it is proved outright (Corollary~\ref{cor:gaussian-known} of the supplementary material).

\subsection{Idealised efficiency and variance identification}\label{sec:idealized-efficiency}

Theorem~\ref{thm:CLT} is stated in a sample-dependent form.  In the idealised benchmark of
Corollary~\ref{cor:idealized-efficiency} below (known precision, stabilised shrinkage),
one has
\[
    A_n\to \left(1+\frac{t_0 N}{n_0}\right)I_p,
\]
with $t_0$ the in-probability limit of $\hat t$ (Assumption~\ref{ass:condT}), which under the regularity conditions of Section~\ref{app:shrinkage} is also the limit of the tuning-scale oracle $t^\star_{\mathrm{tun}}$ (Proposition~\ref{prop:condT-reduction} there).  This yields
\[
    \tau_j^2
    =
    \frac{\sigma^2(n_0+t_0 N)^2}{n_0(n_0+N)}\,(\Sigma^{-1})_{jj}.
\]
Equivalently, the unscaled asymptotic variance of $\tilde\beta_j$ is
\begin{equation}\label{eq:idealized-var}
    \Var(\tilde\beta_j)
    =
    \frac{\sigma^2(n_0+t_0 N)^2}{n_0(n_0+N)^2}\,(\Sigma^{-1})_{jj}.
\end{equation}
This is a \emph{benchmark limit}, the first-order efficiency attainable when the precision is known and the shrinkage has stabilised at $t_0$, not a property of the practical estimator, for which Theorem~\ref{thm:CLT} retains the sample-dependent sandwich form.  Relative to the target-only debiased-Lasso variance $\sigma^2(\Sigma^{-1})_{jj}/n_0$ it gives the following.

\begin{corollary}[{Idealised efficiency and the variance channel}]\label{cor:idealized-efficiency}\label{cor:variance-channel}
In the oracle-precision benchmark $M_2=\Sigma^{-1}$, $M_1=\Sigma^{-1}$, $\hat\Sigma_0\to\Sigma$, $\hat\Sigma_{\mathrm u}\to\Sigma$, and $\hat t\to t_0\in[0,1]$, the limiting variance of Theorem~\ref{thm:CLT} collapses to the form \eqref{eq:idealized-var}, and the idealised variance ratio of $\tilde\beta_j$ relative to the target-only debiased-Lasso variance $\sigma^2(\Sigma^{-1})_{jj}/n_0$ is
\begin{equation}\label{eq:G-ratio}
    G(\kappa,t_0) := \left(\frac{1+t_0\kappa}{1+\kappa}\right)^2, \qquad \kappa := \frac{N}{n_0}.
\end{equation}
In particular {$G(\kappa,t_0)\le 1$ for every $t_0\in[0,1]$, so in this benchmark the DEAL interval is never longer than that of the target-only debiased Lasso:} $t_0=1$ recovers the target-only benchmark, whereas any $t_0<1$ yields a strict first-order variance improvement, $G(\kappa,t_0)<1$.
{An effective-sample-size reading of this idealised variance, under which DEAL behaves as a target-only debiased Lasso with a strictly larger labelled sample, is recorded in Section~\ref{app:shrinkage} of the supplementary material (Corollary~\ref{cor:neff}).}
\end{corollary}

{A worked example of the external-estimator rate that delivers $G(\kappa,t_0)<1$, for the transfer-learning estimator of \citet{li2022transfer}, is given in Remark~\ref{rem:translasso} of the supplementary material.}
The variance-matching part of requirement~(iii) of Theorem~\ref{thm:CLT} is secured at estimated precision by the following rate conditions.

\begin{assumption}[{Rate conditions for variance identification}]\label{ass:rowsparse}

Let $\Sigma:=\Sigma_0$ in the no-shift regime and let $\Omega:=\Sigma^{-1}$ have maximum
row sparsity $s_\Omega:=\max_{1\le m\le p}|\{l\neq m:\Omega_{ml}\neq0\}|$.  The Stage-4
debiasing matrix $M_2$ (either construction of Remark~\ref{rem:jm-scope}) attains,
for the fixed inferential coordinate $j$, the standard nodewise-Lasso $\ell_2$ row rate
\[
    \norm{M_2^\top e_j-\Sigma^{-1}e_j}_2
    =\Op\!\Bigl(\sqrt{s_\Omega\log p/(n_0+N)}\Bigr),
\]
the $A_n$-replacement error vanishes,
\[
    \hat t\,\kappa_n\,\norm{\bigl(M_1^\top\hat\Sigma_{\mathrm u}-I_p\bigr)M_2^\top e_j}_2
    =\op(1),
\]
and the row sparsity satisfies the scaling
\[
    s_\Omega\log p=o\bigl(\sqrt{n_0+N}\bigr).
\]

\end{assumption}

{The first display is the standard row-sparse precision guarantee of \citet{javanmard2014confidence,vandegeer2014asymptotic}, the $\ell_2$ estimation rate of the $j$th precision row.  The second controls the unlabelled-block correction $M_1^\top\hat\Sigma_{\mathrm u}-I_p$ entering the factor $A_n$, damped by the shrinkage: for constructions with bounded operator norm the fluctuation $\norm{(M_1^\top\hat\Sigma_{\mathrm u}-I_p)M_2^\top e_j}_2$ is of order $\sqrt{p/N}$, so the display holds when $\hat t\to_{\mathbb P}0$ at the rate of condition~(T) with $p=o((n_0+N)^{3/2})$, and for a non-vanishing shrinkage limit when $p=o(N)$.  Together with Assumption~\ref{ass:JM} the two displays force the quadratic-form limit $e_j^\top M_2\Sigma M_2^\top e_j\to(\Sigma^{-1})_{jj}$ (Lemma~\ref{lem:m2-efficiency} of the supplementary material) --- the variance-matching ingredient of requirement~(iii) of Theorem~\ref{thm:CLT}; the full stabilisation of $\tau_{j,n}^2$ is verified outright in the known-precision Gaussian refinement (Corollary~\ref{cor:gaussian-known}) and remains a hypothesis for constructions built from the same pooled rows that enter the variance (the reuse obstruction, Remark~\ref{rem:nodewise-iii-hypothesis}).  The price is row sparsity of the precision matrix: $s_\Omega\log p=o(\sqrt{n_0+N})$ is exactly the requirement that the row rate beat $(n_0+N)^{-1/4}$.  The nodewise-Lasso construction of $M_2$---the construction deployed in the numerical study and every application (Remark~\ref{rem:jm-scope})---satisfies the same rates and is treated in Section~\ref{sec:refine}.}

\subsection{Choosing the unlabelled sample size}\label{sec:choose-N}

The central limit theorem of Theorem~\ref{thm:CLT} is valid only while the bias remainder is negligible against the leading stochastic term---requirement~(i), that the composite envelope $\sqrt{n_0+N}\,\bigl(\norm{r(\hat t)}_\infty+\sqrt{\log p/N}\,\norm{\Sigma^{1/2}r(\hat t)}_2\bigr)$ vanish, equivalently that the pseudo-label remainder $\tfrac{N}{n_0+N}\,e_j^\top M_2\hat\Sigma_{\mathrm u}r(\hat t)$ be $o\bigl((n_0+N)^{-1/2}\bigr)$, the order of the leading stochastic term.  The quantity $N$ itself governs this remainder: enlarging the unlabelled sample sharpens the estimated precision and shrinks the variance, but inflates the imputation bias through the same $\sqrt{n_0+N}$ factor.  The size that enters the procedure is chosen from the data rather than set to all of $N_{\mathrm{avail}}$.  We give two devices for this choice.  The first inverts the negligibility requirement directly into a sufficient cap on $N$ (Section~\ref{sec:vbcap}); the second replaces the cap's tolerance by a self-tuning rule that reads the imputation bias in units of standard error and stops where it reaches a fixed level (Section~\ref{sec:vbrule}).

\subsubsection{A sufficient cap by inverting the requirement}\label{sec:vbcap}

The cap solves the negligibility requirement for $N$.  Writing it as $\sqrt{n_0+N}\,b_n\le\gamma_N$, with $b_n$ the residual-bias surrogate of \eqref{eq:bn-gen} and $\gamma_N>0$ a small tolerance, and replacing $b_n$ by a conservative plug-in upper bound, one admits unlabelled rows up to the size at which the shrinkage-discounted bias, magnified by the $\sqrt{n_0+N}$ inference scale, would breach $\gamma_N$.  Let $\hat a_1$ be a plug-in upper bound for $\|\Delta\|_1$ and $\hat a_2$ one for $\|\Sigma^{1/2}\Delta\|_2$, obtained from the external estimator's own error bound (when it is a transfer-learning estimator), a held-out labelled block, or a sample-split target-domain proxy.  Motivated by the sufficient bias conditions of Section~\ref{sec:theory-bias}, this yields the generic-design cap
\[
    N_{\max}^{\mathrm{gen}}
    := \left[\frac{\gamma_N^2}{\bigl\{{\hat{\bar\mu}_1}+(1-\hat t)(1+{\hat{\bar\mu}_1})\bigr\}^2\hat a_1^2}-n_0\right]_+,
\]
and, under Gaussian design, the sharper cap
\[
    N_{\max}^{\mathrm{G}}
    := \left[\frac{\gamma_N^2}{\bigl(1-\hat t+\sqrt{\log p/n_0}\bigr)^2\hat a_2^2}-n_0\right]_+,
\]
where $[x]_+:=\max(x,0)$ {and $\hat{\bar\mu}_1:=\maxnorm{M_1\hat\Sigma_0-I_p}$ is the realised Javanmard--Montanari bound, which does not exceed the tolerance $\bar\mu_1$ on the feasibility event of Assumption~\ref{ass:JM}, so that the sufficient bias conditions of Section~\ref{sec:theory-bias} hold with $\hat{\bar\mu}_1$ in its place}.  The comparative statics follow: the cap falls quadratically in the external bias, $N_{\max}^{\mathrm{gen}}\propto\hat a_1^{-2}$, so a more biased model admits fewer rows; it widens as more of the correction is admitted ($\hat t\to1$); and it tightens with the tolerance $\gamma_N$.  The operating range is then $[1,N_{\mathrm{op}}]$ with
\[
    N_{\mathrm{op}} := \min\{N_{\mathrm{avail}},\,N_{\max}\},
\]
where the admissibility cap $N_{\max}$ is $N_{\max}^{\mathrm{gen}}$ or $N_{\max}^{\mathrm{G}}$ as the design dictates; if the range is empty the procedure reverts to target-only debiased Lasso.  The cap certifies the $N$-increasing entries of the bias requirement, but only through the tolerance $\gamma_N$ and conservative plug-in bounds whose tightness is regime-dependent, so it tends to discard usable unlabelled information; the weighted-$\ell_2$ entry of the Gaussian bias condition of Section~\ref{sec:refine}, the term $\hat t\sqrt{p\log p/(Nn_0)}$ of \eqref{eq:gaussian-bias-cond}, is decreasing in $N$ and is a standing $p$-versus-$n_0$ side condition rather than a cap-driver.  This conservatism motivates the self-tuning rule of Section~\ref{sec:vbrule}.

\subsubsection{The variance-balance rule}\label{sec:vbrule}

The rule instead tracks the bias--variance trade-off, decomposing the standardised error into its competing parts.

\begin{proposition}[Variance decomposition and the variance-balance point]\label{prop:vbalance}
\setlength{\emergencystretch}{3em}%
Suppose Assumptions~\ref{ass:design}, \ref{ass:rsc}, \ref{ass:external-indep} and \ref{ass:JM} hold, with the stacked Lasso tuned as in Theorem~\ref{thm:stacked-lasso} and $\sqrt{n_0+N}\,\mu_2 s\lambda\to0$ (condition~(ii) of Theorem~\ref{thm:CLT}).  Then $\sqrt{n_0+N}\,(\tilde\beta_j-\beta^\star_j)$ admits the decomposition
\begin{equation}\label{eq:lbd-decomp}
    \sqrt{n_0+N}\,(\tilde\beta_j - \beta^\star_j) \;=\; L_j(N) + B_j(N) + R_j(N),
\end{equation}
with
\[
    L_j(N)=\frac{1}{\sqrt{n_0+N}}\,e_j^\top M_2A_nX_0^\top\varepsilon_0,\qquad
    B_j(N)=\sqrt{n_0+N}\,\frac{N}{n_0+N}\,e_j^\top M_2\hat\Sigma_{\mathrm u}\,r(\hat t),
\]
and $R_j(N)=\sqrt{n_0+N}\,e_j^\top(I_p-M_2\hat\Sigma_{\mathrm{stk}})(\hat\beta-\beta^\star)$; the decomposition is an algebraic identity, and $R_j(N)=\op(1)$.  Given the covariates, the tuning data, and the external estimator, $L_j(N)$ is mean zero and $B_j(N)$ is deterministic.  Write $\tau^2_{L,j}(N)$ for the conditional variance of $L_j(N)$ under this conditioning and $\tau^2_{B,j}(N):=B_j(N)^2$.  The ratio $\tau^2_{B,j}(N)/\tau^2_{L,j}(N)$ is invariant to the common $\sqrt{n_0+N}$ scaling and equals the ratio of the squared-bias and leading-variance contributions to the conditional mean-squared error of $\tilde\beta_j$.  In the idealised oracle-precision benchmark of Corollary~\ref{cor:idealized-efficiency} the leading-variance contribution is non-increasing in $N$ while the squared-bias contribution is non-decreasing in $N$, so $\tau^2_{B,j}(N)/\tau^2_{L,j}(N)$ is non-decreasing in $N$ and the balance point defined next is a crossing rather than merely the first $N$ at which the threshold is met.  For a criterion ratio $\varrho\ge 1$ we define the population-level \emph{variance-balance point}
\begin{equation}\label{eq:Nstar-population}
    N^*_j \;:=\; \min\!\Bigl\{N\ge 1\;:\;\tau^2_{B,j}(N)\ge \varrho\,\tau^2_{L,j}(N)\Bigr\},
\end{equation}
the smallest $N$ at which the squared standardised bias reaches $\varrho$ times the leading-noise variance.
\end{proposition}

The monotone ratio of \eqref{eq:Nstar-population} identifies where to stop: using all of $N_{\mathrm{avail}}$ would maximise the bias.  Because the common $\sqrt{n_0+N}$ factor cancels, $\tau^2_{B,j}(N)/\tau^2_{L,j}(N)$ is the squared imputation bias of $\tilde\beta_j$ measured in units of its own standard error.  Requirement~(i) of Theorem~\ref{thm:CLT} is that this standardised bias vanish; holding it instead at a fixed finite level $\varrho$ is the finite-sample analogue, capping the systematic shift at about $\sqrt{\varrho}$ standard errors.  Below $N^*_j$ the chosen level is respected, and beyond it each additional row contributes more bias than it removes noise.  An interior selection---one at which the balance point is reached before the admissible budget is exhausted---trades a controlled bias against width by design.  The central limit theorem is invoked at the selected size only where its bias requirement holds, in particular at saturation, where the admissible budget is exhausted before the balance point is reached.

{The conditional variance of $L_j(N)$ is the sample-dependent $\tau_{j,n}^2$ displayed in Theorem~\ref{thm:CLT}, and its surrogate $\hat\tau^2_{L,j}(N)$ is formed by substituting the quantities the procedure already computes---the debiasing matrices $M_1$, $M_2$, the sample covariances $\hat\Sigma_0$, $\hat\Sigma_{\mathrm u}$, and the noise-variance estimate $\hat\sigma^2$.  For the bias term, the unobservable $r(\hat t)$ enters only through $|e_j^\top M_2\hat\Sigma_{\mathrm u}r(\hat t)|\le\norm{e_j^\top M_2\hat\Sigma_{\mathrm u}}_1\,b_n$, so the bound-form surrogate
$\hat\tau_{B,j}(N):=\sqrt{n_0+N}\,\tfrac{N}{n_0+N}\,\norm{e_j^\top M_2\hat\Sigma_{\mathrm u}}_1\,b_n$
is used, with $b_n$ the surrogate of \eqref{eq:bn-gen}; it is conservative for $|B_j(N)|$.  Both surrogates are explicit in $N$ once these plug-ins are fixed, so the criterion below is evaluated over the candidate grid without refitting, and Stages~2--4 run once at the selected size.}  Aggregating over the index set $J$ of inferential interest, the data-driven choice is
\begin{equation}\label{eq:Nstar-empirical}
    \hat N^* \;:=\; \argmin_{N \in \mathcal N}\;\left|\;\frac{1}{|J|}\sum_{j\in J}\frac{\hat\tau^2_{B,j}(N)}{\hat\tau^2_{L,j}(N)} \;-\; \varrho\;\right|,
    \qquad \mathcal N \subseteq \bigl\{1,\ldots,\min(N_{\mathrm{avail}},N_{\max})\bigr\},
\end{equation}
searched over a coarse grid (for instance, the grid of Section~\ref{app:simdetails}) inside the admissible range of Section~\ref{sec:vbcap}.  The criterion ratio $\varrho\ge 1$ is the only user choice: $\varrho=1$ targets the population variance-balance point~\eqref{eq:Nstar-population} and is the default, while a moderately larger $\varrho$ admits more rows at the cost of a larger standardised bias.  The rule carries no tolerance $\gamma_N$, is invariant under common rescaling of $\Sigma$, requires no held-out labelled data, and reverts to target-only debiased Lasso when no positive $\hat N^*$ lies in the admissible range.

At $\varrho=1$, every Monte Carlo cell of Section~\ref{sec:experiments} and every real-data demonstration of Section~\ref{sec:realdata} selects the unlabelled size by \eqref{eq:Nstar-empirical}.  The admissibility cap of Section~\ref{sec:vbcap} supplies the search range and furnishes the operating size when the surrogates are not formed or the range is empty.

\section{Comparison and robustness}\label{sec:robust}

{
The asymptotic theory of Section~\ref{sec:theory-bias} assumed a correctly specified linear model and an unlabelled design matched to the target distribution.  This section relaxes both assumptions and develops the comparison with the prediction-powered family along two threads.  One establishes validity and robustness: validity for the best linear approximation under model misspecification (Section~\ref{sec:misspec}) and a shift-aware variant for covariate-shifted unlabelled designs (Section~\ref{sec:shift-aware}).  The other compares efficiency: the confidence-interval-length comparison with the prediction-powered family under a linear predictor (Section~\ref{sec:misspec-comparison}), the non-linear-predictor comparison and the complementarity of the two channels (Section~\ref{sec:misspec-nonlinear-labeler}), and the control-variate hybrid (Section~\ref{sec:hybrid}).
}
\subsection{Validity for the best linear approximation}\label{sec:misspec}

The construction of the previous sections operates under the assumption $Y = X^\top\beta^\star + \varepsilon$.  This section drops the linear-truth assumption and analyses DEAL under a non-linear conditional mean $\E[Y\mid X]=\mu(X)$.  The inferential target becomes the population least-squares projection $\beta^\star_{\mathrm{proj}}$ of $Y$ onto the linear span of $X$, which coincides with $\beta^\star$ in the linear-truth specialisation.  Two regimes for the external model are considered in turn: the linear-coefficient form $\hat Y = X^\top\hat\beta_{\mathrm{ext}}$ used in Sections~\ref{sec:method}--\ref{sec:theory-bias} and in Section~\ref{sec:refine} of the supplementary material via Assumption~\ref{ass:external}, and a more general regime in which the model is a measurable function $\hat\mu:\R^p\to\R$ accessed only through its predicted values.  The two regimes share the same inferential target; the general regime is developed in Section~\ref{sec:misspec-nonlinear-labeler}, where its asymptotic-variance form and plug-in interval are shown to match the linear one.

\subsubsection{Projection-parameter reframing}\label{sec:misspec-setup}

\begin{assumption}[Non-linear conditional mean]\label{ass:nonlinear}
The labelled-target sample is generated by $Y = \mu(X) + \varepsilon$ with $X\sim P_0$, $\E[\varepsilon\mid X]=0$, $\varepsilon$ sub-Gaussian conditional on $X$ with conditional sub-Gaussian norm bounded by $\sigma_{\max}$, $\Var(\varepsilon\mid X)=\sigma^2(X)$ uniformly bounded above by $\sigma_{\max}^2$, and $\E_{P_0}[\mu(X)^2]<\infty$.
\end{assumption}

{Recall the population least-squares projection $\beta^\star_{\mathrm{proj}}$ of Definition~\ref{def:estimands}, displayed in \eqref{eq:beta-proj}.}  Define the misspecification residual
\begin{equation}\label{eq:eta-def}
    \eta(X) \;:=\; \mu(X) - X^\top\beta^\star_{\mathrm{proj}},
\end{equation}
which satisfies the orthogonality $\E_{P_0}[X\,\eta(X)]=0$ by the first-order condition characterising the projection.  Under the linear-truth specialisation $\mu(X)=X^\top\beta^\star$, $\beta^\star_{\mathrm{proj}}=\beta^\star$ and $\eta\equiv 0$, so $\beta^\star_{\mathrm{proj}}$ recovers the linear coefficient.

\begin{assumption}[Regularity of the misspecification residual]\label{ass:misspec-reg}
The random variables $\bigl\{X_j\,\eta(X)\bigr\}_{j=1}^p$ are sub-exponential under $P_0$ with a common proxy $\psi_\eta^2<\infty$, and $\eta(X)$ is sub-Gaussian under $P_0$.
\end{assumption}

Assumption~\ref{ass:misspec-reg} is automatic when $\mu$ is Lipschitz in a sub-Gaussian design and is the only new technical condition introduced by the misspecified analysis, which otherwise invokes the precision-row rate of Assumption~\ref{ass:rowsparse} (Section~\ref{sec:theory-bias}) unchanged from the well-specified case.  Throughout this section the inferential target is $\beta^\star_{\mathrm{proj}}$; the parameter $\beta^\star$ in Sections~\ref{sec:method}--\ref{sec:theory-bias} and in Section~\ref{sec:refine} is reinterpreted as $\beta^\star_{\mathrm{proj}}$.

\subsubsection{Validity at the projection parameter}\label{sec:misspec-validity}

The proof of Theorem~\ref{thm:CLT} extends to the misspecified regime upon a single structural observation: the misspecification residual enters the {asymptotic expansion} only through the variance, never through the bias.

{Concretely, the labelled score remains centred: $\E_{P_0}[X(\eta(X)+\varepsilon)]=0$ (Lemma~\ref{lem:misspec-unbiased}, supplementary material, Section~\ref{app:proofs}).}

The pseudo-labels $\tilde f=\tilde X\tilde\beta^{\mathrm{init}}$ are linear in the unlabelled covariates, with the bias-aware initialiser $\tilde\beta^{\mathrm{init}}=\hat\beta_{\mathrm{ext}}+\hat t\,C$ of \eqref{eq:init-bias-aware} as coefficient; since the one-step correction $C$ is built from the labelled inference sample alone, $\eta$ is never evaluated at an unlabelled row, so the unlabelled block contributes no $\eta$ term of its own to the score $X_{\mathrm{stk}}^\top(y_{\mathrm{stk}} - X_{\mathrm{stk}}\beta^\star_{\mathrm{proj}})$ beyond what the labelled block already carries through $C$.  Since the labelled block is itself centred at $\beta^\star_{\mathrm{proj}}$ (Lemma~\ref{lem:misspec-unbiased} of the supplementary material), all bias-control machinery of Sections~\ref{sec:method}--\ref{sec:theory-bias} and of Section~\ref{sec:refine} — the bias-aware shrinkage of Stage~1, the JM cancellation of Stage~4, and the stacked-Lasso $\ell_1$ rate of Theorem~\ref{thm:stacked-lasso} — is unchanged at the population level.  The misspecification residual surfaces only in the variance of the labelled score $X(\eta(X)+\varepsilon)$, which acquires the additive contribution $\eta(X)^2 X X^\top$.
\begin{theorem}[Sandwich-variance CLT under misspecification]\label{thm:CLT-misspec}
Suppose Assumption~\ref{ass:design}(a) and Assumptions~\ref{ass:rsc}, \ref{ass:external}, \ref{ass:external-indep}, \ref{ass:JM}, \ref{ass:condT}, \ref{ass:rowsparse}, \ref{ass:nonlinear}, \ref{ass:misspec-reg} hold.  With $A_n$ as in Theorem~\ref{thm:CLT}, let
\begin{equation}\label{eq:Gamma-fold}
    \hat\Gamma_0 \;:=\; \frac{1}{n_0}\sum_{i=1}^{n_0}\bigl(\eta(X_{0,i})^2 + \sigma^2(X_{0,i})\bigr)\,X_{0,i}X_{0,i}^\top,
    \qquad
    \tau_{j,n}^{2,\mathrm{sand}} \;:=\; \frac{n_0}{n_0+N}\, e_j^\top M_2 A_n\,\hat\Gamma_0\,A_n^\top M_2^\top e_j,
\end{equation}
the misspecified analogs of the homoscedastic labelled-block second moment $\sigma^2\hat\Sigma_0$ and of $\tau_{j,n}^2$ in Theorem~\ref{thm:CLT}.  Fix a coordinate $j$ in the inferential index set, and let $v_j:=(1+t_0N/n_0)\,\Sigma_0^{-1}e_j$ denote the in-probability limit of $A_n^\top M_2^\top e_j$ furnished by Assumptions~\ref{ass:condT} and~\ref{ass:rowsparse}.  Assume that, at $\beta^\star_{\mathrm{proj}}$ in place of $\beta^\star$,
\begin{enumerate}
    \item[(i)--(ii)] conditions (i)--(ii) of Theorem~\ref{thm:CLT} hold verbatim;
    \item[(iii$'$)] $\tau_{j,n}^{2,\mathrm{sand}} \to \tau_j^{2,\mathrm{sand}}\in(0,\infty)$ in probability;
    \item[(iv$'$)] the Lindeberg ratio
    \[
        \frac{\max_{1\le i\le n_0}\bigl|(v_j^\top x_{0,i})\,u_i\bigr|}
        {\bigl\{\sum_{i=1}^{n_0}(v_j^\top x_{0,i})^2\,u_i^2\bigr\}^{1/2}}
        \;\to\;0
        \quad\text{in probability, where } u_i:=\eta(X_{0,i})+\varepsilon_{0,i};
    \]
    \item[(v$'$)] the $\eta$-channel weight-replacement condition $\sqrt{\log p}\,\norm{A_n^\top M_2^\top e_j - v_j}_1 \to 0$ in probability.
\end{enumerate}
Then, for this coordinate $j$,
\begin{equation}\label{eq:CLT-misspec}
    \sqrt{n_0+N}\,\bigl(\tilde\beta_j - \beta^\star_{\mathrm{proj},j}\bigr) \;\overset{\mathrm{d}}{\longrightarrow}\; \mathcal N\!\bigl(0,\;\tau_j^{2,\mathrm{sand}}\bigr).
\end{equation}

\end{theorem}
Condition~(v$'$) is the price of the misspecification residual: because $\eta$ is centred only unconditionally, the proof cannot condition on the JM matrices and instead replaces the realised weights $A_n^\top M_2^\top e_j$ by their limit $v_j$, and (v$'$) asks the $\ell_1$ error of that replacement to vanish faster than the $\sqrt{\log p}$ max-norm fluctuation of the residual score $n_0^{-1/2}\sum_{i=1}^{n_0}X_{0,i}\,\eta(X_{0,i})$.

\begin{remark}[Linear-truth recovery and plug-in intervals]\label{rem:misspec-linear-recovery}\label{cor:misspec-plugin}
Under linear truth $\eta\equiv 0$, $\hat\Gamma_0\to \E[\sigma^2(X)\,XX^\top]$, which under conditional homoscedasticity reduces to $\sigma^2\Sigma_0$, so that $\tau_{j,n}^{2,\mathrm{sand}}=\tau_{j,n}^2$.  Theorem~\ref{thm:CLT-misspec} then recovers Theorem~\ref{thm:CLT} verbatim.  The added contribution $\E[\eta(X)^2\,XX^\top]$ in \eqref{eq:Gamma-fold} is the misspecification-induced inflation of the labelled-block variance, weighted in the aggregate by the labelled fraction $n_0/(n_0+N)$.

Two-sided $(1-\alpha)$ confidence intervals for $\beta^\star_{\mathrm{proj},j}$ are $\tilde\beta_j \pm z_{1-\alpha/2}\sqrt{\hat\tau_j^{2,\mathrm{sand}}/(n_0+N)}$, where $\hat\tau_j^{2,\mathrm{sand}}$ replaces $\hat\Gamma_0$ in \eqref{eq:Gamma-fold} by the empirical-residual plug-in $\frac{1}{n_0}\sum_{i=1}^{n_0}\hat r_i^2\,X_{0,i}X_{0,i}^\top$, with $\hat r_i:=y_{0,i}-X_{0,i}^\top\tilde\beta$.  Under linear truth and conditional homoscedasticity, $\hat\tau_j^{2,\mathrm{sand}}$ is asymptotically equivalent to the homoscedastic plug-in of Section~\ref{sec:theory-bias}; under misspecification the homoscedastic plug-in is generally inconsistent for $\tau_j^{2,\mathrm{sand}}$ --- it may over- or under-cover depending on the correlation between $\eta(X)^2$ and $(v_j^\top X)^2$ --- and only the sandwich form above is consistent.
\end{remark}

\subsection{Shift-aware variant for covariate-shifted unlabelled designs}\label{sec:shift-aware}

The construction of the previous sections assumes that the unlabelled covariates $\tilde X$ are drawn from the target population, so that $\hat\Sigma_{\mathrm u}$ and $\hat\Sigma_0$ both estimate the same $\Sigma$.  In many applications, however, the unlabelled pool is collected from an adjacent population whose covariate distribution differs from the target's, even when the conditional model $Y \mid X$ is preserved.  {We analyse the resulting bias amplification.  Under covariate shift, the variance-balance rule of Section~\ref{sec:vbrule} caps the imputation bias in units of standard error exactly as without shift.  What covariate shift costs is efficiency.  The one-sided modification of the Stage-4 debiasing below removes the bias amplification at its source.  In the regime where the unlabelled pool dominates the labelled sample ($\kappa_n\to\infty$), it restores valid inference at the full augmented sample when $\|\Sigma_{\mathrm u}\Sigma_0^{-1}\|_{\mathrm{op}}=O(1)$, recovering the efficiency the bias ceiling alone would forgo (Corollary~\ref{cor:shift-restore}).}

Throughout this section the labelled target rows $X_{0,i}$ are i.i.d.\ with covariance $\Sigma_0$, while the unlabelled rows $\tilde X_i$ are i.i.d.\ with covariance $\Sigma_{\mathrm u}$, possibly with $\Sigma_{\mathrm u}\neq \Sigma_0$; write $D:=\Sigma_{\mathrm u}-\Sigma_0$.  The conditional model $Y\mid X = X^\top\beta^\star + \varepsilon$ is preserved with $\varepsilon$ independent of $X$, so there is no concept shift.  Assumptions~\ref{ass:design}, \ref{ass:rsc}, and~\ref{ass:JM} are imposed separately on each block.

For each coordinate $j$, the Stage-4 debiased estimator admits the exact decomposition
\begin{equation}\label{eq:bias-shift-decomp}
\begin{gathered}
    \tilde\beta_j-\beta^\star_j
    \;=\;
    S_j
    \;+\;
    \underbrace{e_j^\top\bigl(I_p - M_2 \hat\Sigma_{\mathrm{stk}}\bigr)\bigl(\hat\beta - \beta^\star\bigr)}_{\text{Term A: shrinkage residual}}
    \;+\;
    \underbrace{\frac{N}{n_0+N}\,e_j^\top M_2\,\hat\Sigma_{\mathrm u}\,r(\hat t)}_{\text{Term B: imputation-bias amplification}},
    \\[6pt]
    S_j:=\frac1{n_0+N}\,e_j^\top M_2A_nX_0^\top\varepsilon_0,
\end{gathered}
\end{equation}
with $A_n$ as in Theorem~\ref{thm:CLT} and $\Sigma_{\mathrm{stk}}:=(n_0\Sigma_0 + N\Sigma_{\mathrm{u}})/(n_0+N)$ the population limit of $\hat\Sigma_{\mathrm{stk}}$.  As in the proof of Theorem~\ref{thm:CLT}, $\E[S_j\mid X_{\mathrm{stk}},\text{tuning}]=0$ (Assumption~\ref{ass:external-indep}), so $S_j$ contributes only to the variance and Term~A $+$ Term~B carry the entire bias of $\tilde\beta_j$.  Under $\Sigma_{\mathrm u}=\Sigma_0$, the JM construction \eqref{eq:JM-M2} sets $M_2\hat\Sigma_{\mathrm{stk}}\approx I$ and $M_2\hat\Sigma_{\mathrm u}\approx I$, so Term~A is negligible at the $\sqrt{n_0+N}$ scale and Term~B reduces to the standard imputation bias controlled by $\hat t$.  Under $\Sigma_{\mathrm u}\neq \Sigma_0$, by contrast, $e_j^\top M_2\hat\Sigma_{\mathrm u}$ may have row-$\ell_1$ norm exceeding unity, and Term~B is amplified.

\begin{proposition}[Bias amplification under covariate shift]\label{prop:shift-amp}
Suppose that the hypotheses of Theorem~\ref{thm:stacked-lasso} hold, with $\Sigma_{\mathrm u}$ in place of $\Sigma$ in the surrogate condition \eqref{eq:bn-surrogate} and with Assumptions~\ref{ass:design}, \ref{ass:rsc} and \ref{ass:JM} imposed separately on the labelled and unlabelled blocks, that $M_2$ is the pooled JM matrix \eqref{eq:JM-M2} based on $\hat\Sigma_{\mathrm{stk}}$, and that regime condition~(ii) of Theorem~\ref{thm:CLT} holds.  For each coordinate $j$, write $\rho_j(M_2,\Sigma_{\mathrm u}):=\bigl\|e_j^\top M_2\Sigma_{\mathrm u}\bigr\|_1$.

\textup{(i) Population identity.}  For any $p\times p$ matrix $M_2$, $M_2\Sigma_{\mathrm u} = M_2\Sigma_{\mathrm{stk}} + \tfrac{n_0}{n_0+N}M_2D$, hence
\[
    \rho_j(M_2,\Sigma_{\mathrm u}) \;\le\; \bigl\|e_j^\top M_2\Sigma_{\mathrm{stk}}\bigr\|_1 + \tfrac{n_0}{n_0+N}\bigl\|e_j^\top M_2D\bigr\|_1 .
\]

\textup{(ii) Uniform bound.}  $\rho_j(M_2,\Sigma_{\mathrm u}) \le \mnorm{M_2}\,\mnorm{\Sigma_{\mathrm u}} = \Op(1)$.

\textup{(iii) Non-centring bound.}  With Term~A, Term~B as in \eqref{eq:bias-shift-decomp}, with probability tending to one,
\begin{align*}
    \sqrt{n_0+N}\,\bigl|\text{Term A} + \text{Term B}\bigr|
    &\;\le\;
    \sqrt{n_0+N}\,\frac{N}{n_0+N}\,\mnorm{M_2}\Bigl(\mnorm{\Sigma_{\mathrm u}}\,\|r(\hat t)\|_\infty\\
    &\qquad\qquad\qquad +C\sqrt{\tfrac{\log p}{N}}\,\|\Sigma_{\mathrm u}^{1/2}r(\hat t)\|_2\Bigr) \;+\; \op(1)
\end{align*}
for a constant $C$; by~(ii) this bound is of order $\rho_j(M_2,\Sigma_{\mathrm u})$ at most and, unlike the no-shift case where $\rho_j=1$ exactly (Corollary~\ref{cor:shift-amp-benchmark}(i)), need not be $o(1)$ relative to the no-shift level.
\end{proposition}

{(Benchmark values under oracle and nodewise precision: Corollary~\ref{cor:shift-amp-benchmark} of the supplementary material.)}

By Corollary~\ref{cor:shift-amp-benchmark}(ii) of the supplementary material, the amplification $\rho_{j_0}>1$ at some coordinate $j_0$ is a population-level consequence of $\Sigma_{\mathrm u}\succ\Sigma_0$, independent of the admissibility caps $N_{\max}^{\mathrm{gen}}$ or $N_{\max}^{\mathrm{G}}$ of Section~\ref{sec:vbcap}: capping $N$ neutralises it only by discarding the augmentation, which is what motivates the one-sided substitution below.

Define two precision matrices $\hat M_{\mathrm u}$, $\hat M_{\mathrm p}$ by the JM construction \eqref{eq:JM-M2} (or, under Gaussian design, by nodewise Lasso as in Section~\ref{sec:refine}) applied respectively to the unlabelled block and the pooled stack; the pooled matrix $\hat M_{\mathrm p}$ coincides with the Stage-4 matrix $M_2$ of \eqref{eq:JM-M2}.  Replace the Stage-4 debiasing matrix by
\begin{equation}\label{eq:M2-adapt}
    \hat M_2^{\mathrm{adapt}} \;:=\;
    \begin{cases}
        \hat M_{\mathrm u} & \text{if } \widehat{\mathrm{Shift\text{-}up}} = 1, \\
        \hat M_{\mathrm p} & \text{otherwise},
    \end{cases}
\end{equation}
where the indicator $\widehat{\mathrm{Shift\text{-}up}}\in\{0,1\}$ is a plug-in shift detector defined in \eqref{eq:shift-detector} below.  Under detected shift-up, the substitution forces $\hat M_2^{\mathrm{adapt}}\,\hat\Sigma_{\mathrm u}\approx I$ entrywise, removing the row-$\ell_1$ amplification $\rho_j(M_2,\Sigma_{\mathrm u})$ of Proposition~\ref{prop:shift-amp} from Term~B (Proposition~\ref{prop:shift-bias-reduction} below), at the price of a non-zero Term~A residual, controllable under the strengthened sparsity scaling of Corollary~\ref{cor:shift-restore}---the cost of deploying a precision matrix not feasible for $\hat\Sigma_0$.  Under no shift or shift-down, the substitution is not triggered and the procedure reduces to the pooled construction.  This one-sidedness is essential: applying the substitution under shift-down would inflate Term~A without compensating reduction of Term~B, since shift-down already attenuates Term~B through $\rho_j(M_2,\Sigma_{\mathrm u})<1$.

The substitution rule \eqref{eq:M2-adapt} requires only an indicator of the population event $\Sigma_{\mathrm u}\succ\Sigma_0$ that is operationally accessible from $(\hat\Sigma_0,\hat\Sigma_{\mathrm u},\hat M_{\mathrm p})$ alone.  Motivated by Corollary~\ref{cor:shift-amp-benchmark} --- $\rho_j(\hat M_{\mathrm p},\Sigma_{\mathrm u})=1$ exactly at no shift, $>1$ at some coordinate under shift-up --- a natural plug-in detector tests the empirical row-$\ell_1$ excess on the inferential index set $J$ directly:
\begin{equation}\label{eq:shift-detector}
    \widehat{\mathrm{Shift\text{-}up}} \;:=\; \1\!\Bigl\{\,\max_{j\in J}\,\bigl\|e_j^\top\,\hat M_{\mathrm p}\,\hat\Sigma_{\mathrm u}\bigr\|_1 \;>\; 1 + c_n\,\Bigr\},
    \qquad
    c_n \asymp \sqrt{\frac{\log p}{n_0\wedge N}},
\end{equation}
with $c_n$ set at the standard sub-Gaussian entrywise deviation rate.  {We use \eqref{eq:shift-detector} as a practical trigger for the substitution, not as a device with a proven size and power: a formal Type-I/power guarantee would need a uniform-in-$J$ row-$\ell_1$ consistency rate for $\hat M_{\mathrm p}\hat\Sigma_{\mathrm u}$, which the row-sum control of Assumption~\ref{ass:JM} alone does not supply (the same $\ell_2$-to-$\ell_1$ gap noted in the proof of Corollary~\ref{cor:shift-amp-benchmark}).  We leave a formal analysis of the detector to future work and rely on the empirical validation of Section~\ref{sec:experiments-shift}.}

The following results quantify the effect of the substitution and give the coordinate-wise CLT it enables in the $\kappa_n\to\infty$ regime of Corollary~\ref{cor:shift-restore} (the auxiliary regime, Remark~\ref{rem:shift-restore-regime}).

\begin{assumption}[Unlabelled-block precision-row rate]\label{ass:rowsparse-shift}
Let $\Omega_{\mathrm u}:=\Sigma_{\mathrm u}^{-1}$ have row sparsity $s_{\Omega_{\mathrm u}}$ at the inferential coordinate $j$.  The unlabelled-block JM matrix $\hat M_{\mathrm u}$ of \eqref{eq:M2-adapt} attains the precision-row rate
\[
    \norm{\hat M_{\mathrm u}^\top e_j-\Omega_{\mathrm u} e_j}_2=\Op\!\Bigl(\sqrt{s_{\Omega_{\mathrm u}}\log p/N}\Bigr),
\]
with $s_{\Omega_{\mathrm u}}\log p=o(\sqrt N)$ --- the unlabelled-block analogue of Assumption~\ref{ass:rowsparse}, needed because that assumption is scoped to the no-shift $\Sigma:=\Sigma_0$.
\end{assumption}

\begin{proposition}[Bias reduction under the one-sided substitution]\label{prop:shift-bias-reduction}
Suppose Assumption~\ref{ass:design}(a) and Assumptions~\ref{ass:external-indep}, \ref{ass:JM} hold on the unlabelled block, and that $\hat M_2^{\mathrm{adapt}}=\hat M_{\mathrm u}$.  The substituted imputation-bias term $\mathrm{Term\ B}^{\mathrm u}:=\tfrac{N}{n_0+N}e_j^\top\hat M_{\mathrm u}\hat\Sigma_{\mathrm u}r(\hat t)$ obeys, with probability tending to one,
\[
    \sqrt{n_0+N}\,\bigl|\mathrm{Term\ B}^{\mathrm u}\bigr|
    \;=\;
    \Op\!\Bigl(\sqrt{n_0+N}\,\bigl(\norm{r(\hat t)}_\infty+\sqrt{\tfrac{\log p}{N}}\,\norm{\Sigma_{\mathrm u}^{1/2}r(\hat t)}_2\bigr)\Bigr),
\]
the same order as the no-shift imputation-bias term of Theorem~\ref{thm:CLT}, with no $\rho_j(\hat M_{\mathrm p},\Sigma_{\mathrm u})$-type multiplicative factor.  Since $\rho_j(\hat M_{\mathrm p},\Sigma_{\mathrm u})$ can exceed one at a fixed, non-vanishing population level (Corollary~\ref{cor:shift-amp-benchmark}(ii)), the substitution strictly reduces the imputation-bias order whenever the pooled construction is amplified: the residual-bias surrogate $b_n$ of the variance-balance rule (Section~\ref{sec:vbrule}) is no longer inflated by the shift, and need not cap $\hat N^*$ below the level it would select without shift to hold the same bias-to-standard-error target.
\end{proposition}

\begin{corollary}[Full-sample validity under one-sided substitution]\label{cor:shift-restore}
Suppose that $\Sigma_{\mathrm u}\succ\Sigma_0$ in the Loewner sense with $\|\Sigma_{\mathrm u}\Sigma_0^{-1}\|_{\mathrm{op}}=O(1)$, that $\hat M_2^{\mathrm{adapt}}=\hat M_{\mathrm u}$, and that the hypotheses of Theorem~\ref{thm:stacked-lasso} hold, with $\Sigma_{\mathrm u}$ in place of $\Sigma$ in the surrogate condition \eqref{eq:bn-surrogate}, with Assumptions~\ref{ass:design}, \ref{ass:rsc} and \ref{ass:JM} imposed separately on the labelled and unlabelled blocks and the stacked-Lasso tuning parameter obeying $\lambda\asymp\sqrt{\log p/(n_0+N)}$.  Write $n_{\mathrm{stk}}:=n_0+N$.  Suppose the regime \eqref{eq:regime} holds with $\kappa_n:=N/n_0\to\infty$ in place of its bounded limit there --- the auxiliary regime this result is confined to (Remark~\ref{rem:shift-restore-regime}) --- together with $\hat t\,\kappa_n=\op(1)$, the sparsity conditions $s\sqrt{\log p}=o(\sqrt{\kappa_n})$ and $s\log p=o(\sqrt{n_0})$, and the sharpened $\Sigma_{\mathrm u}$-analogue of regime condition~(i),
\[
    \sqrt{n_{\mathrm{stk}}\kappa_n}\,\bigl(\norm{r(\hat t)}_\infty+\sqrt{\log p/N}\norm{\Sigma_{\mathrm u}^{1/2}r(\hat t)}_2\bigr)\to0
\]
in probability (the sharpening to the $n_{\mathrm{stk}}/\sqrt{n_0}$ normalisation is derived in Remark~\ref{rem:shift-restore-regime}).  Let, with $A_n$ as in Theorem~\ref{thm:CLT},
\[
    \tau_{j,n}^{2,\mathrm u} := \sigma^2\,e_j^\top \hat M_{\mathrm u}A_n\hat\Sigma_0 A_n^\top \hat M_{\mathrm u}^\top e_j,
\]
and suppose further $\tau_{j,n}^{2,\mathrm u}\to\tau_j^{2,\mathrm u}\in(0,\infty)$ in probability, and the $\Sigma_{\mathrm u}$-analogue of the Lindeberg condition~(iv), stated for the array $\{x_{0,i}^\top A_n^\top\hat M_{\mathrm u}^\top e_j\}_{i=1}^{n_0}$, holds.  Then $\tilde\beta_j^{\mathrm{adapt}}$ is asymptotically linear with remainder $\op(\sqrt{n_0}/n_{\mathrm{stk}})$, and
\[
    \frac{n_{\mathrm{stk}}}{\sqrt{n_0}}\,\bigl(\tilde\beta_j^{\mathrm{adapt}}-\beta^\star_j\bigr) \;\overset{\mathrm{d}}{\longrightarrow}\; \mathcal N\!\bigl(0,\;\tau_j^{2,\mathrm u}\bigr).
\]
\end{corollary}

Once the unlabelled pool is of larger order than the labelled sample, debiasing with the unlabelled-block precision is valid at the full augmented sample, at the faster rate $n_{\mathrm{stk}}/\sqrt{n_0}$.

\begin{remark}[Scope of the auxiliary regime and its normalisation]\label{rem:shift-restore-regime}
Corollary~\ref{cor:shift-restore} is stated at $\kappa_n\to\infty$, outside the bounded-$\kappa$ regime \eqref{eq:regime} used elsewhere, because $\hat M_{\mathrm u}$ is infeasible for the labelled block: $I_p-\hat M_{\mathrm u}\hat\Sigma_0$ converges to $\Sigma_{\mathrm u}^{-1}D$, generically $\Theta(1)$ rather than $o(1)$, so Term~A's labelled-block contribution (Step~3 of the proof) is negligible only once the labelled fraction $n_0/n_{\mathrm{stk}}$ of the stacked sample vanishes.  This is a structural limitation of the one-sided substitution, not a proof artefact.  The companion condition $\hat t\kappa_n=\op(1)$ keeps $A_n=I_p+(\hat tN/n_0)\hat\Sigma_{\mathrm u}M_1$ from diverging as $\kappa_n\to\infty$; it is available under the near-perfect-external-estimator regime of Proposition~\ref{prop:condT-regimes}(i) for $\kappa_n$ not growing too fast, where $\hat t\to_{\mathbb P}0$ at a rate that can dominate a suitably slow $\kappa_n$'s growth.

The change of normalisation from $\sqrt{n_{\mathrm{stk}}}$ to $n_{\mathrm{stk}}/\sqrt{n_0}=\sqrt{n_{\mathrm{stk}}}\cdot\sqrt{1+\kappa_n}$ is forced, not a matter of presentation: the leading term of Step~1 of the proof (Section~\ref{app:proofs-robust}) is $\tfrac1{n_{\mathrm{stk}}}\sum_{i=1}^{n_0}(x_{0,i}^\top A_n^\top\hat M_{\mathrm u}^\top e_j)\varepsilon_{0,i}$, a normalised sum of only $n_0$ mean-zero terms; its conditional variance is $\Theta(n_0/n_{\mathrm{stk}}^2)$, so $\sqrt{n_{\mathrm{stk}}}$ times this quantity has variance $\Theta(n_0/n_{\mathrm{stk}})=\Theta(1/(1+\kappa_n))\to0$ under $\kappa_n\to\infty$: the $\sqrt{n_{\mathrm{stk}}}$-scaled limit is degenerate at zero, not the non-degenerate $\mathcal N(0,\tau_j^{2,\mathrm u})$, $\tau_j^{2,\mathrm u}>0$, the corollary asserts.  Multiplying by the additional factor $\sqrt{1+\kappa_n}$ exactly cancels this decay, which is why $\tau_{j,n}^{2,\mathrm u}$ above carries no $n_0/n_{\mathrm{stk}}$ prefactor (unlike the sample-dependent variance $\tau_{j,n}^2$ of Theorem~\ref{thm:CLT}, whose $\sqrt{n_{\mathrm{stk}}}$-scaling is calibrated to the bounded-$\kappa$ regime instead).
\end{remark}

In the idealised benchmark of Section~\ref{sec:theory-bias} ($\hat\Sigma_{\mathrm u}\to\Sigma_{\mathrm u}$, $\hat\Sigma_0\to\Sigma_0$, $M_1\to\Sigma_0^{-1}$, $\hat M_{\mathrm u}\to\Sigma_{\mathrm u}^{-1}$ in the quadratic-form sense licensed by Assumption~\ref{ass:rowsparse-shift}), the hypothesis $\hat t\kappa_n=\op(1)$ forces $A_n\to I_p$, and the limiting variance simplifies to
\[
    \tau_j^{2,\mathrm u}
    \;=\;
    \sigma^2\,
    e_j^\top \Sigma_{\mathrm u}^{-1}\, \Sigma_0\, \Sigma_{\mathrm u}^{-1} e_j,
\]
the $\Sigma_{\mathrm u}\neq\Sigma_0$, $n_{\mathrm{stk}}/\sqrt{n_0}$-scaled analogue of the oracle-precision benchmark \eqref{eq:idealized-var} at the limit $t_0=0$ this regime requires.

The pooled construction $\hat M_{\mathrm p}$ satisfies a different JM-feasibility normalisation from $\hat M_{\mathrm u}$ ($\hat M_{\mathrm p}\hat\Sigma_{\mathrm{stk}}\to I_p$ versus $\hat M_{\mathrm u}\hat\Sigma_{\mathrm u}\to I_p$), so the van~de~Geer optimality of $\Sigma_{\mathrm{stk}}^{-1}$ over the pooled feasible set \citep{vandegeer2014asymptotic} does not apply to $\hat M_{\mathrm u}^\top e_j\to\Sigma_{\mathrm u}^{-1}e_j$, which lies outside that set; we do not assert a variance ordering between $\tau_j^{2,\mathrm u}$ and the pooled variance Theorem~\ref{thm:CLT} would give at $\hat M_{\mathrm p}$.  The substitution is justified here by the bias reduction of Proposition~\ref{prop:shift-bias-reduction} and the coordinate-wise CLT of Corollary~\ref{cor:shift-restore}, not by an efficiency ordering.


{No parametric model for $\Sigma_0$ or $\Sigma_{\mathrm u}$ is imposed anywhere in this analysis---Proposition~\ref{prop:shift-amp}, Corollary~\ref{cor:shift-restore}, the substitution rule \eqref{eq:M2-adapt}, and the detector \eqref{eq:shift-detector} run entirely through the Loewner ordering and row-$\ell_1$ functionals.}

The empirical evidence in Section~\ref{sec:experiments-shift} shows that the substitution rule holds coverage at or near the nominal level across an unlabelled-design covariance-shift envelope of magnitude $|\Delta\rho|\le 0.4$ in the parameter of the first-order autoregressive (AR(1)) covariance.  Corollary~\ref{cor:shift-restore} accounts for this in the auxiliary regime $\kappa_n\to\infty$; an asymptotic analysis of the modified estimator at bounded $\kappa$, together with optimality of the one-sided substitution against alternative policies, is left to future work.

\subsection{Comparison with prediction-powered inference}\label{sec:misspec-comparison}

The DEAL estimator $\tilde\beta$ targets $\beta^\star_{\mathrm{proj}}$ defined in \eqref{eq:beta-proj}.  Three other procedures used as comparators in this section target the same parameter under the same regime conditions.  The target-only debiased Lasso (DL) of \citet{vandegeer2014asymptotic, javanmard2014confidence, zhang2014confidence} estimates $\beta^\star_{\mathrm{proj}}$ from $(X_0, y_0)$ alone under standard sparsity, with no model input.  The prediction-powered procedures of \citet{angelopoulos2023prediction, angelopoulos2023ppipp}, applied to the squared-loss linear-regression estimating equation $\E[X(Y-X^\top\theta)] = 0$, target the population minimiser $\argmin_\theta\E[(Y-X^\top\theta)^2] = \beta^\star_{\mathrm{proj}}$.  All four procedures are therefore asymptotically estimating the same population quantity, and confidence-interval-length comparisons across them are well-defined at $\beta^\star_{\mathrm{proj}}$.

{All comparisons below are at $\beta^\star_{\mathrm{proj}}$ and, unless stated otherwise, in the no-shift regime $\Sigma_{\mathrm u}=\Sigma_0$.  Under a linear external model the prediction-powered family enters the comparison through an empty channel: the rectifier residual lies in the span of the design.  The residual signal that the PPI and PPI++ rectifier is built to extract therefore vanishes, and both reduce asymptotically to target-only debiased Lasso---the high-dimensional analogue of the exact rectifier cancellation of Proposition~\ref{prop:ppi-cancellation}.  This equivalence is stated and proved in Section~\ref{app:ppi-comparison} (Proposition~\ref{prop:ppipp-equivalence}).  We therefore develop the confidence-interval-length results against target-only debiased Lasso, from which the comparisons against PPI and PPI++ under a linear predictor follow immediately; the non-linear predictor, where the two channels genuinely separate, is taken up in Section~\ref{sec:misspec-nonlinear-labeler}.}

\begin{theorem}[Dominance of DEAL at a common unlabelled budget]\label{thm:deal-vs-dl}
Fix a deterministic unlabelled-budget sequence $N=N_{n_0}$ along \eqref{eq:regime}, with $\kappa_n=N/n_0\to\kappa_\infty\in[0,\infty)$, and suppose the assumptions of Theorem~\ref{thm:CLT-misspec} hold at this budget, with the shrinkage limit $\hat t\to_{\mathbb P} t_0\in[0,1]$.  For the target-only comparator, suppose in addition that the labelled-block matrix $M_1$ satisfies $\norm{M_1^\top e_j-\Sigma_0^{-1}e_j}_2=\Op\bigl(\sqrt{s_\Omega\log p/n_0}\bigr)$ and $\sqrt{\log p}\,\norm{M_1^\top e_j-\Sigma_0^{-1}e_j}_1\to_{\mathbb P}0$, the labelled-block analogues of Assumption~\ref{ass:rowsparse} and of condition~(v$'$).  Then the following hold.
\begin{enumerate}
\item[(i)] \emph{Interval-length ratio.}
\begin{equation}\label{eq:deal-vs-dl}
    \frac{\mathrm{CI}_j^{\mathrm{DEAL}}}{\mathrm{CI}_j^{\mathrm{DL}}} \;\longrightarrow_{\mathbb P}\; \frac{1+t_0\kappa_\infty}{1+\kappa_\infty}.
\end{equation}
The limit is strictly less than one for every $t_0<1$ with $\kappa_\infty>0$; with oracle shrinkage $t_0=0$,
\begin{equation}\label{eq:deal-vs-dl-bound}
    \frac{\mathrm{CI}_j^{\mathrm{DEAL}}}{\mathrm{CI}_j^{\mathrm{DL}}} \;\longrightarrow_{\mathbb P}\; \frac{1}{1+\kappa_\infty} \;<\; 1, \qquad \text{with variance ratio } (1+\kappa_\infty)^{-2}.
\end{equation}
\item[(ii)] \emph{Invariance within the no-shift regime.} The limit in \eqref{eq:deal-vs-dl} depends on the data-generating mechanism only through $(\kappa_\infty,t_0)$.  In particular it does not depend on the misspecification residual $\eta$, which enters the asymptotic variances of $\tilde\beta_j$ and of the target-only debiased Lasso only through the common sandwich middle factor $\Gamma_0:=\E_{P_0}[(\eta(X)^2+\sigma^2(X))XX^\top]=\E_{P_0}[\sigma^2(X)XX^\top]+\Gamma_\eta$, the population limit of $\hat\Gamma_0$ in \eqref{eq:Gamma-fold} (with $\Gamma_\eta:=\E_{P_0}[\eta(X)^2XX^\top]$ the misspecification-induced second moment; the noise term reduces to $\sigma^2\Sigma_0$ under conditional homoscedasticity), and in both through the identical quadratic form $(\Sigma_0^{-1}\Gamma_0\Sigma_0^{-1})_{jj}$, which cancels in the ratio.  Consequently \eqref{eq:deal-vs-dl} and \eqref{eq:deal-vs-dl-bound} hold verbatim whenever the two estimators admit the sandwich central limit theorem with this common middle factor and the same limiting precision---in particular under a non-linear external model (Section~\ref{sec:misspec-nonlinear-labeler}).  The invariance is confined to the no-shift regime $\Sigma_{\mathrm u}=\Sigma_0$ assumed throughout this section: the one-sided substitution of Section~\ref{sec:shift-aware} alters the limiting precision under covariate shift, and with it the ratio.
\end{enumerate}
\end{theorem}

\begin{remark}[Interval length at saturation]\label{rem:saturation-scope}
The theorem is stated at a deterministic budget.  When the budget is chosen by the variance-balance rule of Section~\ref{sec:vbrule}, the displays apply at the realised size whenever the rule saturates, $\Pr(\hat N^*=N)\to1$---the regime in which the bias requirement of Theorem~\ref{thm:CLT} is compatible with the selection; at an interior selection the rule holds the standardised bias at a fixed level by design, and no interval-length limit is asserted; the portfolio's interior selections are reported in the final column of Table~\ref{tab:deal-portfolio} in Section~\ref{subsec:realdata-results}.
\end{remark}


Part~(i) is the confidence-interval-scale counterpart of the idealised variance ratio $G(\kappa,t_0)$ of Corollary~\ref{cor:idealized-efficiency}.  The ratio in \eqref{eq:deal-vs-dl} is a first-order limit under a common limiting precision shared by the two estimators, and does not capture the finite-sample sharpening of the pooled precision, which enters only at higher order.

\begin{remark}[Saturation heuristic]\label{rem:saturation-heuristic}
{The following leading-order heuristic indicates when the variance-balance rule saturates at the full unlabelled budget; it is guidance rather than a formal corollary.}
The asymptotic improvement \eqref{eq:deal-vs-dl} reaches the saturation bound \eqref{eq:deal-vs-dl-bound} once the variance-balance rule's $\hat N^*$ saturates at the unlabelled budget $N$ and the bias-aware shrinkage limit is $t_0=0$; 
both occur together in the near-perfect external regime $n_A\ge n_A^{\mathrm{crit}}$ identified below ({Proposition~\ref{prop:condT-regimes}(i) of the supplementary material gives $t_0=0$ in that regime}; saturation $\hat N^*=N$ occurs when the admissible budget is exhausted before the balance point \eqref{eq:Nstar-population} is reached).  Equating the leading-order bias and noise variances at $\hat N^*$ yields, to leading order, the critical external sample size
\begin{equation}\label{eq:nA-crit}
    n_A^{\mathrm{crit}} \;\asymp\; \frac{N^2\,s}{n_0\,Q_j}, \qquad Q_j \;:=\; \bigl(\Sigma_0^{-1}\Gamma_0\Sigma_0^{-1}\bigr)_{jj},
\end{equation}
in which $\Gamma_0$ is the sandwich middle factor of Theorem~\ref{thm:deal-vs-dl}(ii); above this size {the heuristic predicts} $\hat N^*=N$.  {By Theorem~\ref{thm:deal-vs-dl}(ii) the attained CI ratio at any fixed $\hat N^*$ is itself $\eta$-independent; misspecification can influence the achieved efficiency only indirectly, through the dependence of the selected $\hat N^*$ on the noise level entering the balance \eqref{eq:nA-crit}, and we do not claim a definite sign for this secondary effect.}
\end{remark}
\subsection{Non-linear predictors and the complementarity of the two channels}\label{sec:misspec-nonlinear-labeler}

{The general predictor $\hat\mu$ and its linearisation $\hat\beta_{\mathrm{ext}}^{\mathrm{lin}}$ of \eqref{eq:lin-lasso} were introduced with the method in Section~\ref{sec:general-predictors}, where the predictor is split into its design projection $\beta^{\mathrm{proj}}_0[\hat\mu]$ and the orthogonal residual $\nu[\hat\mu]$ of \eqref{eq:lab-projection}.  Here we impose the regularity that linearisation requires, establish validity at the projection parameter, and compare with prediction-powered inference---the residual $\nu[\hat\mu]$ being the channel that separates the two procedures.}

In place of the rate condition of Assumption~\ref{ass:external}, two structural conditions are imposed on the model $\hat\mu${, gathered as clauses (a) and (b) of a single predictor-regularity assumption}.

%
\begin{assumption}[Predictor regularity]\label{ass:lin-proj}\label{ass:nu-reg}
\begin{enumerate}
\item[(a)] \emph{(Linear-projection accuracy of the model.)} There is a sequence $\rho_{\mathrm{lab}}\to 0$ such that
$\bigl\|\delta_{\mathrm{lin}}[\hat\mu]\bigr\|_2 = \Op(\rho_{\mathrm{lab}})$,
where $\delta_{\mathrm{lin}}[\hat\mu]$ is defined by \eqref{eq:lab-projection}.
\item[(b)] \emph{(Regularity of the model's non-linear residual.)} The variables $\bigl\{X_j\,\nu[\hat\mu](X)\bigr\}_{j=1}^p$ are sub-exponential under $P_0$ with a common proxy $\psi_*^2<\infty$, and $\beta^{\mathrm{proj}}_0[\hat\mu]$ is $s_{\mathrm{lin}}$-sparse with $s_{\mathrm{lin}}=O(s)$.
\end{enumerate}
\end{assumption}

{Clause~(a) of Assumption~\ref{ass:lin-proj}} is strictly weaker than $L^2$-consistency $\|\hat\mu-\mu\|_{L^2}=\Op(\rho_{\mathrm{lab}})$, which by Cauchy--Schwarz is sufficient.  It targets only the linear-projection component of the model's prediction error and is invariant to the orthogonal residual $\nu[\hat\mu]$.  {Clause~(b)} ensures the Lasso linearisation \eqref{eq:lin-lasso} concentrates at the standard rate.  In the linear-coefficient specialisation $\hat\mu(X)=X^\top\hat\beta_{\mathrm{ext}}$, $\nu[\hat\mu]\equiv 0$ so the residual {part of clause~(b)} is vacuous; {the two clauses of Assumption~\ref{ass:nu-reg}} then recover Assumption~\ref{ass:external}---the prediction-norm content from {clause~(a)} and the $\ell_1$ content from {clause~(b)'s} $s_{\mathrm{lin}}$-sparsity of $\beta^{\mathrm{proj}}_0[\hat\mu]=\hat\beta_{\mathrm{ext}}$.

{Under the design and predictor-regularity conditions above, the Lasso linearisation \eqref{eq:lin-lasso} recovers $\beta^\star_{\mathrm{proj}}$ in $\ell_2$ at the rate $\rho_{\mathrm{lab}}+\psi_*\sqrt{s_{\mathrm{lin}}\log p/n_{\mathrm{lin}}}$; the formal statement is Lemma~\ref{lem:lin-rate} of the supplementary material (Section~\ref{app:proofs}).}

\begin{corollary}[Sandwich-variance CLT in the non-linear-predictor regime]\label{cor:CLT-nonlin-labeler}
{Suppose the design and debiasing block of Theorem~\ref{thm:CLT} holds with its external-rate condition removed---Assumption~\ref{ass:design}(a) and Assumptions~\ref{ass:rsc}, \ref{ass:external-indep}, \ref{ass:JM}, \ref{ass:condT}---together with the misspecification pair (Assumptions~\ref{ass:nonlinear} and~\ref{ass:misspec-reg}), the precision-row rates (Assumption~\ref{ass:rowsparse}), and the model-regularity condition (both clauses of Assumption~\ref{ass:lin-proj}, in place of Assumption~\ref{ass:external}); and suppose the remaining regime conditions of Theorem~\ref{thm:CLT-misspec} are preserved with the model rate condition strengthened to}
\begin{equation}\label{eq:rate-cond-nonlin}
    \sqrt{n_0+N}\,\Bigl(\rho_{\mathrm{lab}} \;+\; \psi_*\sqrt{s_{\mathrm{lin}}\log p / n_{\mathrm{lin}}}\Bigr) \;=\; o(1).
\end{equation}
Then with $\hat\beta_{\mathrm{ext}}\leftarrow \hat\beta_{\mathrm{ext}}^{\mathrm{lin}}$ in Stages~1--4, the conclusion of Theorem~\ref{thm:CLT-misspec} holds verbatim: for any fixed coordinate $j$,
\[
    \sqrt{n_0+N}\,\bigl(\tilde\beta_j - \beta^\star_{\mathrm{proj},j}\bigr) \;\overset{\mathrm{d}}{\longrightarrow}\; \mathcal N\!\bigl(0,\;\tau_j^{2,\mathrm{sand}}\bigr).
\]
The asymptotic variance $\tau_j^{2,\mathrm{sand}}$, the empirical sandwich middle factor $\hat\Gamma_0$ of \eqref{eq:Gamma-fold}, and the plug-in confidence intervals (Remark~\ref{rem:misspec-linear-recovery}) are unchanged from the linear-predictor case.
\end{corollary}

{With validity established under a non-linear predictor, we turn to interval length. Unlike the linear case, in which the rectifier residual lies in the span of the design, the two channels now separate: the prediction-powered rectifier reads a design-orthogonal residual signal that the variance channel discards.}

{
The correlation this residual signal carries is recorded in the next proposition, {whose development is given in Section~\ref{app:ppi-comparison} and whose proof in Section~\ref{app:proofs}.}
}

\begin{proposition}[PPI++ rectifier under non-linear predictors]
\label{prop:ppipp-nonlinear}
\setlength{\emergencystretch}{3em}%
Under Assumptions~\ref{ass:design}, \ref{ass:rsc}, \ref{ass:external-indep}, \ref{ass:JM}, \ref{ass:nonlinear}, \ref{ass:misspec-reg}{ and~\ref{ass:lin-proj} (both clauses)}, in the no-shift regime $\Sigma_{\mathrm u}=\Sigma_0$, with the sparsity scaling $s\log p=o(\sqrt{n_0})$ and the labelled-block precision conditions of Theorem~\ref{thm:deal-vs-dl} on $M_1$, and with the design-orthogonal residual non-degenerate at coordinate $j$, $e_j^\top\Sigma_0^{-1}\E_{P_0}\bigl[\nu[\hat\mu](X)^2XX^\top\bigr]\Sigma_0^{-1}e_j$ bounded away from zero, the optimally-tuned PPI++ at the squared-loss estimating equation satisfies
\begin{equation}\label{eq:ppipp-nonlin-ratio}
\begin{gathered}
    \frac{\mathrm{CI}_j^{\mathrm{PPI++}}}{\mathrm{CI}_j^{\mathrm{DL}}} \;\longrightarrow_{\mathbb P}\; \sqrt{1 - \frac{\kappa_\infty}{1+\kappa_\infty}\,\rho_j^2[\hat\mu]},\\
    \rho_j^2[\hat\mu] \;:=\; \frac{\bigl(e_j^\top \Sigma_0^{-1}\,\Cov_{P_0}(g_l, g_l^{\hat Y})\,\Sigma_0^{-1} e_j\bigr)^2}{\bigl(e_j^\top \Sigma_0^{-1}\,\Gamma_0\,\Sigma_0^{-1} e_j\bigr)\bigl(e_j^\top \Sigma_0^{-1}\,\Var_{P_0}(g_l^{\hat Y})\,\Sigma_0^{-1} e_j\bigr)},
\end{gathered}
\end{equation}
where $g_l(\theta)=X(X^\top\theta-Y)$ and $g_l^{\hat Y}(\theta)=X(X^\top\theta-\hat\mu(X))$, the squared-loss estimating functions of the labelled response and of the prediction, are evaluated at $\theta=\beta^\star_{\mathrm{proj}}$, and $\rho_j^2[\hat\mu]\in[0,1]$ is the design-orthogonal signal fraction of the predictor at coordinate $j$.
\end{proposition}

{
\begin{theorem}[Complementarity of the residual and variance channels]\label{thm:complementarity}
Fix a coordinate $j$ and a deterministic unlabelled-budget sequence $N=N_{n_0}$ along \eqref{eq:regime} with $\kappa_n=N/n_0\to\kappa_\infty\in(0,\infty)$, and suppose the assumptions of Corollary~\ref{cor:CLT-nonlin-labeler} (including the rate condition \eqref{eq:rate-cond-nonlin}) and of Proposition~\ref{prop:ppipp-nonlinear} hold at this budget, with $\hat t\to_{\mathbb P}t_0\in[0,1]$.  All comparisons are at the shared projection target $\beta^\star_{\mathrm{proj}}$ and the common budget $N$.
\begin{enumerate}
\item[(i)] \emph{The non-linear-predictor regime inherits the conclusions of Theorem~\ref{thm:deal-vs-dl}.}  The linearised procedure $\hat\beta_{\mathrm{ext}}\leftarrow\hat\beta_{\mathrm{ext}}^{\mathrm{lin}}$ obeys the sandwich central limit theorem with the same middle factor $\Gamma_0$ and the same limiting precision $\Sigma_0^{-1}$ as the target-only debiased Lasso; the proof of Theorem~\ref{thm:deal-vs-dl} therefore goes through, and \eqref{eq:deal-vs-dl} and \eqref{eq:deal-vs-dl-bound} hold verbatim.
\item[(ii)] \emph{CI length versus optimally-tuned PPI++.}
\begin{equation}\label{eq:complementarity-ppipp}
    \frac{\mathrm{CI}_j^{\mathrm{DEAL}}}{\mathrm{CI}_j^{\mathrm{PPI++}}} \;\longrightarrow_{\mathbb P}\; \sqrt{\frac{\bigl((1+t_0\kappa_\infty)/(1+\kappa_\infty)\bigr)^2}{1 - \tfrac{\kappa_\infty}{1+\kappa_\infty}\,\rho_j^2[\hat\mu]}},
\end{equation}
with the $\eta$-independent DEAL factor of \eqref{eq:deal-vs-dl}.  With oracle shrinkage $t_0=0$ the right-hand side is at most $(1+\kappa_\infty)^{-1/2}<1$, since $\rho_j^2[\hat\mu]\le1$, and is smaller still whenever $\sigma^2(X)$ is bounded below by a positive constant on a set of positive $P_0$-measure, which forces $\rho_j^2[\hat\mu]<1$; for $t_0>0$ the ratio can lie on either side of one.
\end{enumerate}
\end{theorem}
}

{As with Theorem~\ref{thm:deal-vs-dl}, the statement is at a deterministic budget and applies at the realised size when the variance-balance rule saturates; the common-budget accounting charges both procedures the same $N$ stacking rows; DEAL in addition consumes a linearisation block of $n_{\mathrm{lin}}$ held-out unlabelled rows (Section~\ref{sec:general-predictors}), a single block shared across all coordinates.  The saturation heuristic of Remark~\ref{rem:saturation-heuristic} indicates when the rule saturates.}

{Clause~(ii) carries one qualification under a non-linear predictor: the comparison with optimally-tuned PPI++ is settled in DEAL's favour at oracle shrinkage ($t_0=0$, the saturated regime), but away from it optimally-tuned PPI++ can be shorter.  The reason is a difference in what each channel can reach: the variance channel sees the model only through its design projection $\beta^{\mathrm{proj}}_0[\hat\mu]$ and discards the orthogonal residual $\nu[\hat\mu]$ of \eqref{eq:lab-projection}, whereas the rectifier extracts the part of that residual correlated with the labelled noise, of squared correlation $\rho_j^2[\hat\mu]$ (Proposition~\ref{prop:ppipp-nonlinear}).  This is the natural objection that the procedure can exploit only information already lying in the span of the design; the answer is that the residual signal is recoverable, without disturbing DEAL's guarantees, by a control variate on the labelled block---the construction developed in the next subsection.  Two further developments are deferred to the supplementary material: models that act on auxiliary covariates outside the design (Section~\ref{sec:aux-cov}), and the joint regime in which covariate shift and misspecification act together (Section~\ref{sec:misspec-joint}), the latter with a numerical illustration.}

\subsection{The control-variate hybrid}\label{sec:hybrid}

{
{Here the complementarity of the two approaches is realised in a single construction.} The control-variate channel and the variance channel act on disjoint components of the coordinate error: a control variate on the labelled residual recovers the part of the model orthogonal to the design, the signal the prediction-powered rectifier is built to extract and that DEAL alone discards, while the variance channel routes the unlabelled covariates through $\mathrm{span}(X)$ to sharpen the stacked fit and the precision estimate, a channel the rectifier cannot reach. The two act in parallel and a single estimator runs both. At a common unlabelled budget their interval-length reductions multiply, so the hybrid sharpens DEAL wherever a non-linear predictor carries design-orthogonal signal. The limits are clean at both ends. Switch off the residual channel, as happens automatically when the predictor is linear, and the hybrid returns the DEAL estimator of the preceding sections unchanged (Proposition~\ref{prop:hybrid-var}); switch off the variance channel and it reduces to the pure control variate on the labelled residual, the rectifier mechanism of the prediction-powered family, matching the efficiency of optimally-tuned PPI++ in the infinite-unlabelled limit. With the variance-balance rule reselecting the budget on the control-variated residual, the deployed gains are sub-additive (the reselected $\hat N^*$ is smaller), the residual channel spending part of the variance channel's budget. The scope is deliberately narrow: the construction is developed in theory and illustrated in simulation, while the real-data portfolio routes each model through its linearised projection alone: the control variate is never formed (its weight $\hat\omega_j=0$ by construction) and the hybrid coincides there with DEAL, not run as a separate procedure.
}

{For inference on coordinate $j$, run the procedure on the control-variated response}
\begin{equation}\label{eq:hybrid-response}
   y_0^{\mathrm{cv}} := y_0 - \hat\omega_j\,\hat\nu,
   \qquad
   \hat\nu := \hat\mu - X^\top\hat\beta^{\mathrm{lin}}_{\mathrm{ext}},
\end{equation}
with $\hat\beta^{\mathrm{lin}}_{\mathrm{ext}}$ the linearised external coefficient \eqref{eq:lin-lasso} and $\hat\omega_j$ the tuning-sample weight estimator \eqref{eq:omega-hat-j} of Section~\ref{app:hybrid}, which is measurable with respect to the tuning sample, the covariates, and the external fit---hence, as $\hat t$ and $A_n$ are in Theorem~\ref{thm:CLT}, independent of the inference-stage noise $\{\varepsilon_{0,i}\}$ conditional on the covariates and tuning sample (Assumption~\ref{ass:external-indep})---and consistent for the coordinate-optimal weight $\omega_j^{\mathrm{cv}}$ of the next proposition; the pseudo-labels, stacking, debiasing, and $N$-selection are unchanged. The control variate is formed on the labelled block and estimated on the tuning block, on both of which the model must therefore be evaluable (Assumption~\ref{ass:aux-eval} of the supplementary material).

\begin{proposition}[Hybrid validity and variance]\label{prop:hybrid-var}

Suppose the assumptions of Corollary~\ref{cor:CLT-nonlin-labeler} and Proposition~\ref{prop:ppipp-nonlinear} hold, and, strengthening clause~(b) of Assumption~\ref{ass:lin-proj} for this construction alone, that $\nu[\hat\mu](X)$ is in addition sub-Gaussian under $P_0$. Write $e:=\eta+\varepsilon$, $\Gamma_{e\nu}:=\E_{P_0}[e\,\nu\,XX^\top]$, $\Gamma_\nu:=\E_{P_0}[\nu^2 XX^\top]$, and recall the design-orthogonal signal fraction $\rho_j^2[\hat\mu]$ of \eqref{eq:ppipp-nonlin-ratio}. Let
\[
   \omega_j^{\mathrm{cv}} := \frac{e_j^\top\Sigma_0^{-1}\Gamma_{e\nu}\Sigma_0^{-1}e_j}{e_j^\top\Sigma_0^{-1}\Gamma_\nu\Sigma_0^{-1}e_j}
\]
be the coordinate-$j$-optimal control-variate weight, with the convention $\omega_j^{\mathrm{cv}}:=0$ when the denominator vanishes, and let $\hat\omega_j$ in \eqref{eq:hybrid-response} be as specified there.

Run Stages~1--4 on the control-variated response \eqref{eq:hybrid-response}, with the shrinkage level $\hat t$ selected on this same residual. Then the control-variated estimator $\tilde\beta_j^{\mathrm{cv}}$ obeys the central limit theorem of Corollary~\ref{cor:CLT-nonlin-labeler} at the projection target $\beta^\star_{\mathrm{proj}}$, with the sandwich middle $\Gamma_0$ replaced by $\Gamma_0^{\mathrm{cv}}=\E_{P_0}[(e-\omega_j^{\mathrm{cv}}\nu)^2 XX^\top]$, and a limiting variance related to the sandwich variance DEAL would itself attain \emph{at the same shrinkage limit} by
\begin{equation}\label{eq:hybrid-var}
   \tau_j^{2,\mathrm{cv}} = \bigl(1-\rho_j^2[\hat\mu]\bigr)\,\tau_j^{2,\mathrm{sand}} \;\le\; \tau_j^{2,\mathrm{sand}},
\end{equation}
with equality if and only if $\rho_j^2[\hat\mu]=0$. When the model is linear in $X$---so $\nu[\hat\mu]\equiv0$ and $\rho_j^2[\hat\mu]=0$---the weight $\omega_j^{\mathrm{cv}}=0$ and the procedure reduces to DEAL.

\end{proposition}

Validity is unchanged, and \eqref{eq:hybrid-var} shows the variance never exceeds the DEAL floor \emph{at the shrinkage level the control-variated procedure itself selects}; Corollary~\ref{cor:hybrid-dom} below turns this into dominance over optimally-tuned PPI++ for any shrinkage limit, and, under an explicit shared-shrinkage hypothesis, an exact ratio against separately-deployed DEAL.

\begin{corollary}[Dominance over optimally-tuned PPI++, uniform in $N$]\label{cor:hybrid-dom}
Under the assumptions of Proposition~\ref{prop:hybrid-var} and Theorem~\ref{thm:complementarity}, write $\kappa=\hat N^*/n_0$ and let $\hat t^{\mathrm{cv}}\in[0,1]$ be the control-variated procedure's shrinkage level.

\emph{(i)} Whatever the selected budget $\hat N^*$,
\[
   \frac{\mathrm{CI}_j^{\mathrm{cv}}}{\mathrm{CI}_j^{\mathrm{DL}}}
   \;\longrightarrow_{\mathbb P}\; \sqrt{1-\rho_j^2[\hat\mu]}\;\frac{n_0+\hat t^{\mathrm{cv}}\hat N^*}{n_0+\hat N^*}
   \;\le\; \sqrt{1-\rho_j^2[\hat\mu]}
   \;\le\; \sqrt{1-\tfrac{\kappa}{1+\kappa}\,\rho_j^2[\hat\mu]},
\]
the first inequality by $\hat t^{\mathrm{cv}}\in[0,1]$, the last term being the limit \eqref{eq:ppipp-nonlin-ratio} of $\mathrm{CI}_j^{\mathrm{PPI++}}/\mathrm{CI}_j^{\mathrm{DL}}$. The control-variated estimator therefore dominates optimally-tuned PPI++ \emph{uniformly in $N$}, for any shrinkage limit, removing the sub-saturation reversal of clause~(ii) of Theorem~\ref{thm:complementarity}.

\emph{(ii)} If in addition the control-variated procedure's shrinkage level converges to separately-deployed DEAL's,
\begin{equation}\label{eq:hybrid-shared-t0}
    \hat t^{\mathrm{cv}} \overset{\mathbb P}{\longrightarrow} t_0
\end{equation}
(Remark~\ref{rem:hybrid-shared-t0} below), then at a common unlabelled budget (both at the same $\hat N^*$),
\[
   \frac{\mathrm{CI}_j^{\mathrm{cv}}}{\mathrm{CI}_j^{\mathrm{DEAL}}} \;\longrightarrow_{\mathbb P}\; \sqrt{1-\rho_j^2[\hat\mu]} \;<\; 1 \quad (\rho_j^2[\hat\mu]>0).
\]
\end{corollary}

\begin{remark}[Status of the shared-shrinkage hypothesis]\label{rem:hybrid-shared-t0}
The construction re-selects $(\hat t,\hat N)$ on the control-variated residual (Section~\ref{app:hybrid} of the supplementary material), so its shrinkage limit $\hat t^{\mathrm{cv}}$ need not coincide with separately-tuned DEAL's $t_0$; part~(ii) of Corollary~\ref{cor:hybrid-dom} therefore imposes \eqref{eq:hybrid-shared-t0} as an explicit hypothesis, whose proof is left to future work. Part~(i) and Proposition~\ref{prop:hybrid-var} need no such hypothesis: the dominance over optimally-tuned PPI++ holds for any $\hat t^{\mathrm{cv}}\in[0,1]$, and the variance identity \eqref{eq:hybrid-var} compares the control-variated procedure only to DEAL at the \emph{same} shrinkage limit.
\end{remark}

{The residual channel is genuine and recoverable: a non-linear predictor can carry signal orthogonal to the design, as the component of a queried language model's output that is driven by the prompt rather than by $X$ does, and it is the control variate, not the augmentation, that recovers the part of such a residual correlated with the labelled noise, a gain exhibited on the synthetic illustration of Section~\ref{sec:experiments-hybrid-main}. Proofs, a scalar-weight variant, and an adaptivity statement are given in Section~\ref{app:hybrid}.}

Table~\ref{tab:regime-map} consolidates the regimes in which DEAL attains a confidence-interval-length improvement over DL, recording for each the estimator and precision variant, the operative result, and the limiting CI-length ratio.  The covariate-shift and joint-perturbation regimes are governed by coverage validity rather than by an efficiency ratio, and are summarised separately in Table~\ref{tab:robustness-map}.

\begin{table}[!ht]
\centering
\caption{\textit{Limiting confidence-interval-length ratio of DEAL relative to target-only debiased Lasso (DL), across the regimes in which DEAL attains an efficiency improvement.}  Here $\kappa:=N/n_0$ and $t_0$ is the limit of the bias-aware shrinkage $\hat t$; every ratio is evaluated at the projection parameter $\beta^\star_{\mathrm{proj}}$ and is misspecification-invariant (Theorem~\ref{thm:deal-vs-dl}(ii)); $X_\Omega=(X_0;\tilde X)$ denotes the pooled covariate rows from which the nodewise precision $\hat\Omega$ is estimated.  Results marked $^{\dagger}$ are stated in the supplementary material.}
\label{tab:regime-map}
\small
\begin{tabular}{@{}p{4.4cm}p{3.4cm}p{2.9cm}p{2.6cm}@{}}
\toprule
Regime & Estimator / precision & Main result & Limiting CI ratio \\
\midrule
\multicolumn{4}{@{}l}{\emph{No shift, $\Sigma_{\mathrm u}=\Sigma_0$}}\\
\addlinespace[2pt]
Well-specified, oracle precision $M_2{=}M_1{=}\Sigma^{-1}$ & pooled JM debiasing & Thm.~\ref{thm:CLT}, Cor.~\ref{cor:idealized-efficiency} & $\tfrac{1+t_0\kappa}{1+\kappa}$ \\
\addlinespace[2pt]
Gaussian design, known precision $\Omega=\Sigma^{-1}$ & $M_2{=}M_1{=}\Omega$ & Cor.~\ref{cor:gaussian-known}$^{\dagger}$ & $\tfrac{1+t_0\kappa}{1+\kappa}$ \\
\addlinespace[2pt]
Gaussian design, nodewise-estimated precision & $\hat\Omega$ from $X_\Omega$ & Cor.~\ref{cor:gaussian-nodewise}$^{\dagger}$ & $\tfrac{1+t_0\kappa}{1+\kappa}$ \\
\midrule
\multicolumn{4}{@{}l}{\emph{Misspecified target, projection parameter $\beta^\star_{\mathrm{proj}}$}}\\
\addlinespace[2pt]
Linear external model & pooled JM $+$ sandwich $\hat\Gamma_0$ & Thm.~\ref{thm:CLT-misspec},~\ref{thm:deal-vs-dl} & $\tfrac{1+t_0\kappa}{1+\kappa}$ ($\to\tfrac{n_0}{n_0+N}$ as $t_0\to0$) \\
\addlinespace[2pt]
Non-linear external model & linearisation $\to$ Stages 1--4 & Cor.~\ref{cor:CLT-nonlin-labeler}; Thm.~\ref{thm:deal-vs-dl}(ii) & $\tfrac{1+t_0\kappa}{1+\kappa}$ ($\eta$-indep.) \\
\bottomrule
\end{tabular}
\end{table}

Under a linear or linearised predictor the high-dimensional PPI and PPI++ estimators are asymptotically equivalent to DL (Proposition~\ref{prop:ppipp-equivalence}), so each ratio in Table~\ref{tab:regime-map} holds verbatim with PPI or PPI++ in place of DL, whereas under a non-linear predictor DEAL strictly dominates PPI++ at full saturation ($\hat N^*=N$, $t_0\to0$; {Theorem~\ref{thm:complementarity}(ii)}).

\begin{table}[!ht]
\centering
\caption{\textit{Validity of DEAL under covariate shift and joint perturbation.}  For each perturbation the table records the behaviour of the uncorrected procedure, the corrective device, and the restored guarantee; these regimes are governed by coverage validity rather than by an efficiency ratio.  Here $\delta_\star[\eta]:=\Sigma_\star^{-1}\E_{P_\star}[X\,\eta(X)]$ is the cross-term of shift and misspecification under the marginal $P_\star$ on which the model is linearised, $\delta_{\mathrm u}[\eta]$ its value at the unlabelled marginal, and $\delta_0[\eta]=0$ at the target (Section~\ref{sec:misspec-joint} of the supplementary material).  Results marked $^{\dagger}$ are stated in the supplementary material.}
\label{tab:robustness-map}
\small
\begin{tabular}{@{}p{2.7cm}p{3.5cm}p{3.5cm}p{3.4cm}@{}}
\toprule
Perturbation & Uncorrected behaviour & Corrective device & Restored guarantee \\
\midrule
Covariate shift, $\Sigma_{\mathrm u}\succ\Sigma_0$ & imputation bias inflated by $\rho_j(M_2,\Sigma_{\mathrm u})>1$ (Prop.~\ref{prop:shift-amp}); coverage held by the variance-balance rule (Section~\ref{sec:vbrule}) at a reduced $\hat N^*$ & one-sided substitution $\hat M_2^{\mathrm{adapt}}$~\eqref{eq:M2-adapt} with detector~\eqref{eq:shift-detector} & bias reduced to the no-shift order (Prop.~\ref{prop:shift-bias-reduction}), recovering the efficiency the cap would forgo (Cor.~\ref{cor:shift-restore}) \\
\addlinespace[4pt]
Joint shift $+$ misspecification & cross-term $\delta_{\mathrm u}[\eta]$ breaks the Gaussian limit & target-marginal linearisation on $P_0$, with $\hat M_2^{\mathrm{adapt}}$ at Stage~4 & Gaussian limit preserved iff $\delta_\star[\eta]=0$ (Prop.~\ref{prop:joint-lin}$^{\dagger}$) \\
\bottomrule
\end{tabular}
\end{table}

\subsection{Scope of the guarantees}\label{sec:scope}
Under sub-Gaussian design the coordinate-wise central limit theorem of Section~\ref{sec:stacking} gives valid confidence intervals at the target coefficients, and at the best linear projection when the target is misspecified or the predictor non-linear (Section~\ref{sec:misspec}).  Neither statement requires the external model to be accurate: its error enters only through the bias requirement of Theorem~\ref{thm:CLT}, which the bias-aware shrinkage is constructed to meet.  Against a linear predictor the prediction-powered family collapses to the target-only debiased Lasso (Section~\ref{sec:misspec-comparison}), and the control-variate hybrid dominates optimally-tuned PPI++ uniformly in the budget (Corollary~\ref{cor:hybrid-dom}(i)).

The refinements carry conditions, each stated with its result.  The interval-length dominance over the target-only debiased Lasso is a first-order limit realised at a saturated budget; at an interior selection the guarantee is coverage rather than width (Section~\ref{sec:choose-N}).  Under covariate shift coverage is held at any budget, while the efficiency the variance-balance rule's bias ceiling forgoes is recovered only when the unlabelled pool dominates (Corollary~\ref{cor:shift-restore}).  The exact hybrid-to-DEAL ratio and the tuning-scale optimality of the shrinkage hold under the further conditions of Corollary~\ref{cor:hybrid-dom}(ii) and Section~\ref{sec:proxy-shrinkage}.

\section{Numerical experiments and real-data portfolio}\label{sec:experiments}

We complement the theoretical analysis with a Monte Carlo study of four experiments, followed by the real-data portfolio.  Section~\ref{sec:experiments-power} examines power adaptivity across external sample sizes $n_A$ from $100$ to $2400$ and validates the variance-balance rule of Section~\ref{sec:vbrule}.  Section~\ref{sec:experiments-misspec} examines validity at the projection target $\beta^\star_{\mathrm{proj}}$ under non-linear truth, in the linear-predictor regime of Section~\ref{sec:misspec-validity} and in the non-linear-predictor regime of Section~\ref{sec:misspec-nonlinear-labeler}.  Section~\ref{sec:experiments-shift} examines robustness to covariate shift in the unlabelled covariate distribution and the shift-aware variant of Section~\ref{sec:shift-aware}.  {Section~\ref{sec:experiments-hybrid-main} is the complementarity check of the residual and variance channels.}

\subsection{Simulation design}\label{sec:experiments-design}

The data-generating process is common to all experiments.  DEAL is run exactly as specified in Algorithm~\ref{alg:deal}, whose consolidated statement and Gaussian-design implementation are collected in Section~\ref{sec:implementation} of the supplementary material.  The target distribution is $X_0 \sim N(0,\Sigma_0)$ on $\R^p$ with $p=120$ and $\Sigma_0$ an AR(1) covariance with parameter $\rho_0 = 0.4$.  The regression coefficient $\beta^\star$ is $s$-sparse with $s=6$ and non-zero entries equal to $0.8$ on the index set $\{1,\ldots,6\}$.  The noise is Gaussian, $\varepsilon \sim N(0,1)$.  The labelled target sample has $800$ rows, split into an inference sample of size $n_0=400$ and a tuning subsample of size $n_{\mathrm{tun}}=400$.  Inferential targets are the signal coordinates $J_{\mathrm{signal}} = \{1,2,3\}$ (evaluated at $\beta^\star_j = 0.8$) and the null coordinates $J_{\mathrm{null}} = \{7,8,9\}$ (evaluated at $\beta^\star_j = 0$).

The external estimator $\hat\beta_{\mathrm{ext}}$ is a target-population Lasso fitted on an external sample of size $n_A$, which varies across cells to control external-estimator quality; the Lasso penalty is selected on the external sample by ten-fold cross-validation.  Full protocol details are given in Section~\ref{app:simdetails}.

We consider five estimators.  (i) \textit{DEAL} is the proposed procedure with $N$ chosen by the variance-balance rule \eqref{eq:Nstar-empirical} and $M_2$ estimated by nodewise Lasso on an auxiliary pool of $3000$ unlabelled rows (Remark~\ref{rem:jm-scope}; Sections~\ref{sec:refine} and~\ref{app:simdetails}).  (ii) \textit{DEAL-shift-aware} replaces $M_2$ by the one-sided substitution $\hat M_2^{\mathrm{adapt}}$ of \eqref{eq:M2-adapt} (used in Section~\ref{sec:experiments-shift} only).  (iii) \textit{DL}, target-only debiased Lasso \citep{javanmard2014confidence,vandegeer2014asymptotic,zhang2014confidence}, applied to $(X_0,y_0)$ with no unlabelled augmentation, serves as the high-dimensional benchmark.  (iv)~\textit{PPI} \citep{angelopoulos2023prediction} and (v)~\textit{PPI++} \citep{angelopoulos2023ppipp}, the prediction-powered baselines, are adapted to the high-dimensional regime by building them on the debiased Lasso of (iii)---the labelled-sample estimator the rectifier corrects, which we call the anchor---and using the external Lasso estimator $\hat\beta_{\mathrm{ext}}$ to score the unlabelled rows.  In every experiment PPI and PPI++ receive the same external input as DEAL, the coefficient $\hat\beta_{\mathrm{ext}}$ or its linearisation $\hat\beta^{\mathrm{lin}}_{\mathrm{ext}}$ where the external model is non-linear; in the applications every procedure receives the raw predictions $\hat\mu$ (Section~\ref{subsec:realdata-results}).  The same nodewise-Lasso precision estimation is applied uniformly across DEAL, DEAL-shift-aware, and DL.

Three criteria are reported.  \textit{Empirical coverage} is the fraction of replications in which the $95\%$ confidence interval contains $\beta^\star_j$.  {The coverage panels throughout report this on the signal coordinates $J_{\mathrm{signal}}$, the binding case for inference; coverage on the null coordinates is uniformly conservative---at or above the nominal level in every experiment---and is documented in full in Section~\ref{app:null-coverage}.}  The DL benchmark itself covers at $0.88$--$0.93$ at $n_0 = 400$ across the experiments.  \textit{CI-length ratio} is the ratio of the median signal-coordinate CI length of the procedure to that of the DL benchmark on the same replication.  \textit{Selected unlabelled sample size $\hat N^*$} is the value chosen by the variance-balance rule, averaged across replications. Unless a figure caption states otherwise, Monte Carlo averages are over $R=100$ replications under a fixed master seed, and shaded regions in the simulation figures are $95\%$ Monte-Carlo error bands across replications.

Although the experiments below fix $p=120$ and $s=6$, we verified separately that DEAL's validity and efficiency are stable across the sparsity level and the ambient dimension: for $s\in\{6,12,24\}$ and $p\in\{120,300,600\}$, with $p=600$ exceeding $n_0$, DEAL signal coverage stays within Monte Carlo error of the nominal $0.95$ (uniformly above the DL benchmark) and the DEAL-to-DL CI ratio stays in $0.64$--$0.71$.


\subsection{Power adaptivity and the variance-balance rule}\label{sec:experiments-power}

We vary the external-estimator quality through ten cells of increasing external sample size $n_A$, holding $n_0$ fixed at $400$.  Larger $n_A$ yields a more accurate external estimator and is expected to reduce the residual imputation bias, freeing the variance-balance rule to admit a larger $\hat N^*$ and to deliver a tighter confidence interval.

\begin{figure}[t]
\centering
\includegraphics[width=\textwidth]{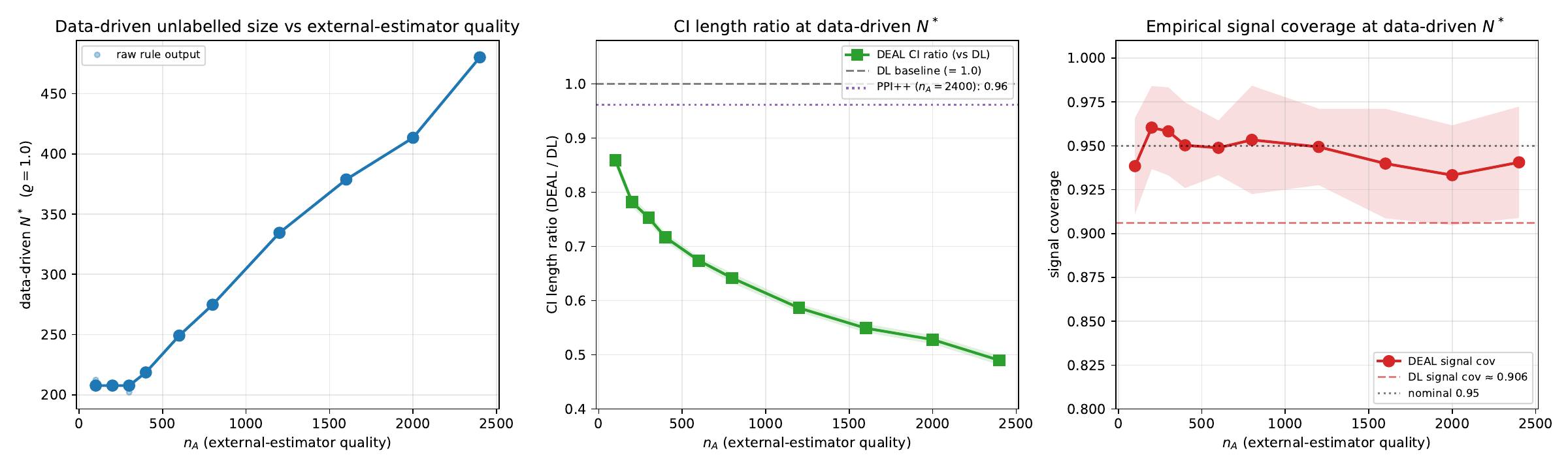}
\caption{\textit{Power adaptivity of DEAL across external-estimator qualities.}  Left: the variance-balance choice $\hat N^*$ versus the external sample size $n_A$; points are the per-cell averages of $\hat N^*$, the line their isotonic fit in $n_A$.  Centre: the median CI-length ratio of DEAL to target-only debiased Lasso (DL) on signal coordinates; horizontal reference lines mark the DL benchmark and the PPI++ benchmark at the highest external-estimator quality ($n_A = 2400$).  Right: empirical signal coverage of DEAL and DL.  Each point is averaged over $R=100$ replications; shaded regions are 95\% Monte-Carlo error bands across replications.  PPI is omitted; it performs similarly to or slightly worse than PPI++.}
\label{fig:study1}
\end{figure}

The empirical findings are summarised in Figure~\ref{fig:study1}.  Across the model-quality range the variance-balance choice $\hat N^*$ rises with model quality to $480$ at $n_A = 2400$ and the DEAL-to-DL CI-length ratio falls from $0.86$ to $0.49$, while signal coverage stays close to the nominal $0.95$.  This ordering matches the adaptivity prediction of {Theorem~\ref{thm:CLT} and Corollary~\ref{cor:idealized-efficiency}}: the leading variance interpolates between the target-only and prediction-assisted benchmarks as the external estimator becomes more accurate.  {The Monte-Carlo band on the CI-length ratio is markedly tighter than the band on coverage.  This is expected: each replication's interval length is itself an average over the labelled sample, a smooth and well-concentrated functional of the estimated variance, and so varies little from one replication to the next, whereas coverage is a binary event for each interval and is intrinsically noisier.}

A direct comparison against the prediction-powered baselines clarifies where the gain comes from.  At the highest external-estimator quality ($n_A = 2400$), PPI++ yields a CI-length ratio of $0.96$, whereas DEAL yields $0.49$.  Routing the augmentation through the variance thus roughly halves the confidence-interval length relative to PPI++, while the bias-aware shrinkage keeps coverage.  This margin reflects the fact that PPI and PPI++ correct only the first-order pseudo-label bias and gain little when the external predictor is already close to the truth, whereas DEAL exploits the unlabelled augmentation in the variance of the debiased estimator rather than only in its mean.


The variance-balance rule of Section~\ref{sec:vbrule}, which drives the selection of $\hat N^*$ above, prescribes it from the data without recourse to a held-out coverage sweep and can be validated directly.\label{sec:experiments-vbrule}  We compare it against an oracle reference $N^{\mathrm{tuned}}$ obtained by an exhaustive grid search that selects the largest $N$ at which empirical coverage on $J_{\mathrm{signal}}$ remains within four percentage points of the DL benchmark.  We report the rule at its default criterion ratio $\varrho=1.0$, the population variance-balance point.

\begin{table}[t]
\centering
\caption{\textit{Empirical validation of the variance-balance rule.}  $N^{\mathrm{tuned}}$ is the coverage-oracle reference, an $N$ chosen with knowledge of the true coefficients: the largest $N$ keeping $J_{\mathrm{signal}}$ coverage within four percentage points of DL.  $\hat N^*_{\varrho=1.0}$ is the rule's recommendation at the default criterion ratio $\varrho=1.0$.  $\hat\gamma_N^{\mathrm{G}}$ is the value of $\gamma_N$ in $N_{\max}^{\mathrm{G}}$ that would render the admissibility cap equal to $N^{\mathrm{tuned}}$.}
\label{tab:vbrule-evidence}
\begin{tabular}{rrrr}
\toprule
$n_A$ & $N^{\mathrm{tuned}}$ & $\hat N^*_{\varrho=1.0}$ & $\hat\gamma_N^{\mathrm{G}}$ \\
\midrule
$100$  & $222$ & $212$ & $7.3$ \\
$400$  & $264$ & $219$ & $6.3$ \\
$800$  & $312$ & $275$ & $5.0$ \\
$1600$ & $450$ & $379$ & $4.1$ \\
$2400$ & $524$ & $480$ & $3.5$ \\
\bottomrule
\end{tabular}
\end{table}

Table~\ref{tab:vbrule-evidence} reports the comparison.  The rule at $\varrho = 1.0$ recovers $N^{\mathrm{tuned}}$ within a factor of $0.83$ to $0.96$ across the range, an undershoot expected from the conservative bound form of the rule's bias term $\hat\tau_{B,j}$ of Section~\ref{sec:vbrule}.  The implied $\hat\gamma_N^{\mathrm{G}}$ that would render the admissibility cap $N_{\max}^{\mathrm{G}}$ equal to $N^{\mathrm{tuned}}$ varies from $7.3$ at $n_A = 100$ to $3.5$ at $n_A = 2400$, a factor of $2.1$ across the model-quality range.  No single fixed value of $\gamma_N$ in either admissibility cap can be simultaneously sharp across this range, illustrating the complementarity of the two devices: the admissibility cap delivers a sufficient condition uniform in the regime, while the variance-balance rule delivers a sharp, regime-adaptive interior point.

\subsection{Inference under model misspecification}\label{sec:experiments-misspec}

We complement the analysis of Section~\ref{sec:misspec} with empirical studies of inference at the projection target $\beta^\star_{\mathrm{proj}}$ when the conditional mean $\mu(X)=\E_{P_0}[Y\mid X]$ does not lie in the linear span of $X$.  Throughout, the data-generating process is
\[
   Y \;=\; X^\top\beta^\star \;+\; \eta(X) \;+\; \varepsilon,\qquad X\sim N(0,\Sigma_0),\qquad \varepsilon\sim N(0,\sigma^2),
\]
with $\Sigma_0$, $\beta^\star$, $\sigma^2$, $n_0$, $n_{\mathrm{tun}}$, $J_{\mathrm{signal}}$, $J_{\mathrm{null}}$ as in Section~\ref{sec:experiments-design}.  In every $\eta$ specification used below, $\E_{P_0}[X\,\eta(X)]=0$ holds either by construction (Hermite forms via Stein's lemma) or up to Monte Carlo error (the frozen non-linear forms via empirical orthogonalisation), so $\beta^\star_{\mathrm{proj}}=\beta^\star$ exactly across the misspecification grid.  Plug-in confidence intervals for all four estimators (PPI and PPI++ being debiased-Lasso-anchored on the same labelled sample) use the sandwich variance estimator of Remark~\ref{rem:misspec-linear-recovery}.

\paragraph{Linear external model under non-linear truth.}

\begin{figure}[t]
\centering
\includegraphics[width=\textwidth]{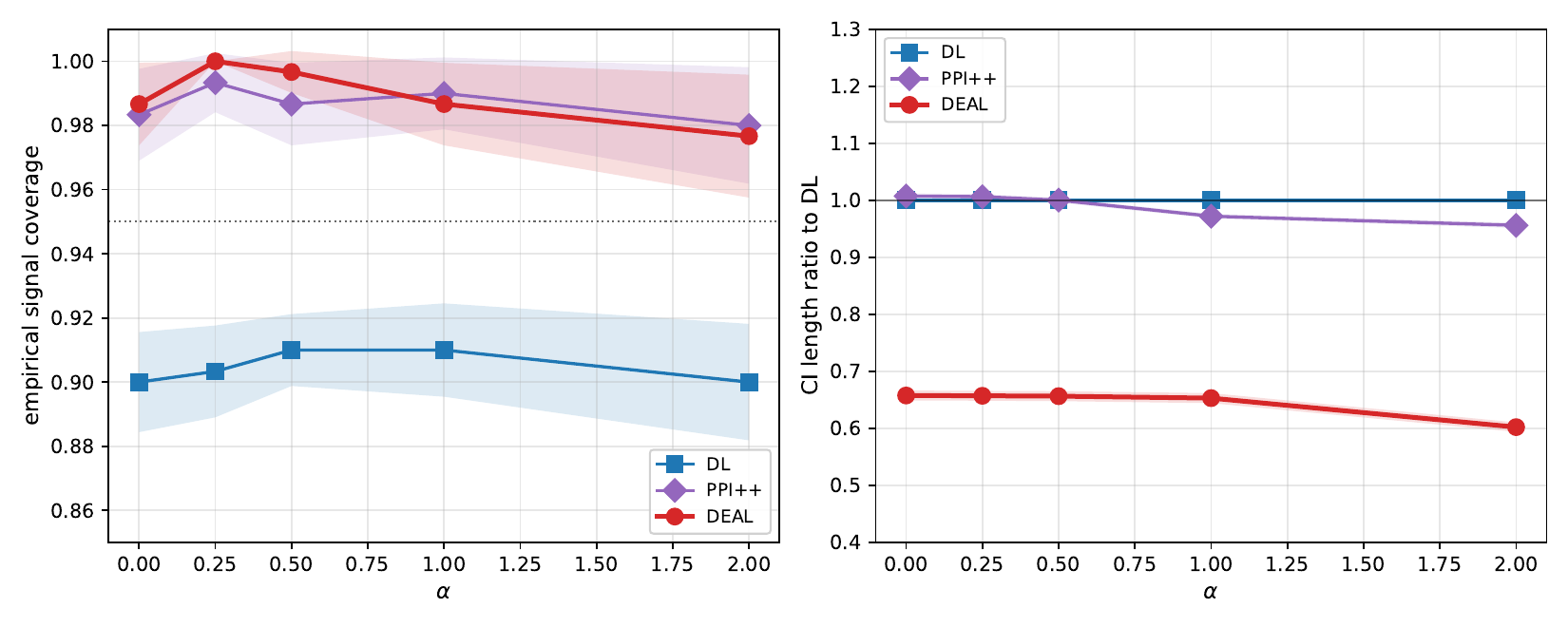}
\caption{\textit{DEAL inference under the oracle linear predictor under non-linear (Hermite) truth.}  Left: empirical signal coverage versus the misspecification strength $\alpha$.  Right: CI-length ratio of each estimator to DL on $J_{\mathrm{signal}}$.  External-estimator coefficient $\hat\beta_{\mathrm{ext}}=\beta^\star$ exactly, $n_0=400$, AR(1) target with $\rho_0=0.4$, $p=120$, $s=6$, $R=100$ replications.  $N$ is chosen per replication by the variance-balance rule, with sandwich variances.  {(PPI omitted as in Figure~\ref{fig:study1}.)}}
\label{fig:misspec-linear-oracle}
\end{figure}

We specialise to the linear predictor $\hat\mu(X)=X^\top\hat\beta_{\mathrm{ext}}$ in the strongest form: $\hat\beta_{\mathrm{ext}} = \beta^\star$ exactly, the oracle linear predictor.  This trivially satisfies Assumption~\ref{ass:external} with $\|\hat\beta_{\mathrm{ext}}-\beta^\star_{\mathrm{proj}}\|_2 = 0$ (since $\beta^\star_{\mathrm{proj}} = \beta^\star$ under the orthogonalised $\eta$), and isolates the misspecification effect from model estimation error so that any departure of the empirical CI ratio from its asymptotic value is attributable to the misspecification residual $\eta$ alone.  The misspecification form is
\[
    \eta(X) \;=\; \alpha\,H_3(X_1), \qquad H_3(z) = z^3 - 3z,
\]
the third probabilist's Hermite polynomial in the first signal coordinate; $\E[X\,\eta(X)] = 0$ follows from Stein's lemma applied to the third moment of $N(0,1)$, and $\Var_{P_0}(\eta) = 6\alpha^2$.  Figure~\ref{fig:misspec-linear-oracle} reports empirical coverage on $J_{\mathrm{signal}}$ and CI-length ratios against DL.

Empirical signal coverage of DEAL and PPI++ holds at or above the nominal $0.95$ across the grid, validating Theorem~\ref{thm:CLT-misspec}.  
The DEAL-to-DL CI ratio is essentially flat in $\alpha$, near $0.66$ and tightening to $0.60$ at $\alpha=2$, in agreement with the $\eta$-independent reduction of {Theorem~\ref{thm:deal-vs-dl}(ii)}; the small residual tightening with $\alpha$ reflects the indirect effect of $\alpha$ on the selected $\hat N^*$ (Remark~\ref{rem:saturation-heuristic}), the ratio at fixed $\hat N^*$ being $\eta$-independent.
  PPI++ remains at DL parity to within a few percent ($0.96$--$1.01$), consistent with Proposition~\ref{prop:ppipp-equivalence}: under a linear predictor the rectifier residual at $\beta^\star_{\mathrm{proj}}$ carries the misspecification residual $\eta$ and cannot tighten below DL, so DEAL alone shortens the interval under non-linear truth.

\paragraph{Linearised oracle predictor under non-linear truth.}

We exercise the non-linear-predictor regime of Section~\ref{sec:misspec-nonlinear-labeler}.  The model $\hat\mu(X) = X^\top\beta^\star + \eta(X)$ is the noiseless conditional mean, paired with the empirical Lasso linearisation \eqref{eq:lin-lasso} executed on an auxiliary covariate-only pool $X_{\mathrm u}^{(\mathrm{lin})}$ of size $n_{\mathrm{lin}} = 2\times 10^4$:
\[
   \hat\beta_{\mathrm{ext}}^{\mathrm{lin}} \;=\; \argmin_{\beta}\;\frac{1}{2 n_{\mathrm{lin}}}\bigl\|\hat\mu(X_{\mathrm u}^{(\mathrm{lin})}) - X_{\mathrm u}^{(\mathrm{lin})}\beta\bigr\|_2^2 + \lambda^{\mathrm{lin}}\|\beta\|_1,
\]
which is then used as $\hat\beta_{\mathrm{ext}}$ in Stages~1--4 of DEAL.  This isolates the linearisation step (Lemma~\ref{lem:lin-rate} {of the supplementary material}) from model-side training error: the model-projection accuracy $\delta_{\mathrm{lin}}[\hat\mu]=0$ by construction, so Assumption~\ref{ass:lin-proj} is met with $\rho_{\mathrm{lab}}=0$ and the model-rate condition \eqref{eq:rate-cond-nonlin} reduces to the Lasso-linearisation rate alone.

To probe robustness across the form of the misspecification, we instantiate $\eta$ in three forms, all rescaled to the common variance $\Var_{P_0}(\eta) \approx \alpha^2$ (exactly $\alpha^2$ for (GB) and (MLP) by their empirical $\sigma_\eta$; for (H) exactly $\alpha^2$ only at $\rho_0=0$) so that $\sigma_{\mathrm{eff}}^2:=\Var_{P_0}(\eta+\varepsilon) \approx 1+\alpha^2$ remains bounded over the $\alpha$ grid:
\begin{itemize}
    \item[(H)] \emph{Hermite.}  $\eta(X) = (\alpha/\sqrt{12})\,(H_3(X_1) + H_3(X_2))$, the symmetrised third Hermite polynomial in the first two signal coordinates.  Orthogonal to $\mathrm{span}(X)$ by Stein's lemma.
    \item[(GB)] \emph{GB-shaped.}  $\eta(X)=(\alpha/\sigma_\eta)\,(\hat\mu_{\mathrm{GB}}(X) - X^\top\hat\beta_{\mathrm{lin}}^{\mathrm{GB}})$, where $\hat\mu_{\mathrm{GB}}$ is a histogram gradient-boosted regressor trained once on a synthetic non-linear target on the signal coordinates $\{0,\ldots,5\}$, and $\hat\beta_{\mathrm{lin}}^{\mathrm{GB}}=\Sigma_0^{-1}\E[X\,\hat\mu_{\mathrm{GB}}(X)]$ is its population linear projection (estimated by Monte Carlo on $5\times 10^4$ fresh draws).  The constant $\sigma_\eta$ is chosen so that $\Var_{P_0}(\eta) = \alpha^2$.
    \item[(MLP)] \emph{MLP-shaped.}  Identical construction with $\hat\mu_{\mathrm{GB}}$ replaced by a frozen single-hidden-layer multilayer perceptron with $32 \to 16$ ReLU activations.
\end{itemize}
Under (GB) and (MLP), $\eta$ is a fixed non-linear function of the first six coordinates whose population linear projection has been removed by construction, so the conditional mean $\mu(X) = X^\top\beta^\star + \eta(X)$ is the sum of a sparse linear function and a non-linear function of the signal coordinates and does not lie in $\mathrm{span}(X)$.

\begin{figure}[t]
\centering
\includegraphics[width=\textwidth]{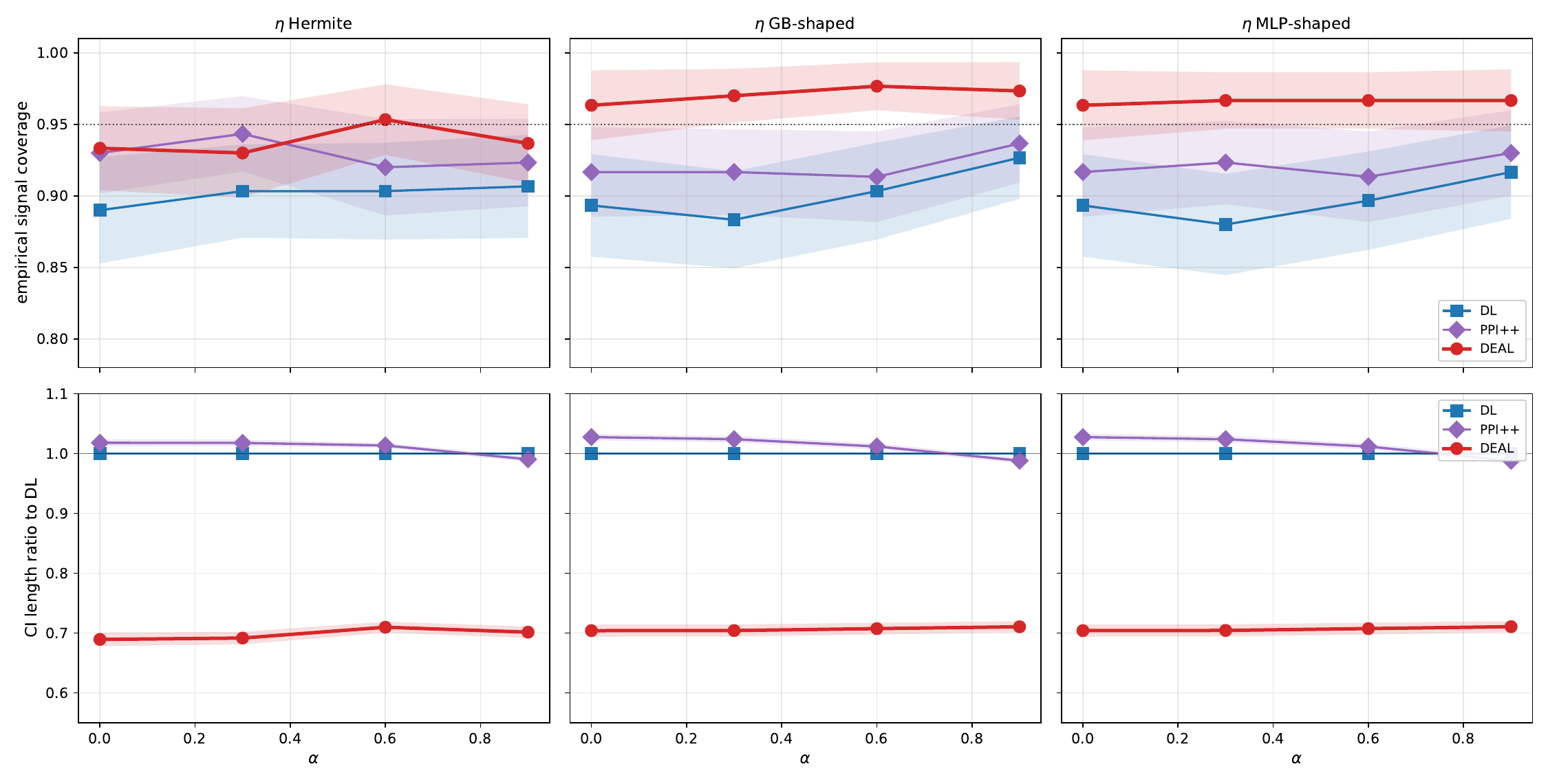}
\caption{\textit{DEAL inference under the linearised oracle predictor across three forms of $\eta$.}  Top row: empirical signal coverage versus the misspecification strength $\alpha$.  Bottom row: CI-length ratio of each estimator to DL on $J_{\mathrm{signal}}$.  Columns correspond to the three $\eta$ specifications: Hermite, GB-shaped, and MLP-shaped.  Reference lines mark the nominal coverage $0.95$ (top) and the DL-parity ratio $1$ (bottom).  External-estimator size $n_A=1600$, $n_0=n_{\mathrm{tun}}=400$, AR(1) target with $\rho_0=0.4$, $p=120$, $s=6$.  {(PPI omitted as in Figure~\ref{fig:study1}.)}}
\label{fig:misspec-anchor}
\end{figure}

Under all three $\eta$ specifications (the anchored-model experiment, Figure~\ref{fig:misspec-anchor}), DEAL coverage tracks the nominal $0.95$ (within $\pm0.06$ across the grid), confirming Corollary~\ref{cor:CLT-nonlin-labeler}.  
The DEAL-to-DL CI ratio is essentially flat in $\alpha$ at $0.69$--$0.71$ and invariant across the three $\eta$ shapes, as Theorem~\ref{thm:deal-vs-dl}(ii) predicts, its hypotheses met here by {Theorem~\ref{thm:complementarity}(i)}.
  
Because the Lasso linearisation renders the model linear, PPI and PPI++ reduce to DL (Proposition~\ref{prop:ppipp-equivalence}) and track parity here.

Together, the linear-predictor and the linearised-oracle-predictor studies cover the two regimes announced in Section~\ref{sec:misspec}.  The linear-predictor study isolates the additive-misspecification axis (Theorem~\ref{thm:CLT-misspec}).  
The linearised-oracle-predictor study isolates the linearisation step (Lemma~\ref{lem:lin-rate} {of the supplementary material}) and the non-linear-predictor validity (Corollary~\ref{cor:CLT-nonlin-labeler}, {Theorem~\ref{thm:complementarity}(i)}), and demonstrates that the rate condition \eqref{eq:rate-cond-nonlin} is empirically met when the external model delivers a noiseless conditional-mean target for the Lasso linearisation.
  In both regimes, the DEAL CI is shorter than DL, PPI, and PPI++ uniformly over $\alpha$ and $\eta$.

\subsection{Robustness to covariate shift in the unlabelled covariate distribution}\label{sec:experiments-shift}

We now examine the covariate-shift regime introduced in Section~\ref{sec:shift-aware}.  The labelled target rows are drawn from the AR(1) covariance with $\rho_0 = 0.4$ as before, while the unlabelled rows are drawn from an AR(1) covariance with parameter $\rho_{\mathrm u}\in\{0,0.1,\ldots,0.8\}$.  The conditional model $Y\mid X$ is unchanged.  Three external-estimator qualities are reported, $n_A\in\{200, 800, 3200\}$, corresponding to poor, medium, and good models.

\begin{figure}[t]
\centering
\includegraphics[width=\textwidth]{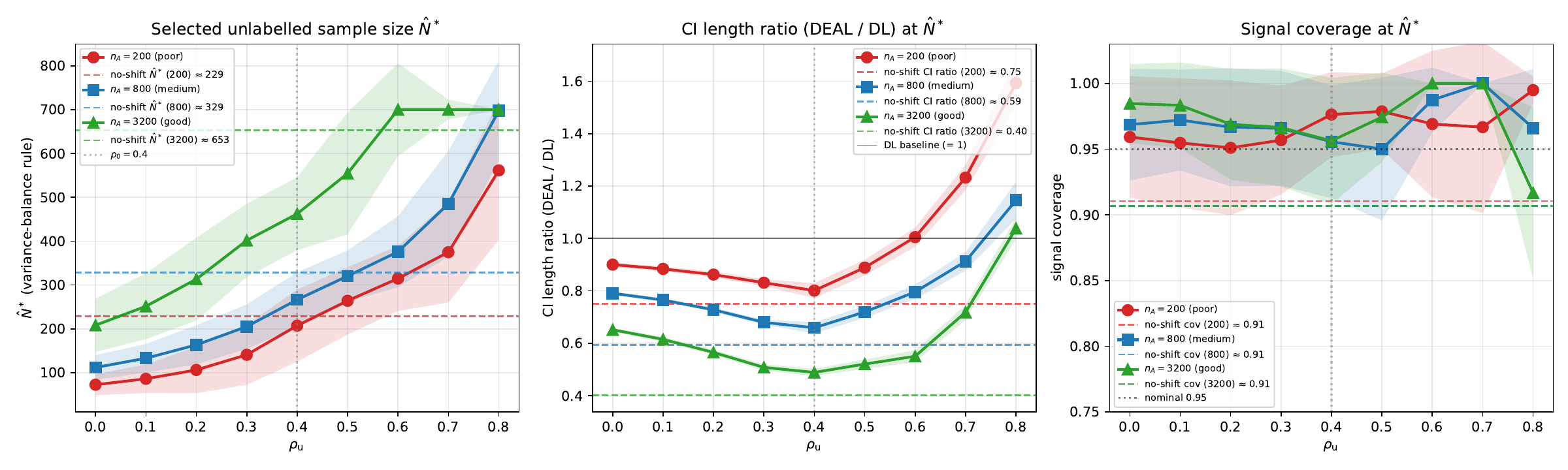}
\caption{\textit{Robustness of DEAL-shift-aware under covariate shift in the unlabelled covariate distribution.}  Left: the variance-balance choice $\hat N^*$ versus the unlabelled-design AR(1) parameter $\rho_{\mathrm u}$.  Centre: CI-length ratio against DL.  Right: empirical signal coverage.  Dashed lines indicate the no-shift reference for each external-estimator quality: DEAL without the shift-aware substitution, at $\rho_{\mathrm u}=\rho_0$.  Each point is averaged over $R=20$ replications; shaded regions are $95\%$ Monte-Carlo error bands across replications.}
\label{fig:study2}
\end{figure}

Empirical signal coverage of DEAL-shift-aware (Figure~\ref{fig:study2}) remains at or near the nominal $0.95$ across the $\rho_{\mathrm u}$ grid.

The CI-length ratio versus DL at $\hat N^*$ is U-shaped in $\rho_{\mathrm u}$, with its minimum at no shift.  Under heavy shift-up at the poorest external-estimator quality ($n_A = 200$, $\rho_{\mathrm u} = 0.8$) the CI-length ratio rises to $1.58$, but at the medium and good external-estimator qualities the ratio under heavy shift remains close to or modestly above the DL benchmark ($1.17$ and $1.08$ respectively): the procedure under shift never under-performs the target-only benchmark by more than the factors just quoted.


\subsection{The hybrid as the complementarity check}\label{sec:experiments-hybrid-main}

{This synthetic illustration is the complementarity check of the theory. It illustrates clause~(ii) of Corollary~\ref{cor:hybrid-dom}---that the design-orthogonal residual channel and the variance channel compound into a product-form dominance---and the observation of Section~\ref{sec:hybrid} that, because the two variance reductions draw on a shared budget, their combined effect on the selected unlabelled budget $\hat N^*$ is sub-additive rather than additive. The setting is synthetic and the model's design-orthogonal component is controlled by construction, so that the design-orthogonal signal fraction can be dialled cleanly. It is a check of the mechanism, not a headline result; every real-data run in the portfolio remains a pure DEAL run.}

\begin{figure}[t]
\centering
\includegraphics[width=\textwidth]{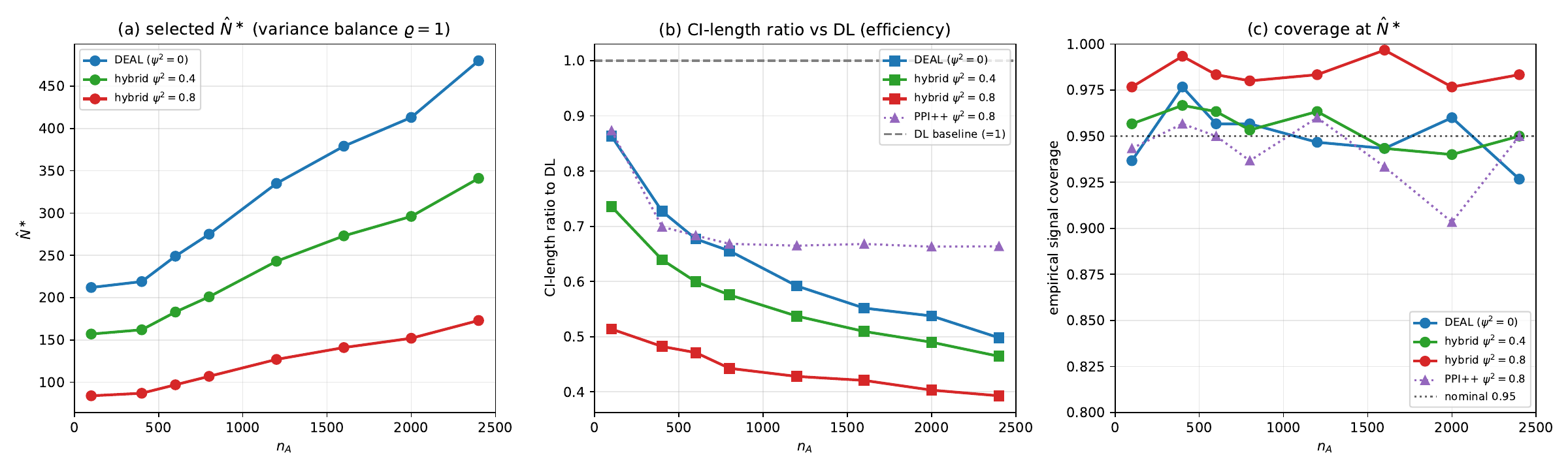}
\caption{\textit{The control-variate hybrid on an auxiliary-covariate model, across the design-orthogonal signal fraction $\psi^2$.} The power-adaptivity design of Figure~\ref{fig:study1} ($p=120$, $s=6$, $n_0=n_{\mathrm{tun}}=400$, AR(1) $\rho_0=0.4$, nodewise precision, cross-validated external model indexed by the external sample size $n_A$, $N$ by the variance-balance rule, $R=100$), with two additions: a design-orthogonal model component $\gamma W$ ($\psi^2=\gamma^2=\rho_j^2[\hat\mu]$ the design-orthogonal signal fraction, total model-residual variance one) and an estimated control-variate weight; no oracle information. The three solid curves are the hybrid at $\psi^2\in\{0,0.4,0.8\}$, the $\psi^2=0$ curve being DEAL; the dotted curve in (b) and (c) is PPI++ at $\psi^2=0.8$. (a)~Selected $\hat N^*$: $212\to480$ at $\psi^2=0$, contracting with $\psi^2$ (to $84\to173$ at $\psi^2=0.8$). (b)~CI-length ratio to debiased Lasso: the $\psi^2=0$ curve reproduces the study ($0.86\to0.50$), larger $\psi^2$ tightening it to $0.51\to0.39$ at $\psi^2=0.8$ and dominating the pure procedure at every quality; PPI++ stays above the hybrid. (c)~Empirical signal coverage at $\hat N^*$: DEAL and the hybrid cover at least $0.92$ at every cell (mostly $0.94$--$1.00$); PPI++, at its native rectifier standard error, covers $0.90$--$0.97$.}
\label{fig:aux-hybrid}
\end{figure}

Figure~\ref{fig:aux-hybrid} summarises the results over $R=100$ replications. At design-orthogonal signal fraction $\psi^2=0$ (the quantity $\rho_j^2[\hat\mu]$ of Section~\ref{sec:misspec-nonlinear-labeler}, written $\psi^2$ in this section and in Figure~\ref{fig:aux-hybrid}) the hybrid matches the confidence-interval efficiency of Section~\ref{sec:experiments-power} to within $0.01$, its CI-length ratio to debiased Lasso falling from $0.86$ to $0.50$ across the model-quality range. Raising $\psi^2$ tightens the interval further at every model quality, to $0.51\to0.39$ at $\psi^2=0.8$: the hybrid dominates the pure procedure throughout, by less than the factor $\sqrt{1-\psi^2}$ of Corollary~\ref{cor:hybrid-dom}(ii), which presumes a common shrinkage limit and a common $\hat N^*$. The selected $\hat N^*$ contracts as $\psi^2$ grows (from $212$ to $84$ at $n_A=100$): the variance and control-variate channels are sub-additive, the residual channel spending part of the variance channel's budget. Optimally-tuned PPI++ improves with $\psi^2$, its rectifier drawing on the same orthogonal component, yet at $\psi^2=0.8$ stays above the hybrid, consistent with the uniform-in-$N$ dominance of {Corollary~\ref{cor:hybrid-dom}}.

\subsection{Portfolio design and reporting protocol}\label{subsec:realdata-overview}\label{sec:realdata}

Five demonstrations are presented, in each of which an external machine-learned predictor is available alongside a moderately sized labelled cohort and a larger pool of unlabelled covariates.    In every demonstration, DEAL is compared against DL, PPI and PPI++ as in Section~\ref{sec:experiments-design}.

A uniform reporting protocol is applied across all five demonstrations, comprising two primary metrics (supplemented in Section~\ref{subsec:realdata-results} by significance counts and, where the noise floor permits, an out-of-sample $R^2$).
\begin{itemize}
\item The \emph{median confidence-interval ratio} for procedure $\pi$ is the median, across the $p$ inferential coordinates, of the ratio of $\pi$'s coordinate-wise confidence-interval half-width to DL's half-width; values below one indicate that $\pi$ tightens against DL.
\item The \emph{bootstrap selection stability} of the anchored discovery set $S_\pi$---the coordinates procedure $\pi$ declares significant on the full inferential cohort---in the spirit of \citet{meinshausen2010stability}: for $B = 20$ half-sample resamples of the inferential cohort, $\pi$ is re-run end-to-end on each subsample and the per-coordinate selection frequency $\hat F_j = \tfrac{1}{B}\sum_b \1\{j \in S_\pi^{(b)}\}$ is averaged across the coordinates in $S_\pi$.  Values close to one indicate that $\pi$'s anchored discoveries are robust to resampling of the inferential rows.
\end{itemize}

Per-demonstration scientific background and data triples are described in Section~\ref{subsec:realdata-demos}; the unified numerical synthesis appears in Section~\ref{subsec:realdata-results}.  Implementation details, including model-training procedures, feature-panel constructions, and side-evidence protocols, are deferred to Section~\ref{sec:realdata-supp}.

\subsection{The five demonstrations: breadth across scientific fields}\label{subsec:realdata-demos}

The five demonstrations span three scientific fields, in each of which the same inferential predicament recurs: a gold-standard response is scarce and expensive, the covariates are cheap and abundant, and an external predictor of the response is available but imperfect.  In astronomy, the response is the citizen-science spiral-arm vote fraction for a galaxy, the covariates are activations of a frozen vision encoder, and the predictor is a fine-tuned morphology head.  In materials chemistry, the response is a measured semiconductor band gap, the covariates are composition descriptors, and the predictor is a first-principles density-functional-theory calculation.  In oncology, the response is the dose response of a targeted agent (measured in patient tumours, xenografts, and cancer cell lines) and the predictor is a cell-line regression model or, in one case, a large language model queried off the shelf.

These three fields trace out a wide statistical envelope: labelled cohorts ranging from $n_0=105$ to $3347$, design dimensions from $p=30$ to $640$, and a model--response correlation $\mathrm{corr}(\hat\mu,y)$ from $0.14$ to $0.87$.  DEAL tightens confidence intervals across this entire range.  
Across these five demonstrations the gain does not order by $\mathrm{corr}(\hat\mu,y)$, which varies alongside $n_0$, $p$ and the unlabelled-cohort size $n_u$: what the variance-balance rule asks of $\hat\mu$ is that its imputation bias be small enough to admit, not that it predict well, and a weakly correlated model can meet that requirement.  The improvement still grows as that bias shrinks (Corollary~\ref{cor:idealized-efficiency}; Section~\ref{sec:experiments-power}).  Table~\ref{tab:realdata-characteristics} records the inferential triple $(n_0,p,n_u)$ and the model for each demonstration; the unified numerical results appear in Section~\ref{subsec:realdata-results} (Table~\ref{tab:deal-portfolio}).

\begin{table}[ht]
\centering
\caption{\textit{Characteristics of the five demonstrations.}  Each row reports the field, the labelled-cohort size $n_0$, the design dimension $p$, the unlabelled-cohort size $n_u$ (the budget $N$ of Section~\ref{sec:setup}), the model--response correlation $\mathrm{corr}(\hat\mu,y)$ on the inferential cohort, and the source of the external model $\hat\mu$.  In every demonstration the unlabelled cohort is drawn from the same distributional context as the labelled cohort, the condition under which the pooled-precision lever is valid; the patient-derived-xenograft demonstration is the case in which this constraint binds (Section~\ref{subsec:pdxe-supp} of the supplementary material).  Abbreviations: BRCA, breast cancer; PDXE, the patient-derived xenograft encyclopedia; PBE, the Perdew--Burke--Ernzerhof density functional.  Numerical results are reported separately in Table~\ref{tab:deal-portfolio}.}
\label{tab:realdata-characteristics}
\footnotesize\setlength{\tabcolsep}{4pt}
\begin{tabular}{l l rrr c >{\raggedright\arraybackslash}p{4.1cm}}
\toprule
& & \multicolumn{3}{c}{Dimensions} & \multicolumn{2}{c}{Predictor} \\
\cmidrule(lr){3-5} \cmidrule(lr){6-7}
Demonstration & Field & $n_0$ & $p$ & $n_u$ & $\mathrm{corr}(\hat\mu,y)$ & Source \\
\midrule
Galaxy morphology  & Astronomy        & $1352$ & $640$ & $2151$     & $0.65$ & Zoobot ConvNeXt head \\
Inorganic band gap & Materials chem.\ & $3347$ & $630$ & $10{,}000$ & $0.87$ & Materials Project PBE \\
BRCA chemoresponse & Oncology         & $226$  & $100$ & $1097$     & $0.23$ & Cancer Cell Line Encyclopedia, paclitaxel \\
PDXE drug response & Oncology         & $105$  & $30$  & $259$      & $0.14$ & Cancer Cell Line Encyclopedia, alpelisib \\
Selumetinib        & Oncology         & $538$  & $80$  & $620$      & $0.48$ & Claude Opus, zero-shot \\
\bottomrule
\end{tabular}
\end{table}

The demonstrations play three roles.  One probes the strong-model frontier: the inorganic band-gap demonstration pairs the strongest model in the portfolio ($\mathrm{corr}=0.87$) with a ten-thousand-compound unlabelled catalogue.

Two probe the weak- and miscalibrated-model regime: the breast-cancer and patient-derived-xenograft demonstrations both use cell-line-to-tissue transfer models that are known to be miscalibrated 
($\mathrm{corr}=0.23$ and $0.14$) yet still tighten; the patient-derived-xenograft demonstration, with the smallest labelled cohort in the panel ($n_0=105$), additionally isolates the same-distributional-context constraint on the unlabelled cohort: a larger but covariate-mismatched pool yields anti-conservative, non-replicating intervals (Section~\ref{subsec:pdxe-supp}).

The fifth demonstration probes the model-modality frontier: the dose response of the MEK1/2 inhibitor selumetinib~\citep{yeh2007biological}, catalogued in the Genomics of Drug Sensitivity in Cancer database~\citep{yang2013genomics}, is imputed by a large language model prompted with a cell-line name and tissue label together with a fixed pathway-prior system prompt (Section~\ref{subsec:llm-oracle-supp}), with no fine-tuning.  
Because the model is queried rather than fitted, it is opaque in the way the opening premise supposes: it is seen only through the scalar values it returns, and the prompt that produced them lies outside the genomic design $X$ altogether.  Beyond the predictor regularity of Assumption~\ref{ass:lin-proj}, no assumption relates it to the covariates: the model is non-linear, and the guarantees of Section~\ref{sec:misspec-nonlinear-labeler} hold at $\beta^\star_{\mathrm{proj}}$ regardless.  
  Scientific background, cohort construction, model-training recipes, and side-evidence protocols for all five demonstrations are collected in Section~\ref{sec:realdata-supp}.

\subsection{Cross-demonstration results}\label{subsec:realdata-results}

The unified portfolio is summarised in Table~\ref{tab:deal-portfolio}.  Across all five demonstrations DEAL delivers narrower confidence intervals than DL, with median ratios ranging from $0.23$ to $0.53$, while PPI++ never materially tightens against DL: its median ratios lie between $0.98$ and $1.10$.  Four of the five demonstrations select an interior budget $\hat N^*<n_u$ (galaxy, band gap, patient-derived xenograft, and selumetinib), the breast-cancer demonstration alone saturating at $\hat N^*=n_u$.  The interval-length limit of Theorem~\ref{thm:deal-vs-dl} is asserted only at saturation.  At the four interior selections the reported tightening is observed rather than guaranteed, coverage secured by the variance-balance rule's bias ceiling (Remark~\ref{rem:saturation-scope}).

DEAL's discoveries also replicate more reliably under resampling.  When the inferential cohort is randomly halved and the analysis is re-run, the fraction of DEAL's originally significant coordinates that come back as significant averages between $0.66$ and $0.97$ across the five demonstrations (the mean-$\hat F$ columns of Table~\ref{tab:deal-portfolio}), against $0.36$--$0.63$ for DL and $0.34$--$0.61$ for PPI++.  Demanding that a discovery survive in at least $80\%$ of resamples, between $48\%$ (selumetinib) and $100\%$ (patient-derived xenograft) of DEAL's coordinates pass, against $0$--$40\%$ for DL and a similar range for PPI++.  DEAL's discoveries are not only more numerous on three of the five demonstrations but substantially more reproducible across the panel.  On the two where it rejects slightly fewer, galaxy and band gap, the coordinates DEAL retains replicate more reliably and carry the stronger out-of-sample signal, the extra DL discoveries being the less reliable of the two sets.  Galaxy's $\hat t = 0$ means the external coefficient entered uncorrected, so no estimate was pulled towards the debiased-Lasso fit; the smaller count instead drops borderline coordinates whose DEAL-to-DL length ratio exceeds one near the significance threshold, non-replicating debiased-Lasso findings that DEAL declines to call.

On the galaxy and band-gap demonstrations (the two whose response admits a non-trivial out-of-sample fit), DEAL's discoveries achieve out-of-sample $R^2_{\mathrm{test}} = 0.42$ and $0.47$ respectively, the highest among the three procedures; on band gap, DEAL is the only procedure whose discoveries generalise at all, with DL and PPI++ each yielding strongly negative held-out $R^2$.

\begin{table}[ht]
\centering
\caption{\textit{Five-demonstration portfolio summary.}  CI-ratio columns report each procedure's median coordinate-wise confidence-interval-length ratio against DL; values below $1$ indicate tightening against DL.  Significance counts are the numbers of coordinates rejecting $\beta^\star_j = 0$ at level $\alpha = 0.05$ for each procedure.  The mean-$\hat F$ columns report bootstrap selection stability on the anchored discovery set: for $B = 20$ half-sample resamples of the inferential cohort, each procedure is re-run end-to-end and the per-coordinate selection frequency $\hat F_j$ is averaged across the coordinates in the procedure's anchored discovery set.  Confidence-interval ratios use DL as the unit reference.  Every procedure receives the raw external model $\hat\mu$; on the selumetinib demonstration PPI++ reads its design-orthogonal component through the rectifier, the genuine two-channel comparison with DEAL, not a linearised surrogate.  PPI is omitted in favour of optimally-tuned PPI++, the sharper prediction-powered baseline, with which it coincides asymptotically under a linear model (Propositions~\ref{prop:ppi-cancellation} and~\ref{prop:ppipp-equivalence}).  The $\hat N^*/n_u$ column reports the fraction of the available unlabelled cohort selected by the variance-balance rule of Section~\ref{sec:vbrule}.}
\label{tab:deal-portfolio}
\footnotesize\setlength{\tabcolsep}{4pt}
\begin{tabular}{l cc ccc ccc c}
\toprule
& \multicolumn{2}{c}{CI ratio vs.\ DL} & \multicolumn{3}{c}{Sig.\ count} & \multicolumn{3}{c}{mean $\hat F$ on anchored set} & \\
\cmidrule(lr){2-3} \cmidrule(lr){4-6} \cmidrule(lr){7-9}
Demonstration & DEAL & PPI++ & DEAL & DL & PPI++ & DEAL & DL & PPI++ & $\hat N^*/n_u$ \\
\midrule
Galaxy             & $0.53$ & $1.02$ & $177$ & $188$ & $179$ & $0.85$ & $0.63$ & $0.61$ & $6\%$ \\
Band gap           & $0.49$ & $1.10$ & $143$ & $209$ & $197$ & $0.88$ & $0.60$ & $0.57$ & $1.6\%$ \\
BRCA chemoresponse & $0.23$ & $1.02$ & $40$  & $12$  & $13$  & $0.97$ & $0.36$ & $0.34$ & $100\%$ \\
PDXE drug response & $0.25$ & $1.06$ & $7$   & $3$   & $3$   & $0.95$ & $0.42$ & $0.38$ & $55\%$ \\
Selumetinib        & $0.49$ & $0.98$ & $29$  & $22$  & $20$  & $0.66$ & $0.62$ & $0.54$ & $36.9\%$ \\
\bottomrule
\end{tabular}
\end{table}

The model's predictive accuracy varies sharply across the portfolio, from $0.14$ on the patient-derived-xenograft demonstration to $0.87$ on band gap.  
An ancillary property of the procedure is its adaptivity to model quality.  If $\hat\mu$ is replaced by a no-information surrogate, the asymptotic theory of Section~\ref{sec:theory-bias} predicts that DEAL's confidence intervals should recover those of DL.  In a no-information check, each model $\hat\mu$ is replaced by a uniform permutation of its outputs, retaining its marginal distribution while breaking its dependence on the response.  The unlabelled cohort is then subsampled to the size $\hat N^*$ prescribed by the variance-balance rule of Section~\ref{sec:vbrule}.  Table~\ref{tab:nosignal-portfolio} reports the resulting CI ratio and significance counts; the small prescriptions $\hat N^*$ it records are the rule's own response to a model that carries no signal.  On the breast-cancer, patient-derived-xenograft, selumetinib, and inorganic-band-gap demonstrations the no-information CI ratio returns to within roughly five percent of unity, confirming the predicted collapse to DL.  
The galaxy demonstration is the lone exception: its no-information CI ratio settles at $1.17$, above unity rather than at it, and its significance count is $160$ against DL's $188$.  The mechanism, specific to the regime $p\approx n_0$, is examined in Section~\ref{subsec:galaxy-supp}.

\begin{table}[ht]
\centering
\caption{\textit{No-information CI ratios under the variance-balance rule of Section~\ref{sec:vbrule}.}  For each demonstration, the model $\hat\mu$ is replaced by a uniform permutation of its outputs; the unlabelled cohort is subsampled to the prescription $\hat N^*$.}
\label{tab:nosignal-portfolio}
\small
\begin{tabular}{l c c c}
\toprule
Demonstration & $\hat N^*$ & CI ratio at $\hat N^*$ & Sig.\ DEAL / DL \\
\midrule
Galaxy                          & $10$  & $1.17$ & $160 / 188$ \\
Band gap                        & $50$   & $0.98$  & $161 / 209$ \\
BRCA chemoresponse              & $48$  & $1.052$ & $12 / 12$ \\
PDXE drug response              & $11$  & $1.025$ & $2 / 3$ \\
Selumetinib                     & $47$  & $0.994$ & $16 / 22$ \\
\bottomrule
\end{tabular}
\end{table}

\section{Discussion}\label{sec:discussion}

This paper began from a negative fact.  In a high-dimensional linear model an external model cannot be made to pay through the mean of a corrected estimator: the rectifier of the prediction-powered family reads only the part of a model's signal lying orthogonal to the design, and against a linear predictor that part is empty (Propositions~\ref{prop:ppi-cancellation} and~\ref{prop:ppipp-equivalence}).  A negative fact of that shape is an instruction about where to look.  If the model's information cannot enter at the mean it must be admitted at the variance, and that is where DEAL admits it: the predictions on the unlabelled covariates enlarge the design that the debiasing step inverts, and a bias-aware shrinkage decides how much of that enlargement is trusted.

The relocation is worth only as much as the safeguard attached to it, and the safeguard, not the enlargement, is the interpretive centre of the construction.  The shrinkage treats the labelled correction as a single noisy reading of an unknown bias and discounts it by a data-driven amount, so that an external model carrying nothing is discarded rather than merely tolerated and the procedure returns the target-only debiased Lasso.  Because the discount is calibrated rather than assumed, the external model need not be vouched for in advance; it is seen only through the values it returns, and may therefore be built on covariates the analysis never records---on text, on images, on a separate assay (Section~\ref{sec:aux-cov} of the supplementary material).  The unlabelled sample size is likewise read off the data rather than fixed by fiat, balancing the imputation bias the external model introduces against the sampling noise the enlargement removes (Section~\ref{sec:choose-N}), and a shift-aware variant of the debiasing step is developed for unlabelled designs drawn under covariate shift (Section~\ref{sec:shift-aware}).  Against the prediction-powered family the relation this produces is complementarity rather than displacement: DEAL reads the variance channel, the rectifier the residual channel, and a hybrid reads both (Section~\ref{sec:hybrid}).  The work belongs to a growing literature on drawing valid inference, and not prediction alone, from model-generated data \citep{keret2025glm, rezaei2025highdim}; what it contributes is the channel through which the model is permitted to act.

It is worth separating what the paper proves from what it shows.  Validity is proved, and proved without reference to the model's quality: the debiased coordinates are asymptotically normal with a variance that adapts to the model's accuracy, and the statement survives the two departures that matter in practice, a misspecified target and a non-linear predictor, both handled at the best linear approximation.  Efficiency is a claim of a different kind, and turns on the external model in the way validity does not: the interval shortens as the model's imputation bias falls.  The length comparison with the target-only debiased Lasso is a first-order limit taken at a saturated unlabelled budget (Theorem~\ref{thm:deal-vs-dl}), and most of the demonstrations select an interior one, where the guarantee is coverage rather than width.  The tightening seen there---median intervals between roughly a quarter and a half of the target-only length---is accordingly an empirical finding rather than a consequence of the theorem.  It is a falsifiable one: replacing each model by a permutation of its own outputs returns the intervals to the target-only benchmark on all but one of the five demonstrations.

The clearest boundary is on the estimand.  Under misspecification DEAL is honest about the best linear approximation of the response, not about the conditional mean itself.  That approximation is well defined and interpretable, the coefficient vector of the closest linear surrogate; it is a summary all the same, and the procedure makes no claim to recover the regression surface.  Where the linear approximation is a poor stand-in, the coordinates reported answer a different question from the one the analyst may have in mind, and whether that substitution is acceptable is a judgement about the science rather than about the interval.  A further boundary concerns what the external model is taken to be.  It is a fixed object, seen only through the values it returns, and the shrinkage is calibrated against a bias that does not move.  An estimator carrying a population drift of its own---a transfer-learning initialiser whose source population has itself shifted \citep{li2022transfer}---would present the shrinkage with a bias that moves with the sample rather than a fixed one, and the rule as stated is not built to track it.  The last boundary is the regime itself.  DEAL is assembled from the apparatus of sparse high-dimensional regression: a debiased-Lasso anchor, an estimated precision matrix, a heteroscedasticity-robust variance for the misspecified case.  The prediction-powered estimators require none of these.  They are general estimating-equation procedures, applicable to any M-estimator at any dimension, and outside the sparse regime they remain the right tool.  What is claimed here is sharpness inside it.

The practical import is that a model no one can audit may still be put to work on a scientific question.  That is the posture the present moment asks for: the external model is increasingly a foundation model queried off the shelf, or a generator of synthetic responses, whose calibration the analyst neither sets nor can readily inspect---one of our demonstrations uses precisely such a model, a large language model prompted zero-shot.  Coverage holds whatever that model returns, and where it is informative the intervals are markedly shorter, in simulation and across demonstrations in astronomy, materials chemistry and oncology.  What makes this possible is the pairing of channel and safeguard, the variance of a debiased estimator rather than the mean of a rectified one, under a shrinkage that reads the model's bias before admitting it; and that pairing is not tied to the Lasso anchor that carries it here.  The hybrid that reads both channels is the nearer step: developed here as theory and exercised once in simulation (Sections~\ref{sec:hybrid} and~\ref{sec:experiments-hybrid-main}), it has still to be formed on real data against a genuinely non-linear model, whose predictive content lies partly outside the recorded design.  Carrying the pairing beyond linear regression, wherever a model of uncertain quality must enter an inference on terms the analyst sets rather than on its own, is the work we would take up after it.

\section*{Data availability statement}
The data underlying the five real-data demonstrations are publicly available from the sources cited in Section~\ref{sec:realdata-supp} of the supplementary material.  Code implementing DEAL and reproducing all simulations and analyses is available from the authors upon reasonable request.

\section*{Acknowledgements}
The authors thank colleagues who commented on earlier versions of this work.

\bibliographystyle{plainnat}
\bibliography{paperref}

\clearpage

\setcounter{section}{0}\setcounter{equation}{0}\setcounter{figure}{0}\setcounter{table}{0}
\setcounter{theorem}{0}\setcounter{proposition}{0}\setcounter{corollary}{0}
\setcounter{lemma}{0}\setcounter{assumption}{0}\setcounter{remark}{0}\setcounter{algorithm}{0}
\renewcommand{\thesection}{S\arabic{section}}
\renewcommand{\theequation}{S.\arabic{equation}}
\renewcommand{\thetheorem}{S\arabic{theorem}}
\renewcommand{\theproposition}{S\arabic{proposition}}
\renewcommand{\thecorollary}{S\arabic{corollary}}
\renewcommand{\thelemma}{S\arabic{lemma}}
\renewcommand{\theassumption}{S\arabic{assumption}}
\renewcommand{\theremark}{S\arabic{remark}}
\renewcommand{\thefigure}{S\arabic{figure}}
\renewcommand{\thetable}{S\arabic{table}}
\setlength{\emergencystretch}{3em}
\renewcommand{\thealgorithm}{S\arabic{algorithm}}

\begin{center}\Large\textbf{Supplementary Material}\end{center}
\medskip

\noindent This supplement collects a notation index, the shrinkage-selector apparatus, the design-specific refinements, three extensions of the procedure, the consolidated algorithm, the comparison with the prediction-powered family, the proofs, the simulation protocols, and the real-data protocols.  Numbering is prefixed by~S; references to the main text are unprefixed.

\section{Notation index}\label{app:notation}

{This section collects the notation used in the main text (Table~\ref{tab:notation}).}

{\small\begin{singlespace}
\begin{longtable}{@{}ll@{}}
\caption{Notation used in the main text.}\label{tab:notation}\\
\toprule
Symbol & Meaning \\
\midrule
\endfirsthead
\caption[]{Notation used in the main text (continued).}\\
\toprule
Symbol & Meaning \\
\midrule
\endhead
\endfoot
\bottomrule
\endlastfoot
\multicolumn{2}{@{}l}{\emph{Data, dimensions, and target}}\\*
$(X_0,y_0)$ & labelled inference sample; $n_0$ rows \\
$\tilde X$ & unlabelled target covariates; $N$ rows \\
$n_0,\;N,\;\kappa:=N/n_0$ & labelled size, unlabelled size, unlabelled-to-labelled ratio \\
$(X_{\mathrm{tun}},y_{\mathrm{tun}}),\; n_{\mathrm{tun}},\; n_A$ & tuning subsample and its size, external-estimator sample size \\
$p,\;s:=\|\beta^\star\|_0$ & ambient dimension, sparsity \\
$\beta^\star,\;\beta^\star_j$ & target regression vector and $j$th coordinate \\
$\beta^\star_{\mathrm{proj}}$ & population least-squares projection (misspecified target) \\
$\varepsilon_0,\;\sigma^2$ & target noise and its variance \\
$\Sigma_0\ (=\Sigma)$ & target covariance; $\Sigma$ in the no-shift regime \\
$J$ & inferential index set \\
\midrule
\multicolumn{2}{@{}l}{\emph{External estimator and DEAL stages}}\\*
$\hat\beta_{\mathrm{ext}}$ & external (possibly transfer-learned) estimator \\
$\Delta:=\hat\beta_{\mathrm{ext}}-\beta^\star$ & external-estimator error \\
$C$ & one-step correction \\
$\hat t\ (\to t_0\in[0,1])$ & bias-aware shrinkage and its limit \\
$\tilde\beta^{\mathrm{init}},\;\tilde f$ & bias-aware initialiser, imputed pseudo-labels \\
$(X_{\mathrm{stk}},y_{\mathrm{stk}}),\;\hat\beta$ & stacked design/response, stacked Lasso \\
$\tilde\beta,\;\lambda$ & final debiased estimator, Lasso penalty \\
\midrule
\multicolumn{2}{@{}l}{\emph{Precision / debiasing}}\\*
$M_1,\;M_2$ & Javanmard--Montanari matrices (Stage 1, Stage 4) \\
$A_n$ & Stage-4 noise-amplification factor (Theorem~\ref{thm:CLT}) \\
$\hat\Sigma_0,\;\hat\Sigma_{\mathrm u}$ & labelled and unlabelled sample covariances \\
$\Omega:=\Sigma^{-1},\;\hat\Omega$ & precision matrix and nodewise estimate \\
$s_\Omega,\;m_\Omega:=n_0+N$ & row sparsity of $\Omega$, covariate rows for $\hat\Omega$ \\
\midrule
\multicolumn{2}{@{}l}{\emph{Efficiency and unlabelled-size rule}}\\*
$G(\kappa,t_0),\;n_{\mathrm{eff}}$ & idealised variance ratio, effective sample size \\
$\tau_{j,n}^2,\;\tau_j^{2,\mathrm{sand}}$ & limiting coordinate variance and its sandwich form \\
$\tau^2_{L,j},\;\tau^2_{B,j}$ & leading-noise variance, squared-bias term \\
$\varrho\ (\ge 1)$ & variance-balance criterion ratio \\
$N^*_j,\;\hat N^*$ & population and empirical variance-balance sizes \\
\midrule
\multicolumn{2}{@{}l}{\emph{Misspecification and model}}\\*
$\mu(X),\;\eta(X)$ & conditional mean, misspecification residual \\
$\hat\mu,\;\nu[\hat\mu]$ & non-linear predictor, design-orthogonal residual \\
$\beta^{\mathrm{proj}}_0[\hat\mu],\;\delta_{\mathrm{lin}}[\hat\mu]$ & model projection and its error \\
$\Gamma_0$ & sandwich middle factor \\
$\rho_j^2[\hat\mu]$ & design-orthogonal signal fraction \\
\midrule
\multicolumn{2}{@{}l}{\emph{Covariate shift}}\\*
$\Sigma_{\mathrm u}$ & unlabelled-block covariance \\
$\rho_j(M_2,\Sigma_{\mathrm u})$ & bias-amplification factor $\|e_j^\top M_2\Sigma_{\mathrm u}\|_1$ \\
$\hat M_{\mathrm u},\;\hat M_{\mathrm p}$ & unlabelled-block and pooled Stage-4 matrices \\
$\hat M_2^{\mathrm{adapt}}$ & shift-aware Stage-4 matrix~\eqref{eq:M2-adapt} \\
\midrule
\multicolumn{2}{@{}l}{\emph{PPI comparators}}\\*
$\hat\beta^{\mathrm{PP}}$ & prediction-powered estimator \\
$\omega\in\R,\;\omega^*_j$ & PPI++ rectifier weight and its variance-minimising value \\
$\omega_j^{\mathrm{cv}},\;\hat\omega_j$ & coordinate-$j$ control-variate weight and its tuning-sample estimator \\
\end{longtable}
\end{singlespace}}

\section{The shrinkage selector: apparatus and consistency}\label{app:shrinkage}

{This section collects the shrinkage-selector apparatus and the efficiency corollary demoted from Sections~\ref{sec:method} and~\ref{sec:theory-bias} of the main text.}

{
\begin{corollary}[Effective sample size of the variance channel]\label{cor:neff}
In the oracle-precision benchmark of Corollary~\ref{cor:idealized-efficiency} ($\hat t\to t_0\in[0,1]$, $\kappa:=N/n_0$), the coordinate-wise asymptotic variance of $\tilde\beta_j$ equals that of a target-only debiased Lasso computed from
\[
    n_{\mathrm{eff}} \;:=\; \frac{n_0(n_0+N)^2}{(n_0+t_0 N)^2} \;=\; n_0\,\frac{(1+\kappa)^2}{(1+t_0\kappa)^2}
\]
i.i.d.\ labelled observations.  For every $t_0<1$ and $\kappa>0$, $n_{\mathrm{eff}}>n_0$; in the limit $t_0\to0$, attained
{in the near-perfect external regime ($a_1=\op(n_0^{-1/2})$; Proposition~\ref{prop:condT-regimes}(i))},
\[
    n_{\mathrm{eff}} \;\longrightarrow\; n_0(1+\kappa)^2 \;=\; \frac{(n_0+N)^2}{n_0},
\]
exceeding even the full labelled-plus-unlabelled count $n_0+N$ for every $\kappa>0$.
\end{corollary}
\begin{proof}
Equate the idealised variance \eqref{eq:idealized-var}, $\Var(\tilde\beta_j)=\sigma^2(n_0+t_0 N)^2(\Sigma^{-1})_{jj}\bigl/\bigl(n_0(n_0+N)^2\bigr)$, to the target-only variance $\sigma^2(\Sigma^{-1})_{jj}/n_{\mathrm{eff}}$ and solve for $n_{\mathrm{eff}}$; the limits and inequalities follow algebraically from $n_0+N=n_0(1+\kappa)$ and $n_0+t_0N=n_0(1+t_0\kappa)$.
\end{proof}
}

{Notation ($\hat B$, $\hat T$, the tuning blocks $(X_{\mathrm{tun},a},y_{\mathrm{tun},a})$ and $(X_{\mathrm{tun},b},y_{\mathrm{tun},b})$, and $M_{1,\mathrm{tun}}$) is as introduced around the definition \eqref{eq:t-hat} of $\hat t$ in Section~\ref{sec:proxy-shrinkage} of the main text.}

The second clause of \eqref{eq:t-hat} makes the selector totally defined and equal to the positive-part form $\hat t=\hat B_+/(\hat B_++\hat T)$.  Since $\hat T>0$ almost surely, $\hat B+\hat T\le0$ forces $\hat B<0$, and on $\{\hat B\le 0,\ \hat B+\hat T>0\}$ the projection already returns $0$.  Without the second clause the ratio would be undefined at $\hat B+\hat T=0$, and the projection would return the \emph{maximal} correction $\hat t=1$ on $\{\hat B<-\hat T\}$, precisely where the bias estimate is most strongly negative.

Moreover $\hat T$ and $\mathrm{tr}(V_{\mathrm{tun}})$ (the trace of the tuning-noise covariance $V_{\mathrm{tun}}$ of Lemma~\ref{lem:tun-regularity} below) are built from the identical random trace $\mathrm{tr}(M_{1,\mathrm{tun}}\hat\Sigma_{\mathrm{tun}}M_{1,\mathrm{tun}}^\top)$, so their ratio is exactly $\hat\sigma^2_{\mathrm{tun}}/\sigma^2$; the consistency of $\hat T$ for $\mathrm{tr}(V_{\mathrm{tun}})$ therefore follows from $\hat\sigma^2_{\mathrm{tun}}\xrightarrow{p}\sigma^2$ alone, with no separate trace-concentration argument.

{The quantitative statements below require operator-norm control of the tuning-stage matrices, which the row-sum clause of the tuning-block Assumption~\ref{ass:JM} does not supply: the row-sum norm bounds only $\mnorm{M_{1,\mathrm{tun}}}$, and $\norm{\hat\Sigma_{\mathrm{tun}}}_{\mathrm{op}}$ is of order $\max(1,p/n_{\mathrm{tun}})$.  It is therefore stated as a separate condition.

\begin{assumption}[Tuning-stage operator-norm regularity]\label{ass:tun-opnorm}
$\norm{M_{1,\mathrm{tun}}}_{\mathrm{op}}+\norm{M^{(a)}_{1,\mathrm{tun}}}_{\mathrm{op}}+\norm{M^{(b)}_{1,\mathrm{tun}}}_{\mathrm{op}}=\Op(1)$.
\end{assumption}

\noindent For a symmetrised precision construction (CLIME-type), the column-sum norm equals the row-sum norm and Assumption~\ref{ass:tun-opnorm} follows from the row clause of Assumption~\ref{ass:JM} via $\norm{M}_{\mathrm{op}}\le\sqrt{\mnorm{M^\top}\,\mnorm{M}}$.}

\begin{lemma}[Tuning-block regularity]\label{lem:tun-regularity}
Under the tuning-block analogues of Assumptions~\ref{ass:design}, \ref{ass:rsc}, \ref{ass:sparsity}, and~\ref{ass:JM}{ and, for part~(b), Assumption~\ref{ass:tun-opnorm}}:
\begin{enumerate}
    \item[(a)] the scaled-Lasso variance estimator is consistent, $\hat\sigma^2_{\mathrm{tun}}\xrightarrow{p}\sigma^2$ \citep{sun2012scaled};
    \item[(b)] the tuning-block noise covariance $V_{\mathrm{tun}}:=\sigma^2 n_{\mathrm{tun}}^{-1}M_{1,\mathrm{tun}}\hat\Sigma_{\mathrm{tun}}M_{1,\mathrm{tun}}^\top$ has divergent effective rank{: $\mathrm{tr}(V_{\mathrm{tun}})/\norm{V_{\mathrm{tun}}}_{\mathrm{op}}\asymp \min(p,n_{\mathrm{tun}})\to\infty$, with $\mathrm{tr}(V_{\mathrm{tun}})\asymp p\sigma^2/n_{\mathrm{tun}}$ and, for $p\ge n_{\mathrm{tun}}$, $\norm{V_{\mathrm{tun}}}_{\mathrm{op}}\asymp\sigma^2 p/n_{\mathrm{tun}}^2$}.
\end{enumerate}
\end{lemma}

{Clause~(a) is a consequence of the maintained assumptions, and clause~(b) of those together with the operator-norm regularity of Assumption~\ref{ass:tun-opnorm}---automatic for a symmetrised (CLIME-type) precision construction, a mild separate condition otherwise. Beyond these, only the following balance condition on the tuning-block bias is additionally required.}

\begin{assumption}[Tuning-block bias balance]\label{ass:eff-rank-tun}
Let $\bm b_{\mathrm{tun}}^{(c)} := -M_{1,\mathrm{tun}}^{(c)}\hat\Sigma_{\mathrm{tun},c}\Delta$ be the bias component of tuning block $c\in\{a,b\}$ (Lemma~\ref{lem:C-decomposition}). We assume that neither block's squared bias energy dominates the sum of the cross-block inner product and the noise trace; that is, for a constant $C_{\mathrm{bal}}<\infty$ that does not depend on $n$,
\begin{equation}\label{eq:tun-balance}
    \norm{\bm b_{\mathrm{tun}}^{(c)}}_2^2 \;\le\; C_{\mathrm{bal}}\bigl(\bigl|\inner{\bm b_{\mathrm{tun}}^{(a)}}{\bm b_{\mathrm{tun}}^{(b)}}\bigr|+\mathrm{tr}(V_{\mathrm{tun}})\bigr)
    \quad\text{for both } c\in\{a,b\}.
\end{equation}
\end{assumption}

\begin{remark}[Interpretation of the balance condition]\label{rem:tun-balance-essence}
This is a regularity condition on the tuning blocks, distinct from and weaker than the inference-scale $\ell_2$ identification discussed in Remark~\ref{rem:t-hat-scope}. At the population level it is trivial: $M_{1,\mathrm{tun}}^{(c)}\Sigma\approx I_p$ gives $\bm b_{\mathrm{tun}}^{(c)}\approx-\Delta$ for both blocks, whence $\inner{\bm b_{\mathrm{tun}}^{(a)}}{\bm b_{\mathrm{tun}}^{(b)}}\approx\norm{\Delta}_2^2\approx\norm{\bm b_{\mathrm{tun}}^{(c)}}_2^2$. Its only content is that the finite-sample residual $\norm{(I_p-M_{1,\mathrm{tun}}^{(c)}\hat\Sigma_{\mathrm{tun},c})\Delta}_2$ not dominate, the sample-level shadow of that population identity; it is irreducible here because the $\ell_2$ residual control of Remark~\ref{rem:t-hat-scope} is unavailable.
\end{remark}

The tuning-scale proxy oracle is the tuning-block analogue of Proposition~\ref{prop:proxy-risk}: writing $B_{\mathrm{tun}} := \inner{\bm{b}_{\mathrm{tun}}^{(a)}}{\bm{b}_{\mathrm{tun}}^{(b)}}$ for the population bias inner product across the two tuning blocks (with $\bm{b}_{\mathrm{tun}}^{(c)}$ the bias component of $C_{\mathrm{tun}}^{(c)}$ given by Lemma~\ref{lem:C-decomposition}) and $V_{\mathrm{tun}}$ for the tuning-noise covariance of Lemma~\ref{lem:tun-regularity}, it is
\[
    t^\star_{\mathrm{tun}} := \Pi_{[0,1]}\!\left(\frac{B_{\mathrm{tun}}}{B_{\mathrm{tun}}+\mathrm{tr}(V_{\mathrm{tun}})}\right)
    \ \text{ on }\{B_{\mathrm{tun}}+\mathrm{tr}(V_{\mathrm{tun}})>0\},
    \qquad
    t^\star_{\mathrm{tun}}:=0
    \ \text{ otherwise},
\]
the same positive-part convention as \eqref{eq:t-hat}.

{The ratio defining $t^\star_{\mathrm{tun}}$ degenerates when $B_{\mathrm{tun}}$ approaches $-\mathrm{tr}(V_{\mathrm{tun}})$; the following condition formalises the non-degenerate-denominator requirement implicit in the consistency argument.

\begin{assumption}[Denominator separation]\label{ass:tun-sep}
There is a constant $c_0\in(0,1]$ such that, with probability tending to one,
\[
    \bigl|B_{\mathrm{tun}}+\mathrm{tr}(V_{\mathrm{tun}})\bigr|\;\ge\;c_0\bigl(|B_{\mathrm{tun}}|+\mathrm{tr}(V_{\mathrm{tun}})\bigr).
\]
\end{assumption}

\begin{lemma}[The separation condition in the named external regimes]\label{lem:tun-sep-derive}
Under the tuning-block analogues of Assumptions~\ref{ass:design}, \ref{ass:rsc}, and~\ref{ass:JM}, and Assumption~\ref{ass:external}, with $n_{\mathrm{tun}}\asymp n_0$ and $N=O(n_0)$, let $\delta_c:=(I_p-M^{(c)}_{1,\mathrm{tun}}\hat\Sigma_{\mathrm{tun},c})\Delta$, $c\in\{a,b\}$.  Then
$B_{\mathrm{tun}}\ge-\tfrac12(\norm{\delta_a}_2^2+\norm{\delta_b}_2^2)$ and
$\norm{\delta_c}_2\le\sqrt p\,\mu_{1,\mathrm{tun}}\norm{\Delta}_1$, so Assumption~\ref{ass:tun-sep} holds, with $c_0$ arbitrarily close to one, whenever
\[
    a_1^2\log p=o(\sigma^2).
\]
This scaling is automatic for a near-perfect external estimator ($a_1=\op(n_0^{-1/2})$); it is implied by the sparsity scaling $s\log p=o(\sqrt{n_0+N})$ of Theorem~\ref{thm:CLT} at the target-only Lasso rate $a_1\asymp s\sigma\sqrt{(\log p)/n_0}/\phi$; and it reads $s_A^2(\log p)^2=o(n_A\sigma^2)$ for an auxiliary-data estimator of effective size $n_A$.
\end{lemma}}

\begin{lemma}[Consistency of the data-driven shrinkage]\label{lem:t-hat-consistency}
Under the tuning-block analogues of Assumptions~\ref{ass:design}, \ref{ass:rsc}, and~\ref{ass:JM}, and Assumption{s}~{\ref{ass:tun-opnorm}, }\ref{ass:eff-rank-tun}{, and~\ref{ass:tun-sep}}, the data-driven shrinkage \eqref{eq:t-hat} is consistent for the tuning-scale proxy oracle, in conditional probability given the tuning covariates and $\hat\beta_{\mathrm{ext}}$:
\[
    \hat t \xrightarrow{p} t^\star_{\mathrm{tun}}.
\]
\end{lemma}

In particular, an asymptotically unbiased external estimator ($\Delta\to0$) drives $t^\star_{\mathrm{tun}}\to0$, so the Stage-1 one-step correction is switched off and the (already accurate) external estimator itself initialises the stacking, which still sharpens the debiased fit through the $N$ unlabelled rows (Corollary~\ref{cor:variance-channel}); a strongly biased one with $B_{\mathrm{tun}}\gg\mathrm{tr}(V_{\mathrm{tun}})$ instead drives $\hat t\xrightarrow{p}1$.

\begin{remark}[Scope of the consistency claim]\label{rem:t-hat-scope}
The selector $\hat t$ is consistent for the \emph{tuning-scale} proxy oracle $t^\star_{\mathrm{tun}}$, which we adopt as the operative selection target throughout.  Identifying $t^\star_{\mathrm{tun}}$ with the \emph{inference-scale} proxy oracle $t^\star$ of Proposition~\ref{prop:proxy-risk} would require strengthening the per-block entrywise bias control $|e_j^\top(I_p-M_{1,\mathrm{tun}}^{(c)}\hat\Sigma_{\mathrm{tun},c})\Delta|\le\mu_{1,\mathrm{tun}}\norm{\Delta}_1$ of Lemma~\ref{lem:bias-envelope} to the $\ell_2$ control $\norm{(I_p-M_{1,\mathrm{tun}}^{(c)}\hat\Sigma_{\mathrm{tun},c})\Delta}_2=\op(\norm{\Delta}_2)$; the only coordinate-wise route gives $\norm{\cdot}_2\le\sqrt p\,\mu_{1,\mathrm{tun}}\norm{\Delta}_1$, which requires $p\,s\,(\log p)/n_{\mathrm{tun}}\to0$, incompatible with the high-dimensional regime $p\gg n_0\ge n_{\mathrm{tun}}$.  The identification $t^\star_{\mathrm{tun}}\to t^\star$ is therefore left open, and the adaptivity guarantee is stated relative to $t^\star_{\mathrm{tun}}$.
\end{remark}

\begin{remark}[Full versus reproducible bias energy]\label{rem:full-vs-reproducible}
The distinction between $t^\star$ and $t^\star_{\mathrm{tun}}$ is one of \emph{which} bias energy enters the oracle, not of splitting as such. The inference-scale oracle $t^\star$ weighs the full squared bias energy $\norm{\bm b}_2^2$; the tuning-scale oracle $t^\star_{\mathrm{tun}}$ weighs the cross-block inner product $\inner{\bm b_{\mathrm{tun}}^{(a)}}{\bm b_{\mathrm{tun}}^{(b)}}$, the part of the bias that reproduces across the two independent tuning halves. The split is what makes the bias energy estimable at all: a single block gives $\E\norm{C}_2^2=\norm{\bm b}_2^2+\mathrm{tr}(V)$, contaminated by the noise trace and biased towards over-correction, whereas two independent halves retain only the reproducible component. The two oracles coincide when the bias $\ell_2$-concentrates, as it does in fixed dimension, and can separate only in the high-dimensional regime, the gap being exactly the quantity controlled by the balance condition of Assumption~\ref{ass:eff-rank-tun}.
\end{remark}

{The consistency of Lemma~\ref{lem:t-hat-consistency} sharpens to a rate.

\begin{proposition}[Estimation rate for the shrinkage selector]\label{prop:that-rate}
Suppose the hypotheses of Lemma~\ref{lem:t-hat-consistency} hold with separation constant $c_0$, that $n_{\mathrm{tun}}\asymp n_0$ with $N=O(n_0)$, and that $\hat\sigma^2_{\mathrm{tun}}$ is the scaled-Lasso estimate with universal penalty $\lambda_0=A\sqrt{2(\log p)/n_{\mathrm{tun}}}$, $A>1$.  Then
\[
    \bigl|\hat t-t^\star_{\mathrm{tun}}\bigr|
    =\Op\!\left(\frac{1}{c_0^{2}}\left[\sqrt{\frac{C_{\mathrm{bal}}}{\min(p,n_{\mathrm{tun}})}}
    +n_{\mathrm{tun}}^{-1/2}+\frac{s\log p}{n_{\mathrm{tun}}}\right]\right).
\]
If moreover $\min(p,n_{\mathrm{tun}})\gg\sqrt{n_0+N}$ (automatic in the regime $p\ge n_{\mathrm{tun}}$) and $s\log p=o(\sqrt{n_0+N})$ (hypothesis~(2) of Theorem~\ref{thm:CLT}), then
$\bigl|\hat t-t^\star_{\mathrm{tun}}\bigr|=\op\bigl((n_0+N)^{-1/4}\bigr)$.
\end{proposition}

\begin{proposition}[Reduction of the shrinkage-rate condition]\label{prop:condT-reduction}
Under the hypotheses of Proposition~\ref{prop:that-rate} and the scalings of its final claim ($\min(p,n_{\mathrm{tun}})\gg\sqrt{n_0+N}$ and $s\log p=o(\sqrt{n_0+N}\,)$), condition~(T) of Assumption~\ref{ass:condT} holds if and only if there is a constant $t_0\in[0,1]$ with
$|t^\star_{\mathrm{tun}}-t_0|=\op\bigl((n_0+N)^{-1/4}\bigr)$, and the limits then coincide.  Both $B_{\mathrm{tun}}$ and $\mathrm{tr}(V_{\mathrm{tun}})$ are functionals of the tuning covariates and of $\Delta$ alone, so the residual condition constrains only the external input and the design.
\end{proposition}

\begin{proposition}[The residual condition at the boundary regimes]\label{prop:condT-regimes}
Write $\rho_n:=B_{\mathrm{tun}}/\mathrm{tr}(V_{\mathrm{tun}})$.  Under the hypotheses of Proposition~\ref{prop:that-rate}, with $p\ge n_{\mathrm{tun}}$ and $s\log p=o(\sqrt{n_0+N}\,)$ (hypothesis~(2) of Theorem~\ref{thm:CLT}):
\begin{enumerate}
    \item[(i)] for a near-perfect external estimator ($a_1=\op(n_0^{-1/2})$), $t^\star_{\mathrm{tun}}=\Op(a_1^2\log p/\sigma^2)=\op\bigl((n_0+N)^{-1/4}\bigr)$, so the residual condition holds with $t_0=0$;
    \item[(ii)] at the target-only Lasso rate $a_1\asymp s\sigma\sqrt{(\log p)/n_0}/\phi$ it holds with $t_0=0$ under the additional scaling $s^2(\log p)^2=o\bigl((n_0+N)^{3/4}\bigr)$;
    \item[(iii)] for an auxiliary-data estimator of effective size $n_A$ it holds with $t_0=0$ provided $s_A^2(\log p)^2\,(n_0+N)^{1/4}=o(n_A\sigma^2)$;
    \item[(iv)] provided $P(\rho_n\ge0)\to1$, it holds with $t_0=1$ if and only if $\rho_n\gg(n_0+N)^{1/4}$, by the identity $1-t^\star_{\mathrm{tun}}=1/(1+\rho_n)$ on $\{\rho_n\ge0\}$;
    \item[(v)] for $t_0\in(0,1)$ it is equivalent to $|\rho_n-t_0/(1-t_0)|=\op\bigl((n_0+N)^{-1/4}\bigr)$, a stabilisation of the exogenous bias-to-noise ratio that the rate bound of Assumption~\ref{ass:external} does not imply; it is retained as an assumption in this regime.
\end{enumerate}
\end{proposition}

\begin{remark}[Scope of the discharge]\label{rem:condT-scope}
Propositions~\ref{prop:that-rate}--\ref{prop:condT-regimes} are stated under the tuning model of Lemma~\ref{lem:C-decomposition}, the scope of Lemma~\ref{lem:t-hat-consistency}.  Under the misspecified model of Section~\ref{sec:misspec} the architecture is unchanged after re-basing: the tuning residual contributes the covariate-measurable component $h^{(c)}:=n_{\mathrm{tun},c}^{-1}M^{(c)}_{1,\mathrm{tun}}X_{\mathrm{tun},c}^\top\eta(X_{\mathrm{tun},c})$, the conditional mean of $\hat B$ given the tuning covariates and $\hat\beta_{\mathrm{ext}}$ equals the cross-block inner product of the re-based biases $\tilde{\bm b}^{(c)}:=-M^{(c)}_{1,\mathrm{tun}}\hat\Sigma_{\mathrm{tun},c}\Delta+h^{(c)}$, $\Delta:=\hat\beta_{\mathrm{ext}}-\beta^\star_{\mathrm{proj}}$, exactly, and bounded conditional noise variance changes constants rather than rates in the variance accounting.  The corresponding extension of Proposition~\ref{prop:that-rate}, with $\hat\sigma^2_{\mathrm{tun}}$ recentred at the total residual variance, is left open, and Assumption~\ref{ass:condT} is retained for the misspecification analysis.
\end{remark}}


\section{Gaussian-design refinements}\label{sec:refine}

This section refines the asymptotic theory of Section~\ref{sec:theory-bias} under Gaussian-design structure on the covariates.  The general theory controls the residual bias $r(\hat t)$ through a max-entry JM
bound and an $\ell_1$-bound on $\Delta$.  Under Gaussian design one can sharpen this step
by exploiting the exact covariance structure.  The key point is that, when the target
precision matrix $\Omega:=\Sigma^{-1}$ is known, the residual takes the form
\[
    r(\hat t) = \Delta - \hat t\,\Omega\hat\Sigma_0\Delta
    = (1-\hat t)\Delta - \hat t\,\Omega(\hat\Sigma_0-\Sigma)\Delta,
\]
so that the stochastic part can be controlled in terms of the prediction norm
$\norm{\Sigma^{1/2}\Delta}_2$ rather than the cruder $\norm{\Delta}_1$.  We first record
the known-precision benchmark and then discuss the case in which $\Omega$ is estimated from
target-domain covariates by nodewise Lasso.

\begin{remark}[Discharge without precision-matrix sparsity]\label{rem:sparsityfree-discharge}
The variance-matching limit of Lemma~\ref{lem:m2-efficiency} is discharged there under row-sparsity of the precision matrix $\Omega=\Sigma^{-1}$.  That restriction can be removed at the price of a stronger design assumption.  Under a Gaussian random design---the rows of $(X_0;\tilde X)$ drawn i.i.d.\ $N(0,\Sigma)$---the estimated convex programme~\eqref{eq:JM-M2} attains the same limit $e_j^\top M_2\Sigma M_2^\top e_j\to(\Sigma^{-1})_{jj}$ with \emph{no} sparsity assumption on $\Omega$, provided the coefficient sparsity obeys $s=o\bigl((n_0+N)/(\log p)^2\bigr)$ \citep{javanmard2018debiasing}.  The row-sparsity of $\Omega$ required in Lemma~\ref{lem:m2-efficiency} is thereby traded for Gaussianity of the design together with a stronger coefficient-sparsity and sample-size requirement; under merely sub-Gaussian rows the row-sparse route of Lemma~\ref{lem:m2-efficiency}(B) remains the operative one.
\end{remark}

\subsection{Known precision matrix}

\begin{assumption}[Gaussian design with known precision]\label{ass:gaussian-known}
Assumption~\ref{ass:design} is strengthened to a Gaussian design: the rows of $(X_0;\tilde X)$ are
i.i.d.\ $N(0,\Sigma)$, where the eigenvalues of $\Sigma$ are bounded away from zero and
infinity.  In addition, the target precision matrix $\Omega:=\Sigma^{-1}$ is row-summable, with maximal $\ell_\infty$ row-sum bounded uniformly in $p$,
\[
    \mnorm{\Omega}=\max_{1\le j\le p}\sum_{k=1}^p|\Omega_{jk}|=O(1).
\]
This precision matrix is known and is used in both
debiasing steps, i.e.
\[
    M_2=\Omega,\qquad M_1=\Omega.
\]
Moreover the target sparsity obeys
\[
    {s=o\!\left(\frac{\sqrt{n_0}}{\log p}\right)}.
\]
\end{assumption}

\begin{remark}[Row-summability of the precision]
The row-summability $\mnorm{\Omega}=O(1)$, equivalently a bounded row-$\ell_1$ norm of the precision, is the standard regularity condition under which a known or estimated debiasing matrix admits the entrywise bias control of the Javanmard--Montanari construction \citep{javanmard2014confidence}.  It is not implied by the bounded-eigenvalue condition alone, and parallels the row-$\ell_1$ control $\mnorm{M_2}=\Op(1)$ that Assumption~\ref{ass:JM} supplies for the estimated debiasing matrix.
\end{remark}

\begin{remark}[On the sparsity scaling]\label{rem:gaussian-sparsity}
The bound $s=o(\sqrt{n_0}/\log p)$ is the $\ell_1$-driven requirement inherited from
condition~(ii) of Theorem~\ref{thm:CLT}; it is what the proof of
Corollary~\ref{cor:gaussian-known} establishes by the deterministic H\"older bound on the
debiasing remainder, exactly as in the generic theory.  The sharper Gaussian control of that
remainder through the prediction norm
$\norm{\Sigma^{1/2}(\hat\beta-\beta^\star)}_2=\Op(\sqrt{s/\phi}\,\lambda)$ would weaken the
requirement to $s=o(n_0/(\log p)^2)$, removing one factor of sparsity; we record this as a
conjecture rather than a theorem, because the requisite concentration step is not available
under the present (single-pool) pipeline (see Remark~\ref{rem:gaussian-sharper-sparsity} in the
proof of Corollary~\ref{cor:gaussian-known}, which records the cross-fitted route).
\end{remark}

\begin{proposition}[Residual-bias bound at known precision]\label{prop:gaussian-r}
Suppose the rows of $X_0$ are i.i.d.\ sub-Gaussian with the eigenvalues of $\Sigma$ bounded away from zero and infinity, that the Stage-1 matrix is $M_1=\Omega:=\Sigma^{-1}$ (as under Assumption~\ref{ass:gaussian-known}, whose Gaussian design is a special case), and that $\Delta$ is independent of $X_0$ (Assumption~\ref{ass:external-indep}).  Let $a_2$ be the prediction-norm rate of Assumption~\ref{ass:external}, a deterministic sequence with
\[
    \Pr\bigl(\norm{\Sigma^{1/2}\Delta}_2 \le a_2\bigr)\to 1.
\]
Then, with probability tending to one,
\[
    \norm{r(\hat t)}_\infty
    \le
    C\Bigl(1-\hat t + \sqrt{\tfrac{\log p}{n_0}}\Bigr)a_2,
    \qquad
    \norm{\Sigma^{1/2}r(\hat t)}_2
    \le
    C\Bigl(1-\hat t + \hat t\sqrt{\tfrac{p}{n_0}}\Bigr)a_2,
\]
for a constant $C$; on the same event, the same argument with $e_j$ in place of $\Omega e_j$ gives $\norm{(\hat\Sigma_0-\Sigma)\Delta}_\infty\le C\sqrt{\log p/n_0}\,a_2$.  Consequently, a sufficient condition for the composite residual-bias requirement~(i) of
Theorem~\ref{thm:CLT} is
\begin{equation}
    \sqrt{n_0+N}\,
    \Bigl(1-\hat t + \sqrt{\tfrac{\log p}{n_0}} + \hat t\sqrt{\tfrac{p\log p}{N n_0}}\Bigr)a_2\to 0
    \quad\text{in probability},
    \label{eq:gaussian-bias-cond}
\end{equation}
the last entry pricing the weighted-$\ell_2$ component of the envelope; it is dominated by $\sqrt{\log p/n_0}$ once $N\gtrsim p$, and is switched off in the limit $\hat t\to0$ when $\hat t$ carries a rate against the $p$-dependent factor (condition~(T) with $p\log p=O(Nn_0\sqrt{n_0+N})$ suffices).
\end{proposition}

The generic surrogate $b_n^{\mathrm{gen}}$ of \eqref{eq:bn-gen} is sharpened under the present design: with $a_2$ the population prediction-norm rate of Proposition~\ref{prop:gaussian-r},
\begin{equation}
    b_n^{\mathrm{G}}
    :=
    C_{\mathrm{G}}\,a_2
    \left(
        1-\hat t + \sqrt{\frac{\log p}{n_0}} + \hat t\sqrt{\frac{p\log p}{N n_0}}
    \right)
    \label{eq:bn-gauss-known}
\end{equation}
is a valid surrogate for a sufficiently large constant $C_{\mathrm{G}}$, in the sense of \eqref{eq:bn-surrogate}: the probability-tending-to-one form of Proposition~\ref{prop:gaussian-r} covers both entries of the composite envelope there, the weighted-$\ell_2$ entry carrying the weight $\sqrt{\log p/N}$ that produces the third term.

\begin{corollary}[Gaussian refinement of the main CLT]\label{cor:gaussian-known}
Suppose that Assumptions~\ref{ass:rsc}, \ref{ass:external},
\ref{ass:external-indep}, and \ref{ass:gaussian-known} hold, that $\hat t\xrightarrow{p}t_0\in[0,1]$ (Lemma~\ref{lem:t-hat-consistency}), that along the regime \eqref{eq:regime} the ratio $\kappa_n:=N/n_0$ converges to $\kappa\in[0,\infty)$, and that the precision-fluctuation condition
\[
    \hat t^{\,2}\,\frac{N\,p}{n_0\,(n_0+N)}\;\longrightarrow_{\mathbb P}\;0
\]
holds (for which $p=o(n_0)$ suffices when $t_0>0$, and at $t_0=0$ Assumption~\ref{ass:condT} together with $p=O(n_0^{3/2})$).  Let $a_2$ be as in Proposition~\ref{prop:gaussian-r} and $b_n=b_n^{\mathrm{G}}$ from \eqref{eq:bn-gauss-known}.  If
\eqref{eq:gaussian-bias-cond} holds and the stacked-Lasso tuning parameter obeys
\[
    \lambda \asymp
    \sigma\sqrt{\frac{\log p}{n_0+N}}
    +
    \frac{N}{n_0+N}\hat t\sqrt{\frac{\log p}{n_0}}
    +
    \frac{N}{n_0+N}b_n^{\mathrm{G}},
\]
then the conclusion of Theorem~\ref{thm:CLT} holds with the limiting variance identified: for each
fixed coordinate $j$,
\[
    \sqrt{n_0+N}\,(\tilde\beta_j-\beta_j^\star)\Rightarrow N(0,\tau_j^2),
    \qquad
    \tau_j^2:=\sigma^2\,\frac{(1+t_0\kappa)^2}{1+\kappa}\,(\Sigma^{-1})_{jj}.
\]
Moreover the first-order variance ratio against the target-only debiased Lasso is
\[
    G(\kappa,t_0)=\left(\frac{1+t_0\kappa}{1+\kappa}\right)^2 .
\]
\end{corollary}

Relative to the generic theory, the Gaussian refinement replaces the $\ell_1$-driven bias
cap by the sharper prediction-norm requirement \eqref{eq:gaussian-bias-cond}.  This is exactly the step that relaxes the auxiliary-sample-size threshold of Remark~\ref{rem:translasso} for high-quality external estimators.

\subsection{Estimated precision matrix by nodewise Lasso}

We now turn to the more realistic case in which $\Omega$ is unknown.  Let
\[
    X_{\Omega}:=
    \begin{pmatrix}
        X_0\\
        \tilde X
    \end{pmatrix}\in\R^{m_{\Omega}\times p},
    \qquad
    m_{\Omega}:=n_0+N,
\]
and estimate $\Omega$ from $X_{\Omega}$ by nodewise Lasso.

\begin{assumption}[Sub-Gaussian design with nodewise-Lasso precision]\label{ass:gaussian-nodewise}
Assumption~\ref{ass:design} is in force --- its noise clause unchanged, its design clause specialised as in Assumption~\ref{ass:gaussian-known} except that the rows are i.i.d.\ sub-Gaussian rather than Gaussian, with the eigenvalues of $\Sigma$ bounded away from zero and infinity --- and the precision matrix $\Omega:=\Sigma^{-1}$, now unknown, retains the row-summability of Assumption~\ref{ass:gaussian-known}:
\[
    \mnorm{\Omega}=O(1).
\]
Let $\hat\Omega$ be the nodewise-Lasso precision estimator of \citet{vandegeer2014asymptotic} built from $X_{\Omega}$: for each $j$, the $j$th row of $\hat\Omega$ is the normalised coefficient vector of the Lasso regression, with penalty $\lambda_{\mathrm{node}}\asymp\sqrt{\log p/m_{\Omega}}$, of the $j$th column of $X_{\Omega}$ on the remaining columns.  Let
\[
    s_{\Omega}:=
    \max_{1\le j\le p}
    \left|
        \{k\neq j:\Omega_{jk}\neq 0\}
    \right|
\]
denote the maximum row sparsity of $\Omega$.  The single pooled estimator $\hat\Omega$ is used both for the Stage-4 matrix $M_2$ and for the Stage-1 matrix $M_1$, in place of the Javanmard--Montanari matrices of \eqref{eq:JM-M1} and \eqref{eq:JM-M2}.  Assume
\[
    s_{\Omega}\frac{\log p}{\sqrt{m_{\Omega}}}\to 0.
\]
\end{assumption}

{The following corollary, promoted from the covariate-shift analysis of Section~\ref{sec:shift-aware}, benchmarks the row-$\ell_1$ amplification factor $\rho_j$ of Proposition~\ref{prop:shift-amp} at the oracle stacked precision --- a free-standing linear-algebra convention needing no distributional assumption --- and at its nodewise-Lasso estimate under this section's Gaussian design.}
\begin{corollary}[Amplification benchmark under oracle and nodewise precision]\label{cor:shift-amp-benchmark}
Take $M_2:=\Sigma_{\mathrm{stk}}^{-1}$, the oracle stacked precision.

\textup{(i)} If $\Sigma_{\mathrm u}=\Sigma_0$, then $\rho_j(M_2,\Sigma_{\mathrm u})=1$ exactly, for every $j$.

\textup{(ii)} If $\Sigma_{\mathrm u}\succ\Sigma_0$ in the Loewner sense, then $\rho_{j_0}(M_2,\Sigma_{\mathrm u})>1$ for at least one coordinate $j_0$.

Under the nodewise-Lasso refinement (Assumption~\ref{ass:gaussian-nodewise}), the empirical estimate $\hat M_2=\hat\Omega_{\mathrm{stk}}$ inherits both conclusions up to an $\op(1)$ remainder: $\rho_j(\hat M_2,\Sigma_{\mathrm u})=1+\op(1)$ under~(i), and $\rho_{j_0}(\hat M_2,\Sigma_{\mathrm u})>1$ persists for large $n_0,N$ under~(ii) at any coordinate $j_0$ where $(\Omega_{\mathrm{stk}}(\Sigma_{\mathrm u}-\Sigma_0))_{j_0j_0}$ is bounded away from zero.
\end{corollary}

The nodewise-Lasso rates of \citet{vandegeer2014asymptotic} require only i.i.d.\ sub-Gaussian rows and a restricted-eigenvalue bound, not Gaussianity.  Under Assumption~\ref{ass:gaussian-nodewise}, the nodewise-Lasso precision estimator contributes the usual additional Gaussian-design remainder of order
\[
    \delta_{\Omega,n}
    :=
    s_{\Omega}\frac{\log p}{\sqrt{m_{\Omega}}},
\]
while the external-initialiser contribution remains controlled by
\eqref{eq:gaussian-bias-cond}.  Sharpening the generic surrogate \eqref{eq:bn-gen} with this additional remainder gives
\begin{equation}
    b_n^{\mathrm{G},\Omega}
    :=
    C_{\mathrm{G},\Omega}\,a_2
    \left(
        1-\hat t + \sqrt{\frac{\log p}{n_0}} + \hat t\sqrt{\frac{p\log p}{N n_0}} + \delta_{\Omega,n}\Bigl(1+\sqrt{\tfrac{p\log p}{N}}\Bigr)
    \right)
    \label{eq:bn-gauss-node}
\end{equation}
for a sufficiently large constant $C_{\mathrm{G},\Omega}$, again in the sense of \eqref{eq:bn-surrogate}: the first three terms are those of $b_n^{\mathrm{G}}$, and the $\delta_{\Omega,n}$ entry carries the precision-estimation error through both components of the composite envelope, the crude $\sqrt p$ conversion pricing its weighted-$\ell_2$ part; both $p$-carrying factors reduce to constants once $N\gtrsim p\log p$.  This leads to the following extension.

\begin{corollary}[Gaussian refinement with estimated precision]\label{cor:gaussian-nodewise}
Suppose that Assumptions~\ref{ass:external}, \ref{ass:external-indep},
\ref{ass:gaussian-nodewise}, \ref{ass:rsc}, and the prediction-norm bias condition
\eqref{eq:gaussian-bias-cond} hold with $a_2$ as in Proposition~\ref{prop:gaussian-r}, that the ratio $\kappa_n=N/n_0$ converges to a finite
limit $\kappa\in(0,\infty)$, and that $\hat t\xrightarrow{p}t_0\in[0,1]$ (Lemma~\ref{lem:t-hat-consistency}).  Suppose further that conditions~(iii)--(iv) of Theorem~\ref{thm:CLT} hold at $M_2=M_1=\hat\Omega$, with the limit in~(iii) equal to the $\tau_j^2$ of Corollary~\ref{cor:gaussian-known}; Remark~\ref{rem:nodewise-iii-hypothesis} below explains why these enter as hypotheses here.  Write
\[
    \delta_{\Omega,n}^{\circ}:=s_{\Omega}\,\frac{\log p}{\sqrt{n_0+N}}
\]
for the nodewise precision-estimation rate (so that $\delta_{\Omega,n}=\delta_{\Omega,n}^{\circ}$; below, $\delta_{\Omega,n}^{\circ}$ also serves as a loose common upper bound for the row-$\ell_1$ precision error, which is smaller by a factor $\sqrt{\log p}$; $\delta_{\Omega,n}^{\circ}\to0$ already holds under Assumption~\ref{ass:gaussian-nodewise}).  If
\[
    s=o\!\left(\frac{\sqrt{n_0}}{\log p}\right),
    \qquad
    \sqrt{n_0+N}\,a_2\,\delta_{\Omega,n}^{\circ}\Bigl(1+\sqrt{\tfrac{p\log p}{N}}\Bigr)\to 0,
\]
and the stacked-Lasso tuning parameter obeys
\[
    \lambda \asymp
    \sigma\sqrt{\frac{\log p}{n_0+N}}
    +
    \frac{N}{n_0+N}\hat t\sqrt{\frac{\log p}{n_0}}
    +
    \frac{N}{n_0+N}b_n^{\mathrm{G},\Omega},
\]
then the conclusion of Corollary~\ref{cor:gaussian-known} --- the central limit theorem with limiting variance $\tau_j^2$ and the ratio $G(\kappa,t_0)$ --- continues to hold with the
surrogate $b_n=b_n^{\mathrm{G},\Omega}$ from \eqref{eq:bn-gauss-node}.
\end{corollary}

\begin{remark}[Why (iii)--(iv) are hypotheses at estimated precision]\label{rem:nodewise-iii-hypothesis}
In Corollary~\ref{cor:gaussian-known} conditions (iii)--(iv) are \emph{verified}: the weight vector there is built from the deterministic $\Omega$ and the unlabelled block, so, conditionally on the tuning sample and the unlabelled covariates, its fluctuation can be integrated out against the labelled rows.  That device is unavailable here: $\hat\Omega$ is computed from the pooled covariate stack $X_\Omega$, which contains the labelled rows $X_0$ and the stacking block itself, so no conditioning separates the weight vector from the quadratic form it is paired with --- the same single-pool reuse obstruction recorded in Remark~\ref{rem:gaussian-sharper-sparsity} for the prediction-norm route.  A nodewise fit cross-fitted on a covariate block reserved from Stages~3--4 would restore the verification at an additional $\Op(\delta_{\Omega,n}^{\circ})$ replacement cost; we do not pursue this, and instead carry (iii)--(iv) as hypotheses, exactly as Theorem~\ref{thm:CLT} itself does for the generic Javanmard--Montanari construction.
\end{remark}

Nodewise Lasso is an $X$-only step: the precision-estimation requirement depends on the total number of target-domain covariates $m_{\Omega}=n_0+N$, not on the number of labelled responses.  A convenient sufficient condition is
\begin{equation}
    n_0+N \gg s_{\Omega}^2(\log p)^2.
    \label{eq:gaussian-N-cond}
\end{equation}
The Gaussian refinement therefore separates the demand on the external estimator from the demand on the debiasing matrix: the initialiser-bias condition is governed by the auxiliary size (for a concrete TransLasso initialiser, the thresholds of Remark~\ref{rem:translasso}), whereas the nodewise step adds the precision-estimation requirement \eqref{eq:gaussian-N-cond}, governed by the pooled covariate count $n_0+N$ alone.  Only the latter is specific to the estimated-precision setting of this subsection.

\subsection{Implementation under Gaussian design}\label{app:gauss-implementation}

Under Gaussian design, the overall four-stage pipeline remains unchanged, except that the practical selection of \(N_{\mathrm{op}}\) should be based on the admissibility cap \(N_{\max}^{\mathrm{G}}\).  The construction of the debiasing matrices then proceeds as follows.  If the target precision matrix \(\Omega=\Sigma^{-1}\) is known, set \(M_2=\Omega\) and \(M_1=\Omega\).  If \(\Omega\) is unknown, estimate it from the target-domain covariates by nodewise Lasso, using
\[
    X_{\Omega}:=\begin{pmatrix}X_0\\ \tilde X\end{pmatrix}\in\R^{(n_0+N)\times p},
\]
where \((\tilde X,N)\) denotes the selected unlabelled subset from Step~4.  In that case one also checks the lower requirement
\[
    n_0+N \gtrsim s_{\Omega}^2(\log p)^2
\]
for stable nodewise-Lasso precision estimation.  In this regime the requirement supplies a lower endpoint for the admissible interval, so the variance-balance grid~\eqref{eq:Nstar-empirical} is restricted to sizes that meet it.  If this lower requirement fails, or if it exceeds the upper admissible size from the admissibility cap, then no admissible unlabelled size supports a theoretically justified augmentation and the procedure reverts to target-only debiased Lasso.  The corresponding Stage-3 tuning parameter should then use the Gaussian surrogate \(b_n^{\mathrm{G}}\) or \(b_n^{\mathrm{G},\Omega}\) in place of the generic bound.

\subsection{TransLasso auxiliary-sample requirements}\label{app:supp}\label{app:supp-translasso}

\begin{remark}[TransLasso as a concrete external estimator]\label{rem:translasso}
A TransLasso-type estimator \citep{li2022transfer} trained on auxiliary labelled data (independent of the
inference-stage target responses) is one instance of the linear external estimator of
Assumption~\ref{ass:external}, and the analysis of Section~\ref{sec:theory-bias} applies with no modification: only a
high-probability $\ell_1$ error bound $\Pr(\norm{\Delta_{\mathrm{TL}}}_1\le a_{\mathrm{TL}})\to1$,
$\Delta_{\mathrm{TL}}:=\hat\beta_{\mathrm{ext}}-\beta^\star$, is required.  By
Lemma~\ref{lem:bias-envelope}, the $\ell_\infty$ entry of the bias requirement (i) of Theorem~\ref{thm:CLT} is controlled once
\begin{equation}
    \sqrt{n_0+N}\,\bigl(\bar\mu_1 + (1-\hat t)(1+\bar\mu_1)\bigr)\,a_{\mathrm{TL}} \to 0,
    \label{eq:translasso-bias-cond}
\end{equation}
with the weighted-$\ell_2$ entry of that requirement covered on the same event once $N\gtrsim p\log p$ (Lemma~\ref{lem:bias-envelope}'s weighted term carries the additional factor $\sqrt{p\log p/N}$),
so a TransLasso initialiser delivers the strict first-order improvement $G(\kappa,t_0)<1$ of
Corollary~\ref{cor:idealized-efficiency} precisely when its coefficient error is small enough
to admit a shrinkage level bounded away from one.  For the canonical rate
$a_{\mathrm{TL}}\asymp s\sqrt{(\log p)/(n_0+n_A)}$ with effective auxiliary size $n_A$, the
idealised benchmark $\hat t\to0$ requires $n_0+n_A \gg s^2(n_0+N)\log p$; under the Gaussian
design of Section~\ref{sec:refine}, where the prediction-norm control of
Proposition~\ref{prop:gaussian-r} replaces the $\ell_1$ cap, this sharpens to
$n_0+n_A \gg s(n_0+N)\log p$, removing one factor of sparsity.  The auxiliary-sample algebra follows.
\end{remark}

Up to constants, $\bar\mu_1\asymp\sqrt{(\log p)/n_0}$ (the in-sample Javanmard--Montanari tolerance of Assumption~\ref{ass:JM}), and three benchmark consequences of the bias condition \eqref{eq:translasso-bias-cond} are then immediate under the canonical TransLasso rate $a_{\mathrm{TL}}\asymp s\sqrt{\log p/(n_0+n_A)}$.
\begin{enumerate}
    \item If $\hat t\to 1$ in probability \emph{with the rate} $\sqrt{n_0+N}\,(1-\hat t)\,a_{\mathrm{TL}}=\op(1)$ (a requirement on $1-\hat t$ that consistency alone does not supply; the envelope of Lemma~\ref{lem:bias-envelope} retains the term $(1-\hat t)(1+\bar\mu_1)a_{\mathrm{TL}}$), it suffices that additionally $\sqrt{n_0+N}\,\bar\mu_1\,a_{\mathrm{TL}}\to 0$, equivalently $s(\log p)\sqrt{n_0+N}/\sqrt{n_0(n_0+n_A)}\to 0$.
    \item For the idealised lower-variance benchmark $\hat t\to 0$, it suffices that $\sqrt{n_0+N}\,a_{\mathrm{TL}}\to 0$, which reduces to $n_0+n_A \gg s^2 (n_0+N)\log p$.
    \item Under the Gaussian design of Section~\ref{sec:refine}, the $\ell_1$ cap of Lemma~\ref{lem:bias-envelope} is replaced by the prediction-norm control \eqref{eq:gaussian-bias-cond} of Proposition~\ref{prop:gaussian-r}, with $\norm{\Sigma^{1/2}\Delta_{\mathrm{TL}}}_2=\Op(a_2)$ at the slow-rate companion $a_2\asymp\sqrt{s\log p/(n_0+n_A)}$.  At the idealised benchmark $\hat t\to 0$ the term $\sqrt{(\log p)/n_0}\,a_2$ is dominated and the $\hat t$-carried third term of \eqref{eq:gaussian-bias-cond} is switched off (outright once $N\gtrsim p$; under the rate of condition~(T) it additionally needs the $p$-versus-$n_0$ side condition $p\log p=O(Nn_0\sqrt{n_0+N})$, automatic in the regime of Corollary~\ref{cor:gaussian-known}), so \eqref{eq:gaussian-bias-cond} reduces to $\sqrt{n_0+N}\,a_2\to 0$, equivalently $n_0+n_A \gg s (n_0+N)\log p$, removing one factor of sparsity relative to item~2.
\end{enumerate}
Combining with the idealised variance ratio of Corollary~\ref{cor:idealized-efficiency}, the method matches the target-only debiased-Lasso variance at first order when $\hat t\to1$, attains $G(\kappa,t_0)<1$ when $\hat t\to t_0\in(0,1)$, and reaches the oracle benchmark $(1+\kappa)^{-2}$ as $\hat t\to0$.

\section{Extensions of the procedure}\label{app:extensions}

Each extension below is stated at the projection parameter $\beta^\star_{\mathrm{proj}}$ and in the notation of Section~\ref{sec:robust} of the main text.

\subsection{The control-variate hybrid: proofs and variants}\label{app:hybrid-supp}\label{app:hybrid}

The control-variate hybrid is stated and motivated in Section~\ref{sec:hybrid} of the main text---the construction on the control-variated response $y_0^{\mathrm{cv}}=y_0-\hat\omega_j\hat\nu$, the validity-and-variance Proposition~\ref{prop:hybrid-var}, and the uniform-in-$N$ dominance Corollary~\ref{cor:hybrid-dom}.  This section supplies their proofs, a scalar-weight variant, and an adaptivity statement; the numerical illustration is reported in Section~\ref{sec:experiments-hybrid-main} of the main text, its protocol in Section~\ref{app:simdetails}.

\noindent\textbf{Construction (control-variated procedure).} For inference on coordinate $j$, run Algorithm~\ref{alg:deal} unchanged, with the labelled response replaced by $y_0^{\mathrm{cv}} = y_0 - \hat\omega_j\,\hat\nu(X,W)$, where $\hat\nu = \hat\mu(X,W) - X^\top\hat\beta^{\mathrm{lin}}_{\mathrm{ext}}$. Pseudo-labels on the unlabelled block are formed from the linearised coefficient $\hat\beta^{\mathrm{lin}}_{\mathrm{ext}}$, never from the raw prediction $\hat\mu$, exactly as in the pure procedure; their bias-aware correction $\hat t\,C$ (Stage~1) is, like every other stage, computed on the control-variated response. Writing, for each tuning-sample unit $i$, $\hat e_i:=y_i-X_i^\top\hat\beta^{\mathrm{lin}}_{\mathrm{ext}}$ for the labelled residual at the linearised coefficient, $\hat\nu_i:=\hat\mu(X_i,W_i)-X_i^\top\hat\beta^{\mathrm{lin}}_{\mathrm{ext}}$ for the control-variate residual, and $\widehat h_i:=(X_i^\top M_1^\top e_j)^2$ for the tuning-sample surrogate of the weighting implicit in $\omega_j^{\mathrm{cv}}$'s defining ratio below, estimate the coordinate weight by the tuning-sample analogue
\begin{equation}\label{eq:omega-hat-j}
    \hat\omega_j \;:=\; \frac{\sum_{i\in\mathrm{tun}} \widehat h_i\,\hat e_i\,\hat\nu_i}{\sum_{i\in\mathrm{tun}} \widehat h_i\,\hat\nu_i^2},
\end{equation}
with the convention $\hat\omega_j:=0$ if the denominator is zero. This adds one requirement to Assumption~\ref{ass:aux-eval}: the model must be evaluable on the labelled inference block and on the tuning block, so that $\hat\nu$ can be formed on both; absent this the construction reduces to the pure procedure.

\begin{proof}[Proof of Proposition~\ref{prop:hybrid-var}]

\emph{Step 0 ($\hat\omega_j$ meets the proposition's hypotheses).} The estimator \eqref{eq:omega-hat-j} is a function of the tuning sample, the external fit $\hat\beta_{\mathrm{ext}}^{\mathrm{lin}}$, and, through the weights $\widehat h_i$, the labelled-block covariates entering $M_1$; conditional on the covariates and the tuning sample it is therefore fixed, and independent of the inference-stage noise $\{\varepsilon_{0,i}\}$ (Assumption~\ref{ass:external-indep}) --- the measurability requirement of the proposition, the same status $\hat t$ and $A_n$ hold in the proof of Theorem~\ref{thm:CLT}.

For consistency, write $m_{1j}:=M_1^\top e_j$, $\Delta_{\mathrm{lin}}:=\hat\beta_{\mathrm{ext}}^{\mathrm{lin}}-\beta^{\mathrm{proj}}_0[\hat\mu]$, and $\Delta:=\hat\beta_{\mathrm{ext}}^{\mathrm{lin}}-\beta^\star_{\mathrm{proj}}$, so that on the tuning block $\hat e_i=e_i-X_i^\top\Delta$ and $\hat\nu_i=\nu_i-X_i^\top\Delta_{\mathrm{lin}}$, with $\norm{\Delta}_2\to_{\mathbb P}0$ (Lemma~\ref{lem:lin-rate}) and $\norm{\Delta_{\mathrm{lin}}}_2\to_{\mathbb P}0$ (its Lasso-estimation component). Conditionally on $(\hat\beta_{\mathrm{ext}}^{\mathrm{lin}},M_1)$ the summands of \eqref{eq:omega-hat-j} are i.i.d.\ across $i\in\mathrm{tun}$, and every factor appearing below --- $X_i^\top m_{1j}$, $e_i$, $\nu_i$, $X_i^\top\Delta$, $X_i^\top\Delta_{\mathrm{lin}}$ --- is sub-Gaussian ($e=\eta+\varepsilon$ by Assumptions~\ref{ass:nonlinear} and~\ref{ass:misspec-reg}, $\nu[\hat\mu]$ by the clause imposed in the proposition, linear forms $X^\top v$ by Assumption~\ref{ass:design}(a)), so all the moments used below are finite. The weak law of large numbers at the fixed weight vector $m_{1j}$ gives $n_{\mathrm{tun}}^{-1}\sum_{i\in\mathrm{tun}}(X_i^\top m_{1j})^2e_i\nu_i\to_{\mathbb P}\E[(X^\top m_{1j})^2e\nu]$ and its $\nu^2$ analogue. Two $\op(1)$ substitutions identify the limits. Replacing $(\hat e_i,\hat\nu_i)$ by $(e_i,\nu_i)$: each substitution error is an average of products carrying at least one factor $X_i^\top\Delta$ or $X_i^\top\Delta_{\mathrm{lin}}$, so Cauchy--Schwarz against $\E[(X^\top v)^2]\le\lambda_{\max}(\Sigma_0)\norm{v}_2^2$ and the finite higher moments of the remaining factors bound it by a constant multiple of $\norm{\Delta}_2\vee\norm{\Delta_{\mathrm{lin}}}_2\to_{\mathbb P}0$. Replacing $m_{1j}$ by $\Sigma_0^{-1}e_j$: $(X^\top m_{1j})^2-(X^\top\Sigma_0^{-1}e_j)^2=(X^\top(m_{1j}-\Sigma_0^{-1}e_j))(X^\top(m_{1j}+\Sigma_0^{-1}e_j))$, and $\norm{m_{1j}-\Sigma_0^{-1}e_j}_2\to_{\mathbb P}0$ by the same row-sparse precision-consistency route Assumption~\ref{ass:rowsparse} supplies for $M_2$ --- the identical convex-programme or nodewise-Lasso construction, applied to the labelled-block precision $\Sigma_0^{-1}$ in place of the stacked one --- so the same Cauchy--Schwarz argument gives an $\op(1)$ cost. The numerator and denominator averages of \eqref{eq:omega-hat-j} therefore converge in probability to $e_j^\top\Sigma_0^{-1}\Gamma_{e\nu}\Sigma_0^{-1}e_j$ and $e_j^\top\Sigma_0^{-1}\Gamma_\nu\Sigma_0^{-1}e_j$, and the continuous-mapping theorem applied to the ratio (or the shared convention at a vanishing limiting denominator) gives $\hat\omega_j\to_{\mathbb P}\omega_j^{\mathrm{cv}}$, a fixed finite constant, hence $\hat\omega_j=\Op(1)$.

\emph{Step 1 (the idealised construction).} Consider first $y_0^{\mathrm{cv,pop}}:=y_0-\omega_j^{\mathrm{cv}}\nu[\hat\mu]$, using the population weight and the population residual, run through Stages~1--4 at the same deployed shrinkage level $\hat t$ and selected budget $\hat N^*$ as the deployed estimator --- $\hat t$ enters the machinery only as a tuning-measurable scalar, and no hypothesis of Corollary~\ref{cor:CLT-nonlin-labeler} requires it to have been selected on the residual the estimator is run on. Because $\E_{P_0}[X\,\nu[\hat\mu]]=0$ and $\E_{P_0}[X\,\eta]=0$ (the first-order conditions of the projections \eqref{eq:lab-projection} and \eqref{eq:beta-proj}), the fixed linear combination $\eta-\omega\nu[\hat\mu]$ remains orthogonal to $X$ in population for every fixed $\omega$: subtracting $\omega\nu[\hat\mu]$ from $y_0$ moves neither the projection estimand $\beta^\star_{\mathrm{proj}}$ nor any term of the expansion of Corollary~\ref{cor:CLT-nonlin-labeler} other than the labelled residual $e\mapsto e-\omega\nu[\hat\mu]$. The effective residual $\eta^{\mathrm{cv}}:=\eta-\omega_j^{\mathrm{cv}}\nu[\hat\mu]$ is sub-Gaussian, with $\norm{\eta^{\mathrm{cv}}}_{\psi_2}\le\norm{\eta}_{\psi_2}+|\omega_j^{\mathrm{cv}}|\,\norm{\nu[\hat\mu]}_{\psi_2}<\infty$ by the Orlicz-norm triangle inequality, Assumption~\ref{ass:misspec-reg}, the sub-Gaussianity of $\nu[\hat\mu]$ imposed above, and $\omega_j^{\mathrm{cv}}$ fixed and finite; likewise $\{X_j\eta^{\mathrm{cv}}\}_{j=1}^p$ is sub-exponential, its proxy bounded by the triangle inequality applied to Assumption~\ref{ass:misspec-reg} and Assumption~\ref{ass:lin-proj}(b). Assumptions~\ref{ass:nonlinear} and~\ref{ass:misspec-reg} therefore transfer verbatim to the pair $(y_0^{\mathrm{cv,pop}},\eta^{\mathrm{cv}})$ in place of $(y_0,\eta)$. The remaining hypothesis of Corollary~\ref{cor:CLT-nonlin-labeler}, Assumption~\ref{ass:condT}, is a rate condition on whatever shrinkage level the tuning-sample selector of Section~\ref{sec:proxy-shrinkage} returns; its content is agnostic to which residual the selector is fed, so we impose it for the level selected on the control-variated tuning residual, with limit $t_0^{\mathrm{cv}}\in[0,1]$ --- not necessarily the pure procedure's own $t_0$, the separate question Remark~\ref{rem:hybrid-shared-t0} addresses. Corollary~\ref{cor:CLT-nonlin-labeler} therefore applies at $\beta^\star_{\mathrm{proj}}$ to the estimator run on $y_0^{\mathrm{cv,pop}}$ at this shrinkage level.

The sandwich middle is therefore $\Gamma_0^{\mathrm{cv}}(\omega) = \Gamma_0 - 2\omega\,\Gamma_{e\nu} + \omega^2\,\Gamma_\nu$. The scalar $e_j^\top\Sigma_0^{-1}\Gamma_0^{\mathrm{cv}}(\omega)\Sigma_0^{-1}e_j$ is a convex quadratic in $\omega$---strictly convex when $e_j^\top\Sigma_0^{-1}\Gamma_\nu\Sigma_0^{-1}e_j>0$, and constant in $\omega$ (trivially minimised everywhere, in particular at $\omega_j^{\mathrm{cv}}:=0$ by the stated convention) otherwise---minimised at $\omega_j^{\mathrm{cv}}$ with value $(e_j^\top\Sigma_0^{-1}\Gamma_0\Sigma_0^{-1}e_j)\,(1-\rho_j^2[\hat\mu])$. That $\rho_j^2[\hat\mu]$ is the ratio \eqref{eq:ppipp-nonlin-ratio}: $g_l = Xe$ gives $\Gamma_0 = \Var_{P_0}(g_l)$, and $g_l^{\hat Y} = X\nu[\hat\mu] + XX^\top\delta_{\mathrm{lin}}[\hat\mu]$ has its in-span term asymptotically negligible, the rate condition \eqref{eq:rate-cond-nonlin} driving $\delta_{\mathrm{lin}}[\hat\mu]\to 0$ (Assumption~\ref{ass:lin-proj}), so $\Cov_{P_0}(g_l,g_l^{\hat Y}) = \Gamma_{e\nu}$ and $\Var_{P_0}(g_l^{\hat Y}) = \Gamma_\nu$ in the limit and the minimised quadratic is $(e_j^\top\Sigma_0^{-1}\Gamma_0\Sigma_0^{-1}e_j)(1-\rho_j^2[\hat\mu])$. Cauchy--Schwarz gives $\rho_j^2[\hat\mu]\in[0,1]$, hence the inequality in \eqref{eq:hybrid-var}; multiplying by the bias-aware normalisation of Corollary~\ref{cor:CLT-nonlin-labeler} yields, for the idealised estimator $\tilde\beta_j^{\mathrm{cv,pop}}$ run on $y_0^{\mathrm{cv,pop}}$,
\[
    \sqrt{n_0+N}\,\bigl(\tilde\beta_j^{\mathrm{cv,pop}}-\beta^\star_{\mathrm{proj},j}\bigr)\overset{\mathrm d}{\longrightarrow}\mathcal N\bigl(0,\tau_j^{2,\mathrm{cv}}\bigr).
\]
For the equality condition and the linear-predictor reduction: when $\rho_j^2[\hat\mu]=0$ with $\nu[\hat\mu]\not\equiv0$, the weighted covariance $e_j^\top\Sigma_0^{-1}\Gamma_{e\nu}\Sigma_0^{-1}e_j$ vanishes while $e_j^\top\Sigma_0^{-1}\Gamma_\nu\Sigma_0^{-1}e_j>0$, so $\omega_j^{\mathrm{cv}}=0$ and $\Gamma_0^{\mathrm{cv}}=\Gamma_0$; for an exactly linear predictor $\nu[\hat\mu]\equiv0$; under exact linearisation recovery ($\lambda^{\mathrm{lin}}=0$ on a full-rank linearisation block) $\hat\nu\equiv0$, $y_0^{\mathrm{cv}}=y_0$ identically, and the estimator coincides with DEAL for any weight, while under the penalised linearisation $\hat\nu=X^\top(\beta^{\mathrm{proj}}_0[\hat\mu]-\hat\beta^{\mathrm{lin}}_{\mathrm{ext}})$ is purely in-span --- the case covered, with $\nu\equiv0$, by the in-span analysis of Step~2.

\emph{Step 2 (from the idealised to the deployed estimator).} The deployed response is $y_0^{\mathrm{cv}}=y_0-\hat\omega_j\hat\nu$ with $\hat\nu=\nu[\hat\mu]-X^\top\Delta_{\mathrm{lin}}$ for $\Delta_{\mathrm{lin}}$ as in Step~0, the linearisation's own estimation error---independent of the inference-stage data by Assumption~\ref{ass:external-indep}, of order $\norm{\Delta_{\mathrm{lin}}}_2=\Op(\psi_*\sqrt{s_{\mathrm{lin}}\log p/n_{\mathrm{lin}}})$ by the proof of Lemma~\ref{lem:lin-rate} (its first, Lasso-estimation term). Decompose
\[
    y_0^{\mathrm{cv}} - y_0^{\mathrm{cv,pop}}
    = (\omega_j^{\mathrm{cv}}-\hat\omega_j)\,\nu[\hat\mu]
    + \hat\omega_j\,X^\top\Delta_{\mathrm{lin}},
\]
and note the imputation-bias (Term~B) contribution of Corollary~\ref{cor:CLT-nonlin-labeler}'s proof is identical for the two responses: it is driven by $r(\hat t)=(I_p-\hat t\,M_1\hat\Sigma_0)\Delta$ (Lemma~\ref{lem:oof-decomposition}), with $\Delta$ as in Step~0 --- response-free given $\hat t$ and the covariates --- while the response enters the pseudo-labels only through the bias-aware correction $\hat t\,C$, the stochastic $\hat tZ_1$ channel of that lemma, score content that the factor $A_n$ carries and the two summands below account for. Term~B is therefore identical for $y_0^{\mathrm{cv}}$ and $y_0^{\mathrm{cv,pop}}$ at the shared $\hat t$ of Step~1.

The first summand is, conditional on the covariates and the tuning sample (under which $\hat\omega_j$ is fixed and independent of the inference-stage noise, Step~0), a fixed scalar multiple of $\nu[\hat\mu]$; by the Step~1 orthogonality argument it perturbs only the labelled residual, not the projection estimand, so the estimator run on $y_0-\hat\omega_j\nu[\hat\mu]$ is asymptotically linear at $\beta^\star_{\mathrm{proj}}$ with sandwich middle $\Gamma_0^{\mathrm{cv}}(\hat\omega_j)$; since $\hat\omega_j\overset{\mathbb P}{\to}\omega_j^{\mathrm{cv}}$ (Step~0) and $\omega\mapsto\Gamma_0^{\mathrm{cv}}(\omega)$ is continuous, $\Gamma_0^{\mathrm{cv}}(\hat\omega_j)\overset{\mathbb P}{\to}\Gamma_0^{\mathrm{cv}}$ and the achieved variance is Slutsky-consistent for $\tau_j^{2,\mathrm{cv}}$; this term affects only the achieved variance, not the centring, at any rate of consistency of $\hat\omega_j$.

The second summand, $\hat\omega_j X^\top\Delta_{\mathrm{lin}}$, lies in the span of $X$: $\E_{P_0}[X\cdot X^\top\Delta_{\mathrm{lin}}]=\Sigma_0\Delta_{\mathrm{lin}}\neq0$ in general, so it shifts the population projection coefficient of the labelled-block response to $\beta^\star_{\mathrm{proj}}+\hat\omega_j\Delta_{\mathrm{lin}}$ exactly, conditional on $(\hat\omega_j,\Delta_{\mathrm{lin}})$. The leading score of Corollary~\ref{cor:CLT-nonlin-labeler}'s proof, evaluated at this shifted coefficient, changes by exactly $\frac{n_0}{n_0+N}e_j^\top M_2A_n\hat\Sigma_0\,\hat\omega_j\Delta_{\mathrm{lin}}$, a purely algebraic consequence of the score's linearity in the response. The stacked-Lasso remainder (Term~A) is controlled by the same oracle-inequality route as Theorem~\ref{thm:stacked-lasso}, now centred at $\beta^\star_{\mathrm{proj}}+\hat\omega_j\Delta_{\mathrm{lin}}$ rather than the $s$-sparse $\beta^\star_{\mathrm{proj}}$; this shifted centre is within $\ell_1$-distance $|\hat\omega_j|\,\norm{\Delta_{\mathrm{lin}}}_1=\Op\bigl(s_{\mathrm{lin}}\sqrt{\log p/n_{\mathrm{lin}}}\bigr)$ of it ($\hat\omega_j=\Op(1)$ by Step~0; the $\ell_1$ rate is the companion to Lemma~\ref{lem:lin-rate}'s $\ell_2$ bound, from the same basic inequality \citep[Theorem~6.1]{buhlmann2011statistics}), which a standard approximate-sparsity extension of the stacked-Lasso oracle inequality absorbs at cost $O(\mu_2\,|\hat\omega_j|\,\norm{\Delta_{\mathrm{lin}}}_1)$ in Term~A, for $\mu_2\asymp\sqrt{\log p/(n_0+N)}$ the tolerance at which \eqref{eq:JM-M2} is tuned; we invoke this extension rather than re-deriving it. At the inference scale this cost is $\sqrt{n_0+N}\,\mu_2\,|\hat\omega_j|\,\norm{\Delta_{\mathrm{lin}}}_1=\Op\bigl(\sqrt{s_{\mathrm{lin}}\log p}\cdot\sqrt{s_{\mathrm{lin}}\log p/n_{\mathrm{lin}}}\bigr)=\op(1)$: the second factor is $o\bigl((n_0+N)^{-1/2}\bigr)$ by the rate condition \eqref{eq:rate-cond-nonlin}, and the first is $o\bigl((n_0+N)^{1/4}\bigr)$ since $s_{\mathrm{lin}}=O(s)$ (Assumption~\ref{ass:lin-proj}(b)) and $s\log p=o(\sqrt{n_0+N})$ (regime condition~(ii) of Theorem~\ref{thm:CLT} at the tuned rates $\mu_2,\lambda\asymp\sqrt{\log p/(n_0+N)}$). Combining,
\[
    \sqrt{n_0+N}\,\bigl(\tilde\beta_j^{\mathrm{cv}} - \beta^\star_{\mathrm{proj},j} - \hat\omega_j(\Delta_{\mathrm{lin}})_j\bigr) \overset{\mathrm d}{\longrightarrow} \mathcal N\bigl(0,\tau_j^{2,\mathrm{cv}}\bigr).
\]
Since $|\hat\omega_j(\Delta_{\mathrm{lin}})_j|\le|\hat\omega_j|\,\norm{\Delta_{\mathrm{lin}}}_2$ and $\hat\omega_j=\Op(1)$ (Step~0), $\sqrt{n_0+N}\,|\hat\omega_j|\,\norm{\Delta_{\mathrm{lin}}}_2=\Op(1)\cdot\Op\bigl(\sqrt{n_0+N}\,\psi_*\sqrt{s_{\mathrm{lin}}\log p/n_{\mathrm{lin}}}\bigr)=\op(1)$ by the rate condition \eqref{eq:rate-cond-nonlin}, already assumed via Corollary~\ref{cor:CLT-nonlin-labeler}. Slutsky's lemma removes this term, giving $\sqrt{n_0+N}(\tilde\beta_j^{\mathrm{cv}}-\beta^\star_{\mathrm{proj},j})\overset{\mathrm d}{\to}\mathcal N(0,\tau_j^{2,\mathrm{cv}})$, the stated conclusion for the deployed estimator.
\end{proof}

\begin{remark}[Scalar weight]\label{rem:hybrid-scalar}
The coordinate weight $\omega_j^{\mathrm{cv}}$ carries the $(X^\top\Sigma_0^{-1}e_j)^2$ weighting implicit in \eqref{eq:ppipp-nonlin-ratio}. A single scalar $\hat\omega = \widehat{\Cov}(\hat e,\hat\nu)/\widehat{\Var}(\hat\nu)$, with $\hat e$ the tuning residual, minimises the \emph{unweighted} residual variance and is a lightweight alternative---the exact analogue of the scalar shrinkage $\hat t$, and the $\widehat h_i\equiv1$ specialisation of \eqref{eq:omega-hat-j}---coinciding with $\omega_j^{\mathrm{cv}}$ when the coordinate weighting $(X^\top\Sigma_0^{-1}e_j)^2$ is itself constant across the design and attaining a reduction no larger than the coordinate-optimal one otherwise. The coordinate weight is what secures the uniform-in-$N$ dominance of Corollary~\ref{cor:hybrid-dom}.
\end{remark}

{Corollary~\ref{cor:hybrid-dom} is stated in Section~\ref{sec:hybrid} of the main text; its proof is given here.}


\begin{proof}[Proof of Corollary~\ref{cor:hybrid-dom}]
Part~(i) uses only $\hat t^{\mathrm{cv}}\in[0,1]$: dividing $\tau_j^{2,\mathrm{cv}}/(n_0+\hat N^*)$, with $\tau_j^{2,\mathrm{cv}}$ from \eqref{eq:hybrid-var}, by $V_j^{\mathrm{DL}}$ and running Steps~2--3 of the proof of Theorem~\ref{thm:deal-vs-dl} at the deployed $(\hat t^{\mathrm{cv}},\hat N^*)$ gives the first limit, whose $\eta$-independent bias-aware factor $(n_0+\hat t^{\mathrm{cv}}\hat N^*)/(n_0+\hat N^*)$ of \eqref{eq:deal-vs-dl} (Theorem~\ref{thm:deal-vs-dl}, licensed in the non-linear-predictor regime by clause~(ii) there and by Theorem~\ref{thm:complementarity}(i)) is at most one; then $\sqrt{1-\rho_j^2[\hat\mu]}\le\sqrt{1-\tfrac{\kappa}{1+\kappa}\rho_j^2[\hat\mu]}$ because $\tfrac{\kappa}{1+\kappa}<1$, the right-hand side being the PPI++ ratio \eqref{eq:ppipp-nonlin-ratio}.  The PPI++ inequality therefore holds for every $\hat N^*$ and every shrinkage limit.  Part~(ii) is where \eqref{eq:hybrid-shared-t0} enters: under it the control-variated procedure and separately-deployed DEAL attain the same limiting $A_n$-sandwich, so the common $(n_0+t_0\hat N^*)/(n_0+\hat N^*)$ factors cancel exactly and \eqref{eq:hybrid-var}'s same-pipeline comparison (Proposition~\ref{prop:hybrid-var}) transfers to the deployed $\mathrm{CI}_j^{\mathrm{DEAL}}$.
\end{proof}

\begin{proposition}[Adaptivity and no first-order harm]\label{prop:hybrid-adapt-app}
Under the assumptions of Proposition~\ref{prop:hybrid-var} and Corollary~\ref{cor:hybrid-dom}, write $\theta_j^{\mathrm{cv,DL}}$ for the estimator obtained by control-variating the target-only debiased Lasso directly (the $t_0=1$ specialisation of the construction, on which no unlabelled augmentation acts) --- the pure control variate. The control-variated estimator $\tilde\beta_j^{\mathrm{cv}}$ is then asymptotically never worse, in coordinate-$j$ interval length, than DEAL at the same shrinkage limit (and than separately-deployed DEAL under \eqref{eq:hybrid-shared-t0}), than $\theta_j^{\mathrm{cv,DL}}$, or than debiased Lasso: at every $\hat N^*$ its coordinate variance $\tau_j^{2,\mathrm{cv}}=(1-\rho_j^2[\hat\mu])\,\tau_j^{2,\mathrm{sand}}\le\tau_j^{2,\mathrm{sand}}$, with equality iff $\rho_j^2[\hat\mu]=0$, and setting $\hat\omega_j=0$ recovers DEAL. It reduces to DEAL when $\rho_j^2[\hat\mu]=0$ (any predictor linear in $X$, for which $\nu[\hat\mu]\equiv0$, and more generally whenever $\nu[\hat\mu]$ is uncorrelated with the labelled residual $e$ in the coordinate-$j$ weighted metric of $\omega_j^{\mathrm{cv}}$'s defining ratio), and its variance coincides with $\theta_j^{\mathrm{cv,DL}}$'s when $t_0=1$.
\end{proposition}
\begin{proof}
This collects Proposition~\ref{prop:hybrid-var} and Corollary~\ref{cor:hybrid-dom}: the variance bound and its equality condition are \eqref{eq:hybrid-var} directly; the comparisons with debiased Lasso and with $\theta_j^{\mathrm{cv,DL}}$ are the first display of Corollary~\ref{cor:hybrid-dom}, whose limiting ratio $\sqrt{1-\rho_j^2[\hat\mu]}\,(n_0+\hat t^{\mathrm{cv}}\hat N^*)/(n_0+\hat N^*)$ is at most $\sqrt{1-\rho_j^2[\hat\mu]}\le1$ for every $\hat N^*$, with the bias-aware factor equal to one exactly at $\hat t^{\mathrm{cv}}=1$, the case of no augmentation in which the ratio's own derivation (Theorem~\ref{thm:deal-vs-dl}) makes $\theta_j^{\mathrm{cv,DL}}$ the relevant comparator.
\end{proof}

\noindent The weight $\hat\omega_j$ requires no change to the variance-balance selection: one first sets $\hat\omega_j$, then applies the rule of Section~\ref{sec:vbrule} to $(\hat t,\hat N)$ on the control-variated residual --- the re-tuning whose own shrinkage limit $\hat t^{\mathrm{cv}}$ Corollary~\ref{cor:hybrid-dom} relates to $t_0$ by the hypothesis \eqref{eq:hybrid-shared-t0} (Remark~\ref{rem:hybrid-shared-t0}). Since $\hat\nu$ is mean-zero against the design ($\E_{P_0}[X\nu]=0$), the control variate lowers $\Gamma_0^{\mathrm{cv}}$ but leaves the imputation bias, and hence validity, untouched. The control variate scales the labelled-noise side of the variance-balance rule's trade-off by $1-\rho_j^2[\hat\mu]$, moving the crossing to a smaller budget, so the selected $\hat N^*$ contracts as $\rho_j^2[\hat\mu]$ grows: the two gains are sub-additive, the residual channel spending part of the augmentation channel's budget.
\subsection{Models on auxiliary covariates}\label{sec:aux-cov}

The residual $\nu[\hat\mu]$ is largest, and the control-variate hybrid of Section~\ref{sec:hybrid} most valuable, when the model draws on predictive content the design covariates cannot reproduce. The canonical instance is a model that acts on covariates outside the design entirely: a measurable map $\hat\mu:\R^p\times\R^q\to\R$, $(x,w)\mapsto\hat\mu(x,w)$, of the design covariates $X\in\R^p$ and \emph{auxiliary} covariates $W\in\R^q$ that need not overlap $X$, frozen conditionally on its training data and seen only through the values $\hat\mu(X_i,W_i)$ it returns on the target units. Overloading \eqref{eq:lab-projection},
\[
   \beta^{\mathrm{proj}}_0[\hat\mu]:=\Sigma_0^{-1}\,\E_{P_0}[X\,\hat\mu(X,W)],
   \qquad
   \nu[\hat\mu](X,W):=\hat\mu(X,W)-X^\top\beta^{\mathrm{proj}}_0[\hat\mu],
\]
with $\E_{P_0}[X\,\nu[\hat\mu]]=0$ by the first-order condition. The auxiliary covariates enter only through this residual: the extra predictive content of $W$ over $X$ falls entirely into $\nu[\hat\mu]$---the design-orthogonal signal the control variate recovers.

The pipeline is run exactly as in Section~\ref{sec:misspec-nonlinear-labeler}: the linearisation \eqref{eq:lin-lasso} regresses the predictions $\hat\mu(X^{(\mathrm{lin})},W^{(\mathrm{lin})})$ on the design covariates $X^{(\mathrm{lin})}$ of an independent target-marginal block, and the resulting $\hat\beta^{\mathrm{lin}}_{\mathrm{ext}}$ plays the role of the external estimator in Stages~1--4. Two features make the reduction exact. First, the linearisation regresses on $X$ alone, so the auxiliary covariates are required only where the model is \emph{evaluated}, never in the stacking or inference algebra, which see the model solely through its scalar outputs. Second, imputing with the projection $X^\top\hat\beta^{\mathrm{lin}}_{\mathrm{ext}}$ rather than the raw prediction discards $\nu[\hat\mu]$ before stacking, which is variance-optimal for the augmentation---feeding a design-orthogonal residual into the pseudo-labels would inject noise the stacked fit cannot use.

The construction adds a single substantive requirement, an observability rather than a modelling condition.

\begin{assumption}[Target-side model evaluability]\label{ass:aux-eval}
The predictions $\hat\mu(X_i,W_i)$ are available on the target-marginal linearisation block, the only block on which the model is evaluated for the pure procedure --- and, where the control-variate hybrid of Section~\ref{sec:hybrid} is deployed, on the labelled inference block and the tuning block as well, so that the control-variate residual $\hat\nu$ and the tuning-sample weight $\hat\omega_j$ of Section~\ref{app:hybrid} can be formed there. The residual family $\{X_j\,\nu[\hat\mu](X,W)\}_{j=1}^p$ is sub-exponential under $P_0$ with proxy $\psi_*^2<\infty$, and $\beta^{\mathrm{proj}}_0[\hat\mu]$ is $s_{\mathrm{lin}}$-sparse with $s_{\mathrm{lin}}=O(s)$.
\end{assumption}

Equivalently, the auxiliary covariates $W$ need be observed on that block alone, or the model's outputs there supplied, since the inference and stacking stages use the linearised coefficient $\hat\beta^{\mathrm{lin}}_{\mathrm{ext}}$ rather than the model itself.  Assumption~\ref{ass:aux-eval} is the auxiliary-covariate reading of Assumptions~\ref{ass:lin-proj}--\ref{ass:nu-reg}: the one genuinely new clause is evaluability of $\hat\mu$ on the target units, and the regularity clause is Assumption~\ref{ass:nu-reg} with $\nu[\hat\mu]$ now absorbing the $W\mid X$ variation, so its proxy $\psi_*$ is in general larger. Under this reading Lemma~\ref{lem:lin-rate} and the sandwich-variance central limit theorem of Corollary~\ref{cor:CLT-nonlin-labeler} hold verbatim, and the interval comparison of Corollary~\ref{cor:deal-vs-dl-nonlin} is unchanged: neither the asymptotic variance $\tau_j^{2,\mathrm{sand}}$ nor the ratio $(n_0+t_0\hat N^*)/(n_0+\hat N^*)$ depends on the auxiliary covariates. Their only footprint is quantitative, through the rate condition \eqref{eq:rate-cond-nonlin}: a larger residual proxy $\psi_*$ enlarges the linearisation error $\psi_*\sqrt{s_{\mathrm{lin}}\log p/n_{\mathrm{lin}}}$, so a model whose predictive power comes largely from covariates outside the design must be linearised on a correspondingly larger auxiliary block to preserve validity. Efficiency, by contrast, is governed as before by the model's linear-projection accuracy $\delta_{\mathrm{lin}}[\hat\mu]$ of Assumption~\ref{ass:lin-proj}, now measuring how well the design covariates alone reproduce the model's target-marginal projection. This governs the design-augmentation channel; the further efficiency carried by the auxiliary covariates is the control-variate factor $\sqrt{1-\rho_j^2[\hat\mu]}$ of the hybrid above.

Two caveats carry over unchanged. The linearisation must be run on covariates from the target marginal $P_0$ (Proposition~\ref{prop:joint-lin}); and if the unlabelled design is itself covariate-shifted, the shift-aware substitution of Section~\ref{sec:shift-aware} composes with the present construction exactly as developed in Section~\ref{sec:misspec-joint}, the auxiliary covariates altering none of that analysis because they have already been projected out.
\subsection{Joint covariate shift and model misspecification}\label{sec:misspec-joint}

This section treats the joint regime in which the unlabelled covariate distribution is shifted relative to the target, $\tilde X\overset{\mathrm{iid}}\sim P_{\mathrm u}$ with $\Sigma_{\mathrm u}\neq\Sigma_0$, \emph{and} the conditional mean is misspecified, $\mu(X)=X^\top\beta^\star+\eta(X)$ with $\eta\in L^2(P_0)$ orthogonal to the target linear span ($\E_{P_0}[X\,\eta(X)]=0$), \emph{and} the model is non-linear so that the linearisation \eqref{eq:lin-lasso} of Section~\ref{sec:misspec-nonlinear-labeler} is in scope.  We show that the shift-aware modification of Section~\ref{sec:shift-aware} and the linearisation step of Section~\ref{sec:misspec-nonlinear-labeler} compose additively, provided that the auxiliary covariates fed to the linearisation are drawn from the target marginal $P_0$ rather than the shifted unlabelled marginal $P_{\mathrm u}$.  The rationale is a population-level cross-term identification, made precise below.

Define the linearisation projection target with respect to an arbitrary marginal $P_\star$ on $\R^p$,
\begin{equation}\label{eq:lin-target-star}
    \beta^{\mathrm{proj}}_\star[\hat\mu] \;:=\; \Sigma_\star^{-1}\,\E_{P_\star}[X\,\hat\mu(X)],
    \qquad \Sigma_\star \;:=\; \E_{P_\star}[X X^\top].
\end{equation}
For the oracle non-linear predictor $\hat\mu(X)=X^\top\beta^\star+\eta(X)$ this evaluates to
\begin{equation}\label{eq:lin-target-explicit}
    \beta^{\mathrm{proj}}_\star[\hat\mu] \;=\; \beta^\star \;+\; \Sigma_\star^{-1}\,\E_{P_\star}[X\,\eta(X)] \;=\; \beta^\star \;+\; \delta_\star[\eta], \qquad
    \delta_\star[\eta] \;:=\; \Sigma_\star^{-1}\,\E_{P_\star}[X\,\eta(X)].
\end{equation}
The misspecification orthogonality $\E_{P_0}[X\,\eta(X)]=0$ gives $\delta_0[\eta]=0$, but $\delta_{\mathrm u}[\eta]$ is in general non-zero whenever $\Sigma_{\mathrm u}\neq\Sigma_0$ and $\eta$ does not lie in a Hermite chaos that is $\Sigma$-orthogonal to the linear span uniformly over Gaussian marginals with standardised second moments.\footnote{For probabilist's Hermite polynomials $H_n$ with $n\ge 2$, $\E_{P_\star}[X_j\,H_n(X_k)]=0$ for any centred Gaussian $P_\star$ with unit-variance marginals, by the orthogonality of Hermite chaoses in $L^2$ of any centred Gaussian measure; see e.g.\ \citet[Ch.\ 1]{nourdin2012normal}.  The Hermite construction in Section~\ref{sec:experiments-misspec} therefore satisfies $\delta_{\mathrm u}[\eta]\equiv 0$ for any $\Sigma_{\mathrm u}$, irrespective of shift.}  In other words, the cross-term $\delta_{\mathrm u}[\eta]$ is the population projection of the misspecification residual onto the linear span of $X$ \emph{in the shifted inner product induced by $\Sigma_{\mathrm u}$}, and, under the Gaussian-marginal parameterisation $P_\star=N(0,\Sigma_\star)$ of Section~\ref{sec:experiments-misspec} and the additional regularity that $\eta$ lies in the Gaussian Sobolev space $\mathbb D^{1,2}(P_0)$ (i.e.\ $\eta$ is weakly differentiable with $\|\nabla\eta\|_{L^2(P_0)}<\infty$), is of order $\|\Sigma_{\mathrm u}-\Sigma_0\|_{\mathrm{op}}$ at first order in the shift, with a constant given by a weighted Gaussian--Sobolev moment of $\eta$ (finite for $\nabla\eta$ of polynomial growth); the expansion is supplied in the proof of Proposition~\ref{prop:joint-lin}.

The relevance to DEAL is the following.  When the linearisation \eqref{eq:lin-lasso} is executed on auxiliary covariates drawn from $P_\star$, the population limit of $\hat\beta_{\mathrm{ext}}^{\mathrm{lin}}$ is $\beta^{\mathrm{proj}}_\star[\hat\mu] = \beta^\star + \delta_\star[\eta]$.  The pseudo-labels imputed on the unlabelled block then take the population form $\tilde X^\top(\beta^\star + \delta_\star[\eta])$, and the population stacked least-squares target shifts by $(N/(n_0+N))\,(\Sigma_{\mathrm{stk}}^{\mathrm{eff}})^{-1}\Sigma_{\mathrm u}\,\delta_\star[\eta]$ relative to $\beta^\star$, where $\Sigma_{\mathrm{stk}}^{\mathrm{eff}}=(n_0/(n_0+N))\Sigma_0+(N/(n_0+N))\Sigma_{\mathrm u}$.  Under the shift-aware substitution $\hat M_2\leftarrow\hat M_{\mathrm u}$ of Section~\ref{sec:shift-aware} and as the unlabelled fraction $N/(n_0+N)\to\pi_{\mathrm u}\in(0,1]$ (a limit distinct from the ratio $\kappa=N/n_0$; $\pi_{\mathrm u}=\kappa/(1+\kappa)$ when $\kappa$ is finite), this propagates to a non-vanishing first-order asymptotic bias of $\sqrt{n_0+N}(\tilde\beta_j-\beta^\star_j)$ proportional to $(\delta_\star[\eta])_j$, breaking the Gaussian limit of Theorem~\ref{thm:CLT-misspec} unless $\delta_\star[\eta]=0$.

\begin{proposition}[Linearisation auxiliary marginal under joint perturbation]\label{prop:joint-lin}
Suppose the regime of Corollary~\ref{cor:CLT-nonlin-labeler} (with the rate condition \eqref{eq:rate-cond-nonlin}) is in force, and that the unlabelled block follows a shifted marginal $\tilde X\sim P_{\mathrm u}$ with $\Sigma_{\mathrm u}\neq\Sigma_0$.  Let $X_0^{(\mathrm{lin})}$ denote a covariate-only sample of size $n_{\mathrm{lin}}$ drawn from the target marginal $P_0$ (independent of the inference labelled and unlabelled samples), and execute the linearisation \eqref{eq:lin-lasso} on $X_0^{(\mathrm{lin})}$ in place of $X_{\mathrm u}^{(\mathrm{lin})}$, producing $\hat\beta_{\mathrm{ext}}^{\mathrm{lin},0}$.  Combine this linearisation with the shift-aware Stage 4 substitution $\hat M_2\leftarrow\hat M_2^{\mathrm{adapt}}$ of Section~\ref{sec:shift-aware}.  Then the conclusion of Corollary~\ref{cor:CLT-nonlin-labeler} holds verbatim,
\[
    \sqrt{n_0+N}\,\bigl(\tilde\beta_j - \beta^\star_{\mathrm{proj},j}\bigr)\;\overset{\mathrm{d}}{\longrightarrow}\;\mathcal N\bigl(0,\,\tau_j^{2,\mathrm{sand}}\bigr),
\]
with the asymptotic variance $\tau_j^{2,\mathrm{sand}}$ evaluated at the realised $\hat M_2^{\mathrm{adapt}}$ (the no-shift form off the detected-shift-up event, a shift-aware sandwich on it) and the same plug-in confidence intervals.
\end{proposition}

\begin{remark}[Necessity of the target-marginal linearisation]\label{rem:joint-lin-necessity}
If instead the linearisation auxiliary $X_{\mathrm u}^{(\mathrm{lin})}$ is drawn from the shifted marginal $P_{\mathrm u}$, the population limit of the DEAL estimator is shifted by $(N/(n_0+N))\,(\Sigma_{\mathrm{stk}}^{\mathrm{eff}})^{-1}\Sigma_{\mathrm u}\,\delta_{\mathrm u}[\eta]$, and $\sqrt{n_0+N}(\tilde\beta_j-\beta^\star_j)$ no longer concentrates at zero unless $\delta_{\mathrm u}[\eta]=0$.  The $\mathbb D^{1,2}(P_0)$ regularity invoked above is what our argument uses: the proof's dilation-difference route bounds the cross-term through $\|\nabla\eta\|$-weighted moments, and that route fails for a general $L^2$ residual (a rapidly oscillating $\eta$ has dilation-difference $L^2$-norm of constant order).  Whether a fully model-free $L^2$-only order holds by some other argument is left open, and no such order is claimed.
\end{remark}

The two corrective devices are therefore independent in their action: Section~\ref{sec:shift-aware} controls the imputation-bias amplification due to $\Sigma_{\mathrm u}\succ\Sigma_0$ at the Stage 4 debiasing step, while Proposition~\ref{prop:joint-lin} controls the cross-term $\delta_{\mathrm u}[\eta]$ due to the joint presence of shift and misspecification at the linearisation step.  Their composition, shift-aware $\hat M_2^{\mathrm{adapt}}$ at Stage 4 and target-marginal linearisation \eqref{eq:lin-lasso} on $X_0^{(\mathrm{lin})}$, preserves the asymptotic Gaussian limit of Corollary~\ref{cor:CLT-nonlin-labeler} under joint perturbation.  The cost of the modification is a slower linearisation rate when $X_0^{(\mathrm{lin})}$ has size smaller than the unlabelled pool $X_{\mathrm u}^{(\mathrm{lin})}$ would have provided; in practice the labelled-target covariate sample is the natural source.  Empirical validation under the AR(1) shift envelope and the GB-shaped misspecification of Section~\ref{sec:experiments-misspec} is reported in Figure~\ref{fig:misspec-joint}.
\clearpage
\section{The DEAL algorithm}\label{sec:implementation}

Algorithm~\ref{alg:deal} assembles the full procedure in one place; it refers to the estimators and tuning parameters defined in the main text rather than restating them.

\begin{algorithm}[t]
\caption{\textit{Debiased External-model-Assisted Lasso (DEAL).}}
\label{alg:deal}
\begin{algorithmic}[1]
\Require Labelled target data, split into a tuning subsample $(X_{\mathrm{tun}},y_{\mathrm{tun}})$ and an inference sample $(X_0,y_0)$ of size $n_0$; unlabelled covariates $\tilde X$ of size $N_{\mathrm{avail}}$; external estimator $\hat\beta_{\mathrm{ext}}$; inferential index set $J$; level $\alpha$.
\State \textbf{Tuning split.} If the labelled data arrive unsplit, reserve the tuning subsample $(X_{\mathrm{tun}},y_{\mathrm{tun}})$ and let $(X_0,y_0)$, of size $n_0$, denote the remaining inference sample.
\State \textbf{Shrinkage level.} Split $(X_{\mathrm{tun}},y_{\mathrm{tun}})$ into two blocks, form $\hat B$ and $\hat T$ from their Javanmard--Montanari corrections, and set $\hat t\in[0,1]$ by~\eqref{eq:t-hat}.
\State \textbf{One-step correction.} Compute $M_1$ by~\eqref{eq:JM-M1} from $X_0$, or by the nodewise construction of Section~\ref{sec:refine} (as in Step~7), and form the one-step correction $C$ of~\eqref{eq:C-def}; the bias-aware initialiser is $\tilde\beta^{\mathrm{init}}=\hat\beta_{\mathrm{ext}}+\hat t\,C$ as in~\eqref{eq:init-bias-aware}.
\State \textbf{Unlabelled sample size.} Form a conservative plug-in bound $\hat a_1$ (or $\hat a_2$ under Gaussian design); on the admissible interval $\bigl[N_{\min},\min(N_{\mathrm{avail}},N_{\max})\bigr]$, select $N$ by the variance-balance rule~\eqref{eq:Nstar-empirical}, where $N_{\min}=1$ unless the Gaussian refinement of Section~\ref{app:gauss-implementation} imposes a larger lower endpoint. If the interval is empty, revert to target-only debiased Lasso and stop. Otherwise draw $N$ rows of $\tilde X$.
\State \textbf{Pseudo-labels.} Impute $\tilde f=\tilde X\,\tilde\beta^{\mathrm{init}}$.
\State \textbf{Stacked Lasso.} Form the stacked data $(X_{\mathrm{stk}},y_{\mathrm{stk}})$, set the penalty $\lambda$ as in Theorem~\ref{thm:stacked-lasso} with the residual-bias surrogate $b_n$ taken, according to the design regime, from~\eqref{eq:bn-gen},~\eqref{eq:bn-gauss-known}, or~\eqref{eq:bn-gauss-node}, and compute $\hat\beta$ by~\eqref{eq:stacked-lasso}.
\State \textbf{Final debiasing.} Construct $M_2$ by~\eqref{eq:JM-M2} or the pooled nodewise Lasso of Section~\ref{sec:refine} (the construction used in the applications)---replaced by the shift-aware $\hat M_2^{\mathrm{adapt}}$ of~\eqref{eq:M2-adapt} under covariate shift in the unlabelled design---and compute $\tilde\beta$ by~\eqref{eq:final-debias}.
\State \textbf{Inference.} For each $j\in J$, report the interval $\tilde\beta_j \pm z_{1-\alpha/2}\,\hat\tau_j/\sqrt{n_0+N}$, with $\hat\tau_j$ the sandwich plug-in standard error of Remark~\ref{rem:misspec-linear-recovery}, reducing to the homoscedastic plug-in of Theorem~\ref{thm:CLT} under linear truth and conditional homoscedasticity.
\Ensure Debiased estimates $\{\tilde\beta_j\}_{j\in J}$ with confidence intervals.
\end{algorithmic}
\end{algorithm}

Steps~1--3 of Algorithm~\ref{alg:deal} realise Stage~1 (the bias-aware initialiser), Step~5 realises Stage~2 (pseudo-label imputation), Step~6 realises Stage~3 (the stacked Lasso), and Step~7 realises Stage~4 (final debiasing); Step~4 selects the unlabelled sample size and Step~8 reports the intervals.

\clearpage
\section[Confidence-interval comparison with the prediction-powered family]{Confidence-interval comparison with the\\ prediction-powered family}\label{app:ppi-comparison}

This section develops the formal confidence-interval comparison of DEAL with the prediction-powered estimators PPI \citep{angelopoulos2023prediction} and PPI++ \citep{angelopoulos2023ppipp}, complementing the projection-parameter analysis of Section~\ref{sec:misspec}; the conclusion is summarised at the close of Section~\ref{sec:misspec-nonlinear-labeler}.  All statements are at the shared projection target $\beta^\star_{\mathrm{proj}}$ and a common unlabelled budget, and their proofs are collected in Section~\ref{app:proofs}.

\subsection{High-dimensional PPI and PPI++ under a linear predictor}\label{app:ppi-linear}

DEAL and the prediction-powered procedures exploit the unlabelled covariate sample through different channels.  The prediction-powered rectifier extracts information from the model's predicted values $\hat Y$: variance reduction over target-only inference is governed by the rectifier residual $\hat Y - X^\top\beta^\star_{\mathrm{proj}}$ and its covariance with the labelled-block influence function.  Whenever $\hat Y = X^\top\hat\beta_{\mathrm{ext}}$ is itself a linear function of $X$, this residual lies in the linear span of $X$ and equals $X^\top(\hat\beta_{\mathrm{ext}} - \beta^\star_{\mathrm{proj}})$, which is $\Op(\sqrt{s/n_A})$ deterministically given $\hat\beta_{\mathrm{ext}}$.  The rectifier then carries no asymptotic information beyond what is already available from target-only inference.

{Under a linear external model both the high-dimensional PPI adaptation and the optimally-tuned PPI++ collapse onto target-only debiased Lasso, as the following records.}
\begin{proposition}[Equivalence of high-dimensional PPI and PPI++ to debiased Lasso]\label{prop:ppipp-equivalence}
Under Assumptions~\ref{ass:design}, \ref{ass:rsc}, \ref{ass:JM}, \ref{ass:external-indep}, \ref{ass:nonlinear}, \ref{ass:misspec-reg}, under the sparsity scaling $s\log p=o(\sqrt{n_0})$ and the labelled-block precision conditions of Theorem~\ref{thm:deal-vs-dl} on $M_1$, in the no-shift regime $\Sigma_{\mathrm u}=\Sigma_0$, with an external estimator consistent at the rate $\sqrt{n_0\,(\log p)/(n_0\wedge N)}\,\norm{\hat\beta_{\mathrm{ext}}-\beta^\star_{\mathrm{proj}}}_1 \to 0$ (a sparse-Lasso external estimator at $\norm{\hat\beta_{\mathrm{ext}}-\beta^\star_{\mathrm{proj}}}_1=\Op(s\sqrt{(\log p)/n_A})$ satisfies this when $n_A\gg s^2(\log p)^2\,n_0/(n_0\wedge N)$), the high-dimensional adaptations of PPI \citep{angelopoulos2023prediction} and the optimally-tuned PPI++ \citep{angelopoulos2023ppipp} satisfy
\[
    \sqrt{n_0}\,\bigl(\hat\theta_j^{\mathrm{PPI++}}(\omega^*) - \hat\theta_j^{\mathrm{DL}}\bigr) = \op(1), \qquad \sqrt{n_0}\,\bigl(\hat\theta_j^{\mathrm{PPI}} - \hat\theta_j^{\mathrm{DL}}\bigr) = \op(1),
\]
for any data-driven $\omega^*$ lying in a fixed bounded interval with probability tending to one (as the variance-minimising weight does under linear truth, $\eta\equiv0$); both estimators are then asymptotically equivalent to target-only debiased Lasso, and the interval-length ratios $\mathrm{CI}_j^{\mathrm{PPI}}/\mathrm{CI}_j^{\mathrm{DL}}$ and $\mathrm{CI}_j^{\mathrm{PPI++}}/\mathrm{CI}_j^{\mathrm{DL}}$ converge to one in probability.
\end{proposition}

This is the high-dimensional asymptotic counterpart of the exact algebraic cancellation of Proposition~\ref{prop:ppi-cancellation}, established under the projection-parameter framework of Assumptions~\ref{ass:nonlinear}--\ref{ass:misspec-reg} without requiring linear truth, for weights confined to a bounded interval.  As in that exact cancellation---which holds for \emph{every} linear predictor---the labelled and unlabelled rectifier blocks share a common leading term that cancels in their difference, leaving only a second-order two-sample covariance fluctuation that a consistent external estimator drives to zero.  The comparators and the $\op(n_0^{-1/2})$ collapse of the rectifier correction are detailed in Section~\ref{app:proofs}, specialising the squared-loss PPI++ derivation of \citet{angelopoulos2023ppipp} to the debiased-Lasso-anchored rectifier of Section~\ref{sec:experiments-design}.

Through this reduction the confidence-interval ratio \eqref{eq:deal-vs-dl} governs the prediction-powered family as well: under a linear predictor $\mathrm{CI}_j^{\mathrm{DEAL}}/\mathrm{CI}_j^{\mathrm{PPI++}}$ and $\mathrm{CI}_j^{\mathrm{DEAL}}/\mathrm{CI}_j^{\mathrm{PPI}}$ obey the same bound as $\mathrm{CI}_j^{\mathrm{DEAL}}/\mathrm{CI}_j^{\mathrm{DL}}$, the PPI and PPI++ ratios to DL being themselves $1+\op(1)$.

\subsection{PPI and PPI++ under a non-linear predictor}\label{app:ppi-nonlinear}

The asymptotic equivalence of PPI and PPI++ to target-only debiased Lasso established in Proposition~\ref{prop:ppipp-equivalence} relies on the rectifier residual $\hat Y - X^\top\beta^\star_{\mathrm{proj}}$ lying in the linear span of $X$.  Under a non-linear predictor this is no longer true: the rectifier residual at $\beta^\star_{\mathrm{proj}}$ contains the model's non-linear component $\nu[\hat\mu](X)$ orthogonal to $X$ in $L^2$, with non-vanishing variance even as the model's linear-projection error $\delta_{\mathrm{lin}}[\hat\mu]$ vanishes.

{In the notation of Proposition~\ref{prop:ppipp-nonlinear} (main text), }$\rho_j^2[\hat\mu]$ is the squared correlation between the labelled-influence and rectifier-influence functions in $\Sigma_0^{-1}$-coordinates; under linear truth ($\eta\equiv0$) and a linear-coefficient predictor $\hat\mu(X)=X^\top\hat\beta_{\mathrm{ext}}$ it vanishes and \eqref{eq:ppipp-nonlin-ratio} reduces to Proposition~\ref{prop:ppipp-equivalence}.

{The comparison of DEAL with optimally-tuned PPI++ under the non-linear predictor is stated as clause~(ii) of Theorem~\ref{thm:complementarity} in the main text: dividing \eqref{eq:ppipp-nonlin-ratio} into the DEAL-to-DL limit of Corollary~\ref{cor:deal-vs-dl-nonlin} yields \eqref{eq:complementarity-ppipp}.  The proof, including the strictness claim at oracle shrinkage, is given in Section~\ref{app:proofs}.}

{For ease of reference, the companion comparison with target-only debiased Lasso under the non-linear predictor---absorbed as clause~(i) of the Complementarity Theorem (Theorem~\ref{thm:complementarity}) of the main text---is restated here in its original form; its proof is given in Section~\ref{app:proofs}.}

\begin{corollary}[The non-linear-predictor regime inherits the conclusions of Theorem~\ref{thm:deal-vs-dl}]\label{cor:deal-vs-dl-nonlin}
Fix a deterministic unlabelled-budget sequence $N=N_{n_0}$ along \eqref{eq:regime} with $\kappa_n=N/n_0\to\kappa_\infty\in[0,\infty)$, and suppose the assumptions of Corollary~\ref{cor:CLT-nonlin-labeler}, including the rate condition \eqref{eq:rate-cond-nonlin}, hold at this budget, with $\hat t\to_{\mathbb P}t_0\in[0,1]$.  Then the linearised procedure $\hat\beta_{\mathrm{ext}}\leftarrow\hat\beta_{\mathrm{ext}}^{\mathrm{lin}}$ obeys the sandwich central limit theorem with the same middle factor $\Gamma_0$ and the same limiting precision $\Sigma_0^{-1}$ as the target-only debiased Lasso.  the proof of Theorem~\ref{thm:deal-vs-dl} therefore goes through, and \eqref{eq:deal-vs-dl} and \eqref{eq:deal-vs-dl-bound} hold verbatim.
\end{corollary}

\begin{remark}[Trade-off between the two procedures]\label{rem:nonlin-tradeoff}
The prediction-powered procedures accept the full unlabelled budget $N$ unconditionally; PPI++'s tuning $\omega^*$ adapts to the model quality through the squared correlation $\rho_j^2[\hat\mu]$.  Their CI-length reduction over target-only debiased Lasso is $\sqrt{1 - (N/(n_0+N))\rho_j^2[\hat\mu]}$, held strictly above the full-data benchmark $\sqrt{n_0/(n_0+N)}$ by the labelled-noise component $\upsilon_{\varepsilon,j}:=(e_j^\top\Sigma_0^{-1})\,\E_{P_0}[\sigma^2(X)XX^\top]\,(\Sigma_0^{-1}e_j)$ in the denominator of $\rho_j^2[\hat\mu]$.

DEAL operates on a different trade-off.  Inference validity at the projection parameter requires the rate condition \eqref{eq:rate-cond-nonlin} of Corollary~\ref{cor:CLT-nonlin-labeler}, which together with the variance-balance rule's saturation {heuristic (Remark~\ref{rem:saturation-heuristic})} constrains the model quality $\rho_{\mathrm{lab}}$, the linearisation set size $n_{\mathrm{lin}}$, the unlabelled budget $N$, and the underlying sparsity $s$.  When this joint condition is met at saturation $\hat N^* = N$, $\tilde\beta$ achieves {the full-data CI reduction $n_0/(n_0+N)$ (variance ratio $(n_0/(n_0+N))^2$)} over target-only debiased Lasso (Corollary~\ref{cor:deal-vs-dl-nonlin}) and strictly dominates PPI and PPI++ (Theorem~\ref{thm:complementarity}(ii)).  When the joint condition is not met --- typically because the model quality is insufficient or the underlying sparsity too weak relative to the unlabelled budget --- the variance-balance rule selects $\hat N^* < N$, DEAL's CI-length reduction is bounded by $n_0/(n_0+\hat N^*)$, and the ratio in Theorem~\ref{thm:complementarity}(ii) can lie on either side of one.

The two procedures are therefore complementary in scope.  PPI and PPI++ are unconditionally applicable to any unlabelled budget but cap their CI-length reduction at $\sqrt{1 - (N/(n_0+N))\rho_j^2[\hat\mu]} \ge \sqrt{n_0/(n_0+N)}$; DEAL achieves {the strictly smaller CI reduction $n_0/(n_0+N)<\sqrt{n_0/(n_0+N)}$} in the regime where its rate condition holds.
\end{remark}

\subsection{Consolidated dominance}\label{sec:deal-dominance}

The linear comparison of Section~\ref{sec:misspec-comparison} and the non-linear comparison above are consolidated by Corollary~\ref{cor:deal-dominance} below.  Its first part is immediate from Proposition~\ref{prop:ppipp-equivalence}, which places PPI and optimally-tuned PPI++ at the target-only debiased-Lasso benchmark, so that the three limiting ratios
\[
  \frac{\mathrm{CI}_j^{\mathrm{DEAL}}}{\mathrm{CI}_j^{\mathrm{DL}}},\;\;
  \frac{\mathrm{CI}_j^{\mathrm{DEAL}}}{\mathrm{CI}_j^{\mathrm{PPI}}},\;\;
  \frac{\mathrm{CI}_j^{\mathrm{DEAL}}}{\mathrm{CI}_j^{\mathrm{PPI++}}}
\]
coincide at $(1+t_0\kappa_\infty)/(1+\kappa_\infty)<1$ for every $t_0<1$; its second part is Theorem~\ref{thm:complementarity}(ii) with $t_0=0$.

{For ease of reference, the consolidated statement is restated here in its original two-part form: part~(i) is the linear-predictor comparison deferred from the main text, and part~(ii) is clause~(ii) of the Complementarity Theorem (Theorem~\ref{thm:complementarity}).}

\begin{corollary}[Dominance over the prediction-powered family]\label{cor:deal-dominance}
Fix a coordinate $j$ and a deterministic common unlabelled budget $N=N_{n_0}$ along \eqref{eq:regime} with $\kappa_n\to\kappa_\infty>0$, work at the shared projection target $\beta^\star_{\mathrm{proj}}$, and suppose $\hat t\to_{\mathbb P}t_0\in[0,1]$.
\begin{enumerate}
\item[(i)] Under the assumptions of Theorem~\ref{thm:deal-vs-dl} and of Proposition~\ref{prop:ppipp-equivalence} with a linear external model, the high-dimensional PPI and optimally-tuned PPI++ estimators are asymptotically equivalent to target-only debiased Lasso with first-order-equal interval lengths, so \eqref{eq:deal-vs-dl} holds with either in place of DL.
\item[(ii)] Under the assumptions of Theorem~\ref{thm:complementarity} with a non-linear external model and oracle shrinkage $t_0=0$, the DEAL intervals are asymptotically strictly shorter than those of optimally-tuned PPI++, the limiting length ratio being at most $(1+\kappa_\infty)^{-1/2}$.
\end{enumerate}
\end{corollary}

Both parts hold at the same unlabelled budget the prediction-powered comparators consume, and the reason is a difference in \emph{which information each procedure extracts}.  The PPI and PPI++ rectifier acts on the first moment of the prediction residual $\tilde f - X\beta^\star$, which is asymptotically null under a linear predictor and recovers only the design-orthogonal component, capped below the full-data reduction, under a non-linear one (Propositions~\ref{prop:ppipp-equivalence} and~\ref{prop:ppipp-nonlinear}).  DEAL instead routes the same covariates through the \emph{variance} of the debiased estimator: the pseudo-labels enter Stages~3--4 as imputed responses that, when the initialiser is accurate enough to drive $\hat t\to0$, are near-noiseless evaluations of the regression surface, so the $N$ unlabelled rows sharpen the stacked fit and the precision estimate while contributing no labelled-noise term to the score.  This is the channel the rectifier cannot reach, and the one that converts a common budget into a strictly shorter interval: it lets DEAL behave as though it had drawn a strictly larger labelled sample (the effective size $n_{\mathrm{eff}}$ of Corollary~\ref{cor:neff}), coinciding with the rectifier-free benchmark under a linear predictor and strictly dominating PPI++ at full saturation under a non-linear one.  The control-variate refinement of Section~\ref{sec:hybrid} removes the saturation restriction in part~(ii), and under a linear predictor coincides with DEAL.

\clearpage
\section{Proofs}\label{app:proofs}

This section collects the proofs of the results stated in the main text and in this supplement, with two groups housed with their statements instead: the control-variate-hybrid results, proved in Section~\ref{app:hybrid-supp}, and Corollary~\ref{cor:neff}, proved where it is stated. Proofs here are grouped by the section in which the result they prove is stated, and follow the order of that section within each group; auxiliary lemmas stated here accompany the result they serve.  Three proofs sit outside their group because they consume, or are consumed by, an adjacent argument: Corollary~\ref{cor:shift-amp-benchmark}, which is read off a display inside the preceding proof, and Proposition~\ref{prop:ppipp-equivalence} and Corollary~\ref{cor:deal-vs-dl-nonlin}, whose displays and hypothesis verifications are used by the proofs that follow them.  Statements are referenced by number and are not restated.

\subsection{Proofs for Sections~\ref{sec:setup} and~\ref{sec:method} (setup and the procedure)}\label{app:proofs-setup}

\begin{proof}[Proof of Proposition~\ref{prop:ppi-cancellation}]
Substituting the linear predictor $f(X)=X\beta_f$ into the prediction-only term of
\eqref{eq:ppi-ols} and using invertibility of $\tilde X^{\top}\tilde X$,
\[
    \big(\tilde X^{\top}\tilde X\big)^{-1}\tilde X^{\top} f(\tilde X)
    = \big(\tilde X^{\top}\tilde X\big)^{-1}\tilde X^{\top}\tilde X\,\beta_f
    = \beta_f,
\]
so the unlabelled contribution reduces to the constant $\beta_f$, free of any randomness in
$\tilde X$. For the rectifier, the linear predictor gives $f(X_0)=X_0\beta_f$, and the
linear model \eqref{eq:target-model} gives $y_0 = X_0\beta^\star + \varepsilon_0$; hence
\[
    \big(X_0^{\top}X_0\big)^{-1} X_0^{\top}\big(f(X_0)-y_0\big)
    = \big(X_0^{\top}X_0\big)^{-1} X_0^{\top}
        \big(X_0\beta_f - X_0\beta^\star - \varepsilon_0\big)
    = (\beta_f-\beta^\star) - \big(X_0^{\top}X_0\big)^{-1} X_0^{\top}\varepsilon_0,
\]
using invertibility of $X_0^{\top}X_0$ in the last step. Combining the two displays in
\eqref{eq:ppi-ols},
\[
    \hat\beta^{\mathrm{PP}}
    = \beta_f - \Big[(\beta_f-\beta^\star) - \big(X_0^{\top}X_0\big)^{-1} X_0^{\top}\varepsilon_0\Big]
    = \beta^\star + \big(X_0^{\top}X_0\big)^{-1} X_0^{\top}\varepsilon_0,
\]
which is precisely the OLS estimator $\hat\beta_{\mathrm{OLS}}$ on the labelled data
$(X_0,y_0)$, and the dependence on $\beta_f$ has dropped out entirely. Since
$\Var(\varepsilon_0\mid X_0)=\sigma^2 I_{n_0}$, the identity gives
$\Var(\hat\beta^{\mathrm{PP}}\mid X_0)=\sigma^2(X_0^{\top}X_0)^{-1}$.
\end{proof}

{The following lemma is stated in prose around \eqref{eq:C-def} in Section~\ref{sec:method} of the main text; its formal statement is reproduced here, immediately before its proof.}

\begin{lemma}[Decomposition of the one-step correction]\label{lem:C-decomposition}
Let $\Delta:=\hat\beta_{\mathrm{ext}} - \beta^\star$.  Define
\[
    \bm{b} := -M_1\hat\Sigma_0\,\Delta\in\R^p,\qquad
    Z_1 := \frac{1}{n_0}M_1 X_0^\top \varepsilon_0\in\R^p.
\]
Then
\[
    C = \bm{b} + Z_1,
\]
and conditionally on $X_0$, $Z_1$ has mean zero and covariance
\[
    V := \Var(Z_1\mid X_0) = \frac{\sigma^2}{n_0} M_1 \hat\Sigma_0 M_1^\top.
\]
\end{lemma}

\begin{proof}[Proof of Lemma~\ref{lem:C-decomposition}]
Write
\[
    y_0 - X_0\hat\beta_{\mathrm{ext}}
    = X_0(\beta^\star - \hat\beta_{\mathrm{ext}}) + \varepsilon_0
    = -X_0\Delta + \varepsilon_0.
\]
Then
\[
    C = \frac{1}{n_0}M_1 X_0^\top (-X_0\Delta + \varepsilon_0)
      = - M_1\hat\Sigma_0\Delta + \frac{1}{n_0}M_1 X_0^\top\varepsilon_0
      = \bm{b} + Z_1.
\]
Conditional mean and covariance of $Z_1$ follow from $\E[\varepsilon_0\mid X_0]=0$ and
$\Var(\varepsilon_0\mid X_0)=\sigma^2 I_{n_0}$.
\end{proof}

\begin{proof}[Proof of Proposition~\ref{prop:proxy-risk}]
The function $R_{\mathrm{proxy}}(t)$ is a quadratic polynomial in $t$ with derivative
\[
    \frac{d}{dt}R_{\mathrm{proxy}}(t)
    = 2(t-1)\norm{\bm{b}}_2^2 + 2t\,\mathrm{tr}(V).
\]
Setting the derivative to zero gives the stated minimiser.  Since the leading coefficient $\norm{\bm b}_2^2+\mathrm{tr}(V)>0$, $R_{\mathrm{proxy}}$ is convex and this critical point is its global minimiser; as it lies in $[0,1]$ it is also the minimiser over $[0,1]$.
\end{proof}

\begin{lemma}[Initialiser decomposition]\label{lem:oof-decomposition}
With $\bm{b} := -M_1\hat\Sigma_0\Delta$ and $Z_1 := n_0^{-1}M_1 X_0^\top\varepsilon_0$ as in Lemma~\ref{lem:C-decomposition}, the bias-aware initialiser satisfies
\[
    \tilde\beta^{\mathrm{init}} - \beta^\star
    = r(\hat t) + \hat t\, Z_1,
\]
with
\begin{equation}
    r(t) := \Delta - t\,M_1\hat\Sigma_0\Delta = (I_p - t\,M_1\hat\Sigma_0)\Delta.
\end{equation}
Conditionally on the covariates, $Z_1$ has mean zero.
\end{lemma}

\begin{proof}[Proof of Lemma~\ref{lem:oof-decomposition}]
The decomposition $C=\bm b+Z_1$ is Lemma~\ref{lem:C-decomposition}.  Multiplying by the
tuning-sample shrinkage level $\hat t$ and adding $\hat\beta_{\mathrm{ext}}-\beta^\star=\Delta$
gives $\tilde\beta^{\mathrm{init}}-\beta^\star=\Delta+\hat t(\bm b+Z_1)=r(\hat t)+\hat t\,Z_1$.
Conditionally on the covariates, $Z_1=n_0^{-1}M_1X_0^\top\varepsilon_0$ has mean zero because $\E[\varepsilon_0\mid X_0]=0$; and since $\hat t$ is measurable with respect to the tuning sample, the covariates, and $\hat\beta_{\mathrm{ext}}$ --- all independent of $\varepsilon_0$ given the covariates (Assumption~\ref{ass:external-indep}) --- $\hat t\,Z_1$ is conditionally mean zero as well.
\end{proof}

\begin{lemma}[Bias envelope]\label{lem:bias-envelope}
Suppose the JM matrix satisfies $\maxnorm{M_1\hat\Sigma_0 - I_p}\le \bar\mu_1$.
Then, for any $t\in[0,1]$,
\[
    \norm{r(t)}_\infty \le \bigl(\bar\mu_1 + (1-t)(1+\bar\mu_1)\bigr)\,\norm{\Delta}_1,
    \qquad
    \norm{r(t)}_1 \le \bigl((1-t) + t\,p\,\bar\mu_1\bigr)\,\norm{\Delta}_1,
\]
and consequently
$\norm{\Sigma^{1/2}r(t)}_2 \le \bigl(\mnorm{\Sigma}\,\norm{r(t)}_1\,\norm{r(t)}_\infty\bigr)^{1/2}$.
\end{lemma}

\subsection{Proofs for Section~\ref{sec:theory-bias} (asymptotic theory)}\label{app:proofs-theory}

\begin{proof}[Proof of Lemma~\ref{lem:bias-envelope}]
Write $r(t)=(I_p - tM_1\hat\Sigma_0)\Delta$ and split
\[
    (I_p - tM_1\hat\Sigma_0)
    =
    (I_p - M_1\hat\Sigma_0) + (1-t)M_1\hat\Sigma_0.
\]
For each coordinate $j$,
\[
    \left|e_j^\top (I_p - M_1\hat\Sigma_0)\Delta\right|
    \le
    \maxnorm{I_p - M_1\hat\Sigma_0}\,\norm{\Delta}_1
    \le \bar\mu_1 \norm{\Delta}_1,
\]
and likewise
\[
    |e_j^\top M_1\hat\Sigma_0\Delta|
    \le
    |e_j^\top\Delta|
    +
    \left|e_j^\top (M_1\hat\Sigma_0-I_p)\Delta\right|
    \le
    (1+\bar\mu_1)\norm{\Delta}_1.
\]
Combining the two bounds gives the stated
$\norm{r(t)}_\infty \le \bigl(\bar\mu_1+(1-t)(1+\bar\mu_1)\bigr)\norm{\Delta}_1$.
For the $\ell_1$ bound, $\norm{r(t)}_1\le(1-t)\norm{\Delta}_1+t\norm{(I_p-M_1\hat\Sigma_0)\Delta}_1$, and each of the $p$ coordinates of $(I_p-M_1\hat\Sigma_0)\Delta$ is bounded by $\bar\mu_1\norm{\Delta}_1$.  The last claim follows from
$r^\top\Sigma r\le\norm{r}_1\norm{\Sigma r}_\infty\le\mnorm{\Sigma}\norm{r}_1\norm{r}_\infty$.
\end{proof}

\begin{lemma}[Unlabelled-block transfer]\label{lem:bias-transfer}
Suppose Assumption~\ref{ass:design} holds, and let $v$ be a random vector in $\R^p$ independent of the unlabelled block $\tilde X$.  Then there is a constant $C$, depending only on the sub-Gaussian constants of Assumption~\ref{ass:design}, such that, with probability tending to one,
\[
    \norm{\hat\Sigma_{\mathrm u} v}_\infty
    \;\le\;
    \mnorm{\Sigma}\,\norm{v}_\infty
    + C\sqrt{\frac{\log p}{N}}\;\norm{\Sigma^{1/2}v}_2 .
\]
In particular, for any $p\times p$ matrix $M$,
$|e_j^\top M\hat\Sigma_{\mathrm u}v| \le \mnorm{M}\,\norm{\hat\Sigma_{\mathrm u}v}_\infty$.
\end{lemma}

\begin{proof}[Proof of Lemma~\ref{lem:bias-transfer}]
Work conditionally on $v$.  For each coordinate $k$,
\[
    e_k^\top\hat\Sigma_{\mathrm u}v
    = e_k^\top\Sigma v
    + \frac{1}{N}\sum_{i=1}^{N}\bigl\{\tilde X_{ik}\,(\tilde X_i^\top v) - e_k^\top\Sigma v\bigr\},
\]
and $|e_k^\top\Sigma v|\le\mnorm{\Sigma}\norm{v}_\infty$.  The summands of the second term are i.i.d., mean zero, and sub-exponential with $\psi_1$-norm at most $C_0\sqrt{\Sigma_{kk}}\,\norm{\Sigma^{1/2}v}_2\le C_0'\norm{\Sigma^{1/2}v}_2$, the product of the $\psi_2$-norms of $\tilde X_{ik}$ and $\tilde X_i^\top v$.  Bernstein's inequality and a union bound over $k\in\{1,\dots,p\}$ give, with probability tending to one,
\[
    \max_{k\le p}\Bigl|\frac{1}{N}\sum_{i=1}^{N}\bigl\{\tilde X_{ik}(\tilde X_i^\top v)-e_k^\top\Sigma v\bigr\}\Bigr|
    \le C\Bigl(\sqrt{\frac{\log p}{N}}+\frac{\log p}{N}\Bigr)\norm{\Sigma^{1/2}v}_2 ,
\]
and $\log p/N\to0$ along \eqref{eq:regime} absorbs the second term into the constant.  The final claim is H\"older's inequality applied to the rows of $M$.
\end{proof}

\begin{lemma}[Effective score bound]\label{lem:score}
Suppose Assumptions~\ref{ass:design}, \ref{ass:external-indep}, and \ref{ass:JM} hold.  Let $\hat t\in[0,1]$.  Then there exist constants $C_1,C_2,C_3>0$, depending only on the constants of those assumptions, such that, with probability tending to one,
\begin{align*}
    \norm{g}_\infty \le{}&
    C_1\sigma\sqrt{\frac{\log p}{n_0+N}}
    + C_2\,\frac{N}{n_0+N}\,\hat t\,\Bigl(1+\sqrt{\tfrac{p}{N}}\Bigr)\sqrt{\frac{\log p}{n_0}}\\
    &+ C_3\,\frac{N}{n_0+N}\,\Bigl(\norm{r(\hat t)}_\infty + \sqrt{\tfrac{\log p}{N}}\,\norm{\Sigma^{1/2}r(\hat t)}_2\Bigr).
\end{align*}
\end{lemma}

\begin{proof}[Proof of Lemma~\ref{lem:score}]
The bound on $\norm{g_{\mathrm{lab}}}_\infty$ follows from standard sub-Gaussian tail bounds and a union bound:
\[
    \norm{g_{\mathrm{lab}}}_\infty \le C_1\sigma\sqrt{\frac{\log p}{n_0+N}}
\]
with probability tending to one.  For $g_{\mathrm{pl,G}}$, write $w:=Z_1=n_0^{-1}M_1 X_0^\top\varepsilon_0$, so that $g_{\mathrm{pl,G}}=\tfrac{N}{n_0+N}\,\hat t\,\hat\Sigma_{\mathrm u}w$; the vector $w$ is built from the labelled block alone and is independent of $\tilde X$.  Conditional on the covariates $X_0$, the tuning data, and the external estimator, $w$ is mean zero with $\norm{w}_\infty=\Op(\sqrt{\log p/n_0})$, by the row-sum bound $\mnorm{M_1}=\Op(1)$ and a sub-Gaussian union bound, and
\begin{gather*}
    \norm{\Sigma^{1/2}w}_2^2
    \le C\norm{w}_2^2 = \Op(p/n_0),
    \qquad\text{since}\\
    \E\bigl[\norm{w}_2^2\mid X_0\bigr]
    = \frac{\sigma^2}{n_0}\,\frac{\mathrm{tr}(M_1 X_0^\top X_0 M_1^\top)}{n_0}
    = \Op\Bigl(\frac{p}{n_0}\Bigr),
\end{gather*}
where $\mathrm{tr}(M_1\hat\Sigma_0 M_1^\top)\le p\max_j m_{1j}^\top\hat\Sigma_0 m_{1j}\le p\,\mnorm{M_1}^2\bigl(\mnorm{\Sigma}+\maxnorm{\hat\Sigma_0-\Sigma}\bigr)=\Op(p)$.  Lemma~\ref{lem:bias-transfer} applied to $v=w$ then gives, with probability tending to one,
\[
    \norm{g_{\mathrm{pl,G}}}_\infty
    \le \frac{N}{n_0+N}\,\hat t\,\Bigl(\mnorm{\Sigma}\,\norm{w}_\infty + C\sqrt{\tfrac{\log p}{N}}\,\norm{\Sigma^{1/2}w}_2\Bigr)
    \le C_2\,\frac{N}{n_0+N}\,\hat t\,\Bigl(1+\sqrt{\tfrac{p}{N}}\Bigr)\sqrt{\frac{\log p}{n_0}}.
\]
Finally, $g_{\mathrm{pl,R}}=\tfrac{N}{n_0+N}\hat\Sigma_{\mathrm u}r(\hat t)$, and $r(\hat t)$ is built from $(M_1,\hat\Sigma_0,\hat t,\Delta)$, none of which involves $\tilde X$ (Assumption~\ref{ass:external-indep}), so Lemma~\ref{lem:bias-transfer} applies with $v=r(\hat t)$:
\begin{align*}
    \norm{g_{\mathrm{pl,R}}}_\infty
    &\le \frac{N}{n_0+N}\Bigl(\mnorm{\Sigma}\,\norm{r(\hat t)}_\infty + C\sqrt{\tfrac{\log p}{N}}\,\norm{\Sigma^{1/2}r(\hat t)}_2\Bigr)\\
    &\le C_3\,\frac{N}{n_0+N}\Bigl(\norm{r(\hat t)}_\infty + \sqrt{\tfrac{\log p}{N}}\,\norm{\Sigma^{1/2}r(\hat t)}_2\Bigr).
\end{align*}
Combining the three pieces completes the proof.
\end{proof}

\begin{proof}[Proof of Theorem~\ref{thm:stacked-lasso}]

On the surrogate event of \eqref{eq:bn-surrogate}, which has probability $1-o(1)$, Lemma~\ref{lem:score} implies, with probability tending to one,
\[
    \norm{g}_\infty \le
    C_1\sigma\sqrt{\frac{\log p}{n_0+N}}
    + C_2\,\frac{N}{n_0+N}\,\hat t\,\Bigl(1+\sqrt{\tfrac{p}{N}}\Bigr)\sqrt{\frac{\log p}{n_0}}
    + C_3\,\frac{N}{n_0+N}\,b_n,
\]
where $g:=\frac{1}{n_0+N}X_{\mathrm{stk}}^\top(y_{\mathrm{stk}}-X_{\mathrm{stk}}\beta^\star)$ is the stacked score.  Taking $K_0\ge 2\max(C_1,C_2,C_3)$ in the definition of $\lambda$ ensures that, on this event,
\begin{equation}
    \lambda \ge 2\norm{g}_\infty.
    \label{eq:lambda-score}
\end{equation}
We complete the proof by the standard basic-inequality and restricted-eigenvalue argument for the Lasso \citep[cf.][Chapter~6]{buhlmann2011statistics}, working in addition on the probability-tending-to-one event of Assumption~\ref{ass:rsc} on which $\hat\Sigma_{\mathrm{stk}}$ obeys the restricted-eigenvalue condition with constant $\phi$.  The intersection of these three probability-tending-to-one events itself has probability tending to one, as claimed in the theorem.

Write $\delta:=\hat\beta-\beta^\star$, $S:=\mathrm{supp}(\beta^\star)$ with $|S|=s$, and abbreviate
\[
    P:=(\hat\beta-\beta^\star)^\top\hat\Sigma_{\mathrm{stk}}(\hat\beta-\beta^\star)
      =\frac{1}{n_0+N}\norm{X_{\mathrm{stk}}\delta}_2^2,
\]
the identity holding because $\hat\Sigma_{\mathrm{stk}}=\frac{1}{n_0+N}X_{\mathrm{stk}}^\top X_{\mathrm{stk}}$.

\emph{Step 1 (basic inequality).}  Since $\hat\beta$ minimises the stacked objective \eqref{eq:stacked-lasso}, evaluating that objective at $\hat\beta$ and at $\beta^\star$ gives
\[
    \frac{1}{2(n_0+N)}\norm{y_{\mathrm{stk}}-X_{\mathrm{stk}}\hat\beta}_2^2
    +\lambda\norm{\hat\beta}_1
    \le
    \frac{1}{2(n_0+N)}\norm{y_{\mathrm{stk}}-X_{\mathrm{stk}}\beta^\star}_2^2
    +\lambda\norm{\beta^\star}_1.
\]
Writing $y_{\mathrm{stk}}-X_{\mathrm{stk}}\hat\beta=(y_{\mathrm{stk}}-X_{\mathrm{stk}}\beta^\star)-X_{\mathrm{stk}}\delta$ and expanding the squared norm, the cross term equals $\frac{1}{n_0+N}\inner{X_{\mathrm{stk}}\delta}{y_{\mathrm{stk}}-X_{\mathrm{stk}}\beta^\star}=\delta^\top g$, so
\begin{equation}
    \tfrac12 P
    \le
    \delta^\top g + \lambda\bigl(\norm{\beta^\star}_1-\norm{\hat\beta}_1\bigr).
    \label{eq:basic}
\end{equation}
By H\"older's inequality and \eqref{eq:lambda-score}, $|\delta^\top g|\le\norm{g}_\infty\norm{\delta}_1\le\tfrac{\lambda}{2}\norm{\delta}_1$.  Decomposing $\delta=\delta_S+\delta_{S^c}$ and using $\beta^\star_{S^c}=0$ with the triangle inequality,
\[
    \norm{\hat\beta}_1
    =\norm{\beta^\star_S+\delta_S}_1+\norm{\delta_{S^c}}_1
    \ge\norm{\beta^\star}_1-\norm{\delta_S}_1+\norm{\delta_{S^c}}_1,
\]
hence $\norm{\beta^\star}_1-\norm{\hat\beta}_1\le\norm{\delta_S}_1-\norm{\delta_{S^c}}_1$.  Substituting both bounds into \eqref{eq:basic} and using $\norm{\delta}_1=\norm{\delta_S}_1+\norm{\delta_{S^c}}_1$,
\begin{equation}
    \tfrac12 P
    \le
    \tfrac{\lambda}{2}\bigl(\norm{\delta_S}_1+\norm{\delta_{S^c}}_1\bigr)
    +\lambda\bigl(\norm{\delta_S}_1-\norm{\delta_{S^c}}_1\bigr)
    =\tfrac{\lambda}{2}\bigl(3\norm{\delta_S}_1-\norm{\delta_{S^c}}_1\bigr).
    \label{eq:master}
\end{equation}

\emph{Step 2 (cone membership and restricted eigenvalue).}  Since $P\ge0$, \eqref{eq:master} forces $3\norm{\delta_S}_1-\norm{\delta_{S^c}}_1\ge0$, i.e.\ $\norm{\delta_{S^c}}_1\le3\norm{\delta_S}_1$.  Thus $\delta$ lies in the cone of Assumption~\ref{ass:rsc}, and on the restricted-eigenvalue event
\[
    P=\delta^\top\hat\Sigma_{\mathrm{stk}}\delta\ge\phi\norm{\delta}_2^2.
\]
By Cauchy--Schwarz, $\norm{\delta_S}_1\le\sqrt{s}\,\norm{\delta_S}_2\le\sqrt{s}\,\norm{\delta}_2\le\sqrt{s}\,\sqrt{P/\phi}$, the compatibility bound
\begin{equation}
    \norm{\delta_S}_1\le\sqrt{\tfrac{s\,P}{\phi}}.
    \label{eq:compat}
\end{equation}

\emph{Step 3 ($\ell_1$ bound).}  Rearranging \eqref{eq:master} as $P+\lambda\norm{\delta_{S^c}}_1\le3\lambda\norm{\delta_S}_1$ and adding $\lambda\norm{\delta_S}_1$ to both sides yields
\begin{equation}
    P+\lambda\norm{\delta}_1\le4\lambda\norm{\delta_S}_1
    \le4\lambda\sqrt{\tfrac{s\,P}{\phi}},
    \label{eq:combined}
\end{equation}
using \eqref{eq:compat}.  Applying the arithmetic--geometric inequality $2ab\le a^2+b^2$ with $a=\sqrt{P}$ and $b=2\lambda\sqrt{s/\phi}$ gives $4\lambda\sqrt{sP/\phi}\le P+4s\lambda^2/\phi$, so the right-hand side of \eqref{eq:combined} is at most $P+4s\lambda^2/\phi$; the term $P$ cancels and
\[
    \norm{\hat\beta-\beta^\star}_1=\norm{\delta}_1\le\frac{4s}{\phi}\lambda,
\]
the first stated bound.

\emph{Step 4 (prediction-error bound).}  From \eqref{eq:master}, discarding the nonpositive term $-\tfrac{\lambda}{2}\norm{\delta_{S^c}}_1\le0$,
\[
    \tfrac12 P\le\tfrac{3\lambda}{2}\norm{\delta_S}_1,
    \qquad\text{i.e.}\qquad
    P\le 3\lambda\norm{\delta_S}_1.
\]
Inserting the compatibility bound \eqref{eq:compat} gives $P\le 3\lambda\sqrt{sP/\phi}$.  If $P=0$ the prediction bound is trivial; otherwise dividing by $\sqrt{P}$ yields $\sqrt{P}\le 3\lambda\sqrt{s/\phi}$, and squaring gives
\[
    (\hat\beta-\beta^\star)^\top\hat\Sigma_{\mathrm{stk}}(\hat\beta-\beta^\star)
    =\frac{1}{n_0+N}\norm{X_{\mathrm{stk}}(\hat\beta-\beta^\star)}_2^2
    \le\frac{9s}{\phi}\lambda^2,
\]
the second stated bound.
\end{proof}

\begin{lemma}[Discharge of the variance-matching limit]\label{lem:m2-efficiency}

In the no-shift regime $\Sigma_{\mathrm{stk}}=\Sigma_0=:\Sigma$ of Theorem~\ref{thm:CLT}, with the eigenvalues of $\Sigma$ bounded away from $0$ and $\infty$ and $\Omega:=\Sigma^{-1}$ row-sparse, fix the inferential coordinate $j$ and let $M_2$ be either \textup{(A)} the convex programme~\eqref{eq:JM-M2} or \textup{(B)} its nodewise-Lasso realisation under i.i.d.\ sub-Gaussian rows \citep{vandegeer2014asymptotic}, each attaining under the row-sparse $\Omega$ the precision-row rate $\norm{M_2^\top e_j-\Sigma^{-1}e_j}_2=\Op\bigl(\sqrt{s_\Omega\log p/(n_0+N)}\bigr)$ (Assumption~\ref{ass:rowsparse}).  Then
\[
    e_j^\top M_2\Sigma M_2^\top e_j \;\longrightarrow\; e_j^\top\Sigma^{-1}e_j=(\Sigma^{-1})_{jj}\in\bigl[\lambda_{\max}(\Sigma)^{-1},\lambda_{\min}(\Sigma)^{-1}\bigr]\subset(0,\infty).
\]
This is the variance-matching limit underlying requirement~\textup{(iii)} of Theorem~\ref{thm:CLT}.

\end{lemma}

\begin{proof}

Write $\xi_j:=M_2^\top e_j-\Omega e_j$, so that $\norm{\xi_j}_2=\op(1)$ by the stated precision-row rate.  The identity $\Omega\Sigma\Omega=\Omega$ is algebraic---$\Omega\Sigma=I_p$ gives $e_j^\top\Omega\Sigma\Omega e_j=(\Omega)_{jj}=(\Sigma^{-1})_{jj}$ for every $\Sigma\succ0$, with no distributional input---so, using $\Sigma\Omega e_j=e_j$,
\[
    e_j^\top M_2\Sigma M_2^\top e_j=(\Sigma^{-1})_{jj}+2\,\xi_j^\top\Sigma\Omega e_j+\xi_j^\top\Sigma\xi_j=(\Sigma^{-1})_{jj}+2(\xi_j)_j+\xi_j^\top\Sigma\xi_j.
\]
The cross term satisfies $|2(\xi_j)_j|\le2\norm{\xi_j}_2$ and the quadratic term $\xi_j^\top\Sigma\xi_j\le\lambda_{\max}(\Sigma)\norm{\xi_j}_2^2$, both $\op(1)$; the limit lies in $[\lambda_{\max}(\Sigma)^{-1},\lambda_{\min}(\Sigma)^{-1}]$ by the spectral bounds, hence is finite and positive.  The argument uses only this algebraic identity and the row rate of Assumption~\ref{ass:rowsparse}, not the variance-matching clause of Assumption~\ref{ass:JM}.

\end{proof}

\begin{proof}[Proof of Theorem~\ref{thm:CLT}]
By Lemma~\ref{lem:oof-decomposition},
\[
    y_{\mathrm{stk}} - X_{\mathrm{stk}}\beta^\star
    =
    \begin{pmatrix}
        \varepsilon_0\\
        \tilde X r(\hat t) + \hat t\,\tilde X Z_1
    \end{pmatrix}.
\]
Using $Z_1=n_0^{-1}M_1 X_0^\top\varepsilon_0$, we obtain
\[
    \tilde X^\top \tilde X Z_1
    =
    \frac{N}{n_0}\hat\Sigma_{\mathrm u} M_1 X_0^\top \varepsilon_0.
\]
Therefore
\[
    \frac{1}{n_0+N}M_2 X_{\mathrm{stk}}^\top (y_{\mathrm{stk}} - X_{\mathrm{stk}}\beta^\star)
    =
    \frac{1}{n_0+N}
    M_2 A_n X_0^\top \varepsilon_0
    + \frac{N}{n_0+N}M_2\hat\Sigma_{\mathrm u} r(\hat t).
\]
As in the standard debiased-Lasso expansion,
\[
    \tilde\beta - \beta^\star
    =
    \frac{1}{n_0+N}
    M_2 A_n X_0^\top \varepsilon_0
    + \frac{N}{n_0+N}M_2\hat\Sigma_{\mathrm u} r(\hat t)
    + (I_p - M_2\hat\Sigma_{\mathrm{stk}})(\hat\beta - \beta^\star).
\]
Hence
\[
    \tilde\beta_j - \beta^\star_j
    =
    \frac{1}{n_0+N}
    e_j^\top M_2 A_n X_0^\top \varepsilon_0
    + R_{j,n},
\]
where
\[
    R_{j,n}
    :=
    \frac{N}{n_0+N}e_j^\top M_2\hat\Sigma_{\mathrm u} r(\hat t)
    + e_j^\top(I_p - M_2\hat\Sigma_{\mathrm{stk}})(\hat\beta - \beta^\star).
\]
By Assumption~\ref{ass:JM},
\[
    \left|e_j^\top(I_p - M_2\hat\Sigma_{\mathrm{stk}})(\hat\beta - \beta^\star)\right|
    \le
    \maxnorm{I_p - M_2\hat\Sigma_{\mathrm{stk}}}\,\norm{\hat\beta-\beta^\star}_1
    \le
    \mu_2 \norm{\hat\beta-\beta^\star}_1.
\]
Combining this with Theorem~\ref{thm:stacked-lasso} yields
\[
    \sqrt{n_0+N}\,
    \left|e_j^\top(I_p - M_2\hat\Sigma_{\mathrm{stk}})(\hat\beta - \beta^\star)\right|
    =
    \Op\!\left(\sqrt{n_0+N}\,\mu_2 s\lambda\right)
    =
    \op(1)
\]
by assumption~(ii).  Likewise,
\begin{align*}
    &\sqrt{n_0+N}\,
    \left|
        \frac{N}{n_0+N}e_j^\top M_2\hat\Sigma_{\mathrm u}r(\hat t)
    \right|\\
    &\qquad\le
    \sqrt{n_0+N}\,\mnorm{M_2}\,
    \Bigl(\mnorm{\Sigma}\,\norm{r(\hat t)}_\infty
    + C\sqrt{\tfrac{\log p}{N}}\,\norm{\Sigma^{1/2}r(\hat t)}_2\Bigr)
    =
    \op(1)
\end{align*}
by Lemma~\ref{lem:bias-transfer}---applicable because $r(\hat t)$ is built from $(M_1,\hat\Sigma_0,\hat t,\Delta)$, none of which involves $\tilde X$ (Assumption~\ref{ass:external-indep})---together with $\mnorm{M_2}=\Op(1)$ and assumption~(i).  Hence $\sqrt{n_0+N}\,R_{j,n}=\op(1)$.

Conditional on the covariates and on the tuning data, the leading term is a sum of
independent mean-zero random variables because $\hat t$ is independent of the
inference-sample noise:
\[
    \frac{1}{\sqrt{n_0+N}}
    e_j^\top M_2 A_n X_0^\top \varepsilon_0
    =
    \frac{1}{\sqrt{n_0+N}}
    \sum_{i=1}^{n_0}
    \bigl(x_{0,i}^\top A_n^\top M_2^\top e_j\bigr)\,\varepsilon_{0,i}.
\]
Its conditional variance is $\tau_{j,n}^2$.  Combined with the i.i.d.\ mean-zero finite-variance noise of Assumption~\ref{ass:design}, Assumption~(iv) supplies the Lindeberg condition for this triangular array, so the Lindeberg--Feller theorem gives conditional asymptotic
normality with variance $\tau_{j,n}^2$.  Assumption~(iii) and Slutsky's lemma complete the proof.
\end{proof}

\begin{proof}[Proof of Corollary~\ref{cor:idealized-efficiency}]
In the benchmark, $M_1=\Sigma^{-1}$ and $\hat\Sigma_{\mathrm u}\to\Sigma$ give
$A_n=I_p+(\hat tN/n_0)\hat\Sigma_{\mathrm u}M_1\to(1+t_0N/n_0)I_p$, and with
$M_2=\Sigma^{-1}$ and $\hat\Sigma_0\to\Sigma$ the sandwich of Theorem~\ref{thm:CLT} becomes
\begin{align*}
    \tau_{j,n}^2
    =\frac{\sigma^2 n_0}{n_0+N}\,e_j^\top M_2A_n\hat\Sigma_0A_n^\top M_2^\top e_j
    &\;\longrightarrow\;
    \frac{\sigma^2 n_0}{n_0+N}\Bigl(1+\frac{t_0N}{n_0}\Bigr)^2 e_j^\top\Sigma^{-1}\Sigma\,\Sigma^{-1}e_j\\
    &=\frac{\sigma^2(n_0+t_0N)^2}{n_0(n_0+N)}\,(\Sigma^{-1})_{jj}=:\tau_j^2 .
\end{align*}
Dividing by $n_0+N$ gives \eqref{eq:idealized-var}, and dividing \eqref{eq:idealized-var} by
$\sigma^2(\Sigma^{-1})_{jj}/n_0$ gives $\{(n_0+t_0N)/(n_0+N)\}^2=G(\kappa,t_0)$ with $\kappa=N/n_0$.
Since $0\le t_0\le1$ and $N>0$, $n_0+t_0N\le n_0+N$ with equality if and only if $t_0=1$, so
$G(\kappa,t_0)\le1$, $G(\kappa,1)=1$, and $G(\kappa,t_0)<1$ for every $t_0<1$.
\end{proof}

\begin{proof}[Proof of Proposition~\ref{prop:vbalance}]
The decomposition \eqref{eq:lbd-decomp}, with the $L_j(N)$, $B_j(N)$, $R_j(N)$ displayed in the statement, is the exact expansion obtained in the proof of Theorem~\ref{thm:CLT}---an algebraic identity given the Stage~1--4 construction.  $R_j(N)=\op(1)$ follows from $\maxnorm{I_p-M_2\hat\Sigma_{\mathrm{stk}}}\le\mu_2$, the $\ell_1$ rate of Theorem~\ref{thm:stacked-lasso}, and condition~(ii), exactly as in that proof; requirement~(i) is not used.  Given the covariates, the tuning data, and the external estimator, $\varepsilon_0$ is the only remaining randomness (Assumptions~\ref{ass:design} and~\ref{ass:external-indep}), so $L_j(N)$ is conditionally mean zero with conditional variance $\tau_{j,n}^2$ and $B_j(N)$ is conditionally deterministic.  Both $L_j(N)$ and $B_j(N)$ carry the common factor $\sqrt{n_0+N}$; since $\tau^2_{L,j}$ and $\tau^2_{B,j}$ are quadratic in this factor, the $(n_0+N)$ cancels in the ratio $\tau^2_{B,j}(N)/\tau^2_{L,j}(N)$, which therefore equals that of the corresponding unscaled leading-variance and squared-bias contributions to the mean-squared error of $\tilde\beta_j$, and it suffices to establish the two monotonicities for those unscaled contributions, which we do in the idealised oracle-precision benchmark.

\emph{Leading contribution.}  In the idealised oracle-precision benchmark of Corollary~\ref{cor:idealized-efficiency}, the leading contribution to $\Var(\tilde\beta_j)$ is, by \eqref{eq:idealized-var},
\[
    V_{L,j}(N) \;=\; \frac{\sigma^2(\Sigma^{-1})_{jj}}{n_0}\,u(N)^2,
    \qquad u(N):=\frac{n_0+t_0 N}{n_0+N}.
\]
Since $u'(N) = n_0(t_0-1)/(n_0+N)^2 \le 0$ for $t_0\in[0,1]$ and $u(N)>0$, the map $N\mapsto V_{L,j}(N)$ is non-increasing.

\emph{Bias contribution.}  The residual imputation bias enters $\tilde\beta_j$ only through the factor $\tfrac{N}{n_0+N}\,e_j^\top M_2\hat\Sigma_{\mathrm u}r(\hat t)$.  In the idealised benchmark the precision is the oracle $\Omega$, the unlabelled covariance is the population $\Sigma$, and the shrinkage sits at its limit $t_0$, so the bias factor $e_j^\top\Omega\Sigma\,r(t_0)=r_j(t_0)$ does not vary with $N$: the unscaled squared-bias contribution is $V_{B,j}(N)=r_j(t_0)^2\,w(N)^2$ with $w(N):=N/(n_0+N)$, and $w'(N)=n_0/(n_0+N)^2>0$, so $N\mapsto V_{B,j}(N)$ is non-decreasing.

\emph{Ratio and balance point.}  As a ratio of a non-decreasing numerator to a positive non-increasing denominator, $V_{B,j}(N)/V_{L,j}(N)=\tau^2_{B,j}(N)/\tau^2_{L,j}(N)$ is non-decreasing in $N$.  Hence for any $\varrho\ge 1$ the set $\{N\ge 1:\tau^2_{B,j}(N)\ge\varrho\,\tau^2_{L,j}(N)\}$ is an up-set, and the variance-balance point $N^*_j$ of \eqref{eq:Nstar-population} is its well-defined minimum; when the set is empty the procedure reverts to target-only debiased Lasso.
\end{proof}

\subsection{Proofs for Section~\ref{app:shrinkage} (the shrinkage selector)}\label{app:proofs-shrinkage}

\begin{proof}[Proof of Lemma~\ref{lem:tun-regularity}]
Part~(a) is the consistency of the scaled-Lasso noise-level estimator under the sub-Gaussian design, restricted-eigenvalue, and sparsity scaling of the tuning-block analogues of Assumptions~\ref{ass:design}--\ref{ass:sparsity} \citep{sun2012scaled}.
{For part~(b) we bound the trace and the operator norm directly; the operator-norm relation $M_{1,\mathrm{tun}}\hat\Sigma_{\mathrm{tun}}=I_p+\op(1)$ is unavailable when $p>n_{\mathrm{tun}}$, where $\hat\Sigma_{\mathrm{tun}}$ is singular.
\emph{Trace.}  Each diagonal entry $e_j^\top M_{1,\mathrm{tun}}\hat\Sigma_{\mathrm{tun}}M_{1,\mathrm{tun}}^\top e_j$ equals the quadratic form $m_{1j}^\top\hat\Sigma_{\mathrm{tun}}m_{1j}$.  By Cauchy--Schwarz in the $\hat\Sigma_{\mathrm{tun}}$-semi-inner product against $e_j$ and the feasibility $|e_j^\top\hat\Sigma_{\mathrm{tun}}m_{1j}-1|\le\mu_{1,\mathrm{tun}}$ of the tuning-block Assumption~\ref{ass:JM},
$m_{1j}^\top\hat\Sigma_{\mathrm{tun}}m_{1j}\ge(1-\mu_{1,\mathrm{tun}})^2/\hat\Sigma_{\mathrm{tun},jj}$ uniformly in $j$, with $\max_j\hat\Sigma_{\mathrm{tun},jj}=\Op(1)$ under the sub-Gaussian design; by H\"older and the same feasibility,
$m_{1j}^\top\hat\Sigma_{\mathrm{tun}}m_{1j}\le\norm{m_{1j}}_1\norm{\hat\Sigma_{\mathrm{tun}}m_{1j}}_\infty\le\mnorm{M_{1,\mathrm{tun}}}\,(1+\mu_{1,\mathrm{tun}})=\Op(1)$.
Summing over $j$, $\mathrm{tr}(V_{\mathrm{tun}})=(\sigma^2/n_{\mathrm{tun}})\sum_{j=1}^p m_{1j}^\top\hat\Sigma_{\mathrm{tun}}m_{1j}\asymp p\,\sigma^2/n_{\mathrm{tun}}$ with probability tending to one.
\emph{Operator norm.}  Unconditionally, $\operatorname{rank}(V_{\mathrm{tun}})\le\min\bigl(p,\operatorname{rank}(\hat\Sigma_{\mathrm{tun}})\bigr)\le\min(p,n_{\mathrm{tun}})$, so $\norm{V_{\mathrm{tun}}}_{\mathrm{op}}\ge\mathrm{tr}(V_{\mathrm{tun}})/\min(p,n_{\mathrm{tun}})\gtrsim\sigma^2\max(1,p/n_{\mathrm{tun}})/n_{\mathrm{tun}}$; conversely, under Assumption~\ref{ass:tun-opnorm} and the sub-Gaussian sample-covariance bound $\norm{\hat\Sigma_{\mathrm{tun}}}_{\mathrm{op}}=\Op(\max(1,p/n_{\mathrm{tun}}))$ \citep{koltchinskii2017concentration},
\[
\norm{V_{\mathrm{tun}}}_{\mathrm{op}}\le(\sigma^2/n_{\mathrm{tun}})\norm{M_{1,\mathrm{tun}}}_{\mathrm{op}}^2\norm{\hat\Sigma_{\mathrm{tun}}}_{\mathrm{op}}=\Op\bigl(\sigma^2\max(1,p/n_{\mathrm{tun}})/n_{\mathrm{tun}}\bigr).
\]
Together, $\mathrm{tr}(V_{\mathrm{tun}})/\norm{V_{\mathrm{tun}}}_{\mathrm{op}}\asymp\min(p,n_{\mathrm{tun}})\to\infty$.}
\end{proof}

\begin{proof}[Proof of Lemma~\ref{lem:t-hat-consistency}]
We argue conditionally on the $\sigma$-field $\mathcal F$ generated by all tuning covariates
and by $\hat\beta_{\mathrm{ext}}$ (hence by $\Delta$); ``$\xrightarrow{p}$'' denotes
convergence in conditional probability, and the tuning regime, the tuning-block analogues of
Assumptions~\ref{ass:design}, \ref{ass:rsc}, and~\ref{ass:JM}, Lemma~\ref{lem:tun-regularity}, and Assumption~\ref{ass:eff-rank-tun} are in
force.  Applying Lemma~\ref{lem:C-decomposition} on each disjoint tuning block gives the
\emph{exact} identities $C_{\mathrm{tun}}^{(c)} = \bm{b}_{\mathrm{tun}}^{(c)} +
Z_{\mathrm{tun}}^{(c)}$, $c\in\{a,b\}$, with $\bm{b}_{\mathrm{tun}}^{(c)}$ $\mathcal
F$-measurable and $Z_{\mathrm{tun}}^{(c)} := n_{\mathrm{tun},c}^{-1}M_{1,\mathrm{tun}}^{(c)}
X_{\mathrm{tun},c}^\top\varepsilon_{\mathrm{tun},c}$ having, conditionally on $\mathcal F$,
mean zero and covariance $V_{\mathrm{tun}}^{(c)}$.  The blocks use disjoint tuning labels, so
by the tuning-block analogue of Assumption~\ref{ass:design} the noise families are
independent and $Z_{\mathrm{tun}}^{(a)}\perp Z_{\mathrm{tun}}^{(b)}$ given $\mathcal F$.

\emph{Conditional mean.}  Expanding the bilinear form
$\hat B = \inner{\bm{b}_{\mathrm{tun}}^{(a)}+Z_{\mathrm{tun}}^{(a)}}
{\bm{b}_{\mathrm{tun}}^{(b)}+Z_{\mathrm{tun}}^{(b)}}$, the three summands carrying a single noise
factor have zero conditional expectation (mean-zero noise, independent across blocks), giving
the exact identity, with \emph{no} proxy approximation,
\[
   \E[\hat B\mid\mathcal F] = \inner{\bm{b}_{\mathrm{tun}}^{(a)}}{\bm{b}_{\mathrm{tun}}^{(b)}}
   =: B_{\mathrm{tun}}.
\]

\emph{Conditional variance.}  By the same independence the four summands of $\hat B$ are
pairwise uncorrelated given $\mathcal F$ (any cross-product is, after conditioning on one
block, an odd linear form in the other block's independent mean-zero noise), so
\[
   \Var(\hat B\mid\mathcal F)
   = \bm{b}_{\mathrm{tun}}^{(a)\top}V_{\mathrm{tun}}^{(b)}\bm{b}_{\mathrm{tun}}^{(a)}
   + \bm{b}_{\mathrm{tun}}^{(b)\top}V_{\mathrm{tun}}^{(a)}\bm{b}_{\mathrm{tun}}^{(b)}
   + \mathrm{tr}\!\bigl(V_{\mathrm{tun}}^{(a)}V_{\mathrm{tun}}^{(b)}\bigr).
\]
{By Assumption~\ref{ass:tun-opnorm} and the sub-Gaussian bound $\norm{\hat\Sigma_{\mathrm{tun},c}}_{\mathrm{op}}=\Op(\max(1,p/n_{\mathrm{tun}}))$,
\[
    \norm{V_{\mathrm{tun}}^{(c)}}_{\mathrm{op}}
    \le \frac{\sigma^2}{n_{\mathrm{tun},c}}\,\norm{M_{1,\mathrm{tun}}^{(c)}}_{\mathrm{op}}^2\,
    \norm{\hat\Sigma_{\mathrm{tun},c}}_{\mathrm{op}}
    = \Op\!\Bigl(\frac{\sigma^2\max(1,p/n_{\mathrm{tun}})}{n_{\mathrm{tun}}}\Bigr).
\]
Set $a_{\mathrm{tun}} := |B_{\mathrm{tun}}| + \mathrm{tr}(V_{\mathrm{tun}}) > 0$; by Lemma~\ref{lem:tun-regularity}, $a_{\mathrm{tun}}\ge\mathrm{tr}(V_{\mathrm{tun}})\ge c\,p\sigma^2/n_{\mathrm{tun}}$ with probability tending to one.  By the $\ell_2$-balance condition of Assumption~\ref{ass:eff-rank-tun}, $\norm{\bm{b}_{\mathrm{tun}}^{(c)}}_2^2\le C_{\mathrm{bal}}\bigl(|B_{\mathrm{tun}}|+\mathrm{tr}(V_{\mathrm{tun}})\bigr)=C_{\mathrm{bal}}\,a_{\mathrm{tun}}$, so the first two terms obey
\[
    \frac{\bm{b}_{\mathrm{tun}}^{(c)\top}V_{\mathrm{tun}}^{(c')}\bm{b}_{\mathrm{tun}}^{(c)}}{a_{\mathrm{tun}}^2}
    \le\frac{\norm{V_{\mathrm{tun}}^{(c')}}_{\mathrm{op}}\,C_{\mathrm{bal}}}{a_{\mathrm{tun}}}
    =\Op\!\Bigl(\frac{C_{\mathrm{bal}}}{\min(p,n_{\mathrm{tun}})}\Bigr),
\]
and the third term, by $\mathrm{tr}(AB)\le\norm{A}_{\mathrm{op}}\mathrm{tr}(B)$ for positive semidefinite $A,B$ and the per-block trace bound of Lemma~\ref{lem:tun-regularity} applied to each block ($\mathrm{tr}(V_{\mathrm{tun}}^{(c)})\asymp p\sigma^2/n_{\mathrm{tun}}$ under the constant block fractions $n_{\mathrm{tun},c}/n_{\mathrm{tun}}$),
\[
    \frac{\mathrm{tr}(V_{\mathrm{tun}}^{(a)}V_{\mathrm{tun}}^{(b)})}{a_{\mathrm{tun}}^2}
    \le\frac{\norm{V_{\mathrm{tun}}^{(a)}}_{\mathrm{op}}\,\mathrm{tr}(V_{\mathrm{tun}}^{(b)})}{\mathrm{tr}(V_{\mathrm{tun}})^2}
    =\Op\!\Bigl(\frac{1}{\min(p,n_{\mathrm{tun}})}\Bigr).
\]
Hence $W_n:=\Var(\hat B\mid\mathcal F)/a_{\mathrm{tun}}^2=\Op\bigl(C_{\mathrm{bal}}/\min(p,n_{\mathrm{tun}})\bigr)=\op(1)$,
and Chebyshev's inequality conditional on $\mathcal F$ gives
$u := (\hat B - B_{\mathrm{tun}})/a_{\mathrm{tun}}\xrightarrow{p}0$.}

\emph{Denominator.}  The statistic
$\hat T = (\hat\sigma_{\mathrm{tun}}^2/n_{\mathrm{tun}})
\mathrm{tr}(M_{1,\mathrm{tun}}\hat\Sigma_{\mathrm{tun}}M_{1,\mathrm{tun}}^\top)$ is the exact
plug-in for $\mathrm{tr}(V_{\mathrm{tun}})$, differing only by the substitution
$\hat\sigma_{\mathrm{tun}}^2$ for $\sigma^2$.  Since $\mathrm{tr}(V_{\mathrm{tun}})>0$
($V_{\mathrm{tun}}$ is positive semidefinite with positive trace under the lower eigenvalue
bound of Assumption~\ref{ass:design} and Assumption~\ref{ass:JM}) and
$\hat\sigma_{\mathrm{tun}}^2\xrightarrow{p}\sigma^2$ (Lemma~\ref{lem:tun-regularity}),
$\hat T/\mathrm{tr}(V_{\mathrm{tun}})=\hat\sigma_{\mathrm{tun}}^2/\sigma^2\xrightarrow{p}1$,
so $v := (\hat T - \mathrm{tr}(V_{\mathrm{tun}}))/a_{\mathrm{tun}}\xrightarrow{p}0$.

{\emph{Ratio.}  Write $\bar N := B_{\mathrm{tun}}/a_{\mathrm{tun}}\in[-1,1]$ and
$\bar D := (B_{\mathrm{tun}}+\mathrm{tr}(V_{\mathrm{tun}}))/a_{\mathrm{tun}}$, so that
$\hat B/(\hat B+\hat T)=(\bar N+u)/(\bar D+u+v)$ and $t^\star_{\mathrm{tun}}=\Pi_{[0,1]}(\bar N/\bar D)$.
By Assumption~\ref{ass:tun-sep}, $|\bar D|\ge c_0$ with probability tending to one, and we argue on the
event $E_n:=\{|u|+|v|\le c_0/2\}$, whose conditional probability tends to one by the convergences
$u,v\xrightarrow{p}0$ established above.  Two branches.  If $\bar D\ge c_0$, the perturbed denominator
satisfies $\bar D+u+v\ge c_0/2$, and because $\Pi_{[0,1]}$ is $1$-Lipschitz,
\[
   \bigl|\hat t - t^\star_{\mathrm{tun}}\bigr|
   \le \left|\frac{\bar N+u}{\bar D+u+v}-\frac{\bar N}{\bar D}\right|
   = \frac{|\bar D\,u - \bar N(u+v)|}{\bar D\,(\bar D+u+v)}
   \le \frac{2|u|+|v|}{c_0\cdot c_0/2}
   \xrightarrow{p}0 ,
\]
using $|\bar N|\le1$ and $\bar D\le1$; for $B_{\mathrm{tun}}\ge0$ one has $\bar D=1$, may take $c_0=1$,
and recovers the bound $4|u|+2|v|$.  If instead $\bar D\le-c_0$, then
$B_{\mathrm{tun}}+\mathrm{tr}(V_{\mathrm{tun}})=\bar D\,a_{\mathrm{tun}}<0$, so
$t^\star_{\mathrm{tun}}=0$ by the positive-part convention; on
$E_n$ the empirical denominator obeys $\hat B+\hat T=a_{\mathrm{tun}}(\bar D+u+v)<0$, so
$\hat t=0$ by the same convention of \eqref{eq:t-hat}, and
$\hat t=0=t^\star_{\mathrm{tun}}$ exactly.  In both branches $\hat t\xrightarrow{p}t^\star_{\mathrm{tun}}$.}  In particular,
an (asymptotically) unbiased external estimator drives $\Delta\to0$, hence
$\bm{b}_{\mathrm{tun}}^{(c)}\to0$ and $B_{\mathrm{tun}}\to0$, so
$\hat t\xrightarrow{p}0=t^\star_{\mathrm{tun}}$ by Lemma~\ref{lem:tun-regularity}, switching off the Stage-1 correction; a strongly biased one with
$B_{\mathrm{tun}}\gg\mathrm{tr}(V_{\mathrm{tun}})$ gives $\hat t\xrightarrow{p}1$.
\end{proof}

{
\begin{proof}[Proof of Lemma~\ref{lem:tun-sep-derive}]
Since $\bm b_{\mathrm{tun}}^{(c)}=-\Delta+\delta_c$, the polarisation identity gives
\[
    B_{\mathrm{tun}}=\inner{-\Delta+\delta_a}{-\Delta+\delta_b}
    =\tfrac14\norm{-2\Delta+\delta_a+\delta_b}_2^2-\tfrac14\norm{\delta_a-\delta_b}_2^2
    \ge-\tfrac12\bigl(\norm{\delta_a}_2^2+\norm{\delta_b}_2^2\bigr).
\]
The entrywise envelope $|e_j^\top\delta_c|\le\mu_{1,\mathrm{tun}}\norm{\Delta}_1$ (the tuning-block form of Lemma~\ref{lem:bias-envelope}) gives $\norm{\delta_c}_2^2\le p\,\mu_{1,\mathrm{tun}}^2a_1^2$ on the event of Assumption~\ref{ass:external}.  With $\mu_{1,\mathrm{tun}}^2\asymp(\log p)/n_{\mathrm{tun}}$ and $\mathrm{tr}(V_{\mathrm{tun}})\ge c\,p\sigma^2/n_{\mathrm{tun}}$ (Lemma~\ref{lem:tun-regularity}), the negative part of $B_{\mathrm{tun}}$ is $\Op\bigl(p\,a_1^2\log p/n_{\mathrm{tun}}\bigr)=\op\bigl(\mathrm{tr}(V_{\mathrm{tun}})\bigr)$ whenever $a_1^2\log p=o(\sigma^2)$.  On $\{B_{\mathrm{tun}}\ge0\}$ the separation holds with $c_0=1$; on $\{B_{\mathrm{tun}}<0\}$, $|B_{\mathrm{tun}}|=\op(\mathrm{tr}(V_{\mathrm{tun}}))$ gives $|B_{\mathrm{tun}}+\mathrm{tr}(V_{\mathrm{tun}})|\ge(1-\op(1))\,\mathrm{tr}(V_{\mathrm{tun}})$ and $|B_{\mathrm{tun}}|+\mathrm{tr}(V_{\mathrm{tun}})\le(1+\op(1))\,\mathrm{tr}(V_{\mathrm{tun}})$, so the separation holds for any fixed $c_0<1$ with probability tending to one.  The regime specialisations insert the respective $a_1$: $a_1=\op(n_0^{-1/2})$ gives $a_1^2\log p=\op(n_0^{-1}\log p)=o(\sigma^2)$ using $\log p=o(n_0)$; $a_1\asymp s\sigma\sqrt{(\log p)/n_0}/\phi$ gives $a_1^2\log p\asymp\sigma^2s^2(\log p)^2/(n_0\phi^2)=o(\sigma^2)$ under $s\log p=o(\sqrt{n_0+N})$ with $n_{\mathrm{tun}}\asymp n_0$ and $N=O(n_0)$; $a_1\asymp s_A\sqrt{(\log p)/n_A}$ gives the stated $s_A^2(\log p)^2=o(n_A\sigma^2)$.
\end{proof}

\begin{proof}[Proof of Proposition~\ref{prop:that-rate}]
The proof of Lemma~\ref{lem:t-hat-consistency} established, conditionally on $\mathcal F$, the exact identity $\E[\hat B\mid\mathcal F]=B_{\mathrm{tun}}$ and the normalised variance bound $W_n=\Var(\hat B\mid\mathcal F)/a_{\mathrm{tun}}^2=\Op\bigl(C_{\mathrm{bal}}/\min(p,n_{\mathrm{tun}})\bigr)$.  Since $W_n$ is $\mathcal F$-measurable, conditional Chebyshev gives $P\bigl(|u|>K\sqrt{W_n}\,\big|\,\mathcal F\bigr)\le K^{-2}$ pointwise, and the tower property makes the bound unconditional, whence $u=\Op\bigl(\sqrt{C_{\mathrm{bal}}/\min(p,n_{\mathrm{tun}})}\bigr)$.  For the denominator, $\hat T/\mathrm{tr}(V_{\mathrm{tun}})=\hat\sigma^2_{\mathrm{tun}}/\sigma^2$ exactly and $\mathrm{tr}(V_{\mathrm{tun}})/a_{\mathrm{tun}}\le1$, so $|v|\le r_\sigma:=|\hat\sigma^2_{\mathrm{tun}}/\sigma^2-1|$; under the tuning-block design, restricted-eigenvalue, and sparsity analogues, the scaled Lasso with the stated penalty attains
\[
    r_\sigma=\Op\!\Bigl(n_{\mathrm{tun}}^{-1/2}+\frac{s\log p}{n_{\mathrm{tun}}}\Bigr)
\]
\citep[Theorem~2]{sun2012scaled}: decomposing $\hat\sigma^2_{\mathrm{tun}}-\sigma^2$ into the noise average, $\Op(\sigma^2n_{\mathrm{tun}}^{-1/2})$ by the central limit theorem with $\E\varepsilon^4<\infty$; the cross term, bounded by $2\norm{n_{\mathrm{tun}}^{-1}X_{\mathrm{tun}}^\top\varepsilon_{\mathrm{tun}}}_\infty\norm{\hat\beta-\beta^\star}_1=\Op(\sigma^2s\log p/n_{\mathrm{tun}})$; and the prediction error, $\Op(\sigma^2 s\log p/n_{\mathrm{tun}})$.  On the event $E_n=\{|u|+|v|\le c_0/2\}$ the two branches of the ratio step in the proof of Lemma~\ref{lem:t-hat-consistency} give $|\hat t-t^\star_{\mathrm{tun}}|\le(4|u|+2|v|)/c_0^2$ on $\{\bar D\ge c_0\}$ and $\hat t=t^\star_{\mathrm{tun}}$ exactly on $\{\bar D\le-c_0\}$, while Assumption~\ref{ass:tun-sep} excludes $|\bar D|<c_0$ with probability tending to one.  Since $|\hat t-t^\star_{\mathrm{tun}}|\le1$ always and $\Pr(E_n^c)\to0$, the contribution of $E_n^c$ is $\op$ of any deterministic null sequence, in particular of $(n_0+N)^{-1/4}$, and the stated rate follows.  For the final claim, with $n_{\mathrm{tun}}\asymp n_0\asymp n_0+N$: the first term is $\op\bigl((n_0+N)^{-1/4}\bigr)$ under $\min(p,n_{\mathrm{tun}})\gg\sqrt{n_0+N}$; the second is $O\bigl((n_0+N)^{-1/2}\bigr)$; the third is $o\bigl((n_0+N)^{-1/2}\bigr)$ under $s\log p=o(\sqrt{n_0+N})$.
\end{proof}

\begin{proof}[Proof of Proposition~\ref{prop:condT-reduction}]
If condition~(T) holds with limit $t_0$, then $|t^\star_{\mathrm{tun}}-t_0|\le|t^\star_{\mathrm{tun}}-\hat t|+|\hat t-t_0|=\op\bigl((n_0+N)^{-1/4}\bigr)$ by Proposition~\ref{prop:that-rate}.  Conversely, if $|t^\star_{\mathrm{tun}}-t_0|=\op\bigl((n_0+N)^{-1/4}\bigr)$, then $|\hat t-t_0|\le|\hat t-t^\star_{\mathrm{tun}}|+|t^\star_{\mathrm{tun}}-t_0|=\op\bigl((n_0+N)^{-1/4}\bigr)$, and $t_0$ is then the in-probability limit of $\hat t$.  The measurability claim is immediate from $\bm b_{\mathrm{tun}}^{(c)}=-M^{(c)}_{1,\mathrm{tun}}\hat\Sigma_{\mathrm{tun},c}\Delta$ and the definition of $V_{\mathrm{tun}}$.
\end{proof}

\begin{proof}[Proof of Proposition~\ref{prop:condT-regimes}]
Throughout, $\norm{\bm b_{\mathrm{tun}}^{(c)}}_2\le\norm{\Delta}_2+\norm{\delta_c}_2\le a_1\bigl(1+\sqrt p\,\mu_{1,\mathrm{tun}}\bigr)$ by the entrywise envelope of the proof of Lemma~\ref{lem:tun-sep-derive} and $\norm{\Delta}_2\le\norm{\Delta}_1\le a_1$; since $p\ge n_{\mathrm{tun}}$ gives $\sqrt p\,\mu_{1,\mathrm{tun}}\gtrsim\sqrt{\log p}\ge1$, the constant is dominated and $\norm{\bm b_{\mathrm{tun}}^{(c)}}_2^2=\Op(a_1^2\,p\,\mu_{1,\mathrm{tun}}^2)$.  So on $\{B_{\mathrm{tun}}\ge0\}$,
\[
    t^\star_{\mathrm{tun}}
    \le\frac{B_{\mathrm{tun}}}{\mathrm{tr}(V_{\mathrm{tun}})}
    \le\frac{\norm{\bm b^{(a)}_{\mathrm{tun}}}_2\norm{\bm b^{(b)}_{\mathrm{tun}}}_2}{\mathrm{tr}(V_{\mathrm{tun}})}
    =\Op\!\Bigl(\frac{a_1^2\,p\,\mu_{1,\mathrm{tun}}^2}{p\sigma^2/n_{\mathrm{tun}}}\Bigr)
    =\Op\!\Bigl(\frac{a_1^2\log p}{\sigma^2}\Bigr),
\]
while on $\{B_{\mathrm{tun}}<0\}$ the separation of Lemma~\ref{lem:tun-sep-derive} keeps the argument non-positive and $t^\star_{\mathrm{tun}}=0$.  (i)~With $a_1=\op(n_0^{-1/2})$, $t^\star_{\mathrm{tun}}=\op(n_0^{-1}\log p)=\op\bigl((n_0+N)^{-1/4}\bigr)$, since $\log p=o\bigl((n_0+N)^{3/4}\bigr)$ under $s\log p=o(\sqrt{n_0+N}\,)$.  (ii)~With $a_1\asymp s\sigma\sqrt{(\log p)/n_0}/\phi$, $t^\star_{\mathrm{tun}}=\Op\bigl(s^2(\log p)^2/(n_0\phi^2)\bigr)$, which is $\op\bigl((n_0+N)^{-1/4}\bigr)$ under the stated $s^2(\log p)^2=o\bigl((n_0+N)^{3/4}\bigr)$.  (iii)~With $a_1\asymp s_A\sqrt{(\log p)/n_A}$, $t^\star_{\mathrm{tun}}=\Op\bigl(s_A^2(\log p)^2/(n_A\sigma^2)\bigr)$, and the stated condition makes this $\op\bigl((n_0+N)^{-1/4}\bigr)$.  In each case the residual condition of Proposition~\ref{prop:condT-reduction} holds with $t_0=0$.  (iv)~On $\{\rho_n\ge0\}$, $1-t^\star_{\mathrm{tun}}=1/(1+\rho_n)$ exactly, so $|t^\star_{\mathrm{tun}}-1|=\op\bigl((n_0+N)^{-1/4}\bigr)$ if and only if $(1+\rho_n)^{-1}=\op\bigl((n_0+N)^{-1/4}\bigr)$, i.e.\ $\rho_n\gg(n_0+N)^{1/4}$.  (v)~For $t_0\in(0,1)$, on $\{\rho_n>-1\}$ the map $\rho\mapsto\rho^+/(1+\rho^+)$ is $1$-Lipschitz, and its inverse $t\mapsto t/(1-t)$ is Lipschitz on a neighbourhood of $t_0<1$, so the residual condition is equivalent to $|\rho_n-t_0/(1-t_0)|=\op\bigl((n_0+N)^{-1/4}\bigr)$.  Assumption~\ref{ass:external} bounds $\norm{\Delta}_1$ from above only and places no stabilisation on the sign-and-alignment structure of $B_{\mathrm{tun}}=\inner{\bm b^{(a)}_{\mathrm{tun}}}{\bm b^{(b)}_{\mathrm{tun}}}$, so this cannot be derived from the maintained assumptions (cf.\ Remark~\ref{rem:t-hat-scope}); it is retained as an assumption in this regime.
\end{proof}
}

\subsection{Proofs for Section~\ref{sec:refine} (Gaussian-design refinements)}\label{app:proofs-gauss}

\begin{proof}[Proof of Proposition~\ref{prop:gaussian-r}]
Because $M_1=\Omega$,
the residual in \eqref{eq:r-oof} simplifies to
\[
    r(\hat t)
    =
    \Delta - \hat t\,\Omega\hat\Sigma_0\Delta
    =
    (1-\hat t)\Delta - \hat t\,\Omega(\hat\Sigma_0-\Sigma)\Delta.
\]
Throughout we argue conditionally on $\Delta$, which is independent of $X_0$ by Assumption~\ref{ass:external-indep}, and on the event $\{\norm{\Sigma^{1/2}\Delta}_2\le a_2\}$, whose probability tends to one by hypothesis; conditionally, $\Delta$ is a fixed vector and the summands below are genuinely centred and independent across $i$.  The deterministic term is bounded by
\[
    \norm{(1-\hat t)\Delta}_\infty
    \le (1-\hat t)\norm{\Delta}_2
    \le C(1-\hat t)\norm{\Sigma^{1/2}\Delta}_2,
\]
using the lower eigenvalue bound on $\Sigma$.  For the stochastic term, for each $j$,
\[
    e_j^\top \Omega(\hat\Sigma_0-\Sigma)\Delta
    =
    \frac{1}{n_0}\sum_{i=1}^{n_0}
    \Bigl\{
        (e_j^\top\Omega X_{0,i})(X_{0,i}^\top \Delta)
        -
        \E[(e_j^\top\Omega X_{0,i})(X_{0,i}^\top \Delta)]
    \Bigr\}.
\]
Under i.i.d.\ sub-Gaussian design, each summand is a centred product of two sub-Gaussian linear forms and is therefore centred sub-exponential, by the Orlicz product bound $\norm{UV}_{\psi_1}\le\norm{U}_{\psi_2}\norm{V}_{\psi_2}$, with scale of order
$\norm{\Sigma^{1/2}\Delta}_2$, the factor $\Var(e_j^\top\Omega X)=(\Sigma^{-1})_{jj}\le\lambda_{\min}(\Sigma)^{-1}$ being bounded.  A Bernstein bound followed by a union bound over
$j\in\{1,\dots,p\}$ gives, with conditional probability at least $1-O(p^{-2})$,
\[
    \norm{\Omega(\hat\Sigma_0-\Sigma)\Delta}_\infty
    \le
    C\sqrt{\frac{\log p}{n_0}}\,\norm{\Sigma^{1/2}\Delta}_2 ;
\]
replacing $\Omega e_j$ by $e_j$ (a sub-Gaussian linear form with variance $\Sigma_{jj}\le\lambda_{\max}(\Sigma)$) in the same argument gives $\norm{(\hat\Sigma_0-\Sigma)\Delta}_\infty\le C\sqrt{\log p/n_0}\,\norm{\Sigma^{1/2}\Delta}_2$ on the same event.  Combining with the deterministic term and intersecting with the $a_2$ event yields the $\ell_\infty$ bound with probability tending to one.

For the weighted-$\ell_2$ bound, $\Sigma^{1/2}r(\hat t)=(1-\hat t)\Sigma^{1/2}\Delta-\hat t\,\Sigma^{1/2}\Omega(\hat\Sigma_0-\Sigma)\Delta$; the first term has norm at most $(1-\hat t)a_2$ on the conditioning event, and for the second, $\norm{\Sigma^{1/2}\Omega(\hat\Sigma_0-\Sigma)\Delta}_2\le C\norm{(\hat\Sigma_0-\Sigma)\Delta}_2$ by the bounded eigenvalues, while conditionally on $\Delta$
\[
    \E\bigl[\norm{(\hat\Sigma_0-\Sigma)\Delta}_2^2\,\big|\,\Delta\bigr]
    =\sum_{k=1}^p \Var\!\Bigl(\tfrac1{n_0}\textstyle\sum_i X_{0,ik}(X_{0,i}^\top\Delta)\Bigr)
    \le \frac{Cp}{n_0}\,\norm{\Sigma^{1/2}\Delta}_2^2
\]
by the bounded fourth moments of the sub-Gaussian rows.  The squared norm concentrates at its conditional mean: its conditional variance is of order $(\E[\norm{(\hat\Sigma_0-\Sigma)\Delta}_2^2\mid\Delta])^2\,(p^{-1}+n_0^{-1})$ (the cross-coordinate covariances contribute $\norm{\mathrm{Cov}}_{\mathrm F}^2\le\mathrm{tr}\cdot\norm{\cdot}_{\mathrm{op}}=O(p\,a_2^4/n_0^2)$ and the fourth-cumulant term carries an extra $n_0^{-1}$), so Chebyshev's inequality applied to $\norm{(\hat\Sigma_0-\Sigma)\Delta}_2^2$ gives $\norm{(\hat\Sigma_0-\Sigma)\Delta}_2\le C'\sqrt{p/n_0}\,a_2$ with probability tending to one at the fixed constant $C'$.  Combining the two terms yields the stated weighted-$\ell_2$ bound.  The sufficiency claim for requirement~(i) of Theorem~\ref{thm:CLT} follows by assembling the composite envelope: the $\ell_\infty$ entry contributes the first two terms of \eqref{eq:gaussian-bias-cond}, and the weighted-$\ell_2$ entry, carried at weight $\sqrt{\log p/N}$, contributes $(1-\hat t)\sqrt{\log p/N}\,a_2$ --- dominated by the $(1-\hat t)$ term since $\log p/N\to0$ --- plus the third term $\hat t\sqrt{p\log p/(Nn_0)}\,a_2$.
\end{proof}

\begin{proof}[Proof of Corollary~\ref{cor:gaussian-known}]

The argument is that of Theorem~\ref{thm:CLT}, specialised to the deterministic precision
choice $M_2=M_1=\Omega:=\Sigma^{-1}$ of Assumption~\ref{ass:gaussian-known}.  Two
features are special to the Gaussian refinement: the feasibility tolerances $\mu_1,\mu_2$ are
realised exactly (Step~1), and the residual-bias remainder is controlled through the
prediction norm of Proposition~\ref{prop:gaussian-r} (Step~4) rather than through the
$\ell_1$ bias envelope of Lemma~\ref{lem:bias-envelope}.  The debiasing remainder, by
contrast, is controlled exactly as in the generic theory, by the deterministic H\"older
bound; this is what fixes the sparsity requirement at $s=o(\sqrt{n_0}/\log p)$ in
Assumption~\ref{ass:gaussian-known}.  We proceed in four steps.

\smallskip
\noindent\emph{Step 1 (the design and JM-type hypotheses of Theorem~\ref{thm:CLT} hold).}
Assumption~\ref{ass:gaussian-known} strengthens Assumption~\ref{ass:design}: Gaussian rows
$N(0,\Sigma)$ are sub-Gaussian, and the eigenvalues of $\Sigma$ are bounded away from $0$
and $\infty$.  Assumption~\ref{ass:rsc} is imposed directly.  It remains to verify that the
deterministic choice $M_1=M_2=\Omega$ satisfies the feasibility and matrix-norm
controls of Assumption~\ref{ass:JM}, which is the only ingredient of the proof of
Theorem~\ref{thm:CLT} not assumed here verbatim.  Since $\Omega\Sigma=I_p$, for the stacked
pool of size $n_0+N$ we have $\Omega\hat\Sigma_{\mathrm{stk}}-I_p=\Omega(\hat\Sigma_{\mathrm{stk}}-\Sigma)$.
For fixed $(j,\ell)$ the entry $e_j^\top\Omega(\hat\Sigma_{\mathrm{stk}}-\Sigma)e_\ell$ is an average over
the $n_0+N$ pooled rows of centred products $(e_j^\top\Omega X)(e_\ell^\top X)$ of two
sub-Gaussian linear forms whose sub-Gaussian norms are controlled by the variances $(\Omega)_{jj}\le\lambda_{\min}(\Sigma)^{-1}$ and
$\Sigma_{\ell\ell}\le\lambda_{\max}(\Sigma)$; by the Orlicz product bound $\norm{UV}_{\psi_1}\le\norm{U}_{\psi_2}\norm{V}_{\psi_2}$ each summand is centred sub-exponential with a
constant scale, so Bernstein's inequality and a union bound over the $p^2$ entries give
\[
    \maxnorm{M_2\hat\Sigma_{\mathrm{stk}}-I_p}
    =\maxnorm{\Omega(\hat\Sigma_{\mathrm{stk}}-\Sigma)}
    =\Op\!\left(\sqrt{\tfrac{\log p}{n_0+N}}\right)
    \asymp \mu_2 .
\]
The identical computation on the labelled block yields
$\maxnorm{M_1\hat\Sigma_0-I_p}=\Op(\sqrt{\log p/n_0})\asymp\bar\mu_1$, so that
$\bar\mu_1\asymp\sqrt{(\log p)/n_0}$ as in the generic theory.  The variance-matching
limit holds here with equality---the oracle special case of Lemma~\ref{lem:m2-efficiency} at $M_2=\Omega$: $e_j^\top M_2\Sigma M_2^\top e_j
=e_j^\top\Omega\Sigma\Omega e_j=(\Omega)_{jj}=(\Sigma^{-1})_{jj}$ exactly, and likewise
$e_j^\top M_1\hat\Sigma_0 M_1^\top e_j=e_j^\top\Omega\hat\Sigma_0\Omega e_j\to(\Sigma^{-1})_{jj}=O(1)$.
Finally the row-sum control
\begin{equation}
    \mnorm{M_1}+\mnorm{M_2}=\Op(1)
    \label{eq:gauss-rowsum}
\end{equation}
holds with $\mnorm{M_1}=\mnorm{M_2}=\mnorm{\Omega}=O(1)$, by the row-summability clause of Assumption~\ref{ass:gaussian-known}.\footnote{The row-sum norm
\eqref{eq:gauss-rowsum} is used only to invoke the score and residual-bias bounds of
Lemma~\ref{lem:score} and Theorem~\ref{thm:CLT}; the debiasing remainder of Step~3 uses only
the entrywise tolerance $\mu_2$ and the $\ell_1$ error of $\hat\beta$.}

\smallskip
\noindent\emph{Step 2 (the debiased expansion and its leading term).}  With
$M_1=M_2=\Omega$ the correction matrix of Theorem~\ref{thm:CLT} becomes
\[
    A_n=I_p+\frac{\hat t\,N}{n_0}\hat\Sigma_{\mathrm u}\,\Omega ,
\]
and the debiased expansion of the proof of Theorem~\ref{thm:CLT} reads
\[
    \tilde\beta_j-\beta^\star_j
    =\frac{1}{n_0+N} e_j^\top\Omega A_n X_0^\top\varepsilon_0
    +R_{j,n},
    \qquad
    R_{j,n}=R_{j,n}^{\mathrm{bias}}+R_{j,n}^{\mathrm{deb}},
\]
where
\[
    R_{j,n}^{\mathrm{bias}}:=\frac{N}{n_0+N}\,e_j^\top\Omega\hat\Sigma_{\mathrm u}r(\hat t),
    \qquad
    R_{j,n}^{\mathrm{deb}}:=e_j^\top(I_p-\Omega\hat\Sigma_{\mathrm{stk}})(\hat\beta-\beta^\star).
\]
The leading term is unchanged from Theorem~\ref{thm:CLT}; we control the two remainders in
turn.

\smallskip
\noindent\emph{Step 3 (debiasing remainder).}  Write $u:=\hat\beta-\beta^\star$.  We bound
$R_{j,n}^{\mathrm{deb}}$ by the deterministic H\"older inequality used in the proof of
Theorem~\ref{thm:CLT}, applied with the \emph{exact} entrywise tolerance established in
Step~1,
\[
    \bigl|R_{j,n}^{\mathrm{deb}}\bigr|
    =\bigl|e_j^\top(I_p-\Omega\hat\Sigma_{\mathrm{stk}})\,u\bigr|
    \le\maxnorm{I_p-\Omega\hat\Sigma_{\mathrm{stk}}}\,\norm{u}_1
    =\Op\!\left(\sqrt{\tfrac{\log p}{n_0+N}}\right)\norm{u}_1 .
\]
By Theorem~\ref{thm:stacked-lasso}, tuned against $b_n=b_n^{\mathrm G}$ as validated in
Step~4, $\norm{u}_1\le(4s/\phi)\lambda=\Op(s\lambda)$, with
$\lambda\asymp\sigma\sqrt{\log p/(n_0+N)}+\tfrac{N}{n_0+N}\hat t\sqrt{\log p/n_0}
+\tfrac{N}{n_0+N}b_n^{\mathrm G}=O(\sqrt{\log p/n_0})$ (each summand is $O(\sqrt{\log p/n_0})$,
the surrogate term being $o((n_0+N)^{-1/2})$ by \eqref{eq:gaussian-bias-cond}).  Hence
\[
    \sqrt{n_0+N}\,\bigl|R_{j,n}^{\mathrm{deb}}\bigr|
    =\Op\!\left(\sqrt{\log p}\,s\,\lambda\right)
    =\Op\!\left(\frac{s\,\log p}{\sqrt{n_0}}\right)
    =\op(1),
\]
the last equality by the sparsity condition $s=o(\sqrt{n_0}/\log p)$ of
Assumption~\ref{ass:gaussian-known}.  This is the verification of condition~(ii) of
Theorem~\ref{thm:CLT} under the Gaussian refinement; the entrywise tolerance $\mu_2$ entering
the bound is realised \emph{exactly} (Step~1) rather than only up to feasibility, which is the
only role the known precision plays here.

{%
\begin{remark}[Sharper sparsity scaling under cross-fitting]\label{rem:gaussian-sharper-sparsity}
The displayed argument establishes the $\ell_1$-driven sparsity requirement
$s=o(\sqrt{n_0}/\log p)$, and this is all we claim under the present single-pool pipeline.  One
would like to weaken it to $s=o(n_0/(\log p)^2)$ via the prediction norm, using
$I_p-\Omega\hat\Sigma_{\mathrm{stk}}=-\Omega(\hat\Sigma_{\mathrm{stk}}-\Sigma)$ to write
$R_{j,n}^{\mathrm{deb}}=-\,e_j^\top\Omega(\hat\Sigma_{\mathrm{stk}}-\Sigma)\,u$ as a centred
sub-exponential average, which would give
$\bigl|R_{j,n}^{\mathrm{deb}}\bigr|=\Op\!\bigl(\sqrt{\log p/(n_0+N)}\,\norm{\Sigma^{1/2}u}_2\bigr)$
with $\norm{\Sigma^{1/2}u}_2=\Op(\sqrt{s/\phi}\,\lambda)$ by
Theorem~\ref{thm:stacked-lasso}, and hence
$\sqrt{n_0+N}\,\bigl|R_{j,n}^{\mathrm{deb}}\bigr|=\Op\!\bigl(\sqrt{s(\log p)^2/n_0}\bigr)=\op(1)$.
This sharper scaling is recorded as a conjecture, not a theorem: the multiplier
$u=\hat\beta-\beta^\star$ is the stacked Lasso \eqref{eq:stacked-lasso} computed from the very
pooled rows that form $\hat\Sigma_{\mathrm{stk}}$, so $u$ is not independent of those rows and
the summands are not conditionally independent given $u$; Assumption~\ref{ass:external-indep}
supplies independence from the inference \emph{noise}, not from the design, and so does not
legitimise conditioning on $u$.  A cross-fitted pipeline---computing the stacked Lasso on a
subsample of the pooled covariates disjoint from those used to form $\hat\Sigma_{\mathrm{stk}}$
in Stage~4---would render $u$ independent of the debiasing rows, after which the centred
sub-exponential bound of Proposition~\ref{prop:gaussian-r} applies verbatim and recovers
$s=o(n_0/(\log p)^2)$.  We do not pursue this refinement here (cf.\
Remark~\ref{rem:gaussian-sparsity}).
\end{remark}%
}

\smallskip
\noindent\emph{Step 4 (residual-bias remainder via Proposition~\ref{prop:gaussian-r}).}
Proposition~\ref{prop:gaussian-r} --- whose hypotheses (sub-Gaussian rows, $M_1=\Omega$, $\Delta$ independent of $X_0$, and the probability-tending-to-one bound at $a_2$) are all in force --- bounds both entries of the composite envelope with probability tending to one, so $b_n^{\mathrm G}$ from \eqref{eq:bn-gauss-known} is a valid surrogate in the sense of \eqref{eq:bn-surrogate}, and the stated tuning level
$\lambda$ is exactly the choice of Theorem~\ref{thm:stacked-lasso} with $b_n=b_n^{\mathrm
G}$, so that theorem applies and validates the $\ell_1$ and prediction-error bounds used in
Step~3.  For the bias remainder itself, Lemma~\ref{lem:bias-transfer} with $M=\Omega$ (and $r(\hat t)$ independent of $\tilde X$) gives
\begin{align*}
    \sqrt{n_0+N}\,\bigl|R_{j,n}^{\mathrm{bias}}\bigr|
    &\le \sqrt{n_0+N}\,\mnorm{\Omega}
        \Bigl(\mnorm{\Sigma}\,\norm{r(\hat t)}_\infty
        + C\sqrt{\tfrac{\log p}{N}}\,\norm{\Sigma^{1/2}r(\hat t)}_2\Bigr)\\
    &=\Op\!\left(\sqrt{n_0+N}\,\Bigl(\norm{r(\hat t)}_\infty+\sqrt{\tfrac{\log p}{N}}\,\norm{\Sigma^{1/2}r(\hat t)}_2\Bigr)\right),
\end{align*}
and substituting the two bounds of Proposition~\ref{prop:gaussian-r} makes the right-hand side
$\Op\bigl(\sqrt{n_0+N}\,a_2\{1-\hat t+\sqrt{\log p/n_0}+\hat t\sqrt{p\log p/(Nn_0)}\}\bigr)
=\op(1)$ by the residual-bias condition \eqref{eq:gaussian-bias-cond}.  This is the
verification of condition~(i) of Theorem~\ref{thm:CLT} under the Gaussian refinement.

\smallskip
\noindent\emph{Conclusion.}  Combining Steps~3--4, $\sqrt{n_0+N}\,R_{j,n}=\op(1)$,
so the asymptotic-linearity representation of Theorem~\ref{thm:CLT} holds with the leading
term
\[
    \frac{1}{\sqrt{n_0+N}}\sum_{i=1}^{n_0}
    \bigl(x_{0,i}^\top A_n^\top\Omega^\top e_j\bigr)\,\varepsilon_{0,i}.
\]
Conditionally on the covariates and the tuning sample, this is a sum of independent
mean-zero terms ($\hat t$, hence $A_n$, being independent of the inference-sample noise by
Assumption~\ref{ass:external-indep}), with conditional variance
\[
    \tau_{j,n}^2=\frac{\sigma^2 n_0}{n_0+N}\,
    e_j^\top\Omega A_n\hat\Sigma_0 A_n^\top\Omega e_j .
\]
Conditions~(iii)--(iv) of Theorem~\ref{thm:CLT} are verified as follows.  Condition~(iv) is scale-invariant in the weight vector: conditionally on the tuning sample and the unlabelled block the weights $x_{0,i}^\top A_n^\top\Omega^\top e_j$ are i.i.d.\ sub-Gaussian linear forms, so the self-normalised Lindeberg ratio is
$\Op(\sqrt{\log n_0/n_0})\to0$ whatever their common (possibly divergent) scale.  For condition~(iii), write, using $\Omega\Sigma\Omega e_j=\Omega e_j$,
\[
    w:=A_n^\top\Omega e_j
    =\Bigl(1+\frac{\hat tN}{n_0}\Bigr)\Omega e_j+\frac{\hat tN}{n_0}\,\Omega\zeta,
    \qquad
    \zeta:=(\hat\Sigma_{\mathrm u}-\Sigma)\Omega e_j,
\]
so that
\[
    \tau_{j,n}^2=\frac{\sigma^2 n_0}{n_0+N}\Bigl[\Bigl(1+\tfrac{\hat tN}{n_0}\Bigr)^2 (\Omega e_j)^\top\hat\Sigma_0\Omega e_j
    +2\Bigl(1+\tfrac{\hat tN}{n_0}\Bigr)\tfrac{\hat tN}{n_0}\,(\Omega e_j)^\top\hat\Sigma_0\Omega\zeta
    +\Bigl(\tfrac{\hat tN}{n_0}\Bigr)^2\zeta^\top\Omega\hat\Sigma_0\Omega\zeta\Bigr].
\]
The main term: $(\Omega e_j)^\top\hat\Sigma_0\Omega e_j\to_{\mathbb P}(\Sigma^{-1})_{jj}$ by the scalar law of large numbers at the fixed vector $\Omega e_j$, and $\tfrac{\sigma^2n_0}{n_0+N}(1+\tfrac{\hat tN}{n_0})^2\to\sigma^2(1+t_0\kappa)^2/(1+\kappa)$ by $\hat t\to_{\mathbb P}t_0$ and $\kappa_n\to\kappa$.  The quadratic fluctuation term: the labelled rows are independent of $(\hat\Sigma_{\mathrm u},\text{tuning})$, so $\E[\zeta^\top\Omega\hat\Sigma_0\Omega\zeta\mid\hat\Sigma_{\mathrm u},\text{tuning}]=\zeta^\top\Omega\Sigma\Omega\zeta\le C\norm{\zeta}_2^2$, and $\E\norm{\zeta}_2^2\le Cp/N$ since each coordinate of $\zeta$ has variance $O(1/N)$ by the bounded fourth moments of the sub-Gaussian rows; hence, by Markov's inequality, the quadratic term is
\[
    \frac{\sigma^2 n_0}{n_0+N}\Bigl(\frac{\hat tN}{n_0}\Bigr)^2\zeta^\top\Omega\hat\Sigma_0\Omega\zeta
    =\Op\!\Bigl(\hat t^{\,2}\,\frac{Np}{n_0(n_0+N)}\Bigr)=\op(1)
\]
by the precision-fluctuation condition of the statement, and the cross term is $\op(1)$ by Cauchy--Schwarz between the main and quadratic terms.  Hence
\[
    \tau_{j,n}^2\;\xrightarrow{\;\mathbb P\;}\;\tau_j^2:=\sigma^2\,\frac{(1+t_0\kappa)^2}{1+\kappa}\,(\Sigma^{-1})_{jj}\in(0,\infty),
\]
which is condition~(iii) with the limit as stated.  The Lindeberg--Feller theorem
and Slutsky's lemma therefore give
\[
    \sqrt{n_0+N}\,(\tilde\beta_j-\beta^\star_j)\Rightarrow N(0,\tau_j^2).
\]

\smallskip
\noindent\emph{First-order variance ratio.}  The unscaled variance of $\tilde\beta_j$ is $\tau_j^2/(n_0+N)=\sigma^2(1+t_0\kappa)^2(\Sigma^{-1})_{jj}/\{n_0(1+\kappa)^2\}$; dividing by the target-only
debiased-Lasso variance $\sigma^2(\Sigma^{-1})_{jj}/n_0$ yields
\[
    G(\kappa,t_0)=\left(\frac{1+t_0\kappa}{1+\kappa}\right)^2 .
\]
This completes the proof.
\end{proof}

\begin{proof}[Proof of Corollary~\ref{cor:gaussian-nodewise}]

The argument is that of Corollary~\ref{cor:gaussian-known}, now with the deterministic
oracle precision $\Omega=\Sigma^{-1}$ replaced everywhere by the nodewise-Lasso estimator
$\hat\Omega$ built from the $X$-only stack
$X_\Omega=(X_0;\tilde X)\in\R^{m_\Omega\times p}$, $m_\Omega=n_0+N$.  We retain the four-step
organisation of that proof and indicate only the modifications produced by the estimation of
$\Omega$.  Write $E:=\hat\Omega-\Omega$ for the precision-estimation error.  Throughout we
work on the high-probability event on which Assumptions~\ref{ass:rsc} and
\ref{ass:gaussian-nodewise} and the external prediction-norm bound
$\norm{\Sigma^{1/2}\Delta}_2=\Op(a_2)$ all hold.

\smallskip
\noindent\emph{Step 0 (nodewise-Lasso rates).}  Under i.i.d.\ sub-Gaussian design with eigenvalues of
$\Sigma$ bounded away from $0$ and $\infty$, the standard nodewise-Lasso analysis
\citep{vandegeer2014asymptotic} applied to the pooled $X$-only sample of size $m_\Omega$
gives, with probability tending to one and uniformly over rows $j$,
\begin{equation}
    \norm{e_j^\top E}_1=\Op\!\left(s_\Omega\sqrt{\tfrac{\log p}{m_\Omega}}\right),
    \qquad
    \maxnorm{\hat\Omega\hat\Sigma_{\mathrm{stk}}-I_p}
    =\Op\!\left(\sqrt{\tfrac{\log p}{m_\Omega}}\right)\asymp\mu_2 ,
    \label{eq:nw-rate}
\end{equation}
where the second display is the KKT feasibility of the nodewise programme, computed on
exactly the stacked design $\hat\Sigma_{\mathrm{stk}}=m_\Omega^{-1}X_\Omega^\top X_\Omega$.
Moreover $\mnorm{\hat\Omega}=\Op(1)$: $\mnorm{\Omega}=O(1)$ by the row-summability clause of
Assumption~\ref{ass:gaussian-nodewise}, and \eqref{eq:nw-rate} controls the deviation,
$\max_j\norm{e_j^\top E}_1=\Op\bigl(s_\Omega\sqrt{\log p/m_\Omega}\bigr)\le\Op(\delta_{\Omega,n}^{\circ})\to 0$,
so $\mnorm{\hat\Omega}\le\mnorm{\Omega}+\max_j\norm{e_j^\top E}_1=O(1)+\op(1)$.  (Row sparsity
and bounded eigenvalues alone would give only $\mnorm{\Omega}=O(\sqrt{s_\Omega})$, which is why
the row-summability is imposed in Assumption~\ref{ass:gaussian-nodewise}, mirroring the same
clause of Assumption~\ref{ass:gaussian-known}.)

\smallskip
\noindent\emph{Step 1 (JM-type hypotheses of Theorem~\ref{thm:CLT}).}  With
$M_1=M_2=\hat\Omega$, the feasibility control
$\maxnorm{M_2\hat\Sigma_{\mathrm{stk}}-I_p}=\mu_2\asymp\sqrt{\log p/(n_0+N)}$ holds by
\eqref{eq:nw-rate}, of the same order as the JM tolerance of \eqref{eq:JM-M2}; the
labelled-block analogue is obtained by combining
$\maxnorm{\Omega(\hat\Sigma_0-\Sigma)}=\Op(\sqrt{\log p/n_0})$ (the Bernstein argument of
Step~1 of Corollary~\ref{cor:gaussian-known}) with
$\maxnorm{E\hat\Sigma_0}\le(\max_j\norm{e_j^\top E}_1)\maxnorm{\hat\Sigma_0}
=\Op(\delta_{\Omega,n}^{\circ})$, giving $\bar\mu_1\asymp\sqrt{\log p/n_0}+\delta_{\Omega,n}^{\circ}$.
The row-sum control \eqref{eq:gauss-rowsum} holds with
$\mnorm{M_1}=\mnorm{M_2}=\mnorm{\hat\Omega}=\Op(1)$ by Step~0.  For the variance-matching
limit of Lemma~\ref{lem:m2-efficiency} (branch~(B)), $e_j^\top\hat\Omega\Sigma\hat\Omega^\top e_j
=e_j^\top\Omega\Sigma\Omega e_j+\Op\bigl(\norm{e_j^\top E}_1\bigr)
=(\Sigma^{-1})_{jj}+\Op(\delta_{\Omega,n}^{\circ})\to(\Sigma^{-1})_{jj}$, since
$\delta_{\Omega,n}^{\circ}\to0$.  Thus the clauses of Assumption~\ref{ass:JM}, together with the variance-matching limit of Lemma~\ref{lem:m2-efficiency}, hold for the
estimated precision, with the one qualification that the $M_1$ tolerance is realised at the enlarged level $\bar\mu_1\asymp\sqrt{\log p/n_0}+\delta_{\Omega,n}^{\circ}$ rather than the sharper $\mu_1\asymp\sqrt{\log p/n_0}$; the proof invokes only this realised tolerance, never the sharper one.

\smallskip
\noindent\emph{Step 2 (debiased expansion).}  The expansion of Step~2 of
Corollary~\ref{cor:gaussian-known} holds verbatim with $\Omega$ replaced by $\hat\Omega$:
\[
    A_n=I_p+\frac{\hat t\,N}{n_0}\hat\Sigma_{\mathrm u}\hat\Omega,
    \qquad
    \tilde\beta_j-\beta^\star_j
    =\frac{1}{n_0+N} e_j^\top\hat\Omega A_n X_0^\top\varepsilon_0
    +R_{j,n}^{\mathrm{bias}}+R_{j,n}^{\mathrm{deb}},
\]
with $R_{j,n}^{\mathrm{bias}}=\tfrac{N}{n_0+N}e_j^\top\hat\Omega\hat\Sigma_{\mathrm u}r(\hat t)$ and
$R_{j,n}^{\mathrm{deb}}=e_j^\top(I_p-\hat\Omega\hat\Sigma_{\mathrm{stk}})(\hat\beta-\beta^\star)$.

\smallskip
\noindent\emph{Step 3 (debiasing remainder).}  We split the JM residual into
its oracle and estimation parts,
\[
    I_p-\hat\Omega\hat\Sigma_{\mathrm{stk}}
    =\underbrace{-\,\Omega(\hat\Sigma_{\mathrm{stk}}-\Sigma)}_{\text{oracle part}}
    \;-\;\underbrace{E\,\hat\Sigma_{\mathrm{stk}}}_{\text{estimation part}} ,
\]
using $\Omega\hat\Sigma_{\mathrm{stk}}-I_p=\Omega(\hat\Sigma_{\mathrm{stk}}-\Sigma)$ and
$\hat\Omega=\Omega+E$.  The oracle part is the debiasing remainder of Step~3 of
Corollary~\ref{cor:gaussian-known} verbatim --- the deterministic H\"older bound at the pooled entrywise tolerance, giving $\sqrt{n_0+N}\,|e_j^\top\Omega(\hat\Sigma_{\mathrm{stk}}-\Sigma)u|=\Op(s\log p/\sqrt{n_0})=\op(1)$ under $s=o(\sqrt{n_0}/\log p)$, with the tuning level $\lambda=O(\sqrt{\log p/n_0})$
validated in Step~4 below --- and the prediction-norm route remains unavailable here for the reason recorded in Remark~\ref{rem:gaussian-sharper-sparsity}.  For the estimation part, pair the row-$\ell_1$ bound of
\eqref{eq:nw-rate} with the $\ell_\infty$ control of $\hat\Sigma_{\mathrm{stk}}u$ furnished by
the stacked-Lasso KKT conditions, namely
$\norm{\hat\Sigma_{\mathrm{stk}}u}_\infty\le\norm{(n_0+N)^{-1}X_{\mathrm{stk}}^\top
(y_{\mathrm{stk}}-X_{\mathrm{stk}}\beta^\star)}_\infty+\lambda=\Op(\lambda)
=\Op(\sqrt{\log p/n_0})$ (the score term is $\Op(\lambda)$ by the choice of $\lambda$ in
Theorem~\ref{thm:stacked-lasso}).  Hence
\[
    \bigl|e_j^\top E\,\hat\Sigma_{\mathrm{stk}}u\bigr|
    \le\norm{e_j^\top E}_1\,\norm{\hat\Sigma_{\mathrm{stk}}u}_\infty
    =\Op\!\left(s_\Omega\sqrt{\tfrac{\log p}{m_\Omega}}\right)\Op\!\left(\sqrt{\tfrac{\log p}{n_0}}\right),
\]
so, using $m_\Omega=n_0+N=(1+\kappa+o(1))n_0$,
\[
    \sqrt{n_0+N}\,\bigl|e_j^\top E\,\hat\Sigma_{\mathrm{stk}}u\bigr|
    =\Op\!\left(s_\Omega\,\frac{\log p}{\sqrt{n_0}}\right)
    =\Op\!\left(\sqrt{1+\kappa}\;\delta_{\Omega,n}^{\circ}\right)
    =\op(1),
\]
the last step because $N/n_0\to\kappa\in(0,\infty)$ is finite and
$\delta_{\Omega,n}^{\circ}\to0$.  Combining the two parts,
$\sqrt{n_0+N}\,|R_{j,n}^{\mathrm{deb}}|=\op(1)$.

\smallskip
\noindent\emph{Step 4 (residual-bias remainder and validity of $b_n^{\mathrm{G},\Omega}$).}
With $M_1=\hat\Omega$ the residual \eqref{eq:r-oof} reduces, using
$\hat\Omega\hat\Sigma_0\Delta=\Delta+\Omega(\hat\Sigma_0-\Sigma)\Delta+E\hat\Sigma_0\Delta$, to
\[
    r(\hat t)
    =\underbrace{(1-\hat t)\Delta-\hat t\,\Omega(\hat\Sigma_0-\Sigma)\Delta}_{\text{oracle part}}
    \;-\;\hat t\,\underbrace{E\,\hat\Sigma_0\Delta}_{\text{estimation part}} .
\]
The oracle part is exactly the residual of Proposition~\ref{prop:gaussian-r}, of order
$\Op\bigl((1-\hat t+\sqrt{\log p/n_0})\,a_2\bigr)$.  For the estimation part, write
$\hat\Sigma_0\Delta=\Sigma^{1/2}(\Sigma^{1/2}\Delta)+(\hat\Sigma_0-\Sigma)\Delta$.  Using
$\norm{e_j^\top E}_1$ from \eqref{eq:nw-rate}, the first piece is
\begin{align*}
\norm{E\Sigma^{1/2}(\Sigma^{1/2}\Delta)}_\infty
&\le(\max_j\norm{e_j^\top E}_1)\,\norm{\Sigma^{1/2}(\Sigma^{1/2}\Delta)}_\infty\\
&\le\Op(\delta_{\Omega,n}^{\circ})\,C\norm{\Sigma^{1/2}\Delta}_2
=\Op(\delta_{\Omega,n}^{\circ})\,a_2,
\end{align*}
where $\norm{\Sigma^{1/2}w}_\infty\le
\norm{\Sigma^{1/2}w}_2\le C\norm{w}_2$ by the bounded eigenvalues; the second piece is
$\Op(\delta_{\Omega,n}^{\circ})\cdot\Op(\sqrt{\log p/n_0}\,a_2)$ by the Bernstein bound for
$\norm{(\hat\Sigma_0-\Sigma)\Delta}_\infty$ recorded in Proposition~\ref{prop:gaussian-r}, of strictly
higher order.  Hence
$\norm{\hat t\,E\hat\Sigma_0\Delta}_\infty=\Op(\delta_{\Omega,n}^{\circ})\,a_2$, and
collecting the two parts,
\[
    \norm{r(\hat t)}_\infty
    =\Op\!\left(a_2\Bigl(1-\hat t+\sqrt{\tfrac{\log p}{n_0}}+\delta_{\Omega,n}^{\circ}\Bigr)\right).
\]
For the weighted-$\ell_2$ entry of the composite envelope, the oracle part is bounded by Proposition~\ref{prop:gaussian-r}, $\Op(a_2\{(1-\hat t)+\hat t\sqrt{p/n_0}\})$, and the estimation part by the crude conversion
\[
    \norm{\Sigma^{1/2}\,\hat t\,E\hat\Sigma_0\Delta}_2
    \le C\sqrt p\,\norm{E\hat\Sigma_0\Delta}_\infty
    \le C\sqrt p\,\bigl(\max_j\norm{e_j^\top E}_1\bigr)\norm{\hat\Sigma_0\Delta}_\infty
    =\Op\!\bigl(\sqrt p\;\delta_{\Omega,n}^{\circ}\,a_2\bigr),
\]
using $\norm{\hat\Sigma_0\Delta}_\infty\le\norm{\Sigma\Delta}_\infty+\norm{(\hat\Sigma_0-\Sigma)\Delta}_\infty=\Op(a_2)$.  Carried at the envelope weight $\sqrt{\log p/N}$, the estimation part contributes $\Op(\delta_{\Omega,n}^{\circ}\sqrt{p\log p/N}\,a_2)$, so the composite envelope is
$\Op\bigl(a_2\{1-\hat t+\sqrt{\log p/n_0}+\hat t\sqrt{p\log p/(Nn_0)}+\delta_{\Omega,n}^{\circ}(1+\sqrt{p\log p/N})\}\bigr)$ --- of the order of the nodewise surrogate \eqref{eq:bn-gauss-node} (recall
$\delta_{\Omega,n}=\delta_{\Omega,n}^{\circ}$).  Since the oracle-part bounds hold with probability tending to one at explicit constants (Proposition~\ref{prop:gaussian-r}) and the estimation-part bounds likewise (\eqref{eq:nw-rate} is a probability-tending-to-one statement), for $C_{\mathrm{G},\Omega}$ large enough the composite envelope is at most $b_n^{\mathrm{G},\Omega}$ with probability tending to one: $b_n^{\mathrm{G},\Omega}$ is a
valid surrogate in the sense of \eqref{eq:bn-surrogate}, and the tuning of Step~3 through
Theorem~\ref{thm:stacked-lasso} is licensed.  The bias remainder is then bounded by Lemma~\ref{lem:bias-transfer} with $M=\hat\Omega$ exactly as in Step~4 of Corollary~\ref{cor:gaussian-known}, with $\mnorm{\hat\Omega}=\Op(1)$ from Step~0 in place of $\mnorm{\Omega}$ ($r(\hat t)$ involves only the labelled-block correction and is independent of $\tilde X$): the oracle contribution is
$\op(1)$ by the residual-bias condition \eqref{eq:gaussian-bias-cond}, and the
estimation contribution $\Op\bigl(\sqrt{n_0+N}\,a_2\,\delta_{\Omega,n}^{\circ}(1+\sqrt{p\log p/N})\bigr)$
is $\op(1)$ by the corresponding hypothesis of the corollary.  Thus
$\sqrt{n_0+N}\,R_{j,n}^{\mathrm{bias}}=\op(1)$, verifying condition~(i) of
Theorem~\ref{thm:CLT}.

\smallskip
\noindent\emph{Conclusion and variance.}  Combining Steps~3--4,
$\sqrt{n_0+N}\,R_{j,n}=\op(1)$, and the asymptotic-linearity representation of
Theorem~\ref{thm:CLT} holds with leading term
$\tfrac{1}{\sqrt{n_0+N}}\sum_{i=1}^{n_0}
(x_{0,i}^\top A_n^\top\hat\Omega^\top e_j)\varepsilon_{0,i}$, a conditional sum of
independent mean-zero terms by Assumption~\ref{ass:external-indep}, with conditional variance
\[
    \tau_{j,n}^2
    =\frac{\sigma^2 n_0}{n_0+N}\,
    e_j^\top\hat\Omega A_n\hat\Sigma_0 A_n^\top\hat\Omega^\top e_j .
\]
Conditions~(iii)--(iv) of Theorem~\ref{thm:CLT} at $M_2=M_1=\hat\Omega$ hold here by hypothesis, with the limit in~(iii) equal to the $\tau_j^2$ of Corollary~\ref{cor:gaussian-known}; Remark~\ref{rem:nodewise-iii-hypothesis} records why they enter as hypotheses rather than being verified --- $\hat\Omega$'s dependence on the pooled covariate stack disables the conditioning device of that corollary's conclusion step.  The
Lindeberg--Feller theorem and Slutsky's lemma therefore give, for each fixed coordinate $j$,
\[
    \sqrt{n_0+N}\,(\tilde\beta_j-\beta^\star_j)\Rightarrow N(0,\tau_j^2),
\]
and dividing the unscaled variance $\tau_j^2/(n_0+N)$ by the target-only debiased-Lasso variance $\sigma^2(\Sigma^{-1})_{jj}/n_0$ reproduces
$G(\kappa,t_0)=\bigl((1+t_0\kappa)/(1+\kappa)\bigr)^2$, exactly as in
Corollary~\ref{cor:gaussian-known}.  This completes the proof.
\end{proof}

\subsection{Proofs for Section~\ref{sec:robust} (comparison and robustness)}\label{app:proofs-robust}

\begin{proof}[Proof of Proposition~\ref{prop:shift-amp}]

\emph{(i) Population identity.}
From $\Sigma_{\mathrm{stk}}=(n_0\Sigma_0+N\Sigma_{\mathrm u})/(n_0+N)=\Sigma_{\mathrm u}-\tfrac{n_0}{n_0+N}D$ we have $\Sigma_{\mathrm u}=\Sigma_{\mathrm{stk}}+\tfrac{n_0}{n_0+N}D$, so $M_2\Sigma_{\mathrm u}=M_2\Sigma_{\mathrm{stk}}+\tfrac{n_0}{n_0+N}M_2D$ for any matrix $M_2$; the stated row-$\ell_1$ bound is the triangle inequality applied to row $j$.

\emph{(ii) Uniform bound.}
For any matrices $A,B$ and index $j$, $\|e_j^\top AB\|_1=\sum_l|\sum_kA_{jk}B_{kl}|\le\sum_k|A_{jk}|\sum_l|B_{kl}|\le\|e_j^\top A\|_1\mnorm{B}$, so $\rho_j(M_2,\Sigma_{\mathrm u})=\|e_j^\top M_2\Sigma_{\mathrm u}\|_1\le\mnorm{M_2}\,\mnorm{\Sigma_{\mathrm u}}$.  Here $\mnorm{M_2}=\Op(1)$ by Assumption~\ref{ass:JM} and $\mnorm{\Sigma_{\mathrm u}}\le C_{\Sigma,1}$ by Assumption~\ref{ass:design}(a) imposed on the unlabelled block.

\emph{(iii) Non-centring bound.}
The expansion of $\tilde\beta-\beta^\star$ in the proof of Theorem~\ref{thm:CLT} is purely algebraic and does not use $\hat\Sigma_{\mathrm u}\to\Sigma_0$; it therefore holds verbatim under shift and gives, for the fixed coordinate $j$,
\begin{equation}\label{eq:shift-amp-exact}
    \tilde\beta_j-\beta^\star_j
    =S_j+\text{Term A}+\text{Term B},
\end{equation}
which is \eqref{eq:bias-shift-decomp}.  Here $\Delta_{\mathrm{init}}:=\tilde\beta^{\mathrm{init}}-\beta^\star=r(\hat t)+\hat t\,Z_1$ (Lemma~\ref{lem:oof-decomposition}), and the stochastic part $\hat t Z_1$ of the imputation propagates, through $A_n$, into $S_j$ alone, so Term~B's non-vanishing content is carried by $r(\hat t)$.  For Term~A, Assumption~\ref{ass:JM} gives $\maxnorm{I_p-M_2\hat\Sigma_{\mathrm{stk}}}\le\mu_2$, so by H\"older's inequality and the stacked-Lasso $\ell_1$ rate,
\[
    \sqrt{n_0+N}\,|\text{Term A}|
    \le\sqrt{n_0+N}\,\mu_2\,\norm{\hat\beta-\beta^\star}_1
    =\Op\!\bigl(\sqrt{n_0+N}\,\mu_2 s\lambda\bigr)=\op(1)
\]
by Theorem~\ref{thm:stacked-lasso} and regime condition~(ii).  For Term~B, Lemma~\ref{lem:bias-transfer} applies with $M=M_2$, $v=r(\hat t)$ (independent of $\tilde X$ by Assumption~\ref{ass:external-indep}) and $\Sigma_{\mathrm u}$ in place of $\Sigma$ --- its proof uses only the sub-Gaussianity of the unlabelled rows --- giving
\[
    \bigl|e_j^\top M_2\hat\Sigma_{\mathrm u}r(\hat t)\bigr|
    \le\mnorm{M_2}\Bigl(\mnorm{\Sigma_{\mathrm u}}\norm{r(\hat t)}_\infty+C\sqrt{\log p/N}\,\norm{\Sigma_{\mathrm u}^{1/2}r(\hat t)}_2\Bigr),
\]
which is the stated bound on Term~B after multiplying by $\sqrt{n_0+N}\,N/(n_0+N)$.  Combining the two bounds completes the proof.
\end{proof}

\begin{proof}[Proof of Corollary~\ref{cor:shift-amp-benchmark}]

\emph{(i)} At $\Sigma_{\mathrm u}=\Sigma_0$, $\Sigma_{\mathrm{stk}}=\Sigma_{\mathrm u}$, so $M_2\Sigma_{\mathrm u}=\Sigma_{\mathrm{stk}}^{-1}\Sigma_{\mathrm{stk}}=I_p$ exactly, and $\rho_j(M_2,\Sigma_{\mathrm u})=\|e_j^\top\|_1=1$ for every $j$.

\emph{(ii)} Write $D:=\Sigma_{\mathrm u}-\Sigma_0\succ0$.  By Proposition~\ref{prop:shift-amp}(i), $M_2\Sigma_{\mathrm u}=I_p+\tfrac{n_0}{n_0+N}H$, $H:=M_2D=\Sigma_{\mathrm{stk}}^{-1}D$, since $M_2\Sigma_{\mathrm{stk}}=I_p$ here.  Factor the symmetric positive-definite $M_2=\Sigma_{\mathrm{stk}}^{-1}=PP^\top$ with $P$ nonsingular.  Then $\operatorname{tr}(H)=\operatorname{tr}(\Sigma_{\mathrm{stk}}^{-1}D)=\operatorname{tr}(P^\top DP)>0$, since $P^\top DP\succ0$ has strictly positive diagonal entries and hence strictly positive trace.  Because $\sum_jH_{jj}=\operatorname{tr}(H)>0$, some coordinate $j_0$ has $H_{j_0j_0}\ge p^{-1}\operatorname{tr}(\Sigma_{\mathrm{stk}}^{-1}D)>0$, so $(M_2\Sigma_{\mathrm u})_{j_0j_0}=1+\tfrac{n_0}{n_0+N}H_{j_0j_0}>1$, and since the row-$\ell_1$ norm dominates any single entry in absolute value, $\rho_{j_0}(M_2,\Sigma_{\mathrm u})\ge(M_2\Sigma_{\mathrm u})_{j_0j_0}>1$.

For the nodewise refinement, $\hat\Omega_{\mathrm{stk}}$ is symmetric positive-definite and converges to $\Sigma_{\mathrm{stk}}^{-1}$ in row-$\ell_1$ norm with an $\op(1)$ remainder (Assumption~\ref{ass:gaussian-nodewise}), so both displays carry through with an additional $\op(1)$ term: $\rho_j(\hat M_2,\Sigma_{\mathrm u})=1+\op(1)$ under~(i), and $(\hat M_2\Sigma_{\mathrm u})_{j_0j_0}=1+\tfrac{n_0}{n_0+N}H_{j_0j_0}+\op(1)>1$ for large $n_0,N$ under~(ii), whenever $n_0/(n_0+N)\to c>0$ and $H_{j_0j_0}$ is bounded away from zero.  We do not claim a uniform-in-$p$ margin or a placement of $j_0$ in the leading eigen-direction of $\Sigma_{\mathrm u}$; locating the amplified coordinate would require an additional spectral-gap and eigenvector-alignment hypothesis on $D$ restricted to $J$, not implied by Assumptions~\ref{ass:design}, \ref{ass:rsc}, \ref{ass:JM}, or~\ref{ass:gaussian-nodewise} and not invoked elsewhere.
\end{proof}

\begin{proof}[Proof of Proposition~\ref{prop:shift-bias-reduction}]
By Assumption~\ref{ass:JM} on the unlabelled block, $\mnorm{\hat M_{\mathrm u}}=\Op(1)$.  Lemma~\ref{lem:bias-transfer}, applied with $M=\hat M_{\mathrm u}$, $v=r(\hat t)$ (independent of $\tilde X$ by Assumption~\ref{ass:external-indep}) and $\Sigma_{\mathrm u}$ in place of $\Sigma$ --- its proof uses only the sub-Gaussianity of the unlabelled rows --- gives
\begin{align*}
    \bigl|e_j^\top\hat M_{\mathrm u}\hat\Sigma_{\mathrm u}r(\hat t)\bigr|
    &\le \mnorm{\hat M_{\mathrm u}}\Bigl(\mnorm{\Sigma_{\mathrm u}}\norm{r(\hat t)}_\infty+C\sqrt{\log p/N}\norm{\Sigma_{\mathrm u}^{1/2}r(\hat t)}_2\Bigr)\\
    &= \Op\!\Bigl(\norm{r(\hat t)}_\infty+\sqrt{\log p/N}\norm{\Sigma_{\mathrm u}^{1/2}r(\hat t)}_2\Bigr),
\end{align*}
using $\mnorm{\Sigma_{\mathrm u}}\le C_{\Sigma,1}$ (Assumption~\ref{ass:design}(a)); multiplying by $\sqrt{n_0+N}\,N/(n_0+N)\le\sqrt{n_0+N}$ gives the stated bound on $\mathrm{Term\ B}^{\mathrm u}$.  This is exactly the bound Lemma~\ref{lem:bias-transfer} gives for the no-shift imputation-bias term in the proof of Theorem~\ref{thm:CLT}, and for the pooled Term~B in the proof of Proposition~\ref{prop:shift-amp}(iii), with $\hat M_{\mathrm u}$ in place of $M_2$ --- the row-sum bound $\mnorm{\hat M_{\mathrm u}}=\Op(1)$ requires no feasibility property, so this route alone does not yet distinguish pooled from substituted.  The distinguishing fact is that $\hat M_{\mathrm u}$, unlike $M_{\mathrm p}$, is feasible for $\hat\Sigma_{\mathrm u}$ directly: the unlabelled-block JM programme gives $\maxnorm{\hat M_{\mathrm u}\hat\Sigma_{\mathrm u}-I_p}\le\mu_{\mathrm u}\asymp\sqrt{\log p/N}$ with probability tending to one (Assumption~\ref{ass:JM}), whereas $M_{\mathrm p}\hat\Sigma_{\mathrm u}-I_p$ carries the additional population-level term $\tfrac{n_0}{n_0+N}M_{\mathrm p}D$ (Proposition~\ref{prop:shift-amp}(i)) that Corollary~\ref{cor:shift-amp-benchmark}(ii) shows is not negligible at some coordinate whenever $\Sigma_{\mathrm u}\succ\Sigma_0$ and $n_0/(n_0+N)$ stays bounded away from zero: at any such fixed, finite budget the substitution removes a term the pooled construction cannot.
\end{proof}

\begin{proof}[Proof of Corollary~\ref{cor:shift-restore}]

Write $M_{\mathrm u}:=\hat M_{\mathrm u}$ and $n_{\mathrm{stk}}:=n_0+N$.  The modified estimator is \eqref{eq:final-debias} with $M_2$ replaced by $M_{\mathrm u}$.

\emph{Step 1 (asymptotic linearity).}
The algebraic expansion in the proof of Theorem~\ref{thm:CLT} does not use any feasibility property of the debiasing matrix; repeating it verbatim with $M_{\mathrm u}$ in place of $M_2$ gives, for the fixed coordinate $j$,
\begin{equation}\label{eq:shift-restore-decomp}
\begin{gathered}
    \tilde\beta^{\mathrm{adapt}}_j - \beta^\star_j
    =
    \frac{1}{n_{\mathrm{stk}}} e_j^\top M_{\mathrm u}A_n X_0^\top\varepsilon_0
    + R_{j,n}^{\mathrm{u}},\\
    R_{j,n}^{\mathrm{u}}
    :=
    \underbrace{e_j^\top(I_p-M_{\mathrm u}\hat\Sigma_{\mathrm{stk}})(\hat\beta-\beta^\star)}_{\text{Term A}}
    +
    \underbrace{\frac{N}{n_{\mathrm{stk}}}e_j^\top M_{\mathrm u}\hat\Sigma_{\mathrm u}\,r(\hat t)}_{\text{Term B}},
\end{gathered}
\end{equation}
with $A_n=I_p+\tfrac{\hat t N}{n_0}\hat\Sigma_{\mathrm u}M_1$ as in Theorem~\ref{thm:CLT}.

Both remainder steps below bound $R_{j,n}^{\mathrm u}$ at the sharper target $\tfrac{n_{\mathrm{stk}}}{\sqrt{n_0}}R_{j,n}^{\mathrm u}=\op(1)$ that Remark~\ref{rem:shift-restore-regime} explains is the one the normalisation $n_{\mathrm{stk}}/\sqrt{n_0}$ requires (rather than the coarser $\sqrt{n_{\mathrm{stk}}}R_{j,n}^{\mathrm u}=\op(1)$, which the same bounds also deliver \emph{a fortiori} since $n_{\mathrm{stk}}/\sqrt{n_0}=\sqrt{n_{\mathrm{stk}}}\cdot\sqrt{1+\kappa_n}\ge\sqrt{n_{\mathrm{stk}}}$).

\emph{Step 2 (Term~B).}
By Assumption~\ref{ass:JM} on the unlabelled block, $\mnorm{M_{\mathrm u}}=\Op(1)$.  Lemma~\ref{lem:bias-transfer}, applied with $M=M_{\mathrm u}$, $v=r(\hat t)$ (independent of $\tilde X$ by Assumption~\ref{ass:external-indep}) and $\Sigma_{\mathrm u}$ in place of $\Sigma$ (its proof uses only the sub-Gaussianity of the unlabelled rows), gives
\[
    |\text{Term B}|
    \le
    \frac{N}{n_{\mathrm{stk}}}\,\mnorm{M_{\mathrm u}}\,
    \Bigl(\mnorm{\Sigma_{\mathrm u}}\,\norm{r(\hat t)}_\infty
    + C\sqrt{\tfrac{\log p}{N}}\,\norm{\Sigma_{\mathrm u}^{1/2}r(\hat t)}_2\Bigr).
\]
Since $N/n_{\mathrm{stk}}\le1$ and $N/\sqrt{n_0}=\sqrt{N\kappa_n}\le\sqrt{n_{\mathrm{stk}}\kappa_n}$,
\begin{align*}
    \frac{n_{\mathrm{stk}}}{\sqrt{n_0}}\,|\text{Term B}|
    &\le
    \frac{N}{\sqrt{n_0}}\,\Op(1)\,
    \Bigl(\norm{r(\hat t)}_\infty
    + C\sqrt{\tfrac{\log p}{N}}\,\norm{\Sigma_{\mathrm u}^{1/2}r(\hat t)}_2\Bigr)\\
    &\le
    \Op(1)\,\sqrt{n_{\mathrm{stk}}\kappa_n}\Bigl(\norm{r(\hat t)}_\infty+\sqrt{\tfrac{\log p}{N}}\norm{\Sigma_{\mathrm u}^{1/2}r(\hat t)}_2\Bigr)
    =\op(1)
\end{align*}
by the sharpened $\Sigma_{\mathrm u}$-analogue of regime condition~(i).

\emph{Step 3 (Term~A).}
$M_{\mathrm u}$ is feasible for $\hat\Sigma_{\mathrm u}$ but not for $\hat\Sigma_0$: only the crude bound $\|e_j^\top(I_p-M_{\mathrm u}\hat\Sigma_0)\|_\infty\le1+\mnorm{M_{\mathrm u}}\maxnorm{\hat\Sigma_0}=\Op(1)$ is available (Assumption~\ref{ass:JM}; $\maxnorm{\hat\Sigma_0}=\Op(1)$ by Assumption~\ref{ass:design}(a)).  Decomposing $\hat\Sigma_{\mathrm{stk}}=\tfrac{n_0}{n_{\mathrm{stk}}}\hat\Sigma_0+\tfrac{N}{n_{\mathrm{stk}}}\hat\Sigma_{\mathrm u}$ and using the unlabelled-block feasibility $\|e_j^\top(I_p-M_{\mathrm u}\hat\Sigma_{\mathrm u})\|_\infty\le\mu_{\mathrm u}\asymp\sqrt{\log p/N}$ (Assumption~\ref{ass:JM}),
\[
    \bigl\|e_j^\top(I_p-M_{\mathrm u}\hat\Sigma_{\mathrm{stk}})\bigr\|_\infty
    \le
    \frac{n_0}{n_{\mathrm{stk}}}\,\Op(1)+\frac{N}{n_{\mathrm{stk}}}\,\mu_{\mathrm u}
    = \Op\!\Bigl(\frac{n_0}{n_{\mathrm{stk}}}\Bigr)+\Op\!\Bigl(\sqrt{\tfrac{\log p}{n_{\mathrm{stk}}}}\Bigr).
\]
By H\"older's inequality and the stacked-Lasso $\ell_1$ rate $\norm{\hat\beta-\beta^\star}_1=\Op(s\sqrt{\log p/n_{\mathrm{stk}}})$ (Theorem~\ref{thm:stacked-lasso}, tuning parameter $\lambda\asymp\sqrt{\log p/n_{\mathrm{stk}}}$),
\[
    |\text{Term A}|
    =
    \Op\!\Bigl(\frac{sn_0\sqrt{\log p}}{n_{\mathrm{stk}}^{3/2}}\Bigr)+\Op\!\Bigl(\frac{s\log p}{n_{\mathrm{stk}}}\Bigr),
\]
so that
\[
    \frac{n_{\mathrm{stk}}}{\sqrt{n_0}}\,|\text{Term A}|
    =
    \Op\!\Bigl(s\sqrt{\log p}\cdot\sqrt{\tfrac{n_0}{n_{\mathrm{stk}}}}\Bigr)+\Op\!\Bigl(\frac{s\log p}{\sqrt{n_0}}\Bigr)
    =
    \Op\!\Bigl(\frac{s\sqrt{\log p}}{\sqrt{1+\kappa_n}}\Bigr)+\Op\!\Bigl(\frac{s\log p}{\sqrt{n_0}}\Bigr)
    =\op(1),
\]
the first term by $s\sqrt{\log p}=o(\sqrt{\kappa_n})$, the second by $s\log p=o(\sqrt{n_0})$.  This is precisely the price of $M_{\mathrm u}$'s infeasibility for $\hat\Sigma_0$: the labelled fraction $n_0/n_{\mathrm{stk}}$ must vanish, and fast enough relative to $\kappa_n$'s growth, to absorb the otherwise-uncontrolled labelled-block row (Remark~\ref{rem:shift-restore-regime}); the pooled $M_{\mathrm p}$, feasible for $\hat\Sigma_{\mathrm{stk}}$, needs no such condition.

\emph{Step 4 (the CLT).}
Combining Steps~2--3, $\tfrac{n_{\mathrm{stk}}}{\sqrt{n_0}}\,R_{j,n}^{\mathrm{u}}=\op(1)$, so multiplying \eqref{eq:shift-restore-decomp} by $n_{\mathrm{stk}}/\sqrt{n_0}$,
\[
    \frac{n_{\mathrm{stk}}}{\sqrt{n_0}}\,\bigl(\tilde\beta^{\mathrm{adapt}}_j-\beta^\star_j\bigr)
    =
    \frac{1}{\sqrt{n_0}}\sum_{i=1}^{n_0}
    \bigl(x_{0,i}^\top A_n^\top M_{\mathrm u}^\top e_j\bigr)\,\varepsilon_{0,i}
    + \op(1).
\]
The leading term is, conditionally on the covariates and the tuning data, a normalised sum of $n_0$ independent mean-zero variables with conditional variance $\tfrac1{n_0}\sum_{i=1}^{n_0}(x_{0,i}^\top A_n^\top M_{\mathrm u}^\top e_j)^2\sigma^2=\tau_{j,n}^{2,\mathrm u}$, since $\hat t$ (hence $A_n$) is tuning-measurable and $M_{\mathrm u}$ is a function of the unlabelled covariates only, both independent of $\{\varepsilon_{0,i}\}$ (Assumption~\ref{ass:external-indep}, the i.i.d.\ structure of Assumption~\ref{ass:design}).  By hypothesis, $\tau_{j,n}^{2,\mathrm u}\to_{\mathbb P}\tau_j^{2,\mathrm u}$ and the $\Sigma_{\mathrm u}$-analogue Lindeberg condition holds for this array, so the Lindeberg--Feller theorem gives conditional asymptotic normality with variance $\tau_{j,n}^{2,\mathrm u}$; Slutsky's lemma with the display above gives the stated CLT.
\end{proof}

{\emph{(Proposition~\ref{prop:ppipp-equivalence} is stated in Section~\ref{app:ppi-comparison}.)}}

\begin{proof}[Proof of Proposition~\ref{prop:ppipp-equivalence}]
\emph{Definition of the high-dimensional comparators.}
We first fix the comparators as constructed in Section~\ref{sec:experiments-design}.  Write $\hat Y(x):=x^\top\hat\beta_{\mathrm{ext}}$ for the linear external model, and for a parameter value $\theta\in\R^p$ define the squared-loss influence (estimating-equation) functions on a covariate--response pair by
\[
    g_l(\theta;X,Y):=X\bigl(X^\top\theta - Y\bigr),
    \qquad
    g_l^{\hat Y}(\theta;X):=X\bigl(X^\top\theta - \hat Y(X)\bigr),
\]
the first evaluated on labelled rows and the second on any covariate row.  Let $\hat\theta^{\mathrm{DL}}$ be the target-only debiased Lasso of \citet{javanmard2014confidence,vandegeer2014asymptotic,zhang2014confidence} computed from $(X_0,y_0)$, with the standard asymptotic-linearity expansion
\begin{equation}\label{eq:dl-lin}
    \hat\theta^{\mathrm{DL}}_j - \beta^\star_{\mathrm{proj},j}
    \;=\;
    -\,\frac{1}{n_0}\sum_{i=1}^{n_0} e_j^\top M_1\, g_l\bigl(\beta^\star_{\mathrm{proj}};X_{0,i},y_{0,i}\bigr)
    \;+\; \op(n_0^{-1/2}),
\end{equation}
which holds for each fixed $j$ under Assumptions~\ref{ass:design}, \ref{ass:JM}, \ref{ass:nonlinear}, \ref{ass:misspec-reg}: the leading term is the JM-debiased labelled score, and the remainder $e_j^\top(I_p-M_1\hat\Sigma_0)(\hat\beta^{\mathrm{Lasso}}_0-\beta^\star_{\mathrm{proj}})$ is controlled by $\maxnorm{I_p-M_1\hat\Sigma_0}\,\|\hat\beta^{\mathrm{Lasso}}_0-\beta^\star_{\mathrm{proj}}\|_1\le\mu_1\,\Op(s\sqrt{\log p/n_0})=\op(n_0^{-1/2})$ by Assumption~\ref{ass:JM} (the $\maxnorm{\cdot}$ factor) and Assumption~\ref{ass:rsc} (the target-Lasso $\ell_1$ rate), exactly as the analogous labelled-block remainder is controlled in the proof of Theorem~\ref{thm:CLT}.  Here $g_l(\beta^\star_{\mathrm{proj}};X_{0,i},y_{0,i})=-X_{0,i}(\eta(X_{0,i})+\varepsilon_{0,i})$, and by Lemma~\ref{lem:misspec-unbiased} it has population mean zero, so \eqref{eq:dl-lin} is a normalised sum of i.i.d.\ mean-zero terms; this is the target-only debiased-Lasso estimating equation referenced in the statement.

The high-dimensional PPI and PPI++ estimators of Section~\ref{sec:experiments-design} are obtained from the same debiased-Lasso anchor by appending a rectifier that scores the linear predictor on the unlabelled rows and debiases the resulting discrepancy with the same JM weighting.  For $\omega\in\R$, the PPI++ point estimate solves the rectified estimating equation
\begin{equation}\label{eq:ppipp-ee}
    0 \;=\;
    \underbrace{\frac{1}{n_0}\sum_{i=1}^{n_0} g_l(\theta;X_{0,i},y_{0,i})}_{\text{labelled anchor}}
    \;+\;\omega\Biggl(
    \underbrace{\frac{1}{N}\sum_{i=1}^{N} g_l^{\hat Y}(\theta;\tilde X_i)}_{\text{unlabelled rectifier}}
    \;-\;
    \underbrace{\frac{1}{n_0}\sum_{i=1}^{n_0} g_l^{\hat Y}(\theta;X_{0,i})}_{\text{labelled rectifier}}
    \Biggr),
\end{equation}
debiased coordinate-wise by left-multiplication with $e_j^\top M_1$; vanilla PPI is the special case $\omega=1$, and PPI++ takes the per-coordinate variance-minimising $\omega=\omega^*_j\in\R$.  Linearising \eqref{eq:ppipp-ee} about $\beta^\star_{\mathrm{proj}}$ exactly as in \eqref{eq:dl-lin} (the labelled-anchor block reproduces \eqref{eq:dl-lin}; the JM-debiased Newton step for the rectifier blocks introduces the same $\op(n_0^{-1/2})$ remainder under Assumptions~\ref{ass:JM} and \ref{ass:nonlinear}) yields the asymptotically linear representation
\begin{equation}\label{eq:ppipp-lin}
    \hat\theta^{\mathrm{PPI++}}_j(\omega^*_j) - \beta^\star_{\mathrm{proj},j}
    \;=\;
    \bigl(\hat\theta^{\mathrm{DL}}_j - \beta^\star_{\mathrm{proj},j}\bigr)
    \;-\;\omega^*_j\,\Bigl(\mathcal R^{\mathrm u}_{j} - \mathcal R^{\mathrm l}_{j}\Bigr)
    \;+\; \op(n_0^{-1/2}),
\end{equation}
where, writing the rectifier residual at $\beta^\star_{\mathrm{proj}}$ as $\hat Y(x)-x^\top\beta^\star_{\mathrm{proj}}=x^\top(\hat\beta_{\mathrm{ext}}-\beta^\star_{\mathrm{proj}})$ and $\Delta=\hat\beta_{\mathrm{ext}}-\beta^\star_{\mathrm{proj}}$ (the re-basing fixed after Assumption~\ref{ass:misspec-reg}),
\begin{align*}
    \mathcal R^{\mathrm u}_{j}
    := \frac{1}{N}\sum_{i=1}^{N} e_j^\top M_1\, g_l^{\hat Y}(\beta^\star_{\mathrm{proj}};\tilde X_i)
     = e_j^\top M_1\Bigl(\tfrac1N\textstyle\sum_{i=1}^N \tilde X_i\tilde X_i^\top\Bigr)(-\Delta),\\
    \mathcal R^{\mathrm l}_{j}
    := \frac{1}{n_0}\sum_{i=1}^{n_0} e_j^\top M_1\, g_l^{\hat Y}(\beta^\star_{\mathrm{proj}};X_{0,i})
     = e_j^\top M_1\,\hat\Sigma_0\,(-\Delta).
\end{align*}

\emph{Step 1: the rectifier residual lies in $\mathrm{span}(X)$.}
Because the model is linear, the residual at $\beta^\star_{\mathrm{proj}}$ is $\hat Y(X)-X^\top\beta^\star_{\mathrm{proj}}=X^\top\Delta$, an exact element of the linear span of the rows of $X$, carrying \emph{no} component orthogonal to $\mathrm{span}(X)$; this is the structural fact underlying the algebraic cancellation of Proposition~\ref{prop:ppi-cancellation} and is the property that fails under a non-linear predictor.  By the hypothesis, $\norm{\Delta}_1=\op\bigl(1/\sqrt{\log p}\bigr)\to0$ (for the sparse-Lasso instance, $\norm{\Delta}_1=\Op(s\sqrt{(\log p)/n_A})$ and $\norm{\Delta}_2=\Op(\sqrt{s/n_A})$).  By Assumption~\ref{ass:external-indep}, $\hat\beta_{\mathrm{ext}}$ (hence $\Delta$) is independent of the inference-stage rows; we argue conditionally on $\hat\beta_{\mathrm{ext}}$ and integrate, treating $\Delta$ as deterministic.

\emph{Step 2: the rectifier correction telescopes and collapses at the $\op(n_0^{-1/2})$ scale.}
The two rectifier blocks of \eqref{eq:ppipp-lin} differ only through the design at which $\Delta$ is read off:
\[
    \mathcal R^{\mathrm u}_{j}-\mathcal R^{\mathrm l}_{j}
    \;=\;
    -\,e_j^\top M_1\bigl(\hat\Sigma_{\mathrm u}-\hat\Sigma_0\bigr)\Delta,
\]
since $\mathcal R^{\mathrm u}_{j}=-e_j^\top M_1\hat\Sigma_{\mathrm u}\Delta$ and $\mathcal R^{\mathrm l}_{j}=-e_j^\top M_1\hat\Sigma_0\Delta$.  Writing $M_1\hat\Sigma_0=I_p+(M_1\hat\Sigma_0-I_p)$ and likewise for $\hat\Sigma_{\mathrm u}$, the leading term $-(\Delta)_j$ is \emph{identical} in $\mathcal R^{\mathrm u}_{j}$ and $\mathcal R^{\mathrm l}_{j}$ and cancels in their difference.  What survives is only the second-order two-sample covariance fluctuation $\hat\Sigma_{\mathrm u}-\hat\Sigma_0$ acting on $\Delta$---the same correction $M_1(\hat\Sigma_{\mathrm u}-\hat\Sigma_0)$ that enters the factor $A_n$ of Theorem~\ref{thm:CLT} (cf.\ Assumption~\ref{ass:rowsparse}).  In the no-shift regime $\maxnorm{\hat\Sigma_{\mathrm u}-\hat\Sigma_0}\le\maxnorm{\hat\Sigma_0-\Sigma_0}+\maxnorm{\hat\Sigma_{\mathrm u}-\Sigma_0}=\Op(\sqrt{(\log p)/(n_0\wedge N)})$ by Assumption~\ref{ass:design}---the slower of the two independent blocks, no relation between $N$ and $n_0$ being assumed---and $\mnorm{M_1}=\Op(1)$ by Assumption~\ref{ass:JM}.  Hence, since $\norm{e_j^\top M_1}_1\le\mnorm{M_1}$ and $\norm{(\hat\Sigma_{\mathrm u}-\hat\Sigma_0)\Delta}_\infty\le\maxnorm{\hat\Sigma_{\mathrm u}-\hat\Sigma_0}\,\norm{\Delta}_1$,
\[
    \bigl|\mathcal R^{\mathrm u}_{j}-\mathcal R^{\mathrm l}_{j}\bigr|
    \;\le\;
    \mnorm{M_1}\,\maxnorm{\hat\Sigma_{\mathrm u}-\hat\Sigma_0}\,\norm{\Delta}_1
    \;=\;
    \Op\!\Bigl(\sqrt{\tfrac{\log p}{n_0\wedge N}}\;\norm{\Delta}_1\Bigr).
\]
Multiplying by the $\sqrt{n_0}$ inference scale, uniformly over $\omega^*_j$ in any fixed bounded interval, as the statement assumes,
\[
    \sqrt{n_0}\,\bigl|\omega^*_j\bigl(\mathcal R^{\mathrm u}_{j}-\mathcal R^{\mathrm l}_{j}\bigr)\bigr|
    \;\le\;
    \sqrt{n_0}\,\bigl|\mathcal R^{\mathrm u}_{j}-\mathcal R^{\mathrm l}_{j}\bigr|
    \;=\;
    \Op\!\bigl(\sqrt{n_0\,(\log p)/(n_0\wedge N)}\;\norm{\Delta}_1\bigr)
    \;=\;
    \op(1)
\]
under the consistency hypothesis $\sqrt{n_0\,(\log p)/(n_0\wedge N)}\,\norm{\Delta}_1\to0$ (for the sparse-Lasso instance, $n_A\gg s^2(\log p)^2\,n_0/(n_0\wedge N)$).  Because the bound is uniform in $\omega^*_j$ over bounded intervals, a data-driven choice confined to such an interval cannot prevent the collapse.  Substituting into \eqref{eq:ppipp-lin} and subtracting \eqref{eq:dl-lin},
\[
    \sqrt{n_0}\,\bigl(\hat\theta^{\mathrm{PPI++}}_j(\omega^*_j)-\hat\theta^{\mathrm{DL}}_j\bigr)
    \;=\;
    \sqrt{n_0}\,\omega^*_j\bigl(\mathcal R^{\mathrm u}_{j}-\mathcal R^{\mathrm l}_{j}\bigr) + \op(1)
    \;=\;
    \op(1).
\]
Vanilla PPI is the special case $\omega^*_j\equiv 1$, and the identical bound gives $\sqrt{n_0}\,(\hat\theta^{\mathrm{PPI}}_j-\hat\theta^{\mathrm{DL}}_j)=\sqrt{n_0}\,(\mathcal R^{\mathrm u}_{j}-\mathcal R^{\mathrm l}_{j})+\op(1)=\op(1)$, the second display.

\emph{Step 3: equivalence of estimating equations and confidence intervals.}
Because the rectifier blocks of \eqref{eq:ppipp-ee} contribute $\op(n_0^{-1/2})$ to the linearisation at $\beta^\star_{\mathrm{proj}}$, the PPI++ estimating equation reduces asymptotically to the labelled-anchor block alone, i.e.\ to the target-only debiased-Lasso estimating equation underlying \eqref{eq:dl-lin}.  Consequently the leading-order influence function of $\hat\theta^{\mathrm{PPI++}}_j(\omega^*_j)$ coincides with that of $\hat\theta^{\mathrm{DL}}_j$, both equal to $-e_j^\top M_1 g_l(\beta^\star_{\mathrm{proj}};\cdot)/\sqrt{n_0}$ with limiting variance $\bigl(\Sigma_0^{-1}\Gamma_0\Sigma_0^{-1}\bigr)_{jj}/n_0$ in the notation of Theorem~\ref{thm:CLT-misspec}, where $\Gamma_0=\E_{P_0}[(\eta^2+\sigma^2)XX^\top]$ under the homoscedastic noise of Assumption~\ref{ass:design}(b); by Slutsky's lemma the plug-in confidence intervals satisfy $\mathrm{CI}^{\mathrm{PPI++}}_j/\mathrm{CI}^{\mathrm{DL}}_j\to1$ and $\mathrm{CI}^{\mathrm{PPI}}_j/\mathrm{CI}^{\mathrm{DL}}_j\to1$.  This is the asymptotic counterpart of the exact algebraic cancellation of Proposition~\ref{prop:ppi-cancellation}, here established under the projection-parameter framework of Assumptions~\ref{ass:nonlinear}--\ref{ass:misspec-reg} without requiring linear truth (for weights confined to a bounded interval), since the argument used only that the model is linear (so the residual lies in $\mathrm{span}(X)$) and never that $\eta\equiv0$.
\end{proof}

\begin{lemma}[Population unbiasedness of the labelled score]\label{lem:misspec-unbiased}
Under Assumptions~\ref{ass:design}, \ref{ass:nonlinear}, \ref{ass:misspec-reg},
\[
    \E_{P_0}\bigl[X\bigl(\eta(X)+\varepsilon\bigr)\bigr] \;=\; 0.
\]
\end{lemma}

\begin{proof}[Proof of Lemma~\ref{lem:misspec-unbiased}]
The proof is immediate: $\E[X\varepsilon]=\E[X\,\E[\varepsilon\mid X]]=0$ by Assumption~\ref{ass:nonlinear}, and $\E[X\,\eta(X)]=0$ by definition of $\beta^\star_{\mathrm{proj}}$ as the population least-squares projection.
\end{proof}

\begin{proof}[Proof of Theorem~\ref{thm:CLT-misspec}]

Throughout, $\beta^\star$ in the proof of Theorem~\ref{thm:CLT} is reinterpreted as the projection parameter $\beta^\star_{\mathrm{proj}}$ of \eqref{eq:beta-proj}, and the labelled-block noise variable $\varepsilon_{0,i}$ of the linear analysis is replaced by the centred-at-projection residual
\[
    u_i \;:=\; y_{0,i} - X_{0,i}^\top\beta^\star_{\mathrm{proj}} \;=\; \eta(X_{0,i}) + \varepsilon_{0,i} ,
\]
using $y_{0,i}=\mu(X_{0,i})+\varepsilon_{0,i}$ and \eqref{eq:eta-def}.  By Lemma~\ref{lem:misspec-unbiased}, $\E_{P_0}[X\,u]=0$, and for any deterministic $v\in\R^p$ the scalar $v^\top X\,u$ has mean zero.  The proof has four steps: (I) the linearisation of Theorem~\ref{thm:CLT} is unchanged at the population level and its remainder is misspecification-invariant; (II) the leading term is a normalised sum of unconditionally mean-zero terms; (III) a triangular-array CLT with the sandwich middle factor; (IV) plug-in consistency of $\hat\Gamma_0$.

\smallskip
\noindent\emph{Step I: linearisation and misspecification-invariance of the remainder.}
The initialiser decomposition (Lemma~\ref{lem:oof-decomposition}) and the debiased-Lasso expansion at the head of the proof of Theorem~\ref{thm:CLT} are algebraic identities in the response, the design, and $\hat\beta_{\mathrm{ext}}$; replacing $y_{0,i}-X_{0,i}^\top\beta^\star_{\mathrm{proj}}$ by $u_i$ and the imputation residual $y_{\mathrm u}-\tilde X\beta^\star_{\mathrm{proj}}$ by its pseudo-label form leaves them intact.  Because the pseudo-labels $\hat y_{\mathrm u}=\tilde X\tilde\beta^{\mathrm{init}}$ are linear in the unlabelled covariates, with the labelled-sample-measurable coefficient $\tilde\beta^{\mathrm{init}}=\hat\beta_{\mathrm{ext}}+\hat t\,C$ of \eqref{eq:init-bias-aware}, $\eta$ is never evaluated at an unlabelled row; only the labelled block acquires the $\eta$ term inside $u_i$, exactly as in the linear case. Tracking the expansion verbatim gives, for the fixed coordinate $j$,
\[
    \tilde\beta_j - \beta^\star_{\mathrm{proj},j}
    =
    \frac{1}{n_0+N}\sum_{i=1}^{n_0}\bigl(x_{0,i}^\top A_n^\top M_2^\top e_j\bigr)\,u_i
    \;+\; R_{j,n},
\]
with $A_n=I_p+\tfrac{\hat t N}{n_0}\hat\Sigma_{\mathrm u}M_1$ as in the linear proof and
\[
    R_{j,n}
    =
    \frac{N}{n_0+N}\,e_j^\top M_2\hat\Sigma_{\mathrm u} r(\hat t)
    + e_j^\top\bigl(I_p - M_2\hat\Sigma_{\mathrm{stk}}\bigr)(\hat\beta-\beta^\star_{\mathrm{proj}}).
\]
The two summands of $R_{j,n}$ are controlled exactly as in the proof of Theorem~\ref{thm:CLT}: the second by Assumption~\ref{ass:JM} ($\maxnorm{I_p-M_2\hat\Sigma_{\mathrm{stk}}}\le\mu_2$) together with the stacked-Lasso $\ell_1$ rate $\norm{\hat\beta-\beta^\star_{\mathrm{proj}}}_1=\Op(s\lambda)$ of Theorem~\ref{thm:stacked-lasso}, giving $\sqrt{n_0+N}\,|e_j^\top(I_p-M_2\hat\Sigma_{\mathrm{stk}})(\hat\beta-\beta^\star_{\mathrm{proj}})|=\Op(\sqrt{n_0+N}\,\mu_2 s\lambda)=\op(1)$ by condition~(ii); the first by Lemma~\ref{lem:bias-transfer} with $\mnorm{M_2}=\Op(1)$ (Assumption~\ref{ass:JM}) and the composite envelope $\sqrt{n_0+N}\,\bigl(\norm{r(\hat t)}_\infty+\sqrt{\log p/N}\,\norm{\Sigma^{1/2}r(\hat t)}_2\bigr)\to0$ (condition~(i)).  It remains to check that Theorem~\ref{thm:stacked-lasso} still delivers the $\ell_1$ rate under misspecification.  Its only input is the score bound of Lemma~\ref{lem:score}, whose single misspecification-sensitive term is the labelled component $g_{\mathrm{lab}}=\tfrac{1}{n_0+N}X_0^\top(\eta(X_0)+\varepsilon_0)$.  Under Assumption~\ref{ass:design}(a) and Assumptions~\ref{ass:nonlinear}, \ref{ass:misspec-reg}, each coordinate $\tfrac1{n_0+N}\sum_i X_{0,i,m}(\eta(X_{0,i})+\varepsilon_{0,i})$ is a centred average (the centring being $\E[X_m\,\eta(X)]+\E[X_m\varepsilon]=0$ by Lemma~\ref{lem:misspec-unbiased}) of independent sub-exponential summands with common proxy $\psi_\eta^2+C\sigma^2$ (the $X_m\varepsilon$ piece sub-exponential as a product of the sub-Gaussian $X_m$ and the now conditionally sub-Gaussian $\varepsilon$ of Assumption~\ref{ass:nonlinear}); a Bernstein bound and a union bound over $m\in\{1,\dots,p\}$ give $\norm{g_{\mathrm{lab}}}_\infty\le C_1(\sigma+\psi_\eta)\sqrt{\log p/(n_0+N)}$ with probability at least $1-p^{-2}$, the same order as in Lemma~\ref{lem:score}.  Of the two pseudo-label score terms, $g_{\mathrm{pl,R}}$ is a function of $(M_1,\hat\Sigma_0,\hat t,\Delta)$ alone and is genuinely unchanged.  The term $g_{\mathrm{pl,G}}=\tfrac{N}{n_0+N}\hat t\,\hat\Sigma_{\mathrm u}Z_1$ is not: under the substitution $\varepsilon_{0,i}\to u_i$, $Z_1=n_0^{-1}M_1X_0^\top\varepsilon_0$ becomes $Z_1=n_0^{-1}M_1X_0^\top u_0=n_0^{-1}M_1X_0^\top\varepsilon_0+n_0^{-1}M_1W$, with $W$ the Step-II object.  Since $\norm{W}_\infty=\Op(\sqrt{n_0\log p})$ (Step~II) and $\mnorm{M_1}=\Op(1)$ (Assumption~\ref{ass:JM}), the extra piece $n_0^{-1}M_1W$ has $\norm{\cdot}_\infty=\Op(\sqrt{\log p/n_0})$, the same order as the original noise piece of $Z_1$ (Lemma~\ref{lem:score}'s proof), so $\norm{Z_1}_\infty=\Op(\sqrt{\log p/n_0})$ persists with $\sigma$ replaced by $\sigma+\psi_\eta$; the bound on $g_{\mathrm{pl,G}}$ via Lemma~\ref{lem:bias-transfer} --- which uses only that $Z_1$ is independent of $\tilde X$, not that it is conditionally mean zero --- is unaffected.  Hence the conclusion of Theorem~\ref{thm:stacked-lasso} holds with $\sigma$ replaced by $\sigma+\psi_\eta$ throughout, and $\sqrt{n_0+N}\,R_{j,n}=\op(1)$.  This is the precise sense in which the remainder is misspecification-invariant.

\smallskip
\noindent\emph{Step II: the leading term and its unconditional centring.}
Write
\[
    S_n \;:=\; \frac{1}{\sqrt{n_0+N}}\sum_{i=1}^{n_0} w_i\,u_i,
    \qquad w_i:=x_{0,i}^\top A_n^\top M_2^\top e_j .
\]
Unlike the linear case, $u_i$ is \emph{not} mean zero conditional on the design, because $\E[u_i\mid X_{0,i}]=\eta(X_{0,i})\neq0$ in general; the centring $\E[X\,u]=0$ of Lemma~\ref{lem:misspec-unbiased} holds only unconditionally.  Consequently the CLT must be carried out unconditionally rather than conditionally on the inference-sample covariates.  We do so by freezing the data-dependent weight matrices at their probability limits.  Under the preserved regime conditions of Theorem~\ref{thm:CLT}, Assumption~\ref{ass:JM}, and Assumption~\ref{ass:rowsparse} (in particular $\hat t\to t_0\in[0,1]$, $\hat\Sigma_{\mathrm u}\to_{\mathbb P}\Sigma$, the precision-row convergence $M_2^\top e_j\to\Sigma^{-1}e_j$ of Assumption~\ref{ass:rowsparse}---whence the variance-matching quadratic-form limit of Lemma~\ref{lem:m2-efficiency}---and $\mnorm{M_1}+\mnorm{M_2}=\Op(1)$ of Assumption~\ref{ass:JM}), there is a deterministic vector
\[
    v_j \;:=\; \Bigl(1+\tfrac{t_0 N}{n_0}\Bigr)\Sigma^{-1}e_j
\]
such that $A_n^\top M_2^\top e_j \to_{\mathbb P} v_j$, and we set $\bar w_i:=v_j^\top x_{0,i}$.  The frozen sum
\[
    \bar S_n \;:=\; \frac{1}{\sqrt{n_0+N}}\sum_{i=1}^{n_0} \bar w_i\,u_i
\]
is a normalised sum of row-wise independent terms $\bar w_i u_i=(v_j^\top X_{0,i})u_i$, each \emph{unconditionally} mean zero by Lemma~\ref{lem:misspec-unbiased} applied to the linear functional $v_j^\top X$, and with finite second moment
\begin{gather*}
    \E\bigl[(v_j^\top X)^2 u^2\bigr]
    = v_j^\top\,\E\!\bigl[(\eta(X)^2+\sigma^2(X))\,XX^\top\bigr]\,v_j
    =: v_j^\top\Gamma_0\,v_j<\infty,\\
    \Gamma_0:=\E\bigl[(\eta(X)^2+\sigma^2(X))XX^\top\bigr],
\end{gather*}
finiteness following from $\E[\eta(X)^2 X_m X_{m'}]<\infty$ (sub-exponential $X_m\eta(X)$, Assumption~\ref{ass:misspec-reg}, with Cauchy--Schwarz) and $\E[\sigma^2(X)X_mX_{m'}]\le\sigma_{\max}^2\Sigma_{mm'}$ (Assumption~\ref{ass:nonlinear}).  The conditional second moment $\E[(\eta(X)+\varepsilon)^2\mid X]=\eta(X)^2+\sigma^2(X)$ produces no $\eta\varepsilon$ cross-term because $\E[\varepsilon\mid X]=0$; $\Gamma_0$ is the population limit of $\hat\Gamma_0$ of \eqref{eq:Gamma-fold} and recurs throughout Steps~III--IV below.

The replacement error $S_n-\bar S_n=\tfrac{1}{\sqrt{n_0+N}}\sum_{i=1}^{n_0}(w_i-\bar w_i)u_i=\tfrac1{\sqrt{n_0+N}}z^\top\sum_{i=1}^{n_0}X_{0,i}u_i$, with $z:=A_n^\top M_2^\top e_j-v_j$ the frozen-weight perturbation, splits via $u_i=\varepsilon_{0,i}+\eta(X_{0,i})$ into a noise part and a residual part, both controlled through $z$.

\emph{The rate for $z$.}  Set $\zeta_n:=\sqrt{s_\Omega\log p/(n_0+N)}$, and recall $A_n^\top M_2^\top e_j=\bigl(I_p+\tfrac{\hat tN}{n_0}M_1^\top\hat\Sigma_{\mathrm u}\bigr)M_2^\top e_j$ and $v_j=\bigl(1+\tfrac{t_0N}{n_0}\bigr)\Sigma^{-1}e_j$.  Writing $\kappa_n:=N/n_0=O(1)$ along \eqref{eq:regime} and using $\hat t\,M_1^\top\hat\Sigma_{\mathrm u}-t_0I_p=\hat t\bigl(M_1^\top\hat\Sigma_{\mathrm u}-I_p\bigr)+(\hat t-t_0)I_p$, decompose
\[
    z
    =\underbrace{\bigl(1+t_0\kappa_n\bigr)\bigl(M_2^\top e_j-\Sigma^{-1}e_j\bigr)}_{\text{(a)}}
    +\underbrace{\kappa_n\hat t\,\bigl(M_1^\top\hat\Sigma_{\mathrm u}-I_p\bigr)M_2^\top e_j}_{\text{(b)}}
    +\underbrace{\kappa_n(\hat t-t_0)\,M_2^\top e_j}_{\text{(c)}} .
\]
Term~(a) is $O(1)\cdot\Op(\zeta_n)=\op(1)$ in $\ell_2$ norm by the first rate of Assumption~\ref{ass:rowsparse}; in particular $\norm{M_2^\top e_j}_2\le\norm{\Sigma^{-1}e_j}_2+\op(1)=\Op(1)$.  Term~(b) has $\ell_2$ norm $\kappa_n\hat t\,\norm{(M_1^\top\hat\Sigma_{\mathrm u}-I_p)M_2^\top e_j}_2$, which is $\op(1)$ directly by the $A_n$-replacement clause of Assumption~\ref{ass:rowsparse}.  Term~(c) has $\ell_2$ norm $\kappa_n|\hat t-t_0|\,\norm{M_2^\top e_j}_2=O(1)\cdot\op(1)\cdot\Op(1)=\op(1)$, using consistency $\hat t\to_{\mathbb P}t_0$ (a fortiori from the shrinkage-rate condition, Assumption~\ref{ass:condT}).  Hence
\[
    \norm{z}_2=\op(1).
\]

\emph{The noise part.}  Condition on $\mathcal G:=\sigma\bigl(X_0,\tilde X,\text{tuning data},\hat\beta_{\mathrm{ext}}\bigr)$, under which $z$ is measurable and, by Assumptions~\ref{ass:nonlinear} and~\ref{ass:external-indep}, $\varepsilon_{0,i}$ remains i.i.d.\ mean zero with conditional variance $\sigma^2(X_{0,i})$, independent of $\mathcal G$.  The noise part $\tfrac1{\sqrt{n_0+N}}\sum_{i=1}^{n_0}(w_i-\bar w_i)\varepsilon_{0,i}=\tfrac1{\sqrt{n_0+N}}z^\top\sum_iX_{0,i}\varepsilon_{0,i}$ is then a sum of conditionally independent mean-zero terms with conditional variance
\[
    \tfrac{1}{n_0+N}\sum_{i=1}^{n_0}(w_i-\bar w_i)^2\sigma^2(X_{0,i})
    \;\le\;
    \sigma_{\max}^2\,\tfrac{n_0}{n_0+N}\,z^\top\hat\Sigma_0z
    \;\le\;
    \sigma_{\max}^2\Bigl(\lambda_{\max}(\Sigma)\norm{z}_2^2+\norm{z}_1^2\maxnorm{\hat\Sigma_0-\Sigma}\Bigr),
\]
using $\hat\Sigma_0=\Sigma+(\hat\Sigma_0-\Sigma)$ and H\"older's inequality on the second piece.  The first summand is $\op(1)$ by $\norm{z}_2=\op(1)$ above; the second is $\op(1)$ because $\maxnorm{\hat\Sigma_0-\Sigma}=\Op(\sqrt{\log p/n_0})$ (sub-Gaussian concentration and a union bound, Assumption~\ref{ass:design}(a)) and $\norm{z}_1=\op(1/\sqrt{\log p})$ by the weight-replacement condition~(v$'$).  Hence the conditional variance is $\op(1)$, and the noise part is $\op(1)$ by conditional Chebyshev.

\emph{The residual part.}  Write $W:=\sum_{i=1}^{n_0}X_{0,i}\,\eta(X_{0,i})\in\R^p$.  By Lemma~\ref{lem:misspec-unbiased}, $\E[X\eta(X)]=0$, so each coordinate of $W$ is a centred sum of $n_0$ i.i.d.\ sub-exponential summands (Assumption~\ref{ass:misspec-reg}); a Bernstein bound and a union bound over $m\in\{1,\dots,p\}$ give the unconditional rate $\norm{W}_\infty=\Op\bigl(\sqrt{n_0\log p}\bigr)$.  The residual part is $z^\top W/\sqrt{n_0+N}$, and by H\"older's inequality, deterministically in $z$,
\[
    \frac{\bigl|z^\top W\bigr|}{\sqrt{n_0+N}}
    \;\le\;
    \norm{z}_1\,\frac{\norm{W}_\infty}{\sqrt{n_0+N}}
    \;=\;
    \norm{z}_1\cdot\Op\!\left(\sqrt{\tfrac{n_0}{n_0+N}\log p}\right)
    \;\le\;
    \sqrt{\log p}\,\norm{z}_1\cdot\Op(1) ,
\]
using $n_0\le n_0+N$; this is $\op(1)$ by condition~(v$'$).  Unlike a discharge that conditions on the JM matrices $M_1,M_2$ --- which, being functions of $X_0$, would destroy the i.i.d.\ law of the labelled rows and the centring $\E[X\eta(X)]=0$ --- this bound uses only the \emph{unconditional} rate for $W$ together with a deterministic H\"older step, valid for the realised (random) $z$ whatever its dependence on $M_1,M_2$.  Combined with the noise part above, $S_n-\bar S_n=\op(1)$, i.e.\ $S_n=\bar S_n+\op(1)$.

\smallskip
\noindent\emph{Step III: triangular-array CLT with the sandwich middle factor.}
Conditions (iii$'$)--(iv$'$) are hypotheses of the theorem; we recall them for the frozen array $\{\bar w_iu_i\}$, verify (iv$'$) from more primitive assumptions, and identify the variance $\tau_j^{2,\mathrm{sand}}$ of hypothesis~(iii$'$) with the exact variance of $\bar S_n$.  Recall $\Gamma_0=\E[(\eta(X)^2+\sigma^2(X))XX^\top]$, the population limit of $\hat\Gamma_0$, from Step~II.  For every $v\in\R^p$,
\begin{align*}
    v^\top\Gamma_0v
    &=\E[\sigma^2(X)(v^\top X)^2]+\E[\eta(X)^2(v^\top X)^2]\\
    &\le\sigma_{\max}^2\lambda_{\max}(\Sigma)\norm{v}_2^2+\bigl(\E[\eta(X)^4]\bigr)^{1/2}\bigl(\E[(v^\top X)^4]\bigr)^{1/2}
    \le C\norm{v}_2^2
\end{align*}
for a constant $C$, using $\sigma^2(X)\le\sigma_{\max}^2$ (Assumption~\ref{ass:nonlinear}), $v^\top X$ sub-Gaussian with variance $v^\top\Sigma v\le\lambda_{\max}(\Sigma)\norm{v}_2^2$ (Assumption~\ref{ass:design}(a), so bounded fourth moment), and $\eta$ sub-Gaussian with bounded fourth moment (Assumption~\ref{ass:misspec-reg}), the last two combined via Cauchy--Schwarz; in particular $\lambda_{\max}(\Gamma_0)\le C<\infty$.

\emph{Verifying~(iv$'$).}  $\bar w_i u_i=(v_j^\top X_{0,i})\,u_i$ with $\norm{v_j}_2=O(1)$, so each $|\bar w_i u_i|$ is a product of a sub-Gaussian factor $v_j^\top X_{0,i}$ (Assumption~\ref{ass:design}(a)) and the residual $u_i=\eta(X_{0,i})+\varepsilon_{0,i}$; by Assumption~\ref{ass:misspec-reg} the term $X_{0,i}^\top v_j\,\eta(X_{0,i})$ is sub-exponential, and $X_{0,i}^\top v_j\,\varepsilon_{0,i}$ is sub-exponential (a product of two sub-Gaussians, $\varepsilon$ now conditionally sub-Gaussian by Assumption~\ref{ass:nonlinear}), each with proxy of order $\norm{v_j}_2^2(\psi_\eta^2+\sigma_{\max}^2)$.  A maximal inequality for the $n_0$ independent sub-exponential variables gives $\max_{i}|\bar w_i u_i|=\Op(\log n_0)$.  The terms $\{\bar w_i^2u_i^2\}_{i=1}^{n_0}$ are i.i.d.\ with mean $v_j^\top\Gamma_0v_j\in(0,\infty)$ and variance bounded uniformly in $n$ (since $\norm{v_j}_2=O(1)$), so by Chebyshev's inequality $\tfrac1{n_0}\sum_{i=1}^{n_0}\bar w_i^2u_i^2=v_j^\top\Gamma_0v_j+\op(1)$, bounded away from $0$ for $n_0$ large; since $(n_0+N)\asymp n_0$ along \eqref{eq:regime}, the Lindeberg ratio of~(iv$'$) is $\Op(\log n_0/\sqrt{n_0})=\op(1)$, so condition~(iv$'$) holds.

\emph{Identifying the limit.}  By the same Chebyshev argument, $v_j^\top\hat\Gamma_0v_j=v_j^\top\Gamma_0v_j+\op(1)$ --- a scalar average over the $n_0$ i.i.d.\ pairs $(X_{0,i},u_i)$, entailing no entrywise convergence claim about the $p\times p$ matrix $\hat\Gamma_0$ itself.  With $z=A_n^\top M_2^\top e_j-v_j=\op(1)$ (Step II) and $\hat\Gamma_0$ positive semi-definite,
\[
    z^\top\hat\Gamma_0z\;\le\;\lambda_{\max}(\Gamma_0)\norm{z}_2^2+\norm{z}_1^2\maxnorm{\hat\Gamma_0-\Gamma_0}\;=\;\op(1),
\]
using $\hat\Gamma_0=\Gamma_0+(\hat\Gamma_0-\Gamma_0)$ and H\"older's inequality on the second piece, $\norm{z}_2=\op(1)$, $\maxnorm{\hat\Gamma_0-\Gamma_0}=\Op(\sqrt{\log p/n_0})$ (a Bernstein and union bound over the sub-exponential entries $(\eta(X)^2+\sigma^2(X))X_mX_{m'}$, Assumptions~\ref{ass:misspec-reg}--\ref{ass:nonlinear}), and $\norm{z}_1=\op(1/\sqrt{\log p})$ by the weight-replacement condition~(v$'$).  Cauchy--Schwarz then gives $\bigl|z^\top\hat\Gamma_0v_j\bigr|\le(z^\top\hat\Gamma_0z)^{1/2}(v_j^\top\hat\Gamma_0v_j)^{1/2}=\op(1)$, so
\[
    \tau_{j,n}^{2,\mathrm{sand}}
    =\tfrac{n_0}{n_0+N}(v_j+z)^\top\hat\Gamma_0(v_j+z)
    =\tfrac{n_0}{n_0+N}\,v_j^\top\hat\Gamma_0v_j+\op(1)
    =\tfrac{n_0}{n_0+N}\,v_j^\top\Gamma_0v_j+\op(1) .
\]
By hypothesis~(iii$'$) the left side converges in probability to $\tau_j^{2,\mathrm{sand}}$, so the deterministic sequence $\tfrac{n_0}{n_0+N}v_j^\top\Gamma_0v_j$ converges to the same limit.  The Lindeberg--Feller theorem, applied to the row-wise i.i.d., unconditionally mean-zero array $\{\bar w_iu_i/\sqrt{n_0+N}\}$ with exact variance $\tfrac{n_0}{n_0+N}v_j^\top\Gamma_0v_j\to\tau_j^{2,\mathrm{sand}}$ and the Lindeberg condition~(iv$'$) just verified, yields
\[
    \bar S_n \;\overset{\mathrm d}{\longrightarrow}\; \mathcal N\!\bigl(0,\;\tau_j^{2,\mathrm{sand}}\bigr) .
\]
Combining $S_n=\bar S_n+\op(1)$ (Step II), $\sqrt{n_0+N}(\tilde\beta_j-\beta^\star_{\mathrm{proj},j})=S_n+\sqrt{n_0+N}\,R_{j,n}$ with $\sqrt{n_0+N}\,R_{j,n}=\op(1)$ (Step I), and Slutsky's lemma with~(iii$'$), we obtain \eqref{eq:CLT-misspec}.

\smallskip
\noindent\emph{Step IV: plug-in consistency of $\hat\Gamma_0$.}
It remains to show $\hat\tau_j^{2,\mathrm{sand}}\to_{\mathbb P}\tau_j^{2,\mathrm{sand}}$, where $\hat\tau_j^{2,\mathrm{sand}}$ replaces $\hat\Gamma_0$ in \eqref{eq:Gamma-fold} by the empirical-residual plug-in $\hat\Gamma_0^{\mathrm{pl}}:=\tfrac1{n_0}\sum_{i=1}^{n_0}\hat r_i^2 X_{0,i}X_{0,i}^\top$, $\hat r_i=y_{0,i}-X_{0,i}^\top\tilde\beta$.  We apply the device of Step III with $\hat\Gamma_0^{\mathrm{pl}}$ --- again positive semi-definite --- in place of $\hat\Gamma_0$: it suffices to show $v_j^\top\hat\Gamma_0^{\mathrm{pl}}v_j\to_{\mathbb P}v_j^\top\Gamma_0v_j$ and $\maxnorm{\hat\Gamma_0^{\mathrm{pl}}}=\Op(1)$, the latter standing in for $\maxnorm{\hat\Gamma_0-\Gamma_0}=\Op(\sqrt{\log p/n_0})$ in the bound on $z^\top\hat\Gamma_0^{\mathrm{pl}}z$ --- boundedness rather than a rate, since it is only ever paired with $\norm{z}_1=\op(1/\sqrt{\log p})$.  Write $\hat r_i = u_i - X_{0,i}^\top\delta$ with $\delta:=\tilde\beta-\beta^\star_{\mathrm{proj}}$, so
\[
    \hat r_i^2 = u_i^2 - 2u_i\,X_{0,i}^\top\delta + (X_{0,i}^\top\delta)^2 .
\]
Weighting by $(v_j^\top X_{0,i})^2$ and averaging,
\[
    v_j^\top\hat\Gamma_0^{\mathrm{pl}}v_j
    =\tfrac1{n_0}\sum_{i=1}^{n_0}(v_j^\top X_{0,i})^2u_i^2
    -2\tfrac1{n_0}\sum_{i=1}^{n_0}(v_j^\top X_{0,i})^2u_i\,X_{0,i}^\top\delta
    +\tfrac1{n_0}\sum_{i=1}^{n_0}(v_j^\top X_{0,i})^2(X_{0,i}^\top\delta)^2 .
\]
The first average is $v_j^\top\Gamma_0v_j+\op(1)$ by the Chebyshev argument of Step III.  The cross term obeys
\[
    \Bigl|\tfrac1{n_0}\sum_{i=1}^{n_0}(v_j^\top X_{0,i})^2u_i\,X_{0,i}^\top\delta\Bigr|
    \le \norm{\delta}_1\,\max_{l}\Bigl|\tfrac1{n_0}\sum_{i=1}^{n_0}(v_j^\top X_{0,i})^2u_i X_{0,i,l}\Bigr|
    = \Op(s\lambda)\cdot\Op(1)=\op(1),
\]
using $\norm{\delta}_1=\Op(s\lambda)$ (Theorem~\ref{thm:stacked-lasso}, Step~I) and a Bernstein-and-union bound over $l\in\{1,\dots,p\}$ for the sub-exponential third-moment averages; the quadratic term is likewise $\le\norm{\delta}_1^2\max_{l,l'}\bigl|\tfrac1{n_0}\sum_{i=1}^{n_0}(v_j^\top X_{0,i})^2X_{0,i,l}X_{0,i,l'}\bigr|=\Op(s^2\lambda^2)=\op(1)$ under the preserved sparsity regime.  Hence
\[
    v_j^\top\hat\Gamma_0^{\mathrm{pl}}v_j\;\overset{\mathbb P}{\longrightarrow}\;v_j^\top\Gamma_0v_j .
\]
For the boundedness of $\maxnorm{\hat\Gamma_0^{\mathrm{pl}}}$: since $\hat\Gamma_0^{\mathrm{pl}}$ is positive semi-definite, $|\hat\Gamma^{\mathrm{pl}}_{0,mm'}|\le(\hat\Gamma^{\mathrm{pl}}_{0,mm}\hat\Gamma^{\mathrm{pl}}_{0,m'm'})^{1/2}$ for every pair $(m,m')$, so it suffices to bound the diagonal.  The three-term expansion above, specialised to $v_j\leftarrow e_m$ and maximised over $m\in\{1,\dots,p\}$ by the same Bernstein-and-union argument, gives $\hat\Gamma^{\mathrm{pl}}_{0,mm}=\tfrac1{n_0}\sum_iu_i^2X_{0,i,m}^2+\op(1)$ uniformly in $m$, with the leading average bounded by $\lambda_{\max}(\Gamma_0)\le C$ (Step III); hence $\max_m\hat\Gamma^{\mathrm{pl}}_{0,mm}=\Op(1)$ and $\maxnorm{\hat\Gamma_0^{\mathrm{pl}}}=\Op(1)$.  Consequently, exactly as in Step III, $z^\top\hat\Gamma_0^{\mathrm{pl}}z\le\norm{z}_1^2\maxnorm{\hat\Gamma_0^{\mathrm{pl}}}=\op(1)$, and Cauchy--Schwarz gives $z^\top\hat\Gamma_0^{\mathrm{pl}}v_j=\op(1)$, so
\[
    \hat\tau_j^{2,\mathrm{sand}}
    =\tfrac{n_0}{n_0+N}(v_j+z)^\top\hat\Gamma_0^{\mathrm{pl}}(v_j+z)
    =\tfrac{n_0}{n_0+N}\,v_j^\top\hat\Gamma_0^{\mathrm{pl}}v_j+\op(1)
    \;\overset{\mathbb P}{\longrightarrow}\;\tau_j^{2,\mathrm{sand}} ,
\]
using $\tfrac{n_0}{n_0+N}v_j^\top\Gamma_0v_j\to\tau_j^{2,\mathrm{sand}}$ from Step III.  The plug-in confidence intervals of Remark~\ref{rem:misspec-linear-recovery} are therefore asymptotically valid.  Under linear truth $\eta\equiv0$ and conditional homoscedasticity $\sigma^2(X)\equiv\sigma^2$, $\Gamma_0=\sigma^2\Sigma$ and $\tau_j^{2,\mathrm{sand}}=\tau_j^2$, recovering Theorem~\ref{thm:CLT} ({the linear-truth recovery clause of Theorem~\ref{thm:CLT-misspec}}).
\end{proof}

\begin{proof}[Proof of Theorem~\ref{thm:deal-vs-dl}]

Throughout, $N=N_{n_0}$ is the deterministic unlabelled budget of the statement, with
$\kappa_n=N/n_0\to\kappa_\infty$.  Both estimators are
asymptotically normal, $\mathrm{CI}_j$ is a fixed multiple $2z_{1-\alpha/2}$ of the
estimated standard error, and the plug-in standard errors are consistent
(Remark~\ref{rem:misspec-linear-recovery} for DEAL; the sandwich plug-in
$\bigl(M_1\hat\Gamma_0 M_1^\top\bigr)_{jj}/n_0$ for DL, with $M_1$ the labelled-block JM matrix---$\hat\Sigma_0$ being singular for $p>n_0$---whose consistency follows by the argument of Step~IV of the proof of Theorem~\ref{thm:CLT-misspec} specialised to the labelled block), so
\begin{equation}\label{eq:ratio-is-sd}
    \frac{\mathrm{CI}_j^{\mathrm{DEAL}}}{\mathrm{CI}_j^{\mathrm{DL}}}
    \;=\;
    \sqrt{\frac{V_j^{\mathrm{DEAL}}}{V_j^{\mathrm{DL}}}}\,\bigl(1+\op(1)\bigr),
\end{equation}
where $V_j^{\mathrm{DEAL}}$ and $V_j^{\mathrm{DL}}$ denote the (unscaled) asymptotic
variances of $\tilde\beta_j$ and of the target-only debiased-Lasso estimator
$\hat\beta_j^{\mathrm{DL}}$, respectively.  It therefore suffices to identify the two
variances and form their ratio.  We work, as in the idealised-benchmark computation of
Section~\ref{sec:theory-bias}, in the no-shift regime $\Sigma_{\mathrm u}=\Sigma_0$ with
the consistent JM substitutions $M_2\to\Sigma_0^{-1}$, $M_1\to\Sigma_0^{-1}$,
$\hat\Sigma_0\to\Sigma_0$, $\hat\Sigma_{\mathrm u}\to\Sigma_0$ licensed by the precision-row
rates of Assumption~\ref{ass:rowsparse} with Lemma~\ref{lem:m2-efficiency} for the
quadratic forms, and by entrywise concentration for the sample covariances; each
substitution incurs a multiplicative $1+\op(1)$ absorbed into \eqref{eq:ratio-is-sd}.

\smallskip
\noindent\textbf{Step 1 (DL sandwich variance).}
The target-only debiased Lasso fits $\beta^\star_{\mathrm{proj}}$ on the $n_0$ labelled rows
$(X_{0,i},y_{0,i})$ with $y_{0,i}=X_{0,i}^\top\beta^\star_{\mathrm{proj}}+\eta(X_{0,i})+\varepsilon_{0,i}$.
By Lemma~\ref{lem:misspec-unbiased} the labelled score has population mean zero at
$\beta^\star_{\mathrm{proj}}$, and the debiased-Lasso linearisation---the target-only specialisation (formally $N=0$) of the
expansion in the proof of Theorem~\ref{thm:CLT-misspec}, which under linear truth reduces
to the standard expansion of \citet{vandegeer2014asymptotic,javanmard2014confidence}---gives
\[
    \sqrt{n_0}\,\bigl(\hat\beta_j^{\mathrm{DL}}-\beta^\star_{\mathrm{proj},j}\bigr)
    =
    \frac{1}{\sqrt{n_0}}\sum_{i=1}^{n_0}
        \bigl(e_j^\top\Sigma_0^{-1}X_{0,i}\bigr)\bigl(\eta(X_{0,i})+\varepsilon_{0,i}\bigr)
    + \op(1).
\]
The summands are i.i.d.\ mean-zero (Lemma~\ref{lem:misspec-unbiased}).  Since $\eta(X)$ is $X$-measurable and $\E[\varepsilon\mid X]=0$ (Assumption~\ref{ass:nonlinear}), $\E[\eta(X)\varepsilon\mid X]=\eta(X)\E[\varepsilon\mid X]=0$, so the labelled residual $\eta(X_{0,i})+\varepsilon_{0,i}$ satisfies
\[
    \E\bigl[(\eta(X)+\varepsilon)^2\,XX^\top\bigr]
    \;=\;
    \E[\eta(X)^2 XX^\top] + \E[\sigma^2(X)\,XX^\top]
    \;=\;
    \Gamma_\eta + \E_{P_0}[\sigma^2(X)\,XX^\top]
    \;=\;\Gamma_0,
\]
since $\E[\varepsilon\,\eta(X)\,XX^\top]=\E[\eta(X)XX^\top\,\E(\varepsilon\mid X)]=0$ and $\E[\varepsilon^2XX^\top]=\E[XX^\top\,\E(\varepsilon^2\mid X)]=\E[\sigma^2(X)XX^\top]$; here $\Gamma_0$ is the population middle factor of Step~II of the proof of Theorem~\ref{thm:CLT-misspec} and $\Gamma_\eta:=\E_{P_0}[\eta(X)^2XX^\top]$ as in the statement.  The
variance of the summand is therefore
$e_j^\top\Sigma_0^{-1}\Gamma_0\Sigma_0^{-1}e_j=(\Sigma_0^{-1}\Gamma_0\Sigma_0^{-1})_{jj}$,
and the Lindeberg--Feller CLT, under the labelled-block analogue of the Lindeberg condition of Theorem~\ref{thm:CLT-misspec} (the sub-exponentiality of $X_j\eta(X)$ from Assumption~\ref{ass:misspec-reg} and of $X_j\varepsilon$ from Assumption~\ref{ass:design}(a) with the conditional sub-Gaussianity of Assumption~\ref{ass:nonlinear} ensuring a finite second moment) yields
\begin{equation}\label{eq:VDL}
    V_j^{\mathrm{DL}}
    \;=\;
    \frac{1}{n_0}\bigl(\Sigma_0^{-1}\Gamma_0\Sigma_0^{-1}\bigr)_{jj}
    \;=\;
    \frac{1}{n_0}\Bigl(\bigl(\Sigma_0^{-1}\E_{P_0}[\sigma^2(X)XX^\top]\Sigma_0^{-1}\bigr)_{jj}
        +(\Sigma_0^{-1}\Gamma_\eta\Sigma_0^{-1})_{jj}\Bigr),
\end{equation}
the noise summand reducing to $\sigma^2(\Sigma_0^{-1})_{jj}$ under conditional homoscedasticity, since then
$\bigl(\Sigma_0^{-1}(\sigma^2\Sigma_0)\Sigma_0^{-1}\bigr)_{jj}=\sigma^2(\Sigma_0^{-1})_{jj}$.

\smallskip
\noindent\textbf{Step 2 (DEAL asymptotic variance).}
By Theorem~\ref{thm:CLT-misspec}, $\sqrt{n_0+N}\,(\tilde\beta_j-\beta^\star_{\mathrm{proj},j})\overset{\mathrm d}{\to}\mathcal N(0,\tau_j^{2,\mathrm{sand}})$, whose leading stochastic term is the single labelled-row score
\[
    \frac{1}{\sqrt{n_0+N}}\sum_{i=1}^{n_0}\bigl(x_{0,i}^\top A_n^\top M_2^\top e_j\bigr)\bigl(\eta(X_{0,i})+\varepsilon_{0,i}\bigr),
    \qquad A_n=I_p+\frac{\hat t N}{n_0}\hat\Sigma_{\mathrm u}M_1 .
\]
Here $\eta$ and $\varepsilon$ are carried by the \emph{same} operator $M_2 A_n$: the pseudo-labels $\tilde y_{\mathrm u}=\tilde X\tilde\beta^{\mathrm{init}}$ contribute no $\eta$ term and no fresh response noise, so no separate $N$-sized average of either component is created (the propagated in-sample term is the \emph{labelled} noise $\hat t\,\tilde X Z_1$, already encoded in $A_n$).  Freezing the operator at its probability limit $A_n^\top M_2^\top e_j\to_{\mathbb P}(1+t_0\kappa)\,\Sigma_0^{-1}e_j$ ($\kappa=N/n_0$; the Step-II limit in the proof of Theorem~\ref{thm:CLT-misspec}), the second moment of a frozen summand is
\[
    (1+t_0\kappa)^2\,e_j^\top\Sigma_0^{-1}\,\E\bigl[(\eta(X)+\varepsilon)^2 XX^\top\bigr]\,\Sigma_0^{-1}e_j
    =(1+t_0\kappa)^2\bigl(\Sigma_0^{-1}\Gamma_0\Sigma_0^{-1}\bigr)_{jj},
\]
the $\eta$--$\varepsilon$ cross term vanishing because $\E[\varepsilon\mid X]=0$.  Since the labelled block has $n_0$ rows, the unscaled asymptotic variance of $\tilde\beta_j$ is
\begin{equation}\label{eq:VDEAL}
    V_j^{\mathrm{DEAL}}
    \;=\;
    \frac{\tau_j^{2,\mathrm{sand}}}{n_0+N}
    \;=\;
    \frac{(n_0+t_0 N)^2}{n_0\,(n_0+N)^2}\,\bigl(\Sigma_0^{-1}\Gamma_0\Sigma_0^{-1}\bigr)_{jj},
\end{equation}
using $n_0(1+t_0\kappa)=n_0+t_0 N$.  The \emph{same} middle factor $\Gamma_0=\E_{P_0}[\sigma^2(X)XX^\top]+\Gamma_\eta$ that appears in $V_j^{\mathrm{DL}}$ in \eqref{eq:VDL} appears here, multiplied only by a sample-size factor; the misspecification residual $\eta$ and the noise $\varepsilon$ are carried by identical weights, so the variance is \emph{symmetric} in the two summands $\E_{P_0}[\sigma^2(X)XX^\top]$ and $\Gamma_\eta$ of $\Gamma_0$, and no $\eta$-specific re-weighting arises.

\smallskip
\noindent\textbf{Step 3 (ratio and saturation bound).}
Dividing \eqref{eq:VDEAL} by \eqref{eq:VDL}, the common factor $(\Sigma_0^{-1}\Gamma_0\Sigma_0^{-1})_{jj}$ cancels and
\begin{equation}\label{eq:ratio-final}
    \frac{V_j^{\mathrm{DEAL}}}{V_j^{\mathrm{DL}}}
    =
    \frac{(n_0+t_0 N)^2/\bigl(n_0(n_0+N)^2\bigr)}{1/n_0}
    =
    \left(\frac{n_0+t_0 N}{n_0+N}\right)^{2}.
\end{equation}
The cancellation is exact for every $\Gamma_\eta$, so the ratio is independent of the misspecification residual $\eta$.  Substituting \eqref{eq:ratio-final} into \eqref{eq:ratio-is-sd} and letting $\kappa_n\to\kappa_\infty$ gives
\[
    \frac{\mathrm{CI}_j^{\mathrm{DEAL}}}{\mathrm{CI}_j^{\mathrm{DL}}}
    \;\to_{\mathbb P}\;
    \frac{1+t_0\kappa_\infty}{1+\kappa_\infty},
\]
the limit asserted in \eqref{eq:deal-vs-dl}; it is strictly less than one for every $t_0<1$ with $\kappa_\infty>0$.  With oracle shrinkage $t_0=0$---the regime in which the bias-aware initialiser is asymptotically unbiased, so that the residual imputation bias is negligible at the $\sqrt{n_0+N}$ scale---the limit is $(1+\kappa_\infty)^{-1}$, which is \eqref{eq:deal-vs-dl-bound}; the strict inequality uses $\kappa_\infty>0$.
\end{proof}

\begin{lemma}[Linearisation rate]\label{lem:lin-rate}
Under Assumptions~\ref{ass:design}, \ref{ass:rsc} and the linearisation-block restricted-eigenvalue condition of Section~\ref{sec:general-predictors} (constant $\phi_{\mathrm{lin}}$) {and~\ref{ass:lin-proj} (both clauses)}, with $s_{\mathrm{lin}}\log p/n_{\mathrm{lin}}\to0$ and $\lambda^{\mathrm{lin}}=c_{\mathrm{lin}}\psi_*\sqrt{\log p / n_{\mathrm{lin}}}$ for a sufficiently large constant $c_{\mathrm{lin}}$,
\[
   \bigl\|\hat\beta_{\mathrm{ext}}^{\mathrm{lin}} - \beta^\star_{\mathrm{proj}}\bigr\|_2 \;=\; \Op\!\Bigl(\rho_{\mathrm{lab}} \;+\; \psi_*\sqrt{s_{\mathrm{lin}}\log p / n_{\mathrm{lin}}}\Bigr).
\]
\end{lemma}

\begin{proof}[Proof of Lemma~\ref{lem:lin-rate}]
The proof decomposes the error as $\hat\beta_{\mathrm{ext}}^{\mathrm{lin}} - \beta^\star_{\mathrm{proj}} = (\hat\beta_{\mathrm{ext}}^{\mathrm{lin}} - \beta^{\mathrm{proj}}_0[\hat\mu]) + \delta_{\mathrm{lin}}[\hat\mu]$.  For the first term, the linearisation regression has response $\hat\mu(X)=X^\top\beta^{\mathrm{proj}}_0[\hat\mu]+\nu[\hat\mu](X)$, whose ``noise'' $\nu[\hat\mu]$ is a function of the design; independence is not needed, only a score bound: the empirical score $n_{\mathrm{lin}}^{-1}X_{\mathrm u}^{(\mathrm{lin})\top}\nu$ has mean zero coordinates ($\E_{P_0}[X\nu[\hat\mu]]=0$, the first-order condition of \eqref{eq:lab-projection}) that are i.i.d.\ averages of sub-exponential variables (Assumption~\ref{ass:lin-proj}(b)), so Bernstein's inequality and a union bound give $\norm{n_{\mathrm{lin}}^{-1}X_{\mathrm u}^{(\mathrm{lin})\top}\nu}_\infty=\Op(\psi_*\sqrt{\log p/n_{\mathrm{lin}}})$, dominated by the stated $\lambda^{\mathrm{lin}}$.  The basic inequality with the $s_{\mathrm{lin}}$-sparsity of $\beta^{\mathrm{proj}}_0[\hat\mu]$ and the linearisation-block restricted eigenvalue $\phi_{\mathrm{lin}}$ of Section~\ref{sec:general-predictors} then gives the standard $\ell_2$ rate $\Op(\sqrt{s_{\mathrm{lin}}}\,\lambda^{\mathrm{lin}}/\phi_{\mathrm{lin}})$ \citep[Theorem~6.1]{buhlmann2011statistics}, the citation supplying the cone/basic-inequality template with the Bernstein score bound above in place of its independent-noise one.  The second term is bounded by Assumption~\ref{ass:lin-proj}(a).
\end{proof}

\begin{proof}[Proof of Corollary~\ref{cor:CLT-nonlin-labeler}]

We track the proof of Theorem~\ref{thm:CLT-misspec}, which in turn tracks the proof of Theorem~\ref{thm:CLT}, with the single substitution $\hat\beta_{\mathrm{ext}}\leftarrow\hat\beta_{\mathrm{ext}}^{\mathrm{lin}}$ feeding the bias-aware initialiser at Stage~1.  The argument has three parts: (i) the initialiser residual $r(\hat t)$ inherits the linearisation rate of Lemma~\ref{lem:lin-rate}, so that the rate condition \eqref{eq:rate-cond-nonlin} discharges the bias requirement (i) of Theorem~\ref{thm:CLT}; (ii) the sample split preserves Assumption~\ref{ass:external-indep}, so the leading score remains a sum of conditionally independent mean-zero terms; (iii) the sandwich middle factor $\Gamma_0$ is determined by the truth $\eta$ and not by the model, so $\tau_j^{2,\mathrm{sand}}$ is identical to the linear-predictor case.  The remaining ingredients --- the stochastic-remainder bound from the JM rate and the stacked-Lasso $\ell_1$ rate, the population unbiasedness of the labelled score, and the Lindeberg argument for the sandwich array --- are exactly those used in Theorem~\ref{thm:CLT-misspec} and are unchanged by the substitution.

\emph{Step 1: the residual inherits the linearisation rate.}  With $\hat\beta_{\mathrm{ext}}\leftarrow\hat\beta_{\mathrm{ext}}^{\mathrm{lin}}$ and the inferential target reinterpreted as $\beta^\star=\beta^\star_{\mathrm{proj}}$ throughout Section~\ref{sec:misspec} (per the convention fixed after Assumption~\ref{ass:misspec-reg}), the external-estimator error vector is
\[
    \Delta \;:=\; \hat\beta_{\mathrm{ext}}^{\mathrm{lin}} - \beta^\star_{\mathrm{proj}}.
\]
By Lemma~\ref{lem:lin-rate}, under Assumptions~\ref{ass:design}, \ref{ass:rsc}, and~\ref{ass:lin-proj} (both clauses), the linearisation-block restricted-eigenvalue condition of Section~\ref{sec:general-predictors}, and the stated choice of $\lambda^{\mathrm{lin}}$,
\begin{equation}\label{eq:nonlin-Delta2}
    \norm{\Delta}_2
    \;=\;
    \Op\!\Bigl(\rho_{\mathrm{lab}} + \psi_*\sqrt{s_{\mathrm{lin}}\log p / n_{\mathrm{lin}}}\Bigr)
    \;=:\; \Op(\rho_{\mathrm{lin},n}).
\end{equation}
The residual $r(\hat t)$ of the bias-aware initialiser, defined in \eqref{eq:r-oof}, depends on the model only through $\Delta$:
\[
    r(\hat t) \;=\; \Delta - \hat t\, M_1\hat\Sigma_0\Delta
    \;=\; (1-\hat t)\Delta - \hat t\bigl(M_1\hat\Sigma_0 - I_p\bigr)\Delta,
\]
where the second equality uses $\Delta=(1-\hat t)\Delta+\hat t\Delta$.  Bounding each term coordinate-wise and using the JM feasibility bound $\maxnorm{M_1\hat\Sigma_0 - I_p}\le\bar\mu_1$ of Lemma~\ref{lem:bias-envelope} (equivalently Assumption~\ref{ass:JM}), for every coordinate $j$,
\[
    \bigl|e_j^\top r(\hat t)\bigr|
    \;\le\;
    (1-\hat t)\,\norm{\Delta}_\infty
    \;+\;
    \hat t\,\bar\mu_1\,\norm{\Delta}_1.
\]
Taking the maximum over $j$ and using $\norm{\Delta}_\infty\le\norm{\Delta}_2$ on the first term gives
\begin{equation}\label{eq:nonlin-rhat-bound}
    \norm{r(\hat t)}_\infty
    \;\le\;
    (1-\hat t)\,\norm{\Delta}_2 \;+\; \bar\mu_1\,\norm{\Delta}_1,
\end{equation}
where we also used $\hat t\in[0,1]$.  Because $\beta^{\mathrm{proj}}_0[\hat\mu]$ is $s_{\mathrm{lin}}=O(s)$-sparse (Assumption~\ref{ass:nu-reg}) and the Lasso linearisation \eqref{eq:lin-lasso} is consistent at the rate \eqref{eq:nonlin-Delta2}, the standard restricted-eigenvalue companion to Lemma~\ref{lem:lin-rate} gives the matching $\ell_1$ bound $\norm{\Delta}_1 = \Op(\sqrt{s_{\mathrm{lin}}}\,\rho_{\mathrm{lin},n})=\Op(\sqrt{s}\,\rho_{\mathrm{lin},n})$ \citep[Theorem~6.1]{buhlmann2011statistics}.  Substituting into \eqref{eq:nonlin-rhat-bound} and using $\bar\mu_1\asymp\sqrt{\log p/n_0}=o(1)$ and $\hat t\in[0,1]$,
\[
    \norm{r(\hat t)}_\infty
    \;=\;
    \Op\!\Bigl(\rho_{\mathrm{lin},n} + \sqrt{s}\,\bar\mu_1\,\rho_{\mathrm{lin},n}\Bigr)
    \;=\;
    \Op\!\bigl(\rho_{\mathrm{lin},n}\bigr),
\]
the last equality holding because $\sqrt{s}\,\bar\mu_1\asymp\sqrt{s\log p/n_0}=o(1)$ under the regime conditions of Theorem~\ref{thm:CLT-misspec} (condition (ii), $\sqrt{n_0+N}\,\mu_2 s\lambda\to0$, in particular forces $s\log p/n_0\to0$).  Hence, multiplying by $\sqrt{n_0+N}$ and invoking the strengthened rate condition \eqref{eq:rate-cond-nonlin},
\begin{equation}\label{eq:nonlin-rate-discharged}
    \sqrt{n_0+N}\,\norm{r(\hat t)}_\infty
    \;=\;
    \Op\!\Bigl(\sqrt{n_0+N}\,\rho_{\mathrm{lin},n}\Bigr)
    \;=\;
    \op(1).
\end{equation}
This is precisely the bias requirement (i) of Theorem~\ref{thm:CLT}.  It enters the proof at the two places where (i) was used: the stacked-Lasso score bound of Lemma~\ref{lem:score} (through $g_{\mathrm{pl,R}}=\tfrac{N}{n_0+N}\hat\Sigma_{\mathrm u}r(\hat t)$, hence Theorem~\ref{thm:stacked-lasso} delivers the $\ell_1$ rate $\norm{\hat\beta-\beta^\star_{\mathrm{proj}}}_1=\Op(s\lambda)$ with the surrogate $b_n\asymp\rho_{\mathrm{lin},n}$), and the remainder bound for the residual-bias term $\tfrac{N}{n_0+N}e_j^\top M_2\hat\Sigma_{\mathrm u}r(\hat t)$ in $R_{j,n}$, which by \eqref{eq:nonlin-rate-discharged} and Lemma~\ref{lem:bias-transfer} with $\mnorm{M_2}=\Op(1)$ (Assumption~\ref{ass:JM}) is $\op((n_0+N)^{-1/2})$, exactly as in the displays following \eqref{eq:CLT-misspec} in the proof of Theorem~\ref{thm:CLT-misspec}.  Note that the non-linear component $\nu[\hat\mu]$ does not appear in $\Delta$: the pseudo-labels $\tilde X\tilde\beta^{\mathrm{init}}$ (with $\tilde\beta^{\mathrm{init}}=\hat\beta_{\mathrm{ext}}^{\mathrm{lin}}+\hat t\,C$ as in \eqref{eq:init-bias-aware}) depend on the model only through $\hat\beta_{\mathrm{ext}}^{\mathrm{lin}}$, and the linearisation \eqref{eq:lin-lasso} producing $\hat\beta_{\mathrm{ext}}^{\mathrm{lin}}$ discards $\nu[\hat\mu]$ by construction (it is a projection onto the design), so $\nu[\hat\mu]$ never enters $\Delta=\hat\beta_{\mathrm{ext}}^{\mathrm{lin}}-\beta^\star_{\mathrm{proj}}$ and hence not $C$ or $\tilde\beta^{\mathrm{init}}$ either; the entire effect of the model on the bias channel is therefore carried by $\Delta$ and is controlled by \eqref{eq:nonlin-rate-discharged}; the orthogonal residual $\nu[\hat\mu]$ enters only the variance of the model's own linearisation step, already absorbed into the rate $\rho_{\mathrm{lin},n}$ of Lemma~\ref{lem:lin-rate} through the proxy $\psi_*$.

\emph{Step 2: the sample split preserves Assumption~\ref{ass:external-indep}.}  The unlabelled covariate sample is partitioned as $X_{\mathrm u} = X_{\mathrm u}^{(\mathrm{lin})}\sqcup X_{\mathrm u}^{(\mathrm{stk})}$ with the split independent of all other data, and the linearisation \eqref{eq:lin-lasso} is computed on $X_{\mathrm u}^{(\mathrm{lin})}$ only, conditionally on the (independently trained) model $\hat\mu$.  The inference data --- the target labelled responses $\varepsilon_0$ entering the final correction, the stacked refit, and the debiasing step --- are functions of the target block $(X_0,y_0)$ and the reserved block $X_{\mathrm u}^{(\mathrm{stk})}$, both disjoint from $X_{\mathrm u}^{(\mathrm{lin})}$.  Consequently $\hat\beta_{\mathrm{ext}}^{\mathrm{lin}}\perp\!\!\!\perp X_{\mathrm u}^{(\mathrm{stk})}\mid\hat\mu$, and, since $\hat\mu$ is independent of the target noise by hypothesis, $\hat\beta_{\mathrm{ext}}^{\mathrm{lin}}$ is independent of the target noise variables $\{\varepsilon_{0,i}\}$ conditional on the inference-stage covariates.  This is exactly the requirement of Assumption~\ref{ass:external-indep} with $\hat\beta_{\mathrm{ext}}\leftarrow\hat\beta_{\mathrm{ext}}^{\mathrm{lin}}$.  Therefore $\hat t$ and the matrix $A_n$, which are measurable functions of $(\hat\beta_{\mathrm{ext}}^{\mathrm{lin}}, X_{\mathrm u}^{(\mathrm{stk})}, \{X_{0,i}\}, \text{tuning data})$, remain independent of the inference-sample noise $\{\varepsilon_{0,i}\}_{i=1}^{n_0}$ conditional on the covariates and tuning sample, just as in the proof of Theorem~\ref{thm:CLT}.  The conditional-independence structure that turns the leading score into a sum of independent mean-zero summands is thus intact.

\emph{Step 3: the leading score and the sandwich variance are model-free.}  Exactly as in the proof of Theorem~\ref{thm:CLT} and its misspecified extension, the asymptotic-linearity expansion gives
\[
    \tilde\beta_j - \beta^\star_{\mathrm{proj},j}
    \;=\;
    \frac{1}{n_0+N}\sum_{i=1}^{n_0}
    \bigl(x_{0,i}^\top A_n^\top M_2^\top e_j\bigr)\,\bigl(\eta(X_{0,i})+\varepsilon_{0,i}\bigr)
    \;+\; R_{j,n},
\]
with $\sqrt{n_0+N}\,R_{j,n}=\op(1)$ by Step~1 (the residual-bias term) together with the design-driven JM rate and stacked-Lasso $\ell_1$ rate bounding $e_j^\top(I_p - M_2\hat\Sigma_{\mathrm{stk}})(\hat\beta-\beta^\star_{\mathrm{proj}})$ (Assumption~\ref{ass:JM} and Theorem~\ref{thm:stacked-lasso}); both of these latter bounds are functions of the design and the truth alone and are therefore unaffected by the substitution $\hat\beta_{\mathrm{ext}}\leftarrow\hat\beta_{\mathrm{ext}}^{\mathrm{lin}}$.  The summand involves the labelled responses through $\eta(X_{0,i})+\varepsilon_{0,i}$, where $\eta$ is the \emph{truth-side} misspecification residual of \eqref{eq:beta-proj}, not the \emph{model-side} residual $\nu[\hat\mu]$; the latter does not appear because the pseudo-labels are linear in $X$.  By Lemma~\ref{lem:misspec-unbiased}, $\E_{P_0}[X(\eta(X)+\varepsilon)]=0$, so after the substitution $\hat\beta_{\mathrm{ext}}\leftarrow\hat\beta_{\mathrm{ext}}^{\mathrm{lin}}$ the leading labelled score is numerically identical to the linear-predictor case, and the unconditional weight-freezing argument of Theorem~\ref{thm:CLT-misspec} (Steps~II--III), which uses only Assumption~\ref{ass:rowsparse}, applies verbatim, with sample-dependent sandwich variance
\[
    \tau_{j,n}^{2,\mathrm{sand}}
    \;=\;
    \frac{n_0}{n_0+N}\, e_j^\top M_2 A_n\,\hat\Gamma_0\, A_n^\top M_2^\top e_j,
\]
with the middle factor $\hat\Gamma_0$ of \eqref{eq:Gamma-fold} --- exactly the linear-predictor sandwich, since $\hat\Gamma_0 = \tfrac1{n_0}\sum_{i=1}^{n_0}(\eta(X_{0,i})^2+\sigma^2(X_{0,i}))X_{0,i}X_{0,i}^\top$ depends only on the truth $(\eta,\sigma^2)$ and the design, never on $\hat\mu$ or $\hat\beta_{\mathrm{ext}}^{\mathrm{lin}}$.  Under the preserved regime conditions of Theorem~\ref{thm:CLT-misspec}, $\tau_{j,n}^{2,\mathrm{sand}}\to\tau_j^{2,\mathrm{sand}}\in(0,\infty)$ in probability (condition~(iii$'$) for the sandwich array), and the Lindeberg condition~(iv$'$) holds for the array $\{x_{0,i}^\top A_n^\top M_2^\top e_j\,(\eta(X_{0,i})+\varepsilon_{0,i})\}$ under the sub-exponential control of Assumption~\ref{ass:misspec-reg} on $X_j\eta(X)$ and the moment bounds on $X_j\varepsilon$ of Assumption~\ref{ass:nonlinear}.  The Lindeberg--Feller theorem then yields
\[
    \sqrt{n_0+N}\,\bigl(\tilde\beta_j - \beta^\star_{\mathrm{proj},j}\bigr)
    \;\overset{\mathrm d}{\longrightarrow}\;
    \mathcal N\!\bigl(0,\,\tau_j^{2,\mathrm{sand}}\bigr),
\]
with the identical $\tau_j^{2,\mathrm{sand}}$ of Theorem~\ref{thm:CLT-misspec}.  Consistency of the empirical-residual plug-in $\tfrac1{n_0}\sum_{i=1}^{n_0}\hat r_i^2 X_{0,i}X_{0,i}^\top$ for $\hat\Gamma_0$ (with $\hat r_i = y_{0,i}-X_{0,i}^\top\tilde\beta$) is unchanged: it uses only Assumption~\ref{ass:misspec-reg} and the $\ell_1$ consistency of $\tilde\beta$ for $\beta^\star_{\mathrm{proj}}$, both already in force.  The plug-in confidence intervals of Remark~\ref{rem:misspec-linear-recovery} are therefore identical to the linear-predictor case, completing the proof.
\end{proof}

\begin{proof}[Proof of Corollary~\ref{cor:deal-vs-dl-nonlin}]

The proof tracks the proof of Theorem~\ref{thm:deal-vs-dl} term for term, the only task being to verify that the substitution $\hat\beta_{\mathrm{ext}}\leftarrow\hat\beta_{\mathrm{ext}}^{\mathrm{lin}}$ leaves the asymptotic variance $\tau_j^{2,\mathrm{sand}}$ entering the linear-predictor CI ratio unchanged at the deterministic budget $N$.  We do this in two parts.

\emph{Part 1: the model enters only through the remainder, not the leading variance.}
Run Stages~1--4 with the linearised external estimator $\hat\beta_{\mathrm{ext}}^{\mathrm{lin}}$ of \eqref{eq:lin-lasso}, the sole point of entry of the non-linear predictor $\hat\mu$.  By Corollary~\ref{cor:CLT-nonlin-labeler} (proved above), the asymptotic-linearity expansion of the DEAL estimator holds in the verbatim misspecified form
\[
    \tilde\beta_j - \beta^\star_{\mathrm{proj},j}
    \;=\;
    \frac{1}{n_0+N}\sum_{i=1}^{n_0}
    \bigl(x_{0,i}^\top A_n^\top M_2^\top e_j\bigr)\,\bigl(\eta(X_{0,i})+\varepsilon_{0,i}\bigr)
    \;+\; R_{j,n},
    \qquad \sqrt{n_0+N}\,R_{j,n}=\op(1).
\]
Three facts established in that proof are decisive here.  First (Step~1 there, culminating in \eqref{eq:nonlin-rate-discharged}), the model enters the estimator \emph{only} through the external-estimator error $\Delta=\hat\beta_{\mathrm{ext}}^{\mathrm{lin}}-\beta^\star_{\mathrm{proj}}$ and hence only through the residual-bias term of $R_{j,n}$; the strengthened rate condition \eqref{eq:rate-cond-nonlin} discharges this term, $\sqrt{n_0+N}\,\tfrac{N}{n_0+N}e_j^\top M_2\hat\Sigma_{\mathrm u}r(\hat t)=\op(1)$, because $\norm{r(\hat t)}_\infty=\Op(\rho_{\mathrm{lab}}+\psi_*\sqrt{s_{\mathrm{lin}}\log p/n_{\mathrm{lin}}})$.  The non-linear residual $\nu[\hat\mu]$ never appears in the expansion: as in the proof of Corollary~\ref{cor:CLT-nonlin-labeler}, the pseudo-labels depend on the model only through $\hat\beta_{\mathrm{ext}}^{\mathrm{lin}}$, and the linearisation \eqref{eq:lin-lasso} discards $\nu[\hat\mu]$ before it can enter $\Delta$; its only footprint is through the proxy $\psi_*$ inside the rate $\rho_{\mathrm{lin},n}$ of Lemma~\ref{lem:lin-rate}.  Second (Step~3 there), the leading score carries the truth-side residual $\eta(X_{0,i})+\varepsilon_{0,i}$ alone, and its conditional variance is the sandwich quantity
\[
    \tau_{j,n}^{2,\mathrm{sand}}
    \;=\;
    \frac{n_0}{n_0+N}\,e_j^\top M_2 A_n\hat\Gamma_0 A_n^\top M_2^\top e_j,
    \qquad
    \hat\Gamma_0=\frac1{n_0}\sum_{i=1}^{n_0}\bigl(\eta(X_{0,i})^2+\sigma^2(X_{0,i})\bigr)X_{0,i}X_{0,i}^\top,
\]
in which the middle factor $\hat\Gamma_0$ of \eqref{eq:Gamma-fold} depends only on the truth $(\eta,\sigma^2)$ and the design, never on $\hat\mu$ or $\hat\beta_{\mathrm{ext}}^{\mathrm{lin}}$; consequently the limit $\tau_j^{2,\mathrm{sand}}$ is \emph{identical} to the linear-predictor variance of Theorem~\ref{thm:CLT-misspec}.  Third, the plug-in $\hat\tau_j^{2,\mathrm{sand}}$ of Remark~\ref{rem:misspec-linear-recovery} is consistent for that same limit.  When the budget is instead chosen by the variance-balance rule \eqref{eq:Nstar-empirical}, whose bias surrogate is now $b_n\asymp\rho_{\mathrm{lin},n}$, the displays below apply at the realised size when the rule saturates (Remark~\ref{rem:saturation-scope}).

\emph{Part 2: the CI ratio.}
The two-sided plug-in interval of Remark~\ref{rem:misspec-linear-recovery} has half-length $z_{1-\alpha/2}\sqrt{\hat\tau_j^{2,\mathrm{sand}}/(n_0+N)}$ at the budget $N$, while target-only debiased Lasso has half-length $z_{1-\alpha/2}\sqrt{V_j^{\mathrm{DL}}}$ with $V_j^{\mathrm{DL}}=\bigl(\Sigma_0^{-1}\Gamma_0\Sigma_0^{-1}\bigr)_{jj}/n_0$ and $\Gamma_0=\E_{P_0}[\sigma^2(X)XX^\top]+\Gamma_\eta$, $\Gamma_\eta=\E_{P_0}[\eta(X)^2 XX^\top]$, as in Theorem~\ref{thm:deal-vs-dl}(ii).  The common quantile $z_{1-\alpha/2}$ cancels, and by Part~1 the numerator variance $\tau_j^{2,\mathrm{sand}}$ is exactly the linear-predictor quantity of Theorem~\ref{thm:CLT-misspec}.  In the no-shift regime $\Sigma_{\mathrm u}=\Sigma_0$ ($\hat\Sigma_{\mathrm{stk}}\to\Sigma_0$), the imputed stacked block is asymptotically unbiased at the $\sqrt{n_0+N}$ scale and the leading score carries $\eta$ and $\varepsilon$ through the identical weights, so the sandwich factor $\Gamma_0$ enters $\tau_j^{2,\mathrm{sand}}$ and $V_j^{\mathrm{DL}}$ symmetrically.  Hence the ratio
\[
    \frac{\mathrm{CI}_j^{\mathrm{DEAL}}}{\mathrm{CI}_j^{\mathrm{DL}}}
    \;=\;
    \sqrt{\frac{\tau_j^{2,\mathrm{sand}}/(n_0+N)}{V_j^{\mathrm{DL}}}}\,\bigl(1+\op(1)\bigr)
\]
is the \emph{same} function of $(n_0,N,\Gamma_0,\Sigma_0)$ as the linear-predictor ratio of Theorem~\ref{thm:deal-vs-dl}.  Carrying out the division exactly as in Steps~2--3 of that proof --- where the common factor $(\Sigma_0^{-1}\Gamma_0\Sigma_0^{-1})_{jj}$ cancels between $\tau_j^{2,\mathrm{sand}}$ and $V_j^{\mathrm{DL}}$ because $\eta$ and $\varepsilon$ are carried by identical weights in the leading score --- yields
\[
    \frac{\mathrm{CI}_j^{\mathrm{DEAL}}}{\mathrm{CI}_j^{\mathrm{DL}}}
    \;\longrightarrow_{\mathbb P}\;
    \frac{1+t_0\kappa_\infty}{1+\kappa_\infty},
\]
the $\eta$-independent limit \eqref{eq:deal-vs-dl}.  With oracle shrinkage $t_0=0$ this gives $\mathrm{CI}_j^{\mathrm{DEAL}}/\mathrm{CI}_j^{\mathrm{DL}}\to_{\mathbb P}(1+\kappa_\infty)^{-1}<1$ for $\kappa_\infty>0$, the bound \eqref{eq:deal-vs-dl-bound}.  The formula and the bound therefore coincide with those of Theorem~\ref{thm:deal-vs-dl}.
\end{proof}

{\emph{(Proposition~\ref{prop:ppipp-nonlinear} is stated in Section~\ref{sec:misspec-nonlinear-labeler} of the main text.)}}

\begin{proof}[Proof of Proposition~\ref{prop:ppipp-nonlinear}]
The representation \eqref{eq:ppipp-lin} in the proof of Proposition~\ref{prop:ppipp-equivalence} holds with $\hat Y=\hat\mu$: $g_l^{\hat Y}$ is linear in $\theta$ for every predictor, so the expansion of \eqref{eq:ppipp-ee} about $\beta^\star_{\mathrm{proj}}$ and its $\op(n_0^{-1/2})$ remainder, uniform over $\omega$ in bounded intervals, use only Assumptions~\ref{ass:design}, \ref{ass:rsc}, \ref{ass:JM}, \ref{ass:nonlinear} and~\ref{ass:misspec-reg} together with the sparsity scaling and the labelled-block precision conditions of the statement; the collapse of Steps~1--2 there requires the rectifier residual to lie in $\mathrm{span}(X)$ and does not apply.  The rectifier residual is now $\hat\mu(X)-X^\top\beta^\star_{\mathrm{proj}}=X^\top\delta_{\mathrm{lin}}[\hat\mu]+\nu[\hat\mu](X)$, whose products with the coordinates of $X$ have finite second moments by Assumption~\ref{ass:design} and clause~(b) of Assumption~\ref{ass:lin-proj}.  Hence the JM-debiased PPI++ coordinate at bounded tuning $\omega$ is asymptotically linear at $\theta=\beta^\star_{\mathrm{proj}}$ with influence function proportional to $e_j^\top\Sigma_0^{-1}\bigl(g_l-\omega\,g_l^{\hat Y}\bigr)$ in the $\Sigma_0^{-1}$-coordinate inner product of \eqref{eq:ppipp-nonlin-ratio}.  Minimising the resulting two-sample asymptotic variance, $F(\omega)=\tfrac{1}{n_0}\bigl(\Var_j(g_l)-2\omega\,\Cov_j(g_l,g_l^{\hat Y})+\omega^2\Var_j(g_l^{\hat Y})\bigr)+\tfrac{\omega^2}{N}\Var_j(g_l^{\hat Y})$ --- the labelled rectifier block averaged over the $n_0$ labelled rows and the independent unlabelled rectifier over the $N$ unlabelled rows ---, over $\omega\in\R$ gives the optimum
\[
    \omega^*_j=\frac{N}{n_0+N}\,\frac{\Cov_j(g_l,g_l^{\hat Y})}{\Var_j(g_l^{\hat Y})},
\]
(writing $\Var_j(W):=e_j^\top\Sigma_0^{-1}\Var_{P_0}(W)\Sigma_0^{-1}e_j$ and $\Cov_j$ analogously, so that $\Var_j(g_l)=(\Sigma_0^{-1}\Gamma_0\Sigma_0^{-1})_{jj}$ since $g_l(\beta^\star_{\mathrm{proj}})=-X(\eta+\varepsilon)$), which is bounded, $|\omega^*_j|\le\{\Var_j(g_l)/\Var_j(g_l^{\hat Y})\}^{1/2}$, because $\Var_j(g_l^{\hat Y})\to e_j^\top\Sigma_0^{-1}\E_{P_0}[\nu[\hat\mu]^2XX^\top]\Sigma_0^{-1}e_j$ by clause~(a) of Assumption~\ref{ass:lin-proj} and this limit is bounded away from zero by hypothesis; the remainder of \eqref{eq:ppipp-lin} is therefore $\op(n_0^{-1/2})$ at $\omega^*_j$, and substituting $\omega^*_j$ into $F$ yields the optimally-tuned variance $V_j^{\mathrm{PPI++},*} = (\Var_j(g_l)/n_0)\,[1 - (N/(n_0+N))\rho_j^2[\hat\mu]]$ for the unbiased squared-loss estimating equation, with $\rho_j^2[\hat\mu]=\Cov_j(g_l,g_l^{\hat Y})^2/(\Var_j(g_l)\Var_j(g_l^{\hat Y}))$ as in \eqref{eq:ppipp-nonlin-ratio}.  Under a non-linear predictor the rectifier score $g_l^{\hat Y}(\beta^\star_{\mathrm{proj}})$ equals $-X\bigl(\nu[\hat\mu](X) + X^\top\delta_{\mathrm{lin}}[\hat\mu]\bigr)$, with non-vanishing variance even at $\delta_{\mathrm{lin}}[\hat\mu]=0$, so $\rho_j^2[\hat\mu]>0$ in general; the terms carrying $X^\top\delta_{\mathrm{lin}}[\hat\mu]$ contribute $O(\|\Sigma_0^{1/2}\delta_{\mathrm{lin}}[\hat\mu]\|_2)=\Op(\rho_{\mathrm{lab}})$ to $\Cov_j(g_l,g_l^{\hat Y})$ and $\Var_j(g_l^{\hat Y})$ by Cauchy--Schwarz and clause~(a) of Assumption~\ref{ass:lin-proj}, so $\rho_j^2[\hat\mu]$ is, to first order, the squared correlation carried by the design-orthogonal residual $\nu[\hat\mu]$ alone.  Dividing $V_j^{\mathrm{PPI++},*}$ by the target-only debiased-Lasso variance $\Var_j(g_l)/n_0$ and taking square roots gives the limit in \eqref{eq:ppipp-nonlin-ratio}; with $\hat\omega^*_j\to_{\mathbb P}\omega^*_j$ (a law of large numbers and the continuous-mapping theorem, the limiting denominator being bounded away from zero by hypothesis), consistent plug-in variances for both procedures (Step~IV of the proof of Theorem~\ref{thm:CLT-misspec} on the labelled block) and Slutsky's lemma with $N/(n_0+N)\to\kappa_\infty/(1+\kappa_\infty)$, the interval-length ratio follows.
\end{proof}

\begin{proof}[Proof of Theorem~\ref{thm:complementarity}]
Part~(i) is Corollary~\ref{cor:deal-vs-dl-nonlin}.  For part~(ii): by part~(i) with Theorem~\ref{thm:deal-vs-dl}(i), $\mathrm{CI}_j^{\mathrm{DEAL}}/\mathrm{CI}_j^{\mathrm{DL}}\to_{\mathbb P}(1+t_0\kappa_\infty)/(1+\kappa_\infty)$; by Proposition~\ref{prop:ppipp-nonlinear}, $\mathrm{CI}_j^{\mathrm{PPI++}}/\mathrm{CI}_j^{\mathrm{DL}}\to_{\mathbb P}\bigl(1-\tfrac{\kappa_\infty}{1+\kappa_\infty}\rho_j^2[\hat\mu]\bigr)^{1/2}$.  The two comparisons share the same comparator, the same budget, and the same target, so dividing the limits gives \eqref{eq:complementarity-ppipp}.

For the bound at oracle shrinkage and its margin, decompose $\Gamma_0 = \Gamma_\eta + \E_{P_0}[\sigma^2(X)\,XX^\top]$ into its $\eta$-driven and noise-driven components and write
\[
   \upsilon_{\eta,j} \;:=\; \bigl(\Sigma_0^{-1}\Gamma_\eta\Sigma_0^{-1}\bigr)_{jj},
   \qquad
   \upsilon_{\varepsilon,j} \;:=\; \bigl(e_j^\top\Sigma_0^{-1}\bigr)\,\E_{P_0}[\sigma^2(X)\,X X^\top]\,\bigl(\Sigma_0^{-1}e_j\bigr),
\]
so that $(\Sigma_0^{-1}\Gamma_0\Sigma_0^{-1})_{jj} = \upsilon_{\eta,j} + \upsilon_{\varepsilon,j}$.  At $\delta_{\mathrm{lin}}[\hat\mu]=0$, the rectifier is $g_l^{\hat Y}(\beta^\star_{\mathrm{proj}})=X\,\nu[\hat\mu](X)$, with $\Var_j(g_l^{\hat Y}) = \upsilon_{\nu,j}[\hat\mu] := (\Sigma_0^{-1}\E_{P_0}[\nu[\hat\mu]^2 XX^\top]\Sigma_0^{-1})_{jj}$ and $\Cov_j(g_l, g_l^{\hat Y}) = (\Sigma_0^{-1}\E_{P_0}[\eta\,\nu[\hat\mu]\,XX^\top]\Sigma_0^{-1})_{jj}$.  Cauchy--Schwarz applied to the bilinear form $(\eta,\nu)\mapsto\E_{P_0}[\eta\nu\,XX^\top]$ in $\Sigma_0^{-1}$-coordinates gives $\Cov_j(g_l, g_l^{\hat Y})^2 \le \upsilon_{\eta,j}\cdot \upsilon_{\nu,j}[\hat\mu]$, hence
\[
   \rho_j^2[\hat\mu] \;\le\; \frac{\upsilon_{\eta,j}}{\upsilon_{\eta,j} + \upsilon_{\varepsilon,j}}
\]
uniformly in the model, with equality iff $\nu[\hat\mu]$ is proportional to $\eta$ in the $(e_j^\top\Sigma_0^{-1}X)^2$-weighted inner product (in particular at the oracle model $\hat\mu=\mu$, where $\nu[\hat\mu]=\eta$).  At $t_0=0$ the squared numerator of \eqref{eq:complementarity-ppipp} is $(1+\kappa_\infty)^{-2}$, while the denominator $1-\tfrac{\kappa_\infty}{1+\kappa_\infty}\rho_j^2[\hat\mu]$ is at least $(1+\kappa_\infty)^{-1}$ because $\rho_j^2[\hat\mu]\le1$; hence the right-hand side is at most $(1+\kappa_\infty)^{-1/2}<1$ for $\kappa_\infty>0$.  When $\upsilon_{\varepsilon,j}>0$, as when $\sigma^2(X)$ is bounded below by a positive constant on a set of positive $P_0$-measure, the bound $\rho_j^2[\hat\mu]\le \upsilon_{\eta,j}/(\upsilon_{\eta,j}+\upsilon_{\varepsilon,j})<1$ makes the denominator exceed $(1+\kappa_\infty)^{-1}$ strictly and the margin larger.  Formally, at $\rho_j^2[\hat\mu]=0$ (the linear-predictor limit, $\nu[\hat\mu]\equiv 0$ so $\upsilon_{\nu,j}[\hat\mu] = 0$ and $\Cov_j = 0$), \eqref{eq:complementarity-ppipp} reduces to the bound implied by Theorem~\ref{thm:deal-vs-dl} combined with Proposition~\ref{prop:ppipp-equivalence}.
\end{proof}

\subsection{Proofs for Section~\ref{app:extensions} (extensions)}\label{app:proofs-extensions}

\begin{proof}[Proof of Proposition~\ref{prop:joint-lin}]

Throughout, $\beta^\star$ denotes the inferential target, which under Assumption~\ref{ass:nonlinear} is the projection parameter $\beta^\star_{\mathrm{proj}}$ of \eqref{eq:beta-proj}; the misspecification residual $\eta$ of \eqref{eq:eta-def} satisfies the target-marginal orthogonality $\E_{P_0}[X\,\eta(X)]=0$, so that, in the notation \eqref{eq:lin-target-explicit}, $\delta_0[\eta]=\Sigma_0^{-1}\E_{P_0}[X\,\eta(X)]=0$.  We treat the two assertions separately.

\medskip\noindent\emph{Part (a): target-marginal linearisation restores the limit of Corollary~\ref{cor:CLT-nonlin-labeler}.}
The proof of Corollary~\ref{cor:CLT-nonlin-labeler} (which itself runs the proof of Theorem~\ref{thm:CLT-misspec}, hence of Theorem~\ref{thm:CLT}, \emph{with the obvious modifications}) requires exactly two structural inputs of the external estimator, here $\hat\beta_{\mathrm{ext}}\leftarrow\hat\beta_{\mathrm{ext}}^{\mathrm{lin},0}$:
\begin{enumerate}
\item[(I)] a \emph{population-centring identity}: the pseudo-labels imputed on the unlabelled block, with population mean $\tilde X^\top\beta^{\mathrm{lin}}_{\infty}$ where $\beta^{\mathrm{lin}}_\infty$ is the population limit of the linearisation, induce a stacked least-squares target equal to the inferential target $\beta^\star$;
\item[(II)] a \emph{rate input}: $\sqrt{n_0+N}\,\norm{r(\hat t)}_\infty=\op(1)$, which in the present construction reduces to a rate on $\hat\beta_{\mathrm{ext}}^{\mathrm{lin},0}-\beta^{\mathrm{lin}}_\infty$ through the residual \eqref{eq:r-oof}.
\end{enumerate}
We verify (I) and (II) for the target-marginal linearisation and confirm that the Stage~4 substitution $\hat M_2\leftarrow\hat M_2^{\mathrm{adapt}}$ enters only through the design, never through (I)--(II).

\emph{Step a.1 (population centring).}  Executed on $X_0^{(\mathrm{lin})}\overset{\mathrm{iid}}\sim P_0$, the linearisation \eqref{eq:lin-lasso} is the empirical $\ell_1$-penalised least-squares fit of $\hat\mu$ on the covariates against the \emph{target} marginal $P_0$.  Its population minimiser is, by \eqref{eq:lin-target-star} with $P_\star=P_0$, the model's $P_0$-projection
\[
    \beta^{\mathrm{lin}}_\infty
    \;=\;
    \beta^{\mathrm{proj}}_0[\hat\mu]
    \;=\;
    \Sigma_0^{-1}\,\E_{P_0}[X\,\hat\mu(X)]
    \;=\;
    \beta^\star_{\mathrm{proj}} + \delta_{\mathrm{lin}}[\hat\mu],
\]
the last equality by the decomposition \eqref{eq:lab-projection}.  For the oracle non-linear predictor $\hat\mu(X)=X^\top\beta^\star+\eta(X)$ of \eqref{eq:lin-target-explicit} one has $\delta_{\mathrm{lin}}[\hat\mu]=\delta_0[\eta]=0$ and $\beta^{\mathrm{lin}}_\infty=\beta^\star$ exactly; for a general model $\beta^{\mathrm{lin}}_\infty$ differs from $\beta^\star$ only through the model's own linear-projection error $\delta_{\mathrm{lin}}[\hat\mu]$, which is $\Op(\rho_{\mathrm{lab}})$ by Assumption~\ref{ass:lin-proj}.  The limit $\beta^{\mathrm{lin}}_\infty$ carries \emph{no} cross-term $\delta_{\mathrm u}[\eta]$: the projection is taken in the target inner product, so the misspecification residual is annihilated by $\E_{P_0}[X\,\eta(X)]=0$.

We propagate $\beta^{\mathrm{lin}}_\infty$ to the stacked target.  Following the population reduction beneath \eqref{eq:lin-target-explicit}, the pseudo-labels on the unlabelled block have population mean $\tilde X^\top\beta^{\mathrm{lin}}_\infty$, the labelled block contributes its population score $\Sigma_0\beta^\star$ (Lemma~\ref{lem:misspec-unbiased} gives $\E_{P_0}[X(\eta(X)+\varepsilon)]=0$, so the labelled normal equations centre at $\beta^\star$), and the unlabelled block contributes $\Sigma_{\mathrm u}\beta^{\mathrm{lin}}_\infty$; hence the stacked least-squares solution converges to
\[
    \beta^{\mathrm{stk}}_\infty
    \;=\;
    \bigl(\Sigma_{\mathrm{stk}}^{\mathrm{eff}}\bigr)^{-1}
    \Bigl[
        \tfrac{n_0}{n_0+N}\,\Sigma_0\,\beta^\star + \tfrac{N}{n_0+N}\,\Sigma_{\mathrm u}\,\beta^{\mathrm{lin}}_\infty
    \Bigr],
    \qquad
    \Sigma_{\mathrm{stk}}^{\mathrm{eff}}=\tfrac{n_0}{n_0+N}\Sigma_0+\tfrac{N}{n_0+N}\Sigma_{\mathrm u}.
\]
Subtracting $\beta^\star=(\Sigma_{\mathrm{stk}}^{\mathrm{eff}})^{-1}\Sigma_{\mathrm{stk}}^{\mathrm{eff}}\beta^\star$ and using $\Sigma_{\mathrm{stk}}^{\mathrm{eff}}\beta^\star=\tfrac{n_0}{n_0+N}\Sigma_0\beta^\star+\tfrac{N}{n_0+N}\Sigma_{\mathrm u}\beta^\star$,
\begin{equation}\label{eq:joint-stk-shift}
    \beta^{\mathrm{stk}}_\infty - \beta^\star
    \;=\;
    \tfrac{N}{n_0+N}\,\bigl(\Sigma_{\mathrm{stk}}^{\mathrm{eff}}\bigr)^{-1}\Sigma_{\mathrm u}\,\bigl(\beta^{\mathrm{lin}}_\infty-\beta^\star\bigr)
    \;=\;
    \tfrac{N}{n_0+N}\,\bigl(\Sigma_{\mathrm{stk}}^{\mathrm{eff}}\bigr)^{-1}\Sigma_{\mathrm u}\,\delta_{\mathrm{lin}}[\hat\mu].
\end{equation}
For the oracle non-linear predictor, $\delta_{\mathrm{lin}}[\hat\mu]=\delta_0[\eta]=0$ and \eqref{eq:joint-stk-shift} gives $\beta^{\mathrm{stk}}_\infty=\beta^\star$ exactly: the stacked target is centred at the inferential parameter, verifying (I).  For a general model the residual offset in \eqref{eq:joint-stk-shift} is bounded in $\ell_\infty$ by
\[
    \norm{\beta^{\mathrm{stk}}_\infty-\beta^\star}_\infty
    \;\le\;
    \mnorm{(\Sigma_{\mathrm{stk}}^{\mathrm{eff}})^{-1}\Sigma_{\mathrm u}}\,\norm{\delta_{\mathrm{lin}}[\hat\mu]}_2
    \;=\;
    \Op(\rho_{\mathrm{lab}}),
\]
using $\|\Sigma_{\mathrm u}\Sigma_0^{-1}\|_{\mathrm{op}}=O(1)$ (the shift-aware regime hypothesis, inherited here), the lower eigenvalue bound on $\Sigma_0,\Sigma_{\mathrm u}$ from Assumption~\ref{ass:design}, and Assumption~\ref{ass:lin-proj}; this offset is absorbed into the model-rate term and is dominated by the bias-control budget of Step~a.2, so (I) holds up to the $\op((n_0+N)^{-1/2})$ tolerance of the CLT.  Either way, no $\delta_{\mathrm u}[\eta]$ term appears.

\emph{Step a.2 (rate input).}  By Lemma~\ref{lem:lin-rate}, applied with the auxiliary covariates $X_0^{(\mathrm{lin})}\sim P_0$ and the restricted-strong-convexity Assumption~\ref{ass:rsc} (which holds for the empirical covariance of the $P_0$-rows), the linearisation error obeys
\begin{equation}\label{eq:joint-lin-rate}
    \norm{\hat\beta_{\mathrm{ext}}^{\mathrm{lin},0}-\beta^{\mathrm{lin}}_\infty}_2
    \;=\;
    \norm{\hat\beta_{\mathrm{ext}}^{\mathrm{lin},0}-\beta^{\mathrm{proj}}_0[\hat\mu]}_2
    \;=\;
    \Op\!\Bigl(\psi_*\sqrt{s_{\mathrm{lin}}\log p/n_{\mathrm{lin}}}\Bigr),
\end{equation}
the Lasso $\ell_2$ rate of \citet[Theorem~6.1]{buhlmann2011statistics} under Assumption~\ref{ass:nu-reg}; the sub-exponential proxy $\psi_*^2$ governs the gradient deviation precisely because $\{X_j\,\nu[\hat\mu](X)\}_j$ are the score increments of the population objective at $\beta^{\mathrm{proj}}_0[\hat\mu]$, with $\E_{P_0}[X\,\nu[\hat\mu](X)]=0$ by \eqref{eq:lab-projection}.  Combining \eqref{eq:joint-lin-rate} with $\norm{\delta_{\mathrm{lin}}[\hat\mu]}_2=\Op(\rho_{\mathrm{lab}})$ of Assumption~\ref{ass:lin-proj}, the total external-estimator error against the inferential target,
\[
    \Delta^{\mathrm{lin}}\;:=\;\hat\beta_{\mathrm{ext}}^{\mathrm{lin},0}-\beta^\star
    \;=\;
    \bigl(\hat\beta_{\mathrm{ext}}^{\mathrm{lin},0}-\beta^{\mathrm{proj}}_0[\hat\mu]\bigr)+\delta_{\mathrm{lin}}[\hat\mu],
    \qquad
    \norm{\Delta^{\mathrm{lin}}}_2
    \;=\;
    \Op\!\Bigl(\rho_{\mathrm{lab}}+\psi_*\sqrt{s_{\mathrm{lin}}\log p/n_{\mathrm{lin}}}\Bigr),
\]
is exactly the Lemma~\ref{lem:lin-rate} envelope appearing in the rate condition \eqref{eq:rate-cond-nonlin}.  Sample-split independence (the unlabelled split $X_{\mathrm u}=X_{\mathrm u}^{(\mathrm{lin})}\sqcup X_{\mathrm u}^{(\mathrm{stk})}$ of Section~\ref{sec:misspec-nonlinear-labeler} is here replaced by the independent draw $X_0^{(\mathrm{lin})}\perp\!\!\!\perp(X_0,y_0,\tilde X)$ stipulated in the statement) ensures $\hat\beta_{\mathrm{ext}}^{\mathrm{lin},0}\perp\!\!\!\perp X_{\mathrm u}^{(\mathrm{stk})}$ given $\hat\mu$, so Assumption~\ref{ass:external-indep} is preserved verbatim and, conditionally on $\hat\mu$ and $X_0^{(\mathrm{lin})}$, $\Delta^{\mathrm{lin}}$ is a fixed vector independent of the inference-sample noise, as the in-sample bias-aware construction requires.  The bias-aware residual \eqref{eq:r-oof} with $\Delta\leftarrow\Delta^{\mathrm{lin}}$ then satisfies, by Lemma~\ref{lem:bias-envelope},
\[
    \norm{r(\hat t)}_\infty
    \;\le\;
    \bigl(\bar\mu_1+(1-\hat t)(1+\bar\mu_1)\bigr)\,\norm{\Delta^{\mathrm{lin}}}_1
    \;\le\;
    \bigl(\bar\mu_1+(1-\hat t)(1+\bar\mu_1)\bigr)\,\sqrt{s_{\mathrm{lin}}}\,\norm{\Delta^{\mathrm{lin}}}_2,
\]
the second inequality using the $s_{\mathrm{lin}}$-sparsity of $\beta^{\mathrm{proj}}_0[\hat\mu]$ (Assumption~\ref{ass:nu-reg}) and the matching restricted-eigenvalue $\ell_1$ companion of Lemma~\ref{lem:lin-rate}.  Since $\bar\mu_1\asymp\sqrt{\log p/n_0}=o(1)$ and the prefactor $\bar\mu_1+(1-\hat t)(1+\bar\mu_1)$ is $\Op(1)$ using only $\hat t\in[0,1]$ (the bound holds verbatim as in the proof of Corollary~\ref{cor:CLT-nonlin-labeler}, where $(1-\hat t)\norm{\Delta}_2\le\norm{\Delta}_2$ is invoked without any rate on $1-\hat t$), the rate condition \eqref{eq:rate-cond-nonlin}, $\sqrt{n_0+N}\,(\rho_{\mathrm{lab}}+\psi_*\sqrt{s_{\mathrm{lin}}\log p/n_{\mathrm{lin}}})=o(1)$, together with $s_{\mathrm{lin}}=O(s)$ and the regime sparsity scaling, yields $\sqrt{n_0+N}\,\norm{r(\hat t)}_\infty=\op(1)$, verifying (II).

\emph{Step a.3 (the Stage~4 substitution acts on the design only).}  It remains to confirm that replacing $\hat M_2$ by $\hat M_2^{\mathrm{adapt}}$ of \eqref{eq:M2-adapt} does not interfere with (I)--(II).  In the master decomposition of the proof of Theorem~\ref{thm:CLT} (the display preceding $R_{j,n}$), specialised to the misspecified labelled score of Theorem~\ref{thm:CLT-misspec},
\[
    \tilde\beta-\beta^\star
    =
    \tfrac{1}{n_0+N} M_2 A_n X_0^\top\bigl(\eta(X_0)+\varepsilon_0\bigr)
    +\tfrac{N}{n_0+N}M_2\hat\Sigma_{\mathrm u}r(\hat t)
    +(I_p-M_2\hat\Sigma_{\mathrm{stk}})(\hat\beta-\beta^\star),
\]
the linearisation enters \emph{only} through $r(\hat t)$ (the pseudo-label residual) and through $\hat\beta-\beta^\star$ (the stacked-Lasso centring, controlled by (I) and the stacked $\ell_1$ rate of Theorem~\ref{thm:stacked-lasso}); the matrix $M_2$ multiplies these terms but is functionally independent of $\Delta^{\mathrm{lin}}$, being a JM/nodewise functional of the covariance estimates $(\hat\Sigma_0,\hat\Sigma_{\mathrm u},\hat\Sigma_{\mathrm{stk}})$ alone.  Under the shift-up event $\{\widehat{\mathrm{Shift\text{-}up}}=1\}$, $\hat M_2^{\mathrm{adapt}}=\hat M_{\mathrm u}$, and Corollary~\ref{cor:shift-restore} controls precisely the two $M_2$-multiplied design terms.  We import only the two \emph{model-free, design-side} bounds of that proposition --- not its variance conclusion, which is stated there under linear truth with limiting variance $\sigma^2(M_{\mathrm u}\Sigma_{\mathrm{stk}}M_{\mathrm u}^\top)_{jj}$ and is replaced here by the sandwich target $\tau_j^{2,\mathrm{sand}}$ of the misspecified setting.  Its Step~3 bounds the shrinkage residual (Term~A) by $\sqrt{n_0+N}\,|e_j^\top(I_p-\hat M_{\mathrm u}\hat\Sigma_{\mathrm{stk}})(\hat\beta-\beta^\star)|=\op(1)$ (in fact the finer $\tfrac{n_0+N}{\sqrt{n_0}}|\cdot|=\op(1)$) under the sparsity conditions that are \emph{hypotheses of Corollary~\ref{cor:shift-restore}} (in force here because that corollary is invoked), and, by Proposition~\ref{prop:shift-bias-reduction}, controls the imputation-bias term $\tfrac{N}{n_0+N}e_j^\top\hat M_{\mathrm u}\hat\Sigma_{\mathrm u}r(\hat t)$ via Lemma~\ref{lem:bias-transfer} at the same order as the no-shift case, since Assumption~\ref{ass:JM} on the unlabelled block gives $\mnorm{\hat M_{\mathrm u}}=\Op(1)$ exactly as $\mnorm{M_2}=\Op(1)$ does there; this is a purely design-side bound, unaffected by the misspecified truth.  Consequently $\tfrac{N}{n_0+N}\,e_j^\top\hat M_{\mathrm u}\hat\Sigma_{\mathrm u}r(\hat t)$ satisfies~(II) exactly as in the no-shift case.  Off the shift-up event, $\hat M_2^{\mathrm{adapt}}=\hat M_{\mathrm p}$ and the analysis is the pooled construction of Corollary~\ref{cor:CLT-nonlin-labeler} verbatim.  In either case the leading labelled score $\tfrac1{\sqrt{n_0+N}}\sum_{i=1}^{n_0}(x_{0,i}^\top A_n^\top M_2^\top e_j)(\eta(X_{0,i})+\varepsilon_{0,i})$ is a normalised sum of mean-zero terms (Lemma~\ref{lem:misspec-unbiased}) whose sandwich middle factor $\hat\Gamma_0$ of \eqref{eq:Gamma-fold} is determined by the truth $(\eta,\sigma^2(\cdot))$ and the design, never by the model.  The middle factor $\hat\Gamma_0$ of the sandwich variance is therefore model- and $\eta$-invariant, and the plug-in standard error of Remark~\ref{rem:misspec-linear-recovery} is consistent for the realised sandwich variance---the no-shift $\tau_j^{2,\mathrm{sand}}$ off the shift-up event, and a shift-aware sandwich $\tau_{j,\mathrm u}^{2,\mathrm{sand}}$ (with $M_{\mathrm u}A_n$ weights) on it.  Combining Steps a.1--a.3 with the Lindeberg--Feller theorem applied to this array (the array conditions of Theorem~\ref{thm:CLT} hold for the sandwich target $\tau_j^{2,\mathrm{sand}}$ as in the proof of Theorem~\ref{thm:CLT-misspec}) gives
\[
    \sqrt{n_0+N}\,\bigl(\tilde\beta_j-\beta^\star_{\mathrm{proj},j}\bigr)\overset{\mathrm d}{\longrightarrow}\mathcal N\bigl(0,\tau_j^{2,\mathrm{sand}}\bigr),
\]
which is the asserted conclusion.  The two correctives are thus independent in action: the linearisation auxiliary $X_0^{(\mathrm{lin})}\sim P_0$ annihilates the cross-term at the level of (I), while $\hat M_2^{\mathrm{adapt}}$ controls the $\Sigma_{\mathrm u}$-amplification at the level of the design-multiplied terms of Step~a.3.

\medskip\noindent\emph{Part (b): the converse, and the first-order order of the cross-term.}
Suppose instead the linearisation auxiliary $X_{\mathrm u}^{(\mathrm{lin})}\sim P_{\mathrm u}$ is used, with $\Sigma_{\mathrm u}\neq\Sigma_0$.  By \eqref{eq:lin-target-star}--\eqref{eq:lin-target-explicit} with $P_\star=P_{\mathrm u}$, the linearisation now converges to
\[
    \beta^{\mathrm{lin}}_\infty
    \;=\;
    \beta^{\mathrm{proj}}_u[\hat\mu]
    \;=\;
    \beta^\star + \delta_{\mathrm u}[\eta] + \delta_{\mathrm{lin},u}[\hat\mu],
    \qquad
    \delta_{\mathrm u}[\eta]:=\Sigma_{\mathrm u}^{-1}\E_{P_{\mathrm u}}[X\,\eta(X)],
\]
where $\delta_{\mathrm{lin},u}[\hat\mu]$ collects the model's own $P_{\mathrm u}$-projection error (of the controlled order $\rho_{\mathrm{lab}}$).  The exact stacked-target identity \eqref{eq:joint-stk-shift}, which used only the population normal equations and is agnostic to the marginal that generated $\beta^{\mathrm{lin}}_\infty$, now reads
\begin{equation}\label{eq:joint-converse-shift}
    \beta^{\mathrm{stk}}_\infty-\beta^\star
    \;=\;
    \tfrac{N}{n_0+N}\,\bigl(\Sigma_{\mathrm{stk}}^{\mathrm{eff}}\bigr)^{-1}\Sigma_{\mathrm u}\,\bigl(\delta_{\mathrm u}[\eta]+\delta_{\mathrm{lin},u}[\hat\mu]\bigr).
\end{equation}
With the unlabelled fraction $N/(n_0+N)\to\pi_{\mathrm u}\in(0,1]$ one has $(\Sigma_{\mathrm{stk}}^{\mathrm{eff}})^{-1}\Sigma_{\mathrm u}\to\Sigma_{\mathrm{stk},\pi_{\mathrm u}}^{-1}\Sigma_{\mathrm u}$, $\Sigma_{\mathrm{stk},\pi_{\mathrm u}}:=(1-\pi_{\mathrm u})\Sigma_0+\pi_{\mathrm u}\Sigma_{\mathrm u}$, so the offset \eqref{eq:joint-converse-shift} is a non-vanishing deterministic vector whenever $\delta_{\mathrm u}[\eta]\neq0$ (the $\delta_{\mathrm{lin},u}[\hat\mu]$ part is $\op(1)$ and contributes nothing at first order).  Carrying \eqref{eq:joint-converse-shift} through the master decomposition of Step~a.3 in place of (I): the term $(I_p-M_2\hat\Sigma_{\mathrm{stk}})(\hat\beta-\beta^\star)$ no longer vanishes after centring, since $\hat\beta\to\beta^{\mathrm{stk}}_\infty\neq\beta^\star$, and the surviving deterministic contribution to $\sqrt{n_0+N}(\tilde\beta_j-\beta^\star_j)$ is
\[
    \sqrt{n_0+N}\,e_j^\top\bigl(\beta^{\mathrm{stk}}_\infty-\beta^\star\bigr)
    \;=\;
    \sqrt{n_0+N}\cdot\tfrac{N}{n_0+N}\,e_j^\top\bigl(\Sigma_{\mathrm{stk}}^{\mathrm{eff}}\bigr)^{-1}\Sigma_{\mathrm u}\,\delta_{\mathrm u}[\eta]\;+\;\op(1),
\]
which diverges in modulus unless $e_j^\top(\Sigma_{\mathrm{stk}}^{\mathrm{eff}})^{-1}\Sigma_{\mathrm u}\,\delta_{\mathrm u}[\eta]=0$; the JM near-identity $M_2\hat\Sigma_{\mathrm{stk}}\approx I$ does not annihilate it, the offset being a genuine displacement of the population target rather than a debiasing remainder.  Hence $\sqrt{n_0+N}(\tilde\beta_j-\beta^\star_j)$ does not concentrate at $0$ and centres instead at the shift $(N/(n_0+N))(\Sigma_{\mathrm{stk}}^{\mathrm{eff}})^{-1}\Sigma_{\mathrm u}\delta_{\mathrm u}[\eta]$ asserted in the statement, unless $\delta_{\mathrm u}[\eta]=0$.  This is the displayed converse.

It remains to establish the first-order order of $\delta_{\mathrm u}[\eta]$ in the shift, by perturbing $\delta_\star[\eta]=\Sigma_\star^{-1}\E_{P_\star}[X\,\eta(X)]$ around $\Sigma_0$, where $\delta_0[\eta]=0$.  This quantitative claim is the one place at which a regularity hypothesis on $\eta$ beyond $\eta\in L^2(P_0)$ is genuinely required: under $\eta\in L^2(P_0)$ alone the dilation difference $\eta\circ T_E-\eta$ can have $L^2(P_0)$-norm of constant order (a rapidly oscillating $\eta$ acquires an $O(1)$ phase shift once $\norm{E}_{\mathrm{op}}$ exceeds the inverse oscillation scale), so no $\norm{E}_{\mathrm{op}}$-rate is available in the $\norm{\eta}_{L^2(P_0)}$-scale.  We therefore establish the order under the Gaussian-marginal parameterisation $P_\star=N(0,\Sigma_\star)$ of Section~\ref{sec:experiments-misspec} together with the additional regularity that $\eta\in\mathbb D^{1,2}(P_0)$ has $\nabla\eta$ of at most polynomial growth (satisfied by the Hermite, GB, and MLP families of Section~\ref{sec:experiments-misspec}; see \citet[Ch.\ 1]{nourdin2012normal})---which bounds the weighted gradient moment \emph{uniformly over a covariance neighbourhood} of $\Sigma_0$, controlling the tilted marginals below---and obtain the order in the weighted-gradient scale $\mathcal G[\eta]:=\bigl(\E_{P_0}[\norm{\nabla\eta(Z)}_2^2\,\norm Z_2^2]\bigr)^{1/2}<\infty$.

Write $\Sigma_{\mathrm u}=\Sigma_0+E$, $E:=\Sigma_{\mathrm u}-\Sigma_0$, and represent $X\sim P_{\mathrm u}$ as $X=T_E Z$ with $Z\sim N(0,\Sigma_0)\sim P_0$ and
\[
    T_E:=\Sigma_{\mathrm u}^{1/2}\Sigma_0^{-1/2}=I+A,\qquad \norm{A}_{\mathrm{op}}=O(\norm{E}_{\mathrm{op}}),
\]
in operator norm, by Lipschitz continuity of the matrix square root (valid since $\Sigma_0$ has eigenvalues bounded away from $0$ and $\infty$ under Assumption~\ref{ass:design} and $\norm{E}_{\mathrm{op}}$ is small).  By the change of variables $X=T_EZ$,
\begin{align*}
    \E_{P_{\mathrm u}}[X\,\eta(X)]
    &\;=\;
    \E_{P_0}\!\bigl[(T_E Z)\,\eta(T_E Z)\bigr]\\
    &\;=\;
    \underbrace{\E_{P_0}[Z\,\eta(Z)]}_{=\,0}
    +
    A\,\underbrace{\E_{P_0}[Z\,\eta(Z)]}_{=\,0}
    +
    \E_{P_0}\!\bigl[Z\,\bigl(\eta(T_EZ)-\eta(Z)\bigr)\bigr]
    +O(\norm{E}_{\mathrm{op}}^2),
\end{align*}
the first two terms vanishing by the target-marginal orthogonality, leaving the displacement term as the leading contribution.  By Cauchy--Schwarz in $L^2(P_0)$, coordinate-wise,
\[
    \bigl|\E_{P_0}[Z_j\,(\eta(T_EZ)-\eta(Z))]\bigr|
    \;\le\;
    \bigl(\E_{P_0}Z_j^2\bigr)^{1/2}\bigl(\E_{P_0}[(\eta(T_EZ)-\eta(Z))^2]\bigr)^{1/2}.
\]
We bound the dilation difference through the gradient.  Since $\eta\in\mathbb D^{1,2}(P_0)$, the fundamental theorem of calculus along the linear path $s\mapsto(I+sA)Z$, $s\in[0,1]$, gives $\eta(T_EZ)-\eta(Z)=\int_0^1\nabla\eta\bigl((I+sA)Z\bigr)^\top AZ\,\mathrm ds$, whence by Minkowski's integral inequality and Cauchy--Schwarz,
\[
    \bigl(\E_{P_0}[(\eta(T_EZ)-\eta(Z))^2]\bigr)^{1/2}
    \;\le\;
    \norm{A}_{\mathrm{op}}\sup_{s\in[0,1]}\Bigl(\E_{P_0}\bigl[\norm{\nabla\eta\bigl((I+sA)Z\bigr)}_2^2\,\norm{Z}_2^2\bigr]\Bigr)^{1/2}.
\]
For each $s\in[0,1]$ the map $Z\mapsto(I+sA)Z$ pushes $P_0=N(0,\Sigma_0)$ forward to $N(0,(I+sA)\Sigma_0(I+sA)^\top)$, mutually absolutely continuous with $P_0$ with log-density $\tfrac12 Z^\top(\Sigma_0^{-1}-\Sigma_s^{-1})Z$ up to an additive constant, where $\Sigma_s:=(I+sA)\Sigma_0(I+sA)^\top$; since $\norm{\Sigma_0^{-1}-\Sigma_s^{-1}}_{\mathrm{op}}=O(\norm{A}_{\mathrm{op}})$ and $\norm Z_2^2=O(p)$ on the moderate-deviation event $\norm Z_2\le C\sqrt p$, the Radon--Nikodym derivative there is $1+O(p\norm{A}_{\mathrm{op}})$ (for $C$ large the complement of the ball contributes a relative $o(1)$, by the uniform integrability of $\norm{\nabla\eta((I+sA)Z)}_2^2\,\norm Z_2^2$ over $s\in[0,1]$ that the polynomial growth of $\nabla\eta$ supplies, Gaussian polynomial moments being finite uniformly over the tilted covariances $\Sigma_s$); this is the legitimate quasi-invariance reweighting of the \emph{gradient norm}, not of a pointwise difference of $\eta$.  The reweighting is therefore $1+o(1)$ under the dimension-controlled shift $p\,\norm{\Sigma_{\mathrm u}-\Sigma_0}_{\mathrm{op}}\to0$---in particular for $p$ fixed with $\norm{E}_{\mathrm{op}}\to0$---under which $O(p\norm{A}_{\mathrm{op}})=o(1)$; the first-order shift order below is asserted in this regime.  The Proposition's forward centring conclusion does not use this order: it rests on the \emph{exact} identity $\delta_0[\eta]=\Sigma_0^{-1}\E_{P_0}[X\,\eta(X)]=0$, which holds for every shift by the target-marginal orthogonality.  Hence, substituting $W=(I+sA)Z$ (whose Jacobian weight $\norm{(I+sA)^{-1}W}_2^2=(1+O(\norm A_{\mathrm{op}}))\norm W_2^2$ is absorbed into the $1+O(p\norm A_{\mathrm{op}})$ factor), $\sup_{s}\E_{P_0}[\norm{\nabla\eta((I+sA)Z)}_2^2\,\norm Z_2^2]=(1+o(1))\,\E_{P_0}[\norm{\nabla\eta(Z)}_2^2\,\norm Z_2^2]=(1+o(1))\,\mathcal G[\eta]^2$.  Combining the three displays,
\[
    \E_{P_{\mathrm u}}[X\eta(X)]=O\!\bigl(\norm{E}_{\mathrm{op}}\,\mathcal G[\eta]\bigr),
\]
and since $\Sigma_{\mathrm u}^{-1}=\Sigma_0^{-1}+O(\norm{E}_{\mathrm{op}})$ has bounded operator norm,
\[
    \delta_{\mathrm u}[\eta]
    \;=\;
    \Sigma_{\mathrm u}^{-1}\,\E_{P_{\mathrm u}}[X\,\eta(X)]
    \;=\;
    O\!\bigl(\norm{\Sigma_{\mathrm u}-\Sigma_0}_{\mathrm{op}}\,\mathcal G[\eta]\bigr),
\]
the asserted first-order order, in the weighted-gradient scale $\mathcal G[\eta]$.  Two limiting cases corroborate it: if $\eta$ lies in a Hermite chaos $\mathcal H_n$, $n\ge2$, that is $\Sigma$-orthogonal to the linear span uniformly over centred Gaussian marginals with standardised second moments (the footnote construction), the displacement term vanishes identically and $\delta_{\mathrm u}[\eta]\equiv0$ for \emph{all} $\Sigma_{\mathrm u}$, consistent with the $O(\cdot)$ bound; and at $\Sigma_{\mathrm u}=\Sigma_0$ the bound returns $\delta_0[\eta]=0$.  Substituting this order into \eqref{eq:joint-converse-shift} shows that the converse shift is itself $O(\norm{\Sigma_{\mathrm u}-\Sigma_0}_{\mathrm{op}}\,\mathcal G[\eta])$, vanishing continuously as the shift is switched off, the quantitative form of the second display of the Proposition.  The oscillation example above shows only that the \emph{dilation-difference route} of this proof cannot deliver an $\norm{E}_{\mathrm{op}}$-rate in the $\norm{\eta}_{L^2(P_0)}$-scale; the $\mathbb D^{1,2}(P_0)$ regularity---with $\nabla\eta$ of polynomial growth, so that $\mathcal G[\eta]<\infty$ uniformly over the tilted marginals---is what licenses the route used here.  Whether a model-free $L^2$-only order holds by a different argument is left open, as noted beneath \eqref{eq:lin-target-explicit}.
\end{proof}

\section{Simulation protocols}\label{app:simdetails}

\paragraph{Common design.}
The labelled-target sample of total size $800$ is split into a tuning subsample of size $n_{\mathrm{tun}} = 400$ used to select the shrinkage level $\hat t$ and an inference subsample of size $n_0 = 400$.  The unlabelled-pool size is determined per cell by the variance-balance rule of Section~\ref{sec:vbrule} on the candidate grid $\mathcal N = \{50, 75, 100, 200, 300, \ldots, 1000\}$.  An auxiliary pool of $3000$ additional unlabelled rows is reserved for nodewise-Lasso precision estimation.  Each cell is replicated $R=100$ times with replication seeds derived deterministically from a master seed, ensuring exact reproducibility.\footnote{Reproduction code, the random seed schedule, and version-pinned package requirements (Python 3.11, NumPy 1.26.4, scikit-learn 1.3.0) are available from the authors upon reasonable request.}

\paragraph{Power adaptivity.}
{The power-adaptivity experiment ranges the external sample size $n_A$ over the grid}
\[
    n_A \in \{100, 200, 300, 400, 600, 800, 1200, 1600, 2000, 2400\}.
\]

\paragraph{Misspecification.}
The linear-oracle-predictor experiment (Figure~\ref{fig:misspec-linear-oracle}) reports $\alpha\in\{0.0,\allowbreak 0.25,\allowbreak 0.5,\allowbreak 1.0,\allowbreak 2.0\}$; the anchored-model experiment (Figure~\ref{fig:misspec-anchor}) sets $n_A = 1600$ and reports $\alpha\in\{0.0,\allowbreak 0.3,\allowbreak 0.6,\allowbreak 0.9\}$.  Both select $N$ per replication by the variance-balance rule, with $R=100$ replications per cell.

\paragraph{Covariate shift.} Each cell of this study is replicated $R=20$ times.
For each cell the variance-balance rule of Section~\ref{sec:vbrule} chooses $\hat N^*$.  The bias surrogate $b_n$ is evaluated at the data-driven plug-in $\hat a_1$ of Section~\ref{sec:vbcap}; no population quantity enters the selection.

Within the AR(1) family, the population covariance is parameterised by a single scalar $\rho$, and the family-free amplification-factor detector \eqref{eq:shift-detector} of Section~\ref{sec:shift-aware} reduces to a one-dimensional comparison along the first super-diagonal.  We therefore implement $\widehat{\mathrm{Shift\text{-}up}}$ in this experiment by the equivalent simplification
\begin{equation}\label{eq:shift-detector-AR1}
    \widehat{\mathrm{Shift\text{-}up}}_{\mathrm{AR(1)}} \;:=\; \1\!\Bigl\{\bar r_{\mathrm u} - \bar r_0 \;>\; \tilde c_n\Bigr\},
    \qquad
    \tilde c_n \asymp \sqrt{(\log p)/(n_0\wedge N)},
\end{equation}
where $\bar r_0$ and $\bar r_{\mathrm u}$ are the means of the first super-diagonal entries of the sample correlation matrices on the labelled and unlabelled blocks respectively.  This is the AR(1)-specific one-dimensional test: within the equal-variance family the general amplification event $\Sigma_{\mathrm u}\succ\Sigma_0$ cannot occur, and its role is taken by the correlation ordering $\rho_{\mathrm u}>\rho_0$, which drives the row-$\ell_1$ amplification $\rho_j(M_2,\Sigma_{\mathrm u})>1$.

\paragraph{Control-variate hybrid.}\label{sec:experiments-hybrid}

We extend the power-adaptivity study to the control-variate hybrid of Section~\ref{sec:hybrid} (Section~\ref{app:hybrid}), which recovers the model's design-orthogonal signal $\rho_j^2[\hat\mu]$ that the pure procedure discards. The design of Section~\ref{sec:experiments-power} is retained unchanged, with two additions: the model carries a design-orthogonal component $\gamma W$, with $W\perp X$ and total model-residual variance normalised to one, so that the design-orthogonal signal fraction is $\psi^2=\gamma^2=\rho_j^2[\hat\mu]$; and the labelled response is control-variated by an externally estimated weight $\hat\omega$. No oracle information enters, and $N$ is chosen by the same variance-balance rule of Section~\ref{sec:vbrule}. We report three model regimes, $\psi^2\in\{0,0.4,0.8\}$; at $\psi^2=0$ the hybrid reduces to DEAL.

Efficiency results for this experiment are reported in Section~\ref{sec:experiments-hybrid-main} of the main text (Figure~\ref{fig:aux-hybrid}).

DEAL and the hybrid cover at least $0.92$ at every cell, and mostly between $0.94$ and $1.00$, coverage rising with $\psi^2$ as $\hat N^*$ contracts and less imputation bias remains for the robust variance to absorb; PPI++, at its native rectifier standard error, covers between $0.90$ and $0.97$. The hybrid inherits DEAL's validity because the control variate is mean-zero against the design.

\subsection{The joint shift-and-misspecification experiment}\label{app:joint-experiment}

We exercise the joint regime of Section~\ref{sec:misspec-joint}, in which the unlabelled covariates are drawn from a shifted marginal $\tilde X\sim N(0,\Sigma_{\mathrm u})$, $\Sigma_{\mathrm u}\neq\Sigma_0$, and the conditional mean is non-linear under the three $\eta$ specifications introduced above: Hermite (H), GB-shaped (GB), and MLP-shaped (MLP).  The DEAL pipeline combines the shift-aware substitution $\hat M_2\leftarrow\hat M_2^{\mathrm{adapt}}$ of Section~\ref{sec:shift-aware} with the target-marginal linearisation of Proposition~\ref{prop:joint-lin}: the auxiliary covariates $X_0^{(\mathrm{lin})}$ are an independent draw of size $n_{\mathrm{lin}} = 10^4$ from the target marginal $P_0$, and the linearisation step uses ordinary least squares as discussed beneath \eqref{eq:lin-lasso} (in this experiment $n_{\mathrm{lin}}/p\approx 83$, so OLS is preferable to the Lasso variant).  We sweep $\rho_{\mathrm u}\in\{0.0, 0.2, 0.4, 0.6, 0.8\}$ at fixed target $\rho_0 = 0.4$, with $\alpha=0.6$, $n_0 = n_{\mathrm{tun}} = 400$, $N=400$, and $R = 100$ replications per cell.

\begin{figure}[t]
\centering
\includegraphics[width=\textwidth]{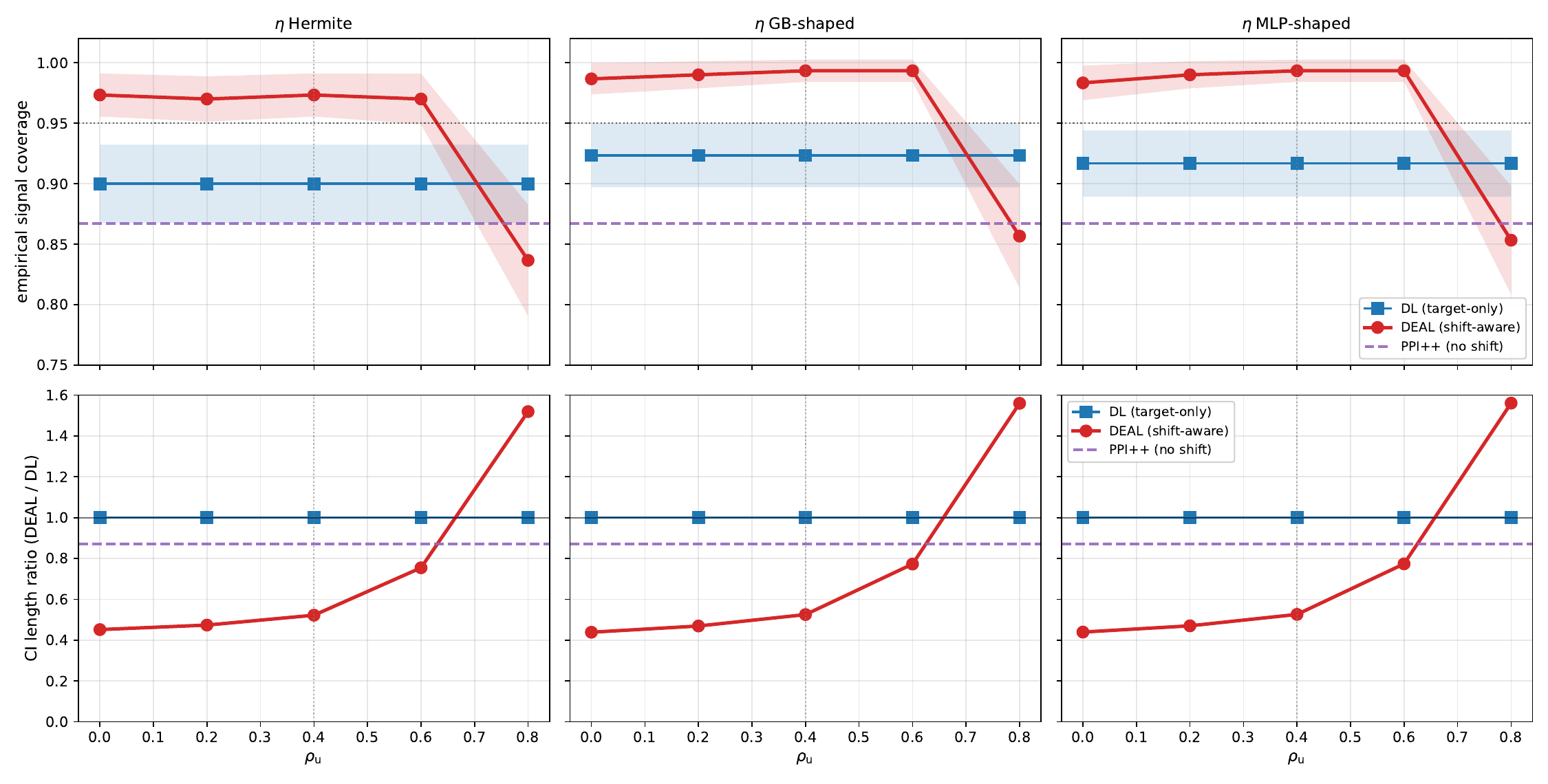}
\caption{\textit{DEAL inference under joint covariate shift and model misspecification.}  Top row: empirical signal coverage versus the unlabelled-design AR(1) parameter $\rho_{\mathrm u}$.  Bottom row: CI-length ratio of DEAL to DL on $J_{\mathrm{signal}}$.  Columns correspond to the three $\eta$ specifications: Hermite (left), GB-shaped (centre), and MLP-shaped (right).  Vertical dotted line marks the no-shift cell $\rho_{\mathrm u} = \rho_0 = 0.4$.  The dashed purple lines mark the no-covariate-shift PPI++ reference---empirical signal coverage in the top row and PPI++-to-DL CI ratio in the bottom row, both $0.87$; PPI++ has no shift-aware variant.  DEAL combines the shift-aware Stage 4 substitution of Section~\ref{sec:shift-aware} with the linearisation of Proposition~\ref{prop:joint-lin}.  $n_0 = n_{\mathrm{tun}} = 400$, $N = 400$, $\alpha = 0.6$, $n_{\mathrm{lin}} = 10^4$, $R = 100$.}
\label{fig:misspec-joint}
\end{figure}

Across the no-shift and moderate-shift cells $\rho_{\mathrm u} \in \{0.0, 0.2, 0.4, 0.6\}$ (Figure~\ref{fig:misspec-joint}), DEAL signal coverage is near nominal ($0.97$--$1.00$) for all three $\eta$ specifications and the DEAL-to-DL CI ratio is $0.44$--$0.78$, shorter than the no-shift PPI++ baseline ($0.87$) until $\rho_{\mathrm u}\approx0.65$, delivering the reduction of Corollary~\ref{cor:deal-vs-dl-nonlin}.  At the extreme cell $\rho_{\mathrm u} = 0.8$ (the boundary of the covariate-shift envelope) coverage drops to $0.84$--$0.86$ and the CI ratio inflates to $1.5$--$1.6$: the shift-aware variant self-protectively widens once its augmentation is no longer beneficial, the failure mode anticipated in Section~\ref{sec:shift-aware}.  The near-identical Hermite, GB-shaped, and MLP-shaped curves across the $\rho_{\mathrm u}$ envelope confirm the population identification of Proposition~\ref{prop:joint-lin}: the target-marginal linearisation eliminates the cross-term $\delta_{\mathrm u}[\eta]$ that would otherwise separate the truth families under joint perturbation.

\subsection{Coverage on the null coordinates}\label{app:null-coverage}
\begin{figure}[t]
\centering
\includegraphics[width=\textwidth]{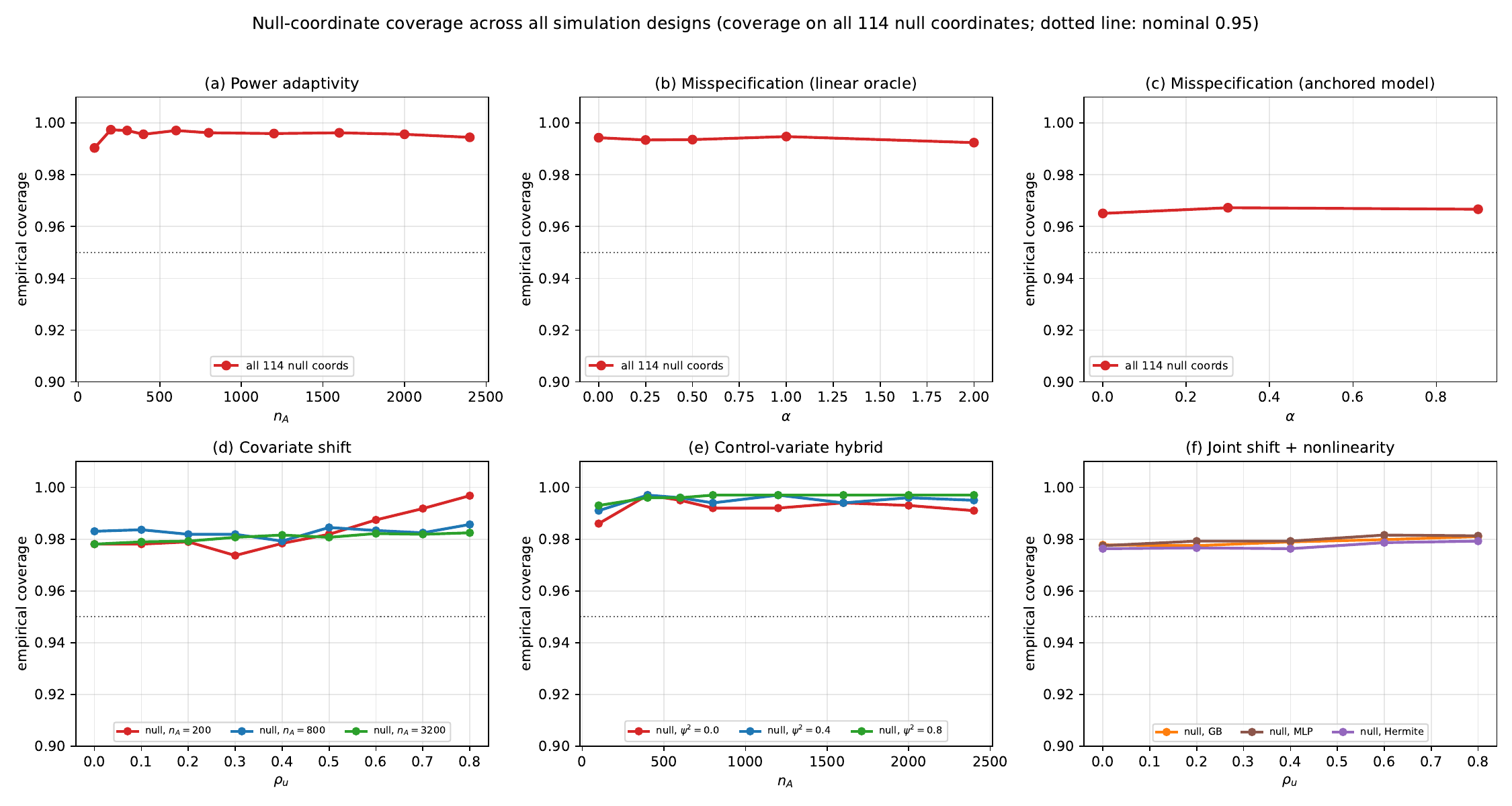}
\caption{{\textit{Empirical coverage on the null coordinates across all simulation designs.}  Each panel plots coverage against the design's sweep variable: (a)~power adaptivity against $n_A$; (b,~c)~misspecification against $\alpha$; (d)~covariate shift against $\rho_{\mathrm u}$, at three external-estimator qualities; (e)~the control-variate hybrid against $n_A$, at three design-orthogonal signal fractions $\psi^2$; (f)~the joint shift-and-non-linearity experiment against $\rho_{\mathrm u}$, for three truth families.  Curves are coverage on all $p-s = 114$ null coordinates; the dotted line marks the nominal $0.95$.  Null coverage is at or above nominal throughout.}}
\label{fig:null-coverage}
\end{figure}

{
The coverage panels of Section~\ref{sec:experiments} report empirical coverage on the signal coordinates, the binding case for inference.  For completeness, Figure~\ref{fig:null-coverage} reports empirical coverage on the null coordinates ($\beta^\star_j = 0$) across all six simulation designs, evaluated on the full complement of $p-s = 114$ null coordinates rather than a subsample.  In every design and at every point of each sweep, null coverage lies at or above the nominal $0.95$, and never below $0.965$: at a true null there is no imputation bias to correct, so the debiased interval is variance-dominated and covers conservatively, most markedly in the power-adaptivity, linear-oracle and hybrid designs (panels~a,~b and~e), where it stays near unity.  The null coordinates therefore carry no coverage risk.}

\section{Real-data protocols}\label{sec:realdata-supp}

\subsection{Galaxy morphology}\label{subsec:galaxy-supp}

\paragraph{Cohort and endpoint.} The inferential triple is assembled from Galaxy Zoo DECaLS volunteer aggregates released alongside \citet{walmsley2022zoobot}.  JPEG cutouts are retrieved from the Legacy Survey cutout service at a fixed pixel scale, normalised, and partitioned into three disjoint cohorts: a model-training (head-only fine-tuning) cohort of $449$ galaxies, an inferential cohort of $1352$ galaxies with paired $(X_i, y_i)$, and an unlabelled cohort of $n_u = 2151$ galaxies retaining only $X$.  Each partition is index-disjoint by survey identifier, eliminating contamination between the model-training cohort and the inferential cohort.  The model-training cohort is the realised count of $449$ after Legacy Survey cutout-availability filtering against an initial request of $800$ galaxies; the inferential and unlabelled cohorts are drawn independently of that filtering, so the realised count bears on the training cohort alone.  The response $y$ is the vote fraction for the spiral-arm question, taken as a continuous variable on $[0,1]$ following the convention in galaxy morphology inference \citep{bamford2009galaxyzoo}.

\paragraph{Feature panel.} The covariate vector $X_i \in \R^{640}$ is the standardised activation from the penultimate layer of the frozen Zoobot ConvNeXt-nano encoder \citep{walmsley2022zoobot}.  The encoder is used as released with its weights held fixed, fine-tuning being confined to the head, so no encoder parameter is fitted on any of the three cohorts.

\paragraph{External model.} The model $\hat\mu$ is a FinetuneableZoobotTree head with canonical architecture (Dropout, Linear, ScaledSigmoid, Dirichlet $\alpha$ parameterisation), trained on the $449$-galaxy head-training cohort using AdamW with learning rate $3 \times 10^{-3}$, weight decay $5 \times 10^{-4}$, dropout $0.5$, batch size $64$, and the Dirichlet negative-log-probability loss of \citet{walmsley2022zoobot}, for $300$ epochs.  The resulting predictor attains $\mathrm{corr}(\hat\mu, y) = 0.65$ on the inferential cohort.

\paragraph{Unlabelled cohort.} The unlabelled cohort is drawn from the same Galaxy Zoo DECaLS release as the inferential cohort and passed through the same frozen encoder, so the two blocks carry covariates on a common panel; each of its galaxies retains only $X$ and carries the value of $\hat\mu$ evaluated on those activations.

\paragraph{Estimator specification.} Because $\hat\mu$ is trained under a Dirichlet likelihood with a scaled-sigmoid output rather than under an OLS criterion, its scale is not calibrated to the linear projection of $y$ onto $X$.  A single $\gamma$-rescaling supplies that calibration at the model boundary, $\hat\mu \mapsto \hat\gamma\, \hat\mu$ with $\hat\gamma = \widehat{\mathrm{cov}}(\hat\mu, y) / \widehat{\mathrm{var}}(\hat\mu)$.  The rescaling preserves the conditional residual structure required for the asymptotic-dominance argument over DL while neutralising the scale gap induced by the non-OLS training criterion.  The sampling variability of $\hat\gamma$ is a standard $\Op(n_0^{-1/2})$ estimation effect that does not enter the reported bias correction on this demonstration, where the data-driven shrinkage returns $\hat t=0$; $\hat\gamma$ is held fixed across the two-block tuning split, so no further data-driven fluctuation enters the selection of $\hat t$.  The rescaling is applied uniformly across DEAL, PPI, and PPI++ so that benchmark comparisons are baseline-tuned at parity.

The point estimator $\hat\beta$ is the standard Lasso on the residualised stacked design; the debiased estimator $\tilde\beta$ employs nodewise-Lasso \citep{vandegeer2014asymptotic} with regularisation constant $c_{\lambda,\mathrm{nw}} = 1.0$, the moderate-$p$ default.  The variance estimator is the stacked-residual sandwich, combining target-cohort residuals with unlabelled-block residuals to capture the contribution of the imputation step.

Selection of $\hat t$ uses an internal split of the inferential cohort of $1352$ galaxies into an inferential block of $1014$ galaxies and a tuning block of $338$ galaxies, so that $\hat t$ is estimated without leakage into the final inferential coordinates.  The two-block tuning split returns $\hat t = 0$ uniformly across the $640$ coordinates: the orthogonal predictor residual is correlated with $y$, but not in the direction that the $\tilde\beta = \hat\beta + \hat t \cdot C$ correction can exploit at the deployment regime $p \approx n_0$.

\paragraph{Side evidence.} Whether the additional coordinates declared significant by DEAL but not by DL carry genuine signal is assessed by held-out sign concordance.  Of the $177$ coordinates declared significant by DEAL and the $188$ by DL, $70$ are common, so the DEAL-extras-over-DL set has cardinality $107$ and the DL-extras-over-DEAL set $118$.  The inferential cohort is split $B = 20$ times at random in the proportion $80/20$; on each split, a univariate OLS refit of $y$ on each coordinate is computed on the held-out $20\%$, and the sign of the refit coefficient is compared with the sign of the DEAL debiased coefficient on the corresponding training $80\%$.  The univariate sign-concordance on the DEAL-extras set is $0.65$, with $p < 10^{-3}$ under the null of concordance $0.5$ via an exact binomial calibration on the $B \cdot 107$ comparisons, pooled as independent trials.

The same splits support an out-of-sample predictive assessment: an OLS refit on the union of each procedure's significant coordinates is evaluated on the held-out $20\%$ portion of each split.  The median $R^2_{\text{test}}$ values are $0.42$ for DEAL, $0.32$ for PPI++, $0.31$ for DL, and $0.17$ for PPI, the last reflecting severe overfitting traceable to a median $\beta$-inflation factor of $1.74$ ($90$th percentile $10.4$).

\paragraph{No-information check.} Under the no-information surrogate of Section~\ref{subsec:realdata-results} (each $\hat\mu$ replaced by a uniform permutation of its outputs), the galaxy CI ratio settles at $1.17$ rather than returning to unity, and the significance count is $160$ against DL's $188$---the lone portfolio exception, in the conservative direction on both counts.  The mechanism is specific to the regime $p\approx n_0$: the variance-balance rule subsamples the unlabelled cohort to a small budget ($\hat N^*=10$), at which the pooled-precision inversion does not fully neutralise the residual no-information unlabelled block, an effect that varies appreciably across random seeds.
\subsection{Inorganic band gap}\label{subsec:bandgap-supp}

\paragraph{Cohort and endpoint.} The labelled triple is constructed from the MatBench \texttt{matbench\_mp\_gap} snapshot \citep{dunn2020benchmarking}, whose raw release contains $4604$ compositions with experimentally measured gaps.  The model $\hat\mu$ is required to be defined on every element of the target sample; the requirement is met by retaining the $n_0 = 3347$ compounds for which a PBE band-gap calculation is available in the Materials Project.  Matching is by reduced formula, computed with \texttt{pymatgen}'s \texttt{Composition.reduced\_formula} so that polymorphs collapse to a single composition; when multiple Materials Project entries share a reduced formula, the entry of lowest formation energy per atom is retained.

\paragraph{Feature panel.} The design $X$ uses a Goldschmidt--Pauling pair-feature panel.  The top two hundred element pairs are selected by frequency of co-occurrence across the $74{,}817$ catalogued Materials Project compositions; each pair $(i,j)$ enters under three weightings constructed from atomic fractions $f_i$, Pauling electronegativities $\chi_i$, and Shannon radii $r_i$.  The Pauling ionic-interaction weighting $(f_i f_j)(\chi_i - \chi_j)^2/4$ captures ionic bond strength, the Goldschmidt size-asymmetry weighting $(f_i + f_j)\min(r_i,r_j)/\max(r_i,r_j)$ captures geometric mismatch, and the dominant-element ionicity weighting $\max(f_i, f_j)\{1 - \exp[-(\chi_i - \chi_j)^2/4]\}$ captures the polarising influence of the majority constituent.  Eight pair-aggregate scalars (mean and dispersion of the three weightings, alongside coordination-count summaries) and twenty-two Magpie elemental-statistic means complete the design at $p = 630$.  The panel is constructed at this width because the nodewise estimator requires a design whose empirical Gram matrix is numerically of full rank: Magpie-only panels at $p = 22$ and $p = 100$ are structurally rank-deficient, with $\mathrm{cond}(X^\top X)$ of order $10^{13}$, whereas the $p = 630$ panel lies within the operating envelope of the nodewise estimator.

\paragraph{External model.} The model is a fixed physical quantity rather than a fitted predictor: $\hat\mu_i$ is the PBE band gap from the Materials Project entry matched to composition $i$.  The PBE calculation~\citep{perdew1996generalized,perdew1983density} is a calibrated computational procedure rather than a statistical fit on $y_0$, though it systematically underestimates experimental gaps in a chemically varying manner \citep{janesko2014beyond}.  It introduces no fit-leakage, and $\hat\mu$ may be applied to the labelled and unlabelled covariates without sample splitting.

\paragraph{Unlabelled cohort.} The unlabelled cohort comprises $n_u = 10{,}000$ compositions drawn uniformly from the $74{,}817$-entry Materials Project catalogue, each carrying its matched PBE model value.

\paragraph{Estimator specification.} DEAL is run with the shrinkage selector of Section~\ref{sec:proxy-shrinkage} and the stacked-residual sandwich variance estimator.  The $\gamma$-rescale at the model boundary is applied to align the model residual scale with the response residual scale.  The nodewise-Lasso regularisation is $c_{\lambda,\mathrm{nw}} = 2.0$, required by the elevated condition number of the panel; the selector returns $\hat t = 0$ at this regime, in agreement with the algebraic prediction that the $X$-orthogonal model residual lies outside the row span of the correction $C$.

\paragraph{Side evidence.} The hold-out generalisation protocol uses $B = 20$ random eighty-twenty splits.  On each split a DEAL analysis is performed on the training portion to obtain a discovery set $S \subseteq \{1, \dots, p\}$, an unpenalised ordinary-least-squares refit is performed on the $|S|$ retained columns, and the coefficient of determination is evaluated on the held-out twenty percent.  With typical discovery-set size $|S| \approx 143$ and test-fold size $n_{\text{test}} \approx 670$, the operative ratio is $n_{\text{test}}/|S| \approx 4.7$, adequate for a stable OLS refit.  The refit is left unpenalised because the diagnostic requires resolution between generalising and non-generalising discovery sets: a ridge refit at $\lambda = |S|$ shrinks signal and noise uniformly, masking the discovery-quality differential and compressing $R^2_{\text{test}}$ for all procedures towards a common positive range.  The resulting median $R^2_{\text{test}}$ is $0.47$ for DEAL, against strongly negative held-out $R^2$ for both DL and PPI++, confirming that only DEAL's discovery set generalises out of sample.
\subsection{Breast cancer neoadjuvant chemoresponse}\label{subsec:brca-supp}

\paragraph{Cohort and endpoint.} The labelled sample $(y_0, X)$ is drawn from GSE25066 \citep{hatzis2011genomic}, restricted to the $226$ patients with a non-missing residual cancer burden (RCB) class assignment~\citep{symmans2017long} and concordant transcriptomic preprocessing.  The response $y_0$ is the ordinal RCB class encoded as $0, 1, 2, 3$ for pathological complete response (pCR, i.e.\ RCB-0), RCB-I, RCB-II and RCB-III respectively.  The ordinal coding is preferred to the binary pCR indicator because it carries strictly more information for the linear-regression target estimand at the same sample size, while preserving the clinically meaningful direction of effect (lower~$y_0$~$=$~better treatment response).

\paragraph{Feature panel.} Expression is log-transformed, quantile-normalised within each cohort, and intersected at the gene-symbol level with the Cancer Cell Line Encyclopedia expression matrix.  The $p = 100$-dimensional panel is obtained by first restricting to the union of the top-$1500$ Cancer Cell Line Encyclopedia genes (by variance) and the gene-symbol intersection, then ranking those genes by absolute correlation with paclitaxel half-maximal inhibitory concentration (IC50) within Cancer Cell Line Encyclopedia alone and retaining the top $100$.  Because the screen depends only on Cancer Cell Line Encyclopedia quantities, it is cohort-independent of GSE25066 and induces no selection bias in the downstream inference on the patient cohort.

\paragraph{External model.} The external model $\hat\mu$ is a cross-validated ridge regression of paclitaxel IC50 on the $p = 1500$-variance-screened gene panel, fit on Cancer Cell Line Encyclopedia cell lines using $\textsc{sklearn.RidgeCV}$ with $n_{\mathrm{train}} = 503$ training rows and a five-fold cross-validation grid for the ridge penalty; the resulting predictor attains a held-out within-Cancer Cell Line Encyclopedia correlation of $0.58$.  Transferred to GSE25066 and restricted to the $p = 100$ panel for the inferential step, the patient-level correlation falls to $\mathrm{corr}(\hat\mu, y) = 0.23$, in line with the published track record of cell-line-to-patient drug-response transfer.

The model residual enters the variance of the DEAL estimator, so the predictor is required to be one that does not over-fit its training cohort.  A multilayer perceptron ($1500 \to 512 \to 128 \to 1$, AdamW with early stopping) fails this requirement under the relatively narrow paclitaxel-response distribution of the Cancer Cell Line Encyclopedia panel: it over-fits the Cancer Cell Line Encyclopedia training cohort, so its residual variance enters the DEAL variance without a compensating reduction in imputation bias.  The cross-validated ridge surrogate meets it.

\paragraph{Unlabelled cohort.} The unlabelled cohort consists of $n_u = 1097$ primary-tumour RNA-seq profiles of The Cancer Genome Atlas breast-cancer cohort (TCGA-BRCA).  The two cohorts share disease site and tissue origin, so the design-side alignment condition $\E_{\mathrm{pool}}[XX^\top] \approx \E_{\mathrm{target}}[XX^\top]$ is plausible by construction.

\paragraph{Estimator specification.} DEAL is run with $\gamma$-rescaling at the model boundary, nodewise-Lasso regularisation $c_{\lambda,\mathrm{nw}} = 0.5$, the shrinkage selector of Section~\ref{sec:proxy-shrinkage} with the stacked-residual sandwich, and a tuning fraction of $0.25$.  The data-driven shrinkage returns $\hat t = \hat B = 0$.

The nodewise-Lasso constant follows the regime rule of Section~\ref{subsec:realdata-design}: with $n_{\mathrm{pool}}/p \approx 13$, where $n_{\mathrm{pool}} = n_0 + n_u$, this demonstration uses $c_{\lambda,\mathrm{nw}} = 0.5$.

\paragraph{Side evidence.} Sign-concordance is computed on the $n = 33$ coordinates returned by DEAL but not by DL.  For each of $B = 20$ random partitions of GSE25066, a held-out OLS coefficient is computed coordinate-wise on the test fold, and the sign of the DEAL point estimate is compared to its sign; the resulting concordance is $0.67$, exceeding the binomial $0.5$ null at $p < 0.05$.
\subsection{Patient-derived xenograft drug response}\label{subsec:pdxe-supp}

\paragraph{Cohort and endpoint.} The patient-derived xenograft encyclopedia (PDXE) \citep{gao2015high} reports tumour-volume trajectories for a panel of patient-derived xenografts (PDXs) under multiple oncology agents.  For each xenograft, the best-average response (BAR) records the minimum-over-time of the time-averaged percent change in tumour volume from baseline.  The target endpoint is BAR for the subset of PDXs treated with alpelisib (BYL719)~\citep{andre2019alpelisib}; after harmonisation with the Cancer Cell Line Encyclopedia expression panel and removal of xenografts lacking matched RNA-seq, $n_0 = 105$ labelled rows remain.  The response is rescaled as $y = \mathrm{BAR}/100$ to place it on a unit-interpretable scale.  Expression features are voom-normalised log counts-per-million (CPM) values, batch-corrected against the Cancer Cell Line Encyclopedia reference panel using ComBat, so that a single ridge model maps across the two platforms.

\paragraph{Feature panel.} The covariate dimension is held at $p = 30$, which keeps the labelled-block design well-conditioned at $n_0 = 105$.  Univariate Pearson correlations between each transcript and response are computed on an independent set of alpelisib-treated Cancer Cell Line Encyclopedia cell lines, disjoint from PDXE, and the $30$ transcripts with the largest $|\mathrm{corr}|$ are retained.  The screen uses only the Cancer Cell Line Encyclopedia side of the data, so the selection is independent of every PDXE sample.

\paragraph{External model.} The model $\hat\mu$ is a cross-validated ridge regression of alpelisib IC50 on the same alpelisib-treated Cancer Cell Line Encyclopedia panel used for feature screening, fit with $\textsc{sklearn.RidgeCV}$ and a five-fold cross-validation grid for the ridge penalty.  Applied to PDXE samples, it has empirical correlation $\mathrm{corr}(\hat\mu, y) = 0.14$: the magnitude of genuine cell-line-to-PDX transfer rather than of an idealised in-domain regressor, so the demonstration exercises the pooled-precision-lever discovery channel.  A linear surrogate is adopted for the reason recorded on the breast-cancer demonstration (Section~\ref{subsec:brca-supp}): a multilayer-perceptron model over-fits the Cancer Cell Line Encyclopedia training cohort.

\paragraph{Unlabelled cohort.} The pooled debiasing matrix $M_2$ approximates the inverse of the stacked-cohort second-moment matrix $\hat\Sigma_{\mathrm{stk}}$, so its consistency for the target precision $\Omega=\Sigma^{-1}$ requires that the labelled and unlabelled covariate distributions share the same population $\Sigma$.  The unlabelled cohort is accordingly composed of the $n_u = 259$ PDXE-internal non-BYL719 xenografts, which are the same biological substrate (xenografted tumour tissue) as the labelled cohort and differ from it only in the drug administered; this is the configuration reported in Section~\ref{subsec:realdata-demos}.  The requirement excludes the larger alternative of TCGA pan-cancer primary tumours ($n_u = 2170$): primary tumour tissue and xenografted tumour tissue do not share a population second-moment matrix at the transcript level (different stromal composition, different selection pressure, different normalisation references), so under that pool $M_2$ is inconsistent for $\Omega$, the confidence intervals computed under $M_2$ understate the true sampling variance, and inference around noise coordinates is anti-conservative.  Same-distributional-context alignment is thus secured at the cost of an order of magnitude in $n_u$.

\paragraph{Estimator specification.} DEAL is run with $\gamma$-rescaling applied at the model boundary, nodewise-Lasso regularisation $c_{\lambda,\mathrm{nw}} = 0.5$, the shrinkage selector of Section~\ref{sec:proxy-shrinkage}, and the stacked-residual sandwich variance estimator.  The shrinkage diagnostic returns $\hat t = \hat B = 0$.

\paragraph{Side evidence.} The principal side-evidence diagnostic is the diff-set sign-concordance test: for each coordinate in the diff-set $\mathcal{D} = \mathcal{S}_{\mathrm{DEAL}} \setminus \mathcal{S}_{\mathrm{DL}}$, the sign of $\tilde\beta_j$ from the full DEAL fit is compared against the sign of $\hat\beta_j$ from an OLS refit on a held-out half of PDXE, repeated over $B = 20$ random splits.  A supplementary $R^2_{\text{test}}$ comparison on the union discovery set is reported alongside it.

\paragraph{Cohort-alignment diagnostic.} The sign-concordance test also registers the second-moment requirement on the unlabelled cohort.  Under the $259$ PDXE-internal non-BYL719 xenografts the median confidence-interval ratio is $0.246$ and the diff-set sign-concordance is $0.75$ univariate, confirming that the discoveries replicate.  Under the TCGA pan-cancer pool the median confidence-interval ratio falls to $0.18$ and $28$ of $30$ coordinates clear the significance threshold, while the diff-set sign-concordance against held-out PDXE rows over $B = 20$ splits is $0.49$, statistically indistinguishable from random.
\subsection{Pan-cancer drug response with a large-language-model oracle}\label{subsec:llm-oracle-supp}

\paragraph{Cohort and endpoint.}  Drug-response data are drawn from the Genomics of Drug Sensitivity in Cancer release 2 (GDSC2), restricted to selumetinib (AZD6244/ARRY-142886).  Among the $1{,}666$ GDSC2 cell-line-by-selumetinib measurements, $538$ are matched to cell lines in the harmonised Cancer Cell Line Encyclopedia expression panel (after dropping the GDSC-tagged \texttt{UNCLASSIFIED} tissue label) and constitute the labelled cohort.  The response $y$ is the area-under-the-dose-response-curve (AUC) on the unit scale.  The covariate matrix is voom-normalised log-RNA-seq expression on the same harmonised Cancer Cell Line Encyclopedia panel, identical to the covariate processing used for the breast-cancer and patient-derived-xenograft demonstrations.

\paragraph{Feature panel.}  The dimension is fixed at $p = 80$.  Genes are ranked by sample variance on the labelled cohort, and the top $80$ are retained.  The ranking uses only the labelled-cohort covariates, with no response information, and is performed before any large-language-model (LLM) call, so the panel is independent of the model.

\paragraph{External model.}  The model $\hat\mu$ is Claude Opus~4.7 (Anthropic) accessed via the standard application programming interface at temperature $T = 0$.  For each cell line, the model is prompted with a one-line specification of the form \texttt{cell\_line=``<NAME>'', tissue=``<TCGA\_DESCRIPTOR>''}; the system prompt provides mitogen-activated protein kinase pathway sensitivity priors (BRAF V600E melanoma and colorectal cancer, NRAS-mutant melanoma, KRAS-mutant carcinoma, NF1-loss tumours) and a five-tier response convention calibrated against published clinical pharmacology.  The model returns a single floating-point AUC prediction per cell line and is applied without fine-tuning, so its output rests on pretraining priors alone.  On the labelled cohort, the empirical correlation is $\mathrm{corr}(\hat\mu, y) = 0.48$ and the Spearman rank correlation is $0.39$ ($p < 10^{-19}$).  After regressing $y$ on the covariate panel $X$ alone, the residual retains empirical correlation $0.20$ with $\hat\mu$, confirming that the LLM contributes information beyond the covariate panel.  As in the breast-cancer and patient-derived-xenograft demonstrations, the model is free of fit-leakage: not having been trained on the inferential cohort's $(X, y)$ tuples, it is independent of the labelled sample, as the asymptotic-coverage statement requires.  Pretraining memorisation of the published responses would weaken that independence; the realised $\mathrm{corr}(\hat\mu, y) = 0.48$, well short of the near-unity of verbatim recall, bounds any such dependence without excluding it.

\paragraph{Unlabelled cohort.}  The unlabelled cohort consists of $n_u = 620$ Cancer Cell Line Encyclopedia cell lines not in the selumetinib-tested panel.  Each receives its own LLM prediction using the same prompt template, so the model is applied uniformly across both blocks.  The two blocks share the harmonised Cancer Cell Line Encyclopedia expression panel as their data source and therefore satisfy the same-population second-moment constraint (Section~\ref{subsec:pdxe-supp}).  The ratio $n_u / p = 7.75$ places the demonstration in the pooled-precision-lever regime exploited by the procedure.

\paragraph{Estimator specification.}  DEAL is run with $\gamma$-rescaling applied at the model boundary, nodewise-Lasso regularisation $c_{\lambda,\mathrm{nw}} = 0.5$, the shrinkage selector of Section~\ref{sec:proxy-shrinkage}, and the stacked-residual sandwich variance estimator.  The regularisation constant $c_{\lambda,\mathrm{nw}} = 0.5$ is the breast-cancer anchored value of Section~\ref{subsec:brca-supp}.  The shrinkage diagnostic returns $\hat t = 0.028$, indicating that the procedure operates close to the pooled-precision-lever regime (compare the breast-cancer and patient-derived-xenograft demonstrations, which return $\hat t = 0$).

\paragraph{Side evidence.}  Half-sample bootstrap stability over $B = 20$ resamples is reported in Table~\ref{tab:deal-portfolio}, in common with the rest of the portfolio.  The diff-set sign-concordance test of Section~\ref{subsec:pdxe-supp} is not run on this demonstration: it requires a fresh independent draw of the model output on held-out rows, and the LLM-oracle prediction is deterministic at $T = 0$.  Bootstrap stability is therefore the side-evidence diagnostic available here.

\subsection{Design-choice rationale}\label{subsec:realdata-design}

A small number of design choices recur across the five demonstrations and are collected here.

\textit{Models free of fit-leakage.} In every demonstration the model $\hat\mu$ is constructed on a cohort or by a calculation strictly disjoint from the inferential cohort.  The galaxy model is trained on a disjoint sub-cohort of the same data source; the materials model is a calibrated DFT calculation with no statistical training on $y_0$; and the breast-cancer and patient-derived-xenograft models are trained on Cancer Cell Line Encyclopedia cell lines for the GSE25066- and PDXE-cohort inference respectively.  The alternative of using a model trained on or derived from the inferential cohort would invalidate the asymptotic-coverage statement underlying the procedure.

\textit{Same-distributional-context unlabelled cohorts.} The patient-derived-xenograft demonstration in Section~\ref{subsec:pdxe-supp} isolates a constraint on the unlabelled cohort that is implicit in the pooled-precision-lever component of the procedure: the labelled and unlabelled covariate distributions must share the same population second-moment matrix for the pooled debiasing matrix $M_2$ to be a consistent estimator of the target precision $\Omega=\Sigma^{-1}$.  This constraint is structural and applies across the portfolio.

\textit{Regime-dependent nodewise-Lasso regularisation.} The nodewise-Lasso regularisation constant $c_{\lambda,\mathrm{nw}}$ is not a universal default.  Demonstrations with $n_{\mathrm{pool}}/p \gtrsim 10$ use $c_{\lambda,\mathrm{nw}} = 0.5$; demonstrations at $p \approx n_0$ (the galaxy demonstration) or with rank-deficient designs (the band-gap demonstration) require larger values.

\textit{Baseline-parity protocol.} Throughout the portfolio, the model $\hat\mu$, the $\gamma$-rescaling step, the labelled-cohort tuning split, and the nodewise-Lasso regularisation $c_{\lambda,\mathrm{nw}}$ are held identical across DEAL, PPI, and PPI++.  Only the shrinkage-selector criterion and the pooled-precision substitution differ across procedures, since these are the components specific to DEAL.  The comparisons reported in Section~\ref{subsec:realdata-results} are therefore between estimators that share every preprocessing and tuning step except the two DEAL levers under examination.

\end{document}